\documentclass[
aps,
prd, 
reprint,
superscriptaddress,
nofootinbib,
]{revtex4-2}

\usepackage{graphicx}
\usepackage[usenames,dvipsnames,table,xcdraw]{xcolor}
\usepackage{xspace}
\usepackage{amsmath,amsfonts,amssymb,amsthm}
\usepackage{mathtools}
\usepackage[varg]{txfonts} 
\usepackage{bm}
\usepackage[utf8]{inputenc}
\usepackage[normalem]{ulem}
\usepackage{makecell}
\usepackage{multirow}
\usepackage{float}
\usepackage{adjustbox}
\usepackage[caption=false]{subfig}
\usepackage{rotating}
\usepackage{dcolumn}
\usepackage{enumitem}
\xdefinecolor{mylinkcolor}{rgb}{0,0,0.7}
\usepackage[
bookmarksnumbered,
bookmarksopen,
bookmarksopenlevel=2,
breaklinks=true,
colorlinks=true,
filecolor=mylinkcolor,
citecolor=mylinkcolor,
linkcolor=mylinkcolor,
urlcolor=mylinkcolor,
menucolor=mylinkcolor,
]{hyperref}

\usepackage{setspace}
\usepackage[hang]{footmisc} 
\renewcommand{\footnoterule}{%
  \kern -5pt  
  \hrule width 1.0 \columnwidth
  \kern 5pt  
}

\allowdisplaybreaks

\def\mr{\mathrm}

\def\di{\mr d}

\DeclareMathOperator{\Order}{\mathcal{O}}
\def\mF{\mathcal{F}}
\def\e{\mathrm{e}}
\def\s{\hspace{0.5pt}}
\def\solarmass{\mathrm{M}_\odot}

\def\NEVENTS{27~}

\DeclareRobustCommand{\textttcustom}[1]{{\normalsize\texttt{#1}}}

\newcommand{\bsout}{\bgroup\markoverwith{\textcolor{blue}{\rule[0.5ex]{2pt}{0.5pt}}}\ULon}

\graphicspath{ {fig/} }

\newcommand{\AEI}{\affiliation{Max Planck Institute for Gravitational Physics (Albert Einstein Institute), Am M\"uhlenberg 1, Potsdam 14476, Germany}}
\newcommand{\Maryland}{\affiliation{Department of Physics, University of Maryland, College Park, MD 20742, USA}}

\newcommand{\Cornell}{\affiliation{Cornell Center for Astrophysics and Planetary Science, Cornell University, Ithaca, New York 14853, USA}}
\newcommand{\Caltech}{\affiliation{Theoretical Astrophysics 350-17, California Institute of Technology, Pasadena, California 91125, USA}}

\newcommand{\UoN}{\affiliation{Nottingham Centre of Gravity \& School of Mathematical Sciences, University of Nottingham, University Park, Nottingham NG7 2RD, United Kingdom}}
\newcommand{\UIB}{\affiliation{Departament de F\'isica, Universitat de les Illes Balears, IAC3 – IEEC, Crta. Valldemossa km 7.5, E-07122 Palma, Spain}}

\begin{document}

\title{
Accurate waveforms for generic planar-orbit binary black holes:\\
The multipolar effective-one-body model \texttt{SEOBNRv6EHM}
}

\author{Aldo Gamboa}
\email{aldo.gamboa@aei.mpg.de}
\AEI

\author{Alessandra Buonanno}
\AEI
\Maryland

\author{Lorenzo Pompili}
\UoN
\AEI

\author{Raffi Enficiaud}
\AEI

\author{Michael Boyle}
\Cornell

\author{Lawrence E. Kidder}
\Cornell

\author{Oliver Long}
\AEI

\author{Peter James Nee}
\AEI

\author{Harald P. Pfeiffer}
\AEI

\author{Antoni Ramos-Buades}
\UIB
\AEI

\author{Mark A. Scheel}
\Caltech


\begin{abstract}
Accurate and computationally efficient waveform models are required to infer the parameters of compact binaries from their gravitational wave (GW) emission.
Among these parameters, orbital eccentricity serves as a smoking gun for dynamical formation channels and must be accounted for to avoid systematic errors in GW analyses.
Here, we present \texttt{SEOBNRv6EHM}, a time-domain, multipolar waveform model for binaries on generic planar orbits, calibrated to quasi-circular (QC) 
numerical-relativity (NR) simulations from the \texttt{SXS} collaboration.
In addition to the dominant $(2,2)$ mode, the model provides the $(2,1)$, $(3,3)$, $(3,2)$, $(4,4)$, and $(4,3)$ multipoles for the full inspiral-merger-ringdown process of coalescing binaries, as well as for dynamical captures and scattering encounters.
The model is built within the effective-one-body (EOB) framework, and it employs novel resummations of the radiation-reaction force and waveform modes.
We validate its accuracy through comparisons against 592 QC, 319 eccentric, one dynamical-capture, and two scattering \texttt{SXS} NR waveforms, and through scattering-angle comparisons against 61 \texttt{SXS} NR simulations.
For QC and small-eccentricity binaries, its accuracy is comparable to previous-generation \texttt{SEOBNRv5} models.
For highly eccentric systems, however, \texttt{SEOBNRv6EHM} attains unprecedented accuracy, with waveform mismatches remaining below or close to $ 2\% $  across the total mass range $ 20-200\, \solarmass $ for eccentricities up to $\sim 0.9$ at 14 periastron passages before merger.
Additionally, \texttt{SEOBNRv6EHM} achieves waveform-generation walltimes that are $ 2 - 8 $ times faster than other state-of-the-art EOB eccentric models, enabling efficient and accurate applications in GW astronomy.
\end{abstract}

\date{\today}

\maketitle



\section{Introduction}
\label{sec:intro}

Nearly 400 gravitational-wave (GW) signals from compact binary coalescences have been observed by the LIGO-Virgo-KAGRA (LVK) collaboration \cite{TheLIGOScientific:2014jea, VIRGO:2014yos, KAGRA:2020tym,LIGOScientific:2016aoc, LIGOScientific:2018mvr, LIGOScientific:2020ibl, LIGOScientific:2021usb, LIGOScientific:2021djp, LIGOScientific:2025slb, LIGOScientific:2025pvj,LIGOScientific:2026wfs,LIGOScientific:2026ctl}.
These GW events are typically associated with binaries composed of black holes (BHs) or neutron stars (NSs) moving on inspiraling quasi-circular (QC) orbits.
Such orbits are astrophysically expected, as GW emission circularizes the binary \cite{Peters:1963ux,Peters:1964zz,Hinder:2007qu,Sperhake:2007gu} over the long lifetime predicted for systems formed in isolation \cite{Bethe:1998bn,Belczynski:2001uc,Belczynski:2014iua,Belczynski:2016obo,Stevenson:2017tfq,Giacobbo:2018etu,Mandel:2018hfr}.

However, astrophysical formation channels also predict a small but non-negligible population of merging binaries with detectable \emph{orbital eccentricity}.
Specifically, the dynamical formation channels are expected to produce eccentric binaries that merge within the sensitive frequency band of current LVK detectors \cite{Sadowski:2007dz,OLeary:2008myb,Rodriguez:2015oxa,Stone:2016wzz,Gondan:2017wzd,Takatsy:2018euo,Fragione:2018vty,Rasskazov:2019gjw,Sedda:2023qlx,Singh:2025ojp} and future GW observatories \cite{Punturo:2010zz,TianQin:2015yph,Reitze:2019iox,LISA:2024hlh}.
These dynamical formation mechanisms are viable in dense stellar environments where different types of gravitational interactions become feasible, including von Zeipel-Kozai-Lidov oscillations in hierarchical triple systems \cite{Zeipel:1910, kozai1962secular, lidov1962evolution,Kimpson:2016dgk,VanLandingham:2016ccd,Hoang:2017fvh,Fragione:2019vgr,Britt:2021dtg,Grishin:2025ofa} and $ n $-body processes \cite{PortegiesZwart:1999nm,Miller:2002pg,Gultekin:2005fd,Samsing:2013kua,Samsing:2017rat,Samsing:2017xmd,Rodriguez:2017pec,Zevin:2018kzq,Arca-Sedda:2018qgq,Tagawa:2020jnc}.
In this way, the presence of eccentricity in GWs serves as a clean discriminator (a \emph{smoking gun}) between the isolated and dynamical formation channels \cite{Breivik:2016ddj,Nishizawa:2016eza,Samsing:2018isx,Zevin:2021rtf,DePorzio:2024cet,Sedda:2026xqr}.

Neglecting orbital eccentricity has widespread consequences across different GW analyses.
A substantial fraction of potential eccentric GW signals may be missed by standard matched-filtering searches relying on QC waveform templates \cite{Brown:2009ng,Huerta:2013qb,Wang:2021qsu,Gadre:2024ndy}, motivating the inclusion of eccentricity in GW search strategies~\cite{LIGOScientific:2019dag,LIGOScientific:2023lpe,Nitz:2019spj,Nitz:2021mzz,Lenon:2021zac,Favata:2021vhw,Pal:2023dyg,Ravichandran:2023qma,Phukon:2024amh,Bhaumik:2024cec,Khalouei:2025ati,Wang:2025yac,Canizares:2026whr}.
Similarly, analyzing data containing an eccentric signal with QC templates can lead to biases in the inferred binary parameters (e.g., masses and spins)~\cite{Favata:2013rwa,Ramos-Buades:2019uvh,OShea:2021faf,Cho:2022cdy,GilChoi:2022waq,Das:2024zib,Divyajyoti:2023rht,Divyajyoti:2025cwq,RoyChowdhury:2026xgb,Yang:2026mam}, as well as in the evidence for gravitational lensing~\cite{Ezquiaga:2020gdt,Mishra:2025dpa}, in the inferred center-of-mass acceleration of the binary~\cite{Tiwari:2025aec,Pathak:2026cik,Roy:2026mco,Roy:2026duh,Pompili:2026tdf}, and in tests of general relativity \cite{Saini:2022igm,Saini:2023rto,Narayan:2023vhm,Gupta:2024gun,Shaikh:2024wyn,Bhat:2022amc,Bhat:2024hyb,Roy:2025xih,Mahapatra:2025cwk,Gupta:2025paz,Wang:2026ibx}.

Several GW events have been identified as potential eccentric candidates;
however, their interpretation remains debated, as different analyses often yield conflicting conclusions.
Some interesting candidates include:
1) GW190701\_203306 \cite{Gupte:2024jfe,Planas:2025jny}, GW200129\_065458 \cite{Gupte:2024jfe,Planas:2025jny}, and GW231223\_032836 \cite{Xu:2025ajj, Gupte:2026whi}, which have support for the eccentric binary black-hole (BBH) hypothesis but is dependent on the specific glitch mitigation method \cite{Gupte:2024jfe,Gupte:2026whi};
2)~GW230712\_090405 \cite{Xu:2025ajj}, whose support for the eccentric aligned-spin BBH hypothesis reduces significantly when compared to the QC precessing-spin hypothesis, thus pointing toward the known degeneracy between eccentricity and spin precession \cite{CalderonBustillo:2020xms,Romero-shaw:2022fbf,Xu:2022zza,Divyajyoti:2023rht,Divyajyoti:2025cwq,Tibrewal:2026jci};
3) GW200208\_222617, associated with an eccentric BBH signal by several studies \cite{Romero-Shaw:2022xko,Gupte:2024jfe,Romero-Shaw:2025vbc,McMillin:2025hof,Planas:2025jny}, but with high false alarm rate;
4) GW190521\_030229, for which some studies find evidence supporting eccentric \cite{Romero-Shaw:2020thy,Gayathri:2020coq} or dynamical-capture \cite{Gamba:2021gap,Lange:2026eqx} BBH hypotheses, while others find no conclusive evidence \cite{Gupte:2024jfe,Iglesias:2022xfc,Ramos-Buades:2023yhy};
and 5) GW200105\_162426, a signal from a neutron-star--black-hole (NSBH) binary with evidence for eccentricity \cite{Fei:2024ruj,Morras:2025xfu,Phukon:2025cky,Tiwari:2025fua}, for which there is a contrasting set of results reported by different follow-up analyses \cite{Planas:2025plq,Kacanja:2025kpr,Jan:2025fps,Clarke:2026cuw}.
The astrophysical implications of these potential eccentric detections have been discussed, e.g., in Refs.~\cite{Romero-Shaw:2025otx,Morras:2026mrv,Gupte:2026whi,Zeeshan:2026pga,Malagon:2026uev}.

Disagreements concerning the presence of eccentricity in GW events highlight the need to develop \emph{accurate} waveform models (and better eccentric parameter-estimation techniques).
Numerical relativity (NR) provides the most accurate eccentric GW templates by directly solving Einstein's field equations \cite{Hinder:2008kv,Mroue:2010re,Lewis:2016lgx,Hinder:2017sxy,Huerta:2019oxn,Habib:2019cui,Ramos-Buades:2019uvh,Ramos-Buades:2022lgf,Healy:2022wdn,Ferguson:2023vta,Bonino:2024xrv,Trenado:2025ccf}.
However, producing enough NR simulations to satisfy the demand required by GW analyses is a highly expensive computational task, which is further complicated by the two additional parameters needed to characterize an eccentric binary \cite{Clarke:2022fma, Carullo:2023kvj,Carullo:2024smg,Nee:2025zdy,Wang:2023wol,Knapp:2024yww}. 
Only recently, enough NR simulations have been produced to allow for the construction of NR surrogate waveform models for eccentric nonspinning BBHs \cite{Islam:2021mha,Nee:2025nmh,Ravichandran:2026iec}.
Nevertheless, it remains crucial to complement NR simulations with semi-analytical waveform models.

Different approaches have been taken to develop eccentric waveform models for comparable-mass binaries.
Using post-Newtonian (PN) results for eccentric binaries \cite{damour1985general,Damour:1988mr,schafer1993second,wex1995second,Memmesheimer:2004cv,Cho:2021oai,Junker:1992kle,Gopakumar:1997bs,Gopakumar:2001dy,Damour:2004bz,Konigsdorffer:2006zt,Arun:2007rg,Arun:2007sg,Arun:2009mc,Mishra:2015bqa,Boetzel:2019nfw,Ebersold:2019kdc,Henry:2023tka,Boetzel:2017zza,Paul:2022xfy,Bhattacharyya:2025ehc,Morras:2025nbp}, Refs.~\cite{Yunes:2009yz,Cornish:2010cd,ShapiroKey:2010cnz,Huerta:2014eca,Loutrel:2017fgu,Tanay:2016zog,Tanay:2019knc,Tiwari:2020hsu,Tiwari:2019jtz,Moore:2018kvz,Moore:2019xkm,Sridhar:2024zms,Phurailatpam:2025ppf} built inspiral-only eccentric models, including spin-precession effects \cite{Klein:2018ybm,Klein:2021jtd,Arredondo:2024nsl,Morras:2025nlp,Morras:2026fho}.
Based on such results and information from the merger-ringdown of QC NR simulations, Refs.~\cite{Huerta:2016rwp,Huerta:2017kez,Hinder:2017sxy,Islam:2024tcs,Ramos-Buades:2019uvh,Chattaraj:2022tay,Manna:2024ycx,Paul:2024ujx} constructed hybrid inspiral-merger-ringdown (IMR) eccentric models.
Alternative methods use empirical properties of NR waveforms to construct eccentric templates from existing QC models \cite{Setyawati:2021gom,Wang:2023ueg,Islam:2024rhm,Islam:2024zqo,Islam:2024bza,Islam:2025rjl,Islam:2025bhf,Islam:2026blk}.
Computationally efficient eccentric waveform approximants have been constructed within the \texttt{IMRPhenom} family of models~\cite{Planas:2025feq,Ramos-Buades:2026kbq}.
Work on small-mass-ratio eccentric waveform modeling has been done, e.g., in Refs.~\cite{Hinderer:2008dm,Hughes:2021exa,Chua:2020stf,Katz:2021yft,Speri:2023jte,Chapman-Bird:2025xtd,Becker:2025zzw}.

Another approach to build eccentric waveforms is based on the effective-one-body (EOB) formalism~\cite{Buonanno:1998gg,Buonanno:2000ef,Damour:2000we,Damour:2001tu,Buonanno:2005xu,Buonanno:2006ui}.
EOB models combine analytical results of the two-body problem [from the
PN, post-Minkowskian, and gravitational self-force (GSF) approximations] with calibrations to NR simulations to produce highly accurate waveforms.
Two families of EOB models exist in the literature, 
\texttt{SEOBNR} (e.g., Refs.~\cite{Bohe:2016gbl,Cotesta:2018fcv,Ossokine:2020kjp,Cotesta:2020qhw,Mihaylov:2021bpf,Pompiliv5,Khalilv5,RamosBuadesv5,VandeMeentv5,Mihaylovv5,Julie:2024fwy,Pompili:2024yec,Buonanno:2024byg,Haberland:2025luz,Leather:2025nhu,Pompili:2025cdc,Estelles:2025zah,Nishimura:2026nse,Pompili:2025rhz,Haberland:2026xvj,FooInPrep,EstellesInPrep})
and \texttt{TEOBResumS} (e.g., Refs.~\cite{Nagar:2018zoe,Nagar:2019wds,Nagar:2020pcj,Gamba:2021ydi,Riemenschneider:2021ppj,Albertini:2021tbt,Gamba:2023mww}),
and both have been extended to model eccentric binaries (independent efforts have been made in Refs.~\cite{Hinderer:2017jcs,Manzini:2025gjx}).
Particularly, the \texttt{TEOBResumS} family has done studies to improve eccentric waveforms both in the comparable-mass and test-mass limits \cite{Chiaramello:2020ehz,Albanesi:2021rby,Albanesi:2022ywx,Albanesi:2022xge,Albanesi:2023bgi,Nagar:2020xsk,Nagar:2021gss,Nagar:2021xnh,Placidi:2021rkh,Nagar:2023zxh,Placidi:2023ofj,Albertini:2023aol,Andrade:2023trh,Nagar:2024dzj,Nagar:2024oyk,Grilli:2024lfh,Albanesi:2024xus,Chiaramello:2024unv,Fontbute:2025vdv,Albanesi:2026qtx,Gamba:2026fqa}, including extensions to spin-precession effects for BBHs \cite{Gamba:2024cvy}, for NSs \cite{Gonzalez:2025xba,Albanesi:2025txj}, and in parametrized deviations of general relativity \cite{Chiaramello:2025bhi}.
The model encapsulating all these developments is collectively known as \texttt{TEOBResumS-Dal\'i}.

For the \texttt{SEOBNR} family, past eccentric models include \texttt{SEOBNRE}~\cite{Cao:2017ndf,Liu:2019jpg,Liu:2021pkr,Liu:2023dgl} (extended to spin-precessing BBHs~\cite{Liu:2023ldr}), \texttt{SEOBNRv4EHM}~\cite{Khalil:2021txt,Ramos-Buades:2021adz}, and \texttt{SEOBNRv5EHM}~\cite{Gamboa:2024imd,Gamboa:2024hli} (extended to a machine-learning surrogate model \cite{Shi:2025paa}).
Additional test-mass-limit studies have been done in Refs.~\cite{Faggioli:2024ugn,Islam:2024vro,Islam:2025wci,Faggioli:2025hff,Faggioli:2026alx}.
\texttt{SEOBNRv5EHM} models aligned-spin BBHs on \emph{bound} eccentric orbits, and is about an order of magnitude more accurate than \texttt{SEOBNRv4EHM} and \texttt{TEOBResumS-Dal\'i}.
In the large eccentricity limit, however, the performance of these models degrades, with mismatches close to or above $ 20\% $ for highly eccentric, long NR waveforms.
Extending the applicability of \texttt{SEOBNRv5EHM} to generic planar orbits and improving its performance across parameter space motivated the development of a new model.

\begin{figure*}
\hspace{-5pt}
\includegraphics[width=0.95\linewidth]{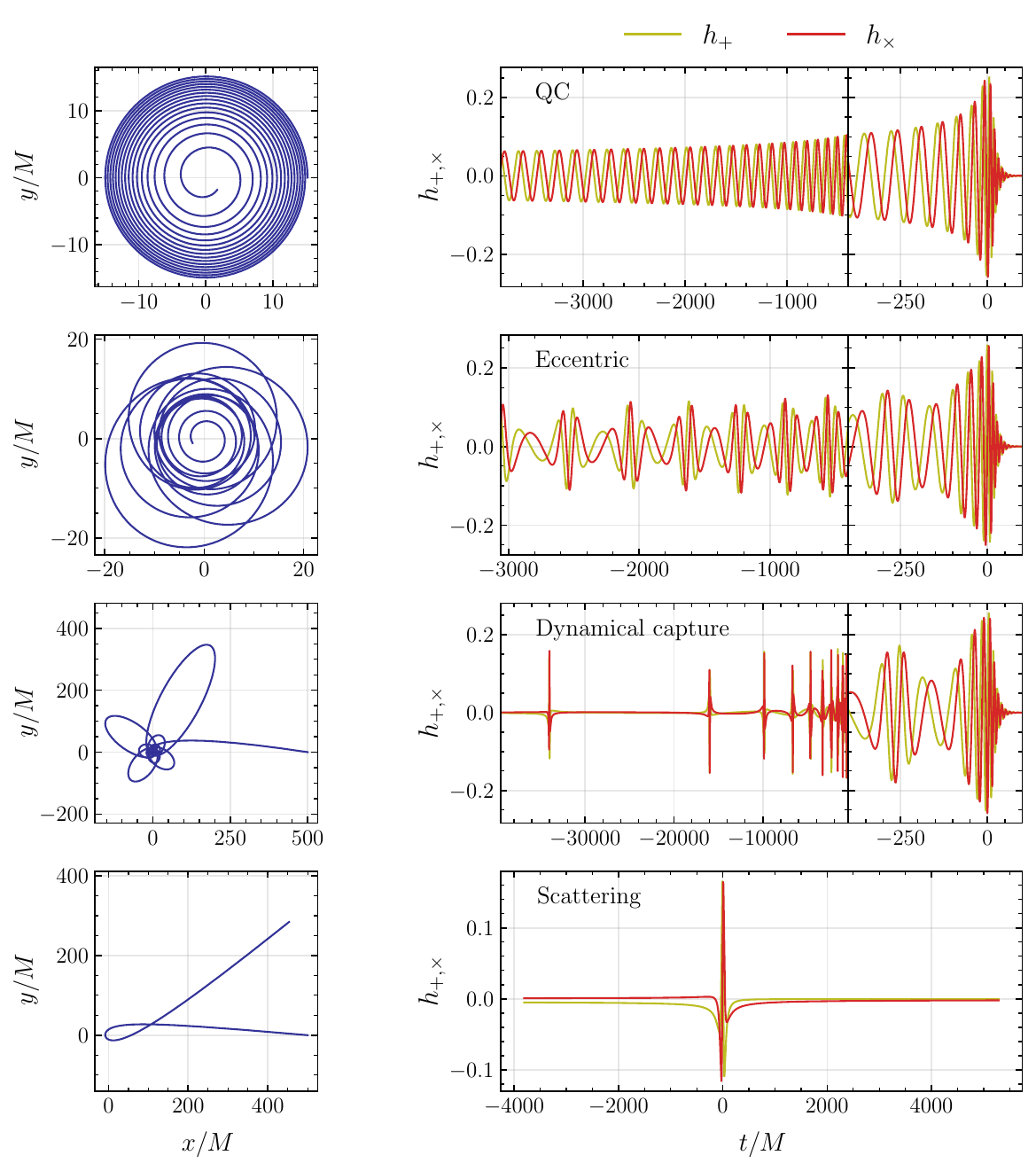}
\vspace{-5pt}
\caption{
EOB dynamics (left panels) and the associated GW emission (right panels), both in geometric units, for different orbital configurations of a BBH system with mass ratio $ q = 1 $, dimensionless spin components $ \chi_1 = 0.3 $ and $ \chi_2 = -0.1 $, and inclination $ \iota = 0 $, as predicted by \texttt{SEOBNRv6EHM}.
The first row shows a QC binary initialized at an orbital frequency of $ M \Omega = 0.017$.
The second row depicts an eccentric binary initialized at an orbit-averaged orbital frequency of $ \langle M \Omega \rangle =  0.015$, eccentricity $ e = 0.4 $, and relativistic anomaly $ \zeta = 0 $.
The third row shows a dynamical capture with initial energy $ E/M = 1.000001 $, total angular momentum $ J/M^2 = 1.225 $, and initial separation $ r/M = 500 $.
Finally, the fourth row displays a scattering encounter with the same total angular momentum and initial separation as the dynamical capture, but with a larger initial energy, $E/M = 1.001$.
}
\label{fig:capture}
\end{figure*}

By leveraging novel results for the EOB radiation-reaction (RR) force and waveform modes, we develop \texttt{SEOBNRv6EHM},
a new time-domain, multipolar waveform model that is capable of modeling the dynamics and GW emission of BBHs on generic planar orbits (see Fig.~\ref{fig:capture}).
IMR, dynamical-capture, and scattering signals are modeled with the multipoles $( \ell, |m| ) \in \{(2, 2),\, (3, 3),\, (2, 1),\, (4, 4),\, (3, 2),\, (4, 3)\}$, with respect to a $-2$ spin-weighted spherical harmonic decomposition.
\texttt{SEOBNRv6EHM} is calibrated only using QC simulations from the \texttt{SXS} catalog \cite{Scheel:2025jct}, and follows the same methods for constructing IMR waveforms as \texttt{SEOBNRv5EHM}, including a QC prescription for the merger-ringdown phase.

\texttt{SEOBNRv6EHM} has improved accuracy, particularly for high eccentricities.
We quantify the accuracy of \texttt{SEOBNRv6EHM} by comparing its predictions against waveforms and scattering angles extracted from \texttt{SXS} NR simulations.
For QC and low-eccentricity binaries, \texttt{SEOBNRv6EHM} shows the same overall accuracy as \texttt{SEOBNRv5EHM}, but removes high-mismatch outliers that are present for \texttt{SEOBNRv5} models.
For bound eccentric binaries, \texttt{SEOBNRv6EHM} is the \emph{first and only} model whose mismatches stay below $ \sim 2 \% $ for \emph{all} eccentricity values when comparing against 319 eccentric NR waveforms.
This represents an order of magnitude improvement in accuracy for the highest eccentricities with respect to \texttt{SEOBNRv5EHM}.
For all eccentricities, \texttt{SEOBNRv6EHM} is overall an order of magnitude more accurate than \texttt{TEOBResumS-Dal\'i}.
For unbound binaries and a dynamical capture, the waveforms and scattering angles predicted by \texttt{SEOBNRv6EHM} are in better agreement with NR results than those of \texttt{TEOBResumS-Dal\'i}.

\texttt{SEOBNRv6EHM} has a better computational performance and has no inherent parameter space restrictions.
Thanks to code optimizations and a simpler analytical structure, \texttt{SEOBNRv6EHM} 
has a speed comparable to that of
\texttt{SEOBNRv5HM} in the QC-orbit limit, and achieves faster waveform-evaluation walltimes by factors of $ \sim 2 - 8 $ with respect to \texttt{SEOBNRv5EHM} and \texttt{TEOBResumS-Dal\'i}, depending on the binary configuration.
In Bayesian parameter estimation, this translates into an overall speedup of a factor of three to five relative to \texttt{SEOBNRv5EHM}, as shown in the companion manuscript~\cite{Pompili:2026yxq}.
Moreover, \texttt{SEOBNRv6EHM} introduces a new set of equations of motion that substantially enlarges the accessible parameter space with respect to \texttt{SEOBNRv5EHM}.

%


\subsection*{Executive summary}

We summarize the main elements of this work, and point to the figures, tables, and sections where they can be found:
\begin{itemize}
\item
\emph{Model.}
\texttt{SEOBNRv6EHM} is a time-domain, multipolar, aligned-spin EOB model for BBHs on generic planar orbits, covering the full inspiral-merger-ringdown of QC and eccentric binaries, as well as dynamical captures and scattering encounters (see Fig.~\ref{fig:capture}).
The equations defining the model are summarized in Sec.~\ref{sec:overview}, and its input parameters are listed in Table~\ref{tab:input_values}.
\item
\emph{New RR force.}
We derive a new RR force resummation for generic planar orbits [Eqs.~\eqref{eq:RR_force_proposal}], in which eccentricity corrections are resummed with a sigmoid function [Eq.~\eqref{eq:F_ecc_resum}], and we select a RR gauge [Eq.~\eqref{eq:ab_gauge}] that guarantees regular expressions even for highly eccentric binaries (see Fig.~\ref{fig:ab_gauges}).
This is presented in Sec.~\ref{sec:RR_force}.
\item
\emph{Initial conditions.}
Prescriptions for the initial conditions of bound orbits, parametrized by an orbit-averaged frequency, an eccentricity, and a radial phase (either the relativistic or the mean anomaly), and of unbound orbits, parametrized by the initial separation, energy, and angular momentum, are given in Sec.~\ref{sec:ics}.
\item
\emph{Waveform modes.}
We introduce a new factorization of the waveform modes for generic planar orbits [Eq.~\eqref{eq:model_fact}], describe the NR-informed corrections applied to the late inspiral, and outline the procedure for attaching the merger-ringdown phase, all in Sec.~\ref{sec:waveforms}.
\item
\emph{Calibration.}
The model employs ten calibration parameters, determined from 425 QC \texttt{SXS} NR waveforms through a nested-sampling analysis.
A summary of the calibration parameters is given in Sec.~\ref{sec:calibration_summary}, and the calibration procedure is detailed in Sec.~\ref{sec:calibration}.
%
\item
\emph{Accuracy for QC binaries.}
Against 592 QC \texttt{SXS} NR waveforms, \texttt{SEOBNRv6EHM} has an accuracy comparable to that of \texttt{SEOBNRv5EHM}, while removing high-mismatch outliers (see Figs.~\ref{fig:mismatch_spaghetti_qc} and \ref{fig:violins_NRsur}, and Sec.~\ref{sec:comparison_qc}).
\item
\emph{Accuracy for eccentric binaries.}
Against 319 eccentric \texttt{SXS} NR waveforms, \texttt{SEOBNRv6EHM} achieves a median of $ \sim 0.02 \% $ for the maximum mismatches over the total mass range $ [20, 200] \, \solarmass $, with mismatches remaining below or close to $ 2\% $ for all eccentricities (up to $ e _{ \text{gw}} \sim 0.9 $).
This represents an improvement of up to an order of magnitude over \texttt{SEOBNRv5EHM} at high eccentricities, and of about an order of magnitude over \texttt{TEOBResumS-Dal\'i} across all eccentricities (see Figs.~\ref{fig:mismatch_spaghetti_ecc}, \ref{fig:mismatch_distributions_ecc}, and \ref{fig:sxs_2527}).
This is discussed in Sec.~\ref{sec:comparison_ecc}.
\item
\emph{Accuracy for generic planar orbits.}
Comparisons against NR waveforms from one dynamical capture and two scattering encounters (see Fig.~\ref{fig:sxs_4000}), and against scattering angles from 61 NR simulations (see Fig.~\ref{fig:scattering_angles}), show improved agreement with respect to \texttt{TEOBResumS-Dal\'i} (see Sec.~\ref{sec:comparison_gen}).
\item
\emph{Computational performance.}
\texttt{SEOBNRv6EHM} has a speed comparable to that of the QC model \texttt{SEOBNRv5HM} in the zero-eccentricity limit, and is $ \sim 2-8 $ times faster than \texttt{SEOBNRv5EHM} and \texttt{TEOBResumS-Dal\'i} for eccentric configurations (see Figs.~\ref{fig:benchmarks_seobnr} and \ref{fig:benchmarks_param_space}, and Sec.~\ref{sec:comp_perf}).
\item
\emph{Robustness.}
The model has no inherent parameter-space restrictions, apart from systems with very high eccentricity near merger, and its waveforms vary smoothly under changes of the input parameters (see Figs.~\ref{fig:initial_separation}--\ref{fig:smoothness_generic}, and Sec.~\ref{sec:robustness}).
\item
\emph{Parameter estimation.}
The performance of \texttt{SEOBNRv6EHM} in Bayesian parameter estimation, including injection-recovery studies with synthetic NR signals and the analysis of \NEVENTS GW events, is presented in the companion manuscript~\cite{Pompili:2026yxq}.
\end{itemize}

We conclude in Sec.~\ref{sec:conclusions} and discuss future work.
We leave secondary results to appendices:
Appendix~\ref{sec:PN_expressions} provides expressions for various quantities derived in this work, and Appendix~\ref{sec:fits} presents the fits employed for the calibration parameters of \texttt{SEOBNRv6EHM}.


\subsection*{Notation}

We use geometric units in which Newton's constant and the speed of light equal one, $G \!=\! c \!= \! 1$.
In PN expansions, the $n$PN order refers to terms of order $v^{2 n} \epsilon^{2n}$, where $v$ is the binary’s relative velocity, and $ \epsilon = 1 $ is a power-counting parameter.

The component masses are labeled as $m_1$ and $m_2$ such that $m_1 \geq \nolinebreak m_2$, and we define the following combinations:
\begin{equation}
\begin{gathered}
M\equiv m_1 + m_2, \qquad \mu \equiv \frac{m_1m_2}{M}, \qquad \nu \equiv \frac{\mu}{M},  \\ 
\delta \equiv\frac{m_1 - m_2}{M},  \qquad q \equiv \frac{m_1}{m_2}, \qquad X_i \equiv \frac{m_i}{M}, \quad i \in \{1,2\}.
\end{gathered}
\end{equation}

The total energy of the binary is denoted as $ E $, from which we define the relative Lorentz factor $ \gamma $ as
\begin{equation}
\gamma
\equiv
\frac{E^2 - m_1^2 - m_2^2}{2 \s m_1 \s m_2}
=
1 + \frac{E^2 / M^2 - 1}{2\, \nu}.
\end{equation}

The orbital angular momentum vector of the binary is denoted as $ \bm{L} $, with magnitude $ L $ and direction $ \bm{l} = \bm{L}/L $.

The impact parameter for unbound binaries is defined as
\begin{equation}
\frac{b}{M}
\equiv
\frac{L/(\s \mu \s M) \, \, E/M }{ \sqrt{\gamma^2 - 1}} .
\end{equation}

The total angular momentum vector of the binary is
\begin{equation}
\label{eq:}
\bm{J} \equiv \bm{L} + \bm{S}_1 + \bm{S}_2.
\end{equation}
where $\bm{S}_1$ and $\bm{S}_2$ are the component spin vectors, with magnitudes $ S_1 $ and $ S_2 $, respectively.

We work with aligned-spin binaries whose spins $\bm{S}_1$ and $\bm{S}_2$ are always parallel to $ \bm{L} $.
Under this assumption, we define the dimensionless spin components as
\begin{equation}
 \label{eq:comp_spins}
\chi_i
\equiv \frac{\bm{S}_i \cdot \bm{l}}{m_i^2}
= \pm \frac{S_i}{m_i^2}
\equiv \frac{a_i}{m_i} \s,  \qquad i \in \{1,2\},
\end{equation}
which take values in the range $(-1,1)$, and where positive spins point to the same direction as $ \bm{L} $.

We define the following combinations of the spins:
\begin{equation}
\label{eq:spinComb}
\begin{gathered}
\chi_\text{S} \equiv \frac{\chi_1 + \chi_2}{2}, \qquad \chi_\text{A} \equiv \frac{\chi_1 - \chi_2}{2},
\\
\chi_{\mathrm{eff}} \equiv \frac{m_1 \, \chi_1 + m_2 \, \chi_2}{m_1+m_2},
\quad
a_\pm \equiv a_1 \pm a_2 = m_1 \, \chi_1 \pm m_2 \, \chi_2 \s .	
\end{gathered}
\end{equation}

We work in the binary's center of mass, and denote the relative separation and momentum vectors as $\bm{r}$ and $\bm{p}$, with magnitudes $r$ and $p$, respectively, such that
\begin{equation}
p^2 = p_r^2 + \frac{L^2}{r^2}, \quad
p_r= \bm{n}\cdot\bm{p}, \quad
\bm{L}=\bm{r}\times\bm{p},
\end{equation}
where $\bm{n}\equiv\bm{r}/r$.
The unit vector perpendicular to $\bm l$ and $\bm n$ is denoted $\bm \lambda \equiv \bm l \times \bm n$.
The aligned-spin hypothesis allows us to restrict the motion to equatorial orbits.
Therefore, we use polar-coordinates phase-space variables $(r,\phi,p_r,p_\phi)$, where the angular momentum reduces to $L =p_\phi$.

We denote the far-zone energy and angular momentum fluxes by $\Phi_E$ and $\Phi_J$, respectively, and the RR force by $\bm{\mF}$, with radial and azimuthal components, $\mF_r$ and $\mF_\phi$, such that
\begin{equation}
\bm{\mF} = \mF_r \, \bm n +  \frac{\mF_\phi}{r} \s \bm \lambda.
\end{equation}

In Appendix~\ref{sec:PN_expressions} and the Supplemental Material~\cite{SEOBNRv6EHM:SupplementalMaterial}, we use dimensionless quantities to simplify notation.
To restore units, one can use the following replacements:
\begin{equation}
\label{eq:dimlessVars}
\begin{gathered}
t \to \frac{t}{M},
\quad r \to \frac{r}{M},
\quad p \to \frac{p}{\mu},
\quad p_r \to \frac{p_r}{\mu},
\\
p_\phi \to \frac{p_\phi}{\mu M},
\quad \dot p_r \to \frac{\dot p_r}{\nu},
\quad E \to \frac{E}{\mu},
\quad \Omega \to M\Omega,
\\
\mF_r \to  \frac{\mF_r}{\nu},
\quad \mF_\phi \to  \frac{\mF_\phi}{\nu M},
\quad \Phi_E \to \frac{\Phi_E}{\nu},
\quad \Phi_J \to \frac{\Phi_J}{\nu M}.
\end{gathered}
\end{equation}
%


\section{Overview of the \textttcustom{SEOBNR\lowercase{v}6EHM} model}
\label{sec:overview}

Waveform generation in EOB models proceeds in two main steps.
First, the binary dynamics is computed using the EOB formalism.
Then, the waveform is constructed from analytical expressions evaluated on the dynamics and, for coalescing binaries, it is smoothly stitched to a phenomenological merger-ringdown ansatz.
Here, we summarize how these steps are done within the \texttt{SEOBNRv6EHM} model.


\subsection{EOB dynamics}
\label{sec:overview_dynamics}

The EOB formalism \cite{Buonanno:1998gg,Buonanno:2000ef,Damour:2000we,Damour:2001tu,Buonanno:2005xu,Buonanno:2006ui} is an analytical approach that maps the gravitational two-body dynamics onto the \emph{effective dynamics} of a test mass or a test spin in a deformed Schwarzschild or Kerr background, with the deformation being parametrized by the symmetric mass ratio $ \nu $ and the spins of the binary $ a_1 $ and $ a_2 $.
The effective Hamiltonian $H_\text{eff}$ governing the dynamics of the effective particle is related to the two-body Hamiltonian $H_\text{EOB}$ via the energy map
\begin{equation}
\label{eq:overview_Heob}
H_\text{EOB} = M \sqrt{1 + 2 \nu \left( \frac{H_\text{eff}}{\mu} - 1 \right)} \,.
\end{equation}
We use the aligned-spin effective Hamiltonian derived in Ref.~\cite{Khalilv5}, which was employed in the \texttt{SEOBNRv5HM}~\cite{Pompiliv5} and \texttt{SEOBNRv5EHM}~\cite{Gamboa:2024imd, Gamboa:2024hli} waveform models.
This Hamiltonian is applicable to binaries on generic planar orbits, and reduces exactly to the Hamiltonian of a test mass in a Kerr background in the limit $\nu \to 0$.
It resums the full 4PN information and most of the 5PN nonspinning contributions, as detailed in Ref.~\cite{Khalilv5}.
Moreover, its accuracy is improved by including two parameters, $ a_6 $ and $ \hat d _{ \text{SO}} $, at high PN orders, which are calibrated against QC NR simulations (see Sec.~\ref{sec:calibration}).

The back-reaction of GW emission on the binary dynamics is modeled by incorporating a RR force into Hamilton's equations of motion.
This RR force is derived from \emph{balance laws} that relate the GW fluxes of energy and angular momentum to the corresponding losses of the system.
To improve its accuracy, the RR force in EOB models is \emph{resummed} using various physically and numerically motivated techniques.

In \texttt{SEOBNRv6EHM}, we use a new expression of the RR force, whose azimuthal $ \mF_\phi $ and radial $ \mF_r $ components are given by
\begin{subequations}
\label{eq:overview_RR_force}
\begin{align}
\mF_\phi &=
\mF_\phi^\text{modes}\, \mF_\phi^\text{ecc},
\\
\mF_r &= \frac{p_{r}}{p_\phi} \ \mF_r^\text{modes} \,\mF_r^\text{ecc},
\label{eq:overview_RR_force_radial}
\\
\mF_\phi^\text{modes} &= -\frac{M}{8 \pi }
\left( 
P^0_5[M \Omega _{ \text{c}}]^{4/3}
+ \frac{3 M}{4 \s r } \dot r^2
\right) 
\sum_{\ell=2}^8 \sum_{m=1}^{\ell} m^2 \frac{\big | \hat d_\mathcal L \s h_{\ell m}^\text{F,\,qc} \big |^2}{(M\Omega)^{1/3}}\!,
\label{eq:overview_F_modes_phi}
\\
\mF_r^\text{modes} &= \frac{M}{8 \pi}
\left( 
\frac{2}{3} \, P^0_5[M \Omega _{ \text{c}}]^{4/3}
+ \frac{M }{2 \s r } \dot r^2
\right) 
\sum_{\ell=2}^8 \sum_{m=1}^{\ell} m^2  \frac{\big | \hat d_\mathcal L \s h_{\ell m}^\text{F,\,qc} \big |^2}{(M \Omega)^{1/3}}\!,
\label{eq:overview_F_modes_r}
\\
\mF_\phi^\text{ecc} &=
\frac{2}{1 + \exp \left[\mF_{\phi}^{\text{arg exp}} \left(r, \dot r, \dot p_{r_*, \s\text{cons}} \right) \right]},
\label{eq:overview_F_ecc_phi}
\\
\mF_r^\text{ecc} &=
\frac{2}{1 + \exp \left[\mF_{r}^{\text{arg exp}} \left(r, \dot r, \dot p_{r_*, \s\text{cons}} \right) \right]}.
\label{eq:overview_F_ecc_r}
\end{align}
\end{subequations}
In the following, we summarize and motivate each element that enters into these equations.

The RR force \eqref{eq:overview_RR_force} employs the EOB dynamical variables $ (r, \phi, p_{r_*}, p_\phi) $, along with their time derivatives $ \dot r $, $ \dot \phi \equiv \Omega $, and $ \dot p_{r_*} $, where $ p_{r_*} $ is the radial momentum conjugate to the tortoise coordinate $r_*$ defined by $\di r / \di r_* \equiv \xi(r)$, where $\xi(r)$ is given by Eq.~(44) of Ref.~\cite{Khalilv5}, and it is related to the canonical radial momentum $ p_r $ through the relation $ p_{r_*} \!=  p_r \s \s \xi(r) $.
The tortoise radial momentum $p_{r_*}$ is employed since it improves the stability of the dynamics near the event horizon~\cite{Damour:2007xr,Pan:2009wj}.
The conservative part of $ \dot p_{r_*} $ is also used, and it is given by
\begin{equation}
\dot p_{r_*,\s\text{cons}}
\equiv
- \xi(r) \frac{\partial H_\text{EOB}}{\partial r}(r,p_{r_*},p_\phi).
\end{equation}

The RR force \eqref{eq:overview_RR_force} is constructed to mimic the structure of the expressions employed in QC EOB models, while keeping the correct eccentricity contributions up to a given PN order.
This motivates the use of $ \dot r $ and $ \dot p_{r_*, \s\text{cons}} $ in the parametrization of the eccentricity contributions, since they satisfy $ \dot r \to 0 $ and $ \dot p_{r_*, \s\text{cons}} \to 0 $ in the circular-orbit limit (using the conservative part of $ \dot p_{r_*} $ is convenient since it is directly given by the Hamiltonian).
Additionally, we introduce the orbital frequency for circular orbits, $ \Omega _{ \text{c}} $, whose PN expansion in terms of the relative separation $ r $ is given by [see Eq.~\eqref{eq:Omega_c}],
\begin{equation}
\label{eq:Omega_c_overview}
\Omega _{ \text{c}} = \frac{M^{1/2}}{r^{3/2}}
+ \frac{1}{c^2}\frac{M^{3/2} }{r^{5/2}} \frac{\nu}{2}
+ \mathcal O \s\s \bigg( \frac{1}{c^3} \bigg),
\end{equation}
to which we apply a $ (0,5) $ Pad\'e resummation, $ P^0_5[\Omega _{ \text{c}}] $, to improve the agreement with the true orbital frequency $ \Omega = \dot \phi $ when applied to QC binaries.
With this parametrization, the terms inside the parentheses in Eqs.~\eqref{eq:overview_F_modes_phi} and \eqref{eq:overview_F_modes_r} lead to the correct leading-order contribution to the RR force for generic planar orbits, in a specific gauge which we set based on the numerical behavior of the RR force in the high-velocity regime.

The eccentricity corrections $ \mF_\phi^\text{ecc} $ and $ \mF_r^\text{ecc} $ ensure the correct PN expansion of the RR force for generic planar orbits beyond the leading order, and are resummed with sigmoids in Eqs.~\eqref{eq:overview_F_ecc_phi} and \eqref{eq:overview_F_ecc_r} to limit their growth during strong-field, high-velocity configurations, such as the plunge of coalescing binaries.
In \texttt{SEOBNRv6EHM}, we employ the 1PN part of $ \mF_{\phi}^{\text{arg exp}} $ and $ \mF_{r}^{\text{arg exp}} $, which can be read from Eqs.~\eqref{eq:F_argexp}.

The remaining element in the RR force \eqref{eq:overview_RR_force} is the dimensionless luminosity distance $ \hat d_{\mathcal L} \equiv d_{\mathcal L} / M $ multiplying the factorized waveform modes for QC orbits, $ h_{\ell m}^\text{F,\,qc} $, given by
\begin{equation}
\label{eq:overview_h_F_qc}
h_{\ell m}^\text{F,\,qc} = h_{\ell m}^\text{N,\,qc}  \s \hat{S}_\text{eff} \, T_{\ell m}^\text{qc} \, f_{\ell m}^\text{qc} \, \e^{i \delta_{\ell m}^\text{qc}},
\end{equation}
where each of these factors is defined in Sec.~\ref{sec:factorized_modes}, and they depend on the binary dynamics through the quantities $ (\phi, \s \Omega, H _{ \text{EOB}}, p_\phi) $.
This is a resummation of the modes widely employed in EOB models.
In \texttt{SEOBNRv6EHM}, we introduce additional parameters in the $ f_{\ell m}^\text{qc} $ factors.
Some of these are calibrated against a nonspinning, equal-mass QC NR simulation, while others approximate the GSF contributions incorporated in the \texttt{SEOBNRv5HM} model \cite{VandeMeentv5,Leather:2025nhu} (see Sec.~\ref{sec:calibration}).
We note that the modes in Eq.~\eqref{eq:overview_h_F_qc} are \emph{only} used in the computation of the RR force.
The modes returned by \texttt{SEOBNRv6EHM} are explained in the next subsection.

The Hamiltonian \eqref{eq:overview_Heob} and RR force \eqref{eq:overview_RR_force} determine the binary's evolution through Hamilton's equations of motion
\begin{subequations}
\label{eq:EOM}
\begin{align}
\dot{r} &= \xi(r) \frac{\partial H_\text{EOB}}{\partial p_{r_*}}(r,p_{r_*},p_\phi),  \\
\Omega &= \frac{\partial H_\text{EOB}}{\partial p_\phi}(r,p_{r_*},p_\phi), \label{eq:phidot_2} \\
\dot{p}_{r_*} &= - \xi(r) \frac{\partial H_\text{EOB}}{\partial r}(r,p_{r_*},p_\phi) + \xi(r) \, \mF_r,
\label{eq:dot_prstar}
\\
\dot{p}_\phi &= \mF_\phi,
\end{align}
\end{subequations}
These equations are integrated numerically from a set of initial conditions $ (r_0, \phi_0, p_{r_*,\s 0}, p_{\phi, \s 0}) $ which is determined from the binary's intrinsic parameters (see Sec.~\ref{sec:ics}).
These are the masses $ m_1 $ and $ m_2 $, the dimensionless spin components $ \chi_1 $ and $ \chi_2 $ in the $ z$-direction, as well as three quantities uniquely characterizing the orbit, summarized in Table \ref{tab:input_values}.

For bound binaries, we parametrize the orbit with the dimensionless orbit-averaged orbital angular frequency $ \langle M \Omega \rangle $, and two eccentric parameters, $ e $ and $ \zeta $, defined within a \emph{Keplerian parametrization} of the orbit \cite{darwin1959gravity},\footnote{
In the small-mass-ratio literature, the parametrization~\eqref{eq:Kep_param} is known as the \emph{quasi-Keplerian} parametrization (e.g., see Ref.~\cite{Lynch:2024hco}).
However, in the PN literature, ``quasi-Keplerian'' is a name associated with another parametrization for eccentric orbits, in which different definitions of eccentricity appear (for example, the \emph{time eccentricity}, $ e_t $) \cite{damour1985general,Damour:1988mr,schafer1993second,wex1995second,Memmesheimer:2004cv,Cho:2021oai}.
\emph{Darwin variables} is another name for the parameters entering Eq.~\eqref{eq:Kep_param}.
}
\begin{equation}
\label{eq:Kep_param}
\frac{r}{M} = \frac{p}{1 + e \cos \zeta},
\end{equation}
where $ e $ is the \emph{Keplerian eccentricity}, $ \zeta $ is a radial phase parameter (known as the \emph{relativistic anomaly}, which reduces to the \emph{true anomaly} in the Newtonian limit), and $ p $ is the semi-latus rectum which can be obtained from $ (\langle M \Omega \rangle, e, \zeta) $ \cite{Gamboa:2024imd}.

For generic planar orbits (including scattering encounters and dynamical captures), we instead use the initial separation $ r $, energy $ E/M $, and total angular momentum $ J / M^2  $, given by
\begin{align}
E/M
&=
H _{ \text{EOB}}/M,
\\
J/M^2
&=
\nu \, p_\phi/(\mu M) + \chi_1 \s X_1^2 + \chi_2 \s X_2^2 .
\end{align}
\begin{table}

\renewcommand{\arraystretch}{1.5}
\begin{tabular}{ c  c  c }

\hline
\hline

& Parameter & Description
\\[2pt]

\hline
\hline

& $m_{1,\s 2}$ \, $ (m_1 \geq m_2) $
& mass components
\\
& $\chi_{1,\s 2}$ \, $ (\s |\s \chi_{1,\s2}| < 1 )$
&
\makecell[cc]{dimensionless spin\\ components }
\\[8pt]

\hline
\\[-12pt]

\multirow{3.5}{*}{\makecell{bound\\orbits}}
& $ \langle M \Omega \rangle $ &
\makecell[cc]{dimensionless orbit-averaged\\orbital frequency}
\\
& $e$ & Keplerian eccentricity
\\[1pt]
& $\zeta \, $ or $ \, l $ & \makecell[cc]{relativistic anomaly\\or mean anomaly}
\\[8pt]

\hline

\multirow{3.1}{*}{\makecell{generic\\orbits}}
& $r$ & relative separation
\\[-1pt]
& $E/M$ & energy
\\[-1pt]
& $J/M^2$ & total angular momentum
\\[3pt]

\hline
\hline
\end{tabular}

\caption{
Input parameters of the \texttt{SEOBNRv6EHM} model characterizing a unique eccentric, aligned-spin BBH system.
}
\label{tab:input_values}
\end{table}
%


\subsection{EOB waveforms}
\label{sec:overview_waveforms}

The plus and cross polarizations of GWs, $ h_+ $ and $ h_\times $, are modeled through the \emph{waveform modes} $ h_{\ell m} $, which are the expansion coefficients of a decomposition in terms of $-2$ spin-weighted spherical harmonics, $ _{-2}Y_{\ell,\s m}$.
Specifically,
\begin{equation}
h_+-i \s h_\times =  \sum^{\infty}_{\ell =2} \sum_{m=-\ell }^{\ell }{} \!_{-2} Y_{\ell,\s m}(\iota,\s \varphi)\, h_{\ell m},
\label{eq:gw_polarizations_decomp}
\end{equation}
where $ \iota $ and $ \varphi $ are the inclination and azimuthal angles, respectively, of the line of sight measured in the source frame.

In \texttt{SEOBNRv6EHM}, we decompose the waveform modes as
\begin{equation}
\label{eq:h_lm_imr}
\begin{aligned}
&h_{\ell m} =
\left\{
\begin{array}{ll}
h_{\ell m}^{\text{F}}, & \text{unbound orbits}, \\
h_{\ell m}^{\text{insp-plunge}} = N_{\ell m} \, h_{\ell m}^{\text{F}}, 
& \text{bound orbits, } t \le t_{\text{match}}, \\
h_{\ell m}^{\text{merger-RD}}\hspace{-23.5pt}, 
& \text{bound orbits, } t \ge t_{\text{match}},
\end{array}
\right.
\\[6pt]
&(\ell, |m|) = \{(2,2), (3,3), (2,1), (4,4), (3,2), (4,3)\},
\end{aligned}
\end{equation}
where $ h_{\ell m}^{\text {F}} $ are the \emph{factorized} modes resumming analytical PN results, $ N_{\ell m} $ are \emph{nonquasicircular corrections} which enhance the late-inspiral waveform of binary coalescences using \emph{input values} calibrated against QC NR waveforms, $ h_{\ell m}^{\text {merger-RD }} $ is the merger-ringdown part of the waveform, modeled with a phenomenological ansatz for QC binaries which employs the least-damped quasinormal mode of the remnant BH, and $ t_{\text {match }} $ is the peak of the $(2,2)$ mode for eccentric binaries, which inherits a calibration against QC NR simulations.
All the elements in Eq.~\eqref{eq:h_lm_imr} are described in Sec.~\ref{sec:waveforms}.
Here, we highlight the new factorization employed in \texttt{SEOBNRv6EHM}.

Specifically, we factorize the EOB modes as
\begin{equation}
\label{eq:overview_model_fact}
h_{\ell m}^\mathrm{F}  = h_{\ell m}^\mathrm{F,\,hyb}(\Omega, r, \dot r, \dot p_{r_*, \s\text{cons}}) \, h_{\ell m}^\mathrm{ecc} \left(r, \dot r, \dot p_{r_*, \s\text{cons}} \right),
\end{equation}
where $ h_{\ell m}^\mathrm{F,\,hyb} $ is defined analogously to Eq.~\eqref{eq:overview_h_F_qc}, but it employs a new leading-order (Newtonian) prefactor valid for generic planar orbits, and $ h_{\ell m}^\mathrm{ecc} $ are eccentricity corrections that ensure the PN expansion of the eccentric modes is accurate beyond leading order.
These corrections are used in PN-expanded form up to relative-1PN order, and are given in Eq.~\eqref{eq:modes_ecc_corr}.

The novel Newtonian prefactor is written as
\begin{equation}
h_{\ell m}^{\rm N}
=
\frac{\nu M}{d_{\mathcal L}} n_{\ell m} \, c_{\ell+\epsilon_{\ell m}}(\nu) \, \hat {h}_{\ell m} \s Y_{\ell-\epsilon_{\ell m},-m}\left(\frac{\pi}{2},\phi\right),
\end{equation}
where the definition of these factors is given in Sec.~\ref{sec:factorized_modes}, and we note that the $ \hat {h}_{\ell m} $ functions generalize the corresponding expressions employed in QC EOB models.

The factorization~\eqref{eq:overview_model_fact} allows for an accurate recovery of the QC expressions employed in \texttt{SEOBNRv5} models, while keeping the correct eccentricity contributions beyond the leading order.
As with the RR force \eqref{eq:overview_RR_force}, this is achieved thanks to the use of the parameters $ \big(P^0_5[\Omega _{ \text{c}}], \dot r, \dot p_{r_*, \s\text{cons}} \big) $.

To compute the waveforms, the factorized modes \eqref{eq:overview_model_fact} are evaluated on the binary dynamics obtained from the numerical integration of the equations of motion \eqref{eq:EOM}.
For coalescing binaries, these modes are supplemented with the nonquasicircular corrections $ N _{ \ell m} $, and are smoothly stitched to a QC merger-ringdown, as indicated in Eq.~\eqref{eq:h_lm_imr}.
This procedure yields highly accurate waveform modes for the entire inspiral-merger-ringdown evolution of bound binaries, or for the scattering process of unbound binaries.


\section{Radiation-reaction force for binaries on generic planar orbits}
\label{sec:RR_force}

The RR force plays a fundamental role in producing accurate waveform templates, as it determines the binary's phase to which GW detectors are highly sensitive.
In this section, we motivate and derive a new resummation of the EOB RR force applicable for binaries moving on generic planar orbits.


\subsection{The PN-expanded RR force}
\label{sec:PN_RR}

The RR force can be derived from first-principle calculations or through a set of \emph{flux-balance} relations.
The former approach relies on direct integration of the retarded field generated by the binary \cite{Damour:1981bh,Damour:1981ntn,damour1982probleme,Schaefer:1985vxb,Kopeikin:1985,Blanchet:1998vx}, or on matching procedures between the near-zone and wave-zone fields \cite{Blanchet:1996vx,Jaranowski:1996nv,Pati:2002ux,Konigsdorffer:2003ue,Nissanke:2004er,Itoh:2009rz,Blanchet:2024loi,Blanchet:2026suq}.
In contrast, the flux-balance method \cite{Iyer:1993xi,Iyer:1995rn,Gopakumar:1997ng} requires a consistency between the (gauge-dependent) local energy and angular momentum losses by the system, $ \dot E _{ \text{sys}} $ and $ \dot J _{ \text{sys}} $, and the corresponding (gauge-invariant) GW fluxes of energy and angular momentum at infinity, $ \Phi_E $ and  $ \Phi_J $.
These quantities can be related by introducing a pair of \emph{Schott terms}, $ E _{ \text{Schott}}  $ and $ J _{ \text{Schott}} $, which represent additional gauge-dependent contributions arising due to the interaction of the system with the radiation field, as in the theory of electromagnetism \cite{Schott:1915aa}. 

To determine the RR force, we start by writing the flux-balance equations for generic planar orbits as~\cite{Bini:2012ji}
\begin{subequations}
\label{eq:fluxbalance}
\begin{align}
\dot E _{ \text{sys}} + \dot E _{ \text{Schott}} + \Phi_E &= 0,
 \\
\dot J _{ \text{sys}} + \dot J _{ \text{Schott}} + \Phi_J &= 0.
\end{align}
\end{subequations}
Then, from the equations of motion \eqref{eq:EOM} we get the relations $\dot E_{\text{sys}} = \nolinebreak \dot r \, \mathcal F_r  + \nolinebreak\Omega \, \mathcal F_\phi$ and $\dot J_{\text{sys}} = \mF_\phi$, which lead to
\begin{subequations}
\label{eq:forces_schott}
\begin{align}
\mF_\phi &= - \dot J _{ \text{Schott}} - \Phi_J ,
\label{eq:fphischott}
 \\
\mathcal F_r  &=
\frac{
-\Phi_E - \dot E _{ \text{Schott}}
+ \Omega  \, ( \Phi_J + \dot J _{ \text{Schott}})
}{\dot r}.
\label{eq:frschott}
\end{align}
\end{subequations}

Afterward, one substitutes expressions for the fluxes (e.g., the 3PN EOB fluxes derived in Ref.~\cite{Gamboa:2024imd}), and proposes ans\"atze for the Schott terms whose time derivatives are obtained by substituting back the equations of motion~\eqref{eq:EOM}.
The Schott terms are parametrized by arbitrary \emph{gauge constants} $ \{ \alpha_i \} $ and $ \{ \beta_i \} $, such that $ J _{ \text{Schott}}(r, p_r, p_\phi, \{ \alpha_i \} ) $ and $ E _{ \text{Schott}}(r, p_r, p_\phi, \{ \beta_i \} ) $.
In this way, an expansion in small velocities leads to a set of linear equations for $ \{ \alpha_i\} $ and $ \{ \beta_i\} $, whose solution can be expressed in terms of a reduced set of independent gauge constants.
This ultimately yields generic, gauge-dependent PN expressions for the RR force components in terms of EOB coordinates.

At leading order (2.5PN order), there are two independent gauge constants, commonly denoted by $ \alpha $ and $ \beta $, which allow the RR force components to be written as \cite{Gamboa:2024imd}:
\begin{subequations}
\label{eq:RR_LO_rprL}
\begin{align}
\mathcal{F}^\text{LO}_\phi &=
\frac{8 M^3 \nu^2}{5 \, r^3} \frac{ p_\phi}{\mu}
\left[
(2 \alpha + 1) \s \frac{p_r^2}{\mu^2}
- (\alpha + 2) \s  \frac{p_\phi^2}{\mu^2 r^2}
+ (\alpha - 2) \s \frac{M}{r}
\right],
\label{eq:Fphi_LO_rprL}
\\
\mathcal{F}^\text{LO}_r &=
\frac{8 \nu^2 M^3}{15 \, r^3} \frac{p_r}{\mu}
\Bigg[
6 \s (\alpha -\beta + 2)\,  \frac{p_r^2}{\mu^2}
- 3 \s ( \alpha - 3 \beta -1)\s  \frac{p_\phi^2}{\mu^2 r^2}
\nonumber
\\ 
&\hspace{55pt} 
+ (9 \alpha -9 \beta + 17) \frac{M}{r}
\Bigg].
\label{eq:Fr_LO_rprL}
\end{align}
\end{subequations}
For later convenience, we rewrite Eqs.~\eqref{eq:RR_LO_rprL} as
\begin{subequations}
\label{eq:RR_LO}
\begin{align}
\mathcal{F}^\text{LO}_\phi &=
-\frac{32 \nu^2 M^2  \Omega}{5}
\left[
\frac{M^2}{r^2}
- \frac{2 \alpha + 1}{4}  \s \frac{M \dot r^2}{r} 
+ \frac{\alpha + 2}{4}\s \frac{M \dot p_r }{\mu}
\right],
\label{eq:Fphi_LO}
\\
\mathcal{F}^\text{LO}_r &=
-\frac{32 \nu^2 M^2 \Omega }{5}
\frac{p_r}{p_\phi}
\!\left[
\frac{- 3 \alpha -10 }{6} \frac{M^2}{r^2}
- \frac{\alpha - \beta + 2}{2} \s  \frac{M \dot r^2}{ r}
\right.
\nonumber
\\ 
&\hspace{73pt} \left.
+ \frac{\alpha - 3 \beta -1}{4}\s \frac{M \dot p_r }{\mu}
\right],
\label{eq:Fr_LO}
\end{align}
\end{subequations}
where we have used the Newtonian relations, $ \Omega = p_\phi/(\s \mu r^2) $, $ \dot r = \nolinebreak p_r / \mu $, and $ p_{\phi}/(\s\mu M) = \sqrt{r/M + \dot p_r \s r^3 / (\s \mu M^2)  } $.
The circular-orbit limit is easily obtained by taking $ M^2/r^2 = \nolinebreak (M\Omega)^{4/3}$, $ \dot r =\nolinebreak 0 $, and $ \dot p_r = 0 $, leading to the well-known expression:
\begin{equation}
\mathcal{F}^\text{LO, qc}_\phi =
-\frac{32 \nu^2 M (M \Omega)^{7/3}}{5}.
\end{equation}

At higher PN orders, the RR force increases in complexity, with a growing number of independent constants.\footnote{
In Ref.~\cite{Gamboa:2024imd}, the calculation was extended up to the relative 3PN order.
At 2.5PN order one needs to take into account the back reaction of the leading order RR effects, which are associated to the constants $ \alpha $ and $ \beta $.
}
However, a suitable gauge condition can be used to determine the values of these constants.
Here, we propose the condition
\begin{subequations}
\label{eq:RRforce_QCgauge_gen}
\begin{align}
\mF_\phi ^{ \text{qc}}
&=
- \frac{\Phi_E^\text{qc}}{\Omega} ,
\\
\mF_r ^{ \text{qc}}
&=
\frac{- 3 \alpha -10 }{6} \,
\frac{p_{r}}{p_\phi} \mF_\phi^\text{qc},
\label{eq:Frqcgauge}
\end{align}
\end{subequations}
where ``qc'' denotes the QC-orbit limit of the corresponding expressions.
This condition keeps $ \alpha $ and $ \beta $ arbitrary, is consistent with the leading-order expressions \eqref{eq:RR_LO}, and generalizes the gauges employed in QC EOB models.
In particular, the gauge employed in \texttt{SEOBNRv5} models \cite{Pompiliv5,Gamboa:2024hli} can be recovered by setting $ \alpha = -16/3 $.

By using the gauge condition \eqref{eq:RRforce_QCgauge_gen}, the same ans\"atze for the Schott terms given in Eqs.~(64) and (65) of Ref.~\cite{Gamboa:2024imd}, and the energy and angular momentum fluxes derived therein,\footnote{
These fluxes are composed of instantaneous and tail contributions.
The instantaneous terms are valid for generic planar orbits, while the tail contributions are derived in an expansion in $ p_r $.
The original tail contributions to the fluxes are given in an orbit-averaged eccentricity expansion.
By employing a $ p_r $-expanded ansatz in terms of EOB variables, the orbit-averaged tail contributions can be transformed into non-averaged expressions.
This transformation makes it possible to derive expressions for the RR force that depend only on EOB variables.
}
we obtain complete 3PN expressions of the RR force for binaries on generic planar orbits, as a function of the leading order gauge constants $ \alpha $ and $ \beta $.
For reference, the 1PN expressions are given in Eq.~\eqref{eq:RRforces_EOB_1PN} of Appendix~\ref{sec:PN_expressions}, while the complete 3PN expressions, as well as the Schott terms, are provided in the Supplemental Material~\cite{SEOBNRv6EHM:SupplementalMaterial}.
Such expressions generalize the ones given in Eqs.~(66) of Ref.~\cite{Gamboa:2024imd}, which can be obtained by taking $ \alpha = -16/3 $ and $ \beta = -13/2 $.


\subsection{A new resummation of the RR force}
\label{sec:new_RR}

PN-expanded expressions perform poorly at high velocities, making them insufficient for modeling the plunge or the periastron passages of binaries with moderate to high eccentricities.
For this reason, EOB models use resummations to improve the robustness and accuracy of PN expressions.
Here, we propose a new resummation of the RR force that enables accurate modeling of binaries on generic planar orbits.


\subsubsection{Motivation}
\label{sec:new_RR_motivation}

Standard resummations of the EOB RR force for QC binaries follow the work from Refs.~\cite{Damour:2007xr,Damour:2007yf,Damour:2008gu,Pan:2010hz}.
In particular, for \texttt{SEOBNRv5} models, the resummation reads \cite{Pompiliv5}:
\begin{subequations}
\label{eq:fact_RRforce_QC}
\begin{align}
\mF_\phi ^{ \text{F, qc}}&= -\frac{\Omega}{8 \pi} \sum_{\ell=2}^8 \sum_{m=1}^{\ell} m^2\big | d_\mathcal L \s h_{\ell m}^{\text{F, qc}}\big|^2 \!, 
\label{eq:fact_RRforce_QC_az}
 \\
\mF_r ^{ \text{F, qc}}&= \frac{p_{r}}{p_\phi} \mF_\phi^\text{F, qc} \!, 
\end{align}
\end{subequations}
where $h_{\ell m}^{\text{F, qc}}$ are the factorized QC modes given by
\begin{equation}
\label{eq:qcFactModesoriginal}
h_{\ell m}^\text{F, qc} = h_{\ell m}^\text{N,\,qc}  \, \hat{S}_\text{eff} \, T_{\ell m}^\text{qc} \, f_{\ell m}^\text{qc} \, \e^{i \delta_{\ell m}^\text{qc}},
\end{equation}
with all factors taking their standard definitions (see Sec.~\ref{sec:factorized_modes}), most of which depend directly on $ \Omega $.

Proposing eccentric RR force resummations that reduce to such standard EOB expressions for QC orbits has been the prevailing approach in the literature \cite{Chiaramello:2020ehz,Khalil:2021txt,Gamboa:2024imd}.
This strategy allows one to leverage the extensive developments achieved for QC binaries.
However, imposing this requirement introduces various challenges.
For example, QC resummations are given in a particular RR gauge and rely heavily on the orbital frequency $ \Omega $.
As a result, it becomes difficult to accommodate the additional eccentric degrees of freedom and the RR force inherent gauge dependence.

To motivate our resummation proposal, let us first consider the generic planar-orbit expressions for the energy and angular momentum fluxes from the waveform modes~\cite{Ruiz:2007yx,Lousto:2007mh}
\begin{subequations}
\begin{align}
\Phi_E^\text{gw}
&=
\frac{1}{16 \pi}\sum^{\infty}_{\ell=2} \sum_{m=-\ell}^{\ell} \big |d _{ \mathcal L} \s \dot{h}_{\ell m}(t) \big|^2 ,
\\
\Phi_J^\text{gw}
&=
\frac{1}{16 \pi}\sum^{\infty}_{\ell=2} \sum_{m=-\ell}^{\ell} m \, \Im \left[ d _{ \mathcal L}^2 \s \dot{h}^*_{\ell m}(t) \, h_{\ell m}(t) \right] ,
\label{eq:PhiJ_gw}
\end{align}
\end{subequations}
where the symbol $ ^* $ denotes the complex conjugate and $ \Im $ the imaginary part.
Next, we decompose the modes as
\begin{equation}
h _{ \ell m} = \nolinebreak A _{ \ell m} \, e^{- i m \phi} ,
\end{equation}
with $ \phi $ the azimuthal orbital phase.
Then, one obtains
\begin{equation}
\label{eq:genflux}
\Phi_J^\text{gw}
=
\frac{1}{16 \pi}\sum^{\infty}_{\ell=2} \sum_{m=-\ell}^{\ell} \left[m^2 \Omega - m \s \Im \left( \frac{\dot A _{ \ell m}}{A _{ \ell m}} \right) \right] \big | d _{ \mathcal L} \s h _{ \ell m} \big |^2 .
\end{equation}

From this relation, it is straightforward to see that Eq.~\eqref{eq:fact_RRforce_QC_az} is an incomplete representation of the eccentric RR force.
Indeed, for QC, aligned-spin binaries, one has $ h _{ \ell\, -m} = \nolinebreak (-1)^\ell h _{ \ell m}^*$ and $ A _{ \ell m} ^{ \text{qc}} = A _{ \ell m} ^{ \text{qc}}(\Omega) $, so $ \Im( \dot A _{ \ell m}^{ \text{qc}} / A _{ \ell m}^{ \text{qc}} ) = \nolinebreak \mathcal O (1/c^{10}) $.\footnote{
Imaginary terms enter the dominant amplitude $ A_{22} ^{ \text{qc}} $ at 2.5PN order;
combined with the 2.5PN leading-order contribution to $ \dot\Omega $, they produce a 5PN correction [e.g., see Eqs.~(5.3b) and (6.17) of Ref.~\cite{Blanchet:2023sbv}].
}
Thus,
\begin{equation}
\label{eq:PhiJ_qc}
\Phi_J ^{ \text{qc}} \approx \frac{ \Omega}{8 \pi} \sum^{\infty}_{\ell=2} \sum_{m=1}^{\ell}  m^2 \, \big | d _{ \mathcal L} \s h _{ \ell m}^{ \text{qc}} \big |^2,
\end{equation}
as employed in the QC RR force~\eqref{eq:fact_RRforce_QC_az}, with $ \mF_\phi^{ \text{qc}}  = - \Phi_J^{ \text{qc}}  $.
In contrast, for eccentric binaries, the term $ \Im( \dot A _{ \ell m} / A _{ \ell m} )$ has contributions at leading order, thereby making Eq.~\eqref{eq:PhiJ_qc} an incomplete formula for the eccentric angular momentum flux.
Additionally, the Schott terms encoding the gauge-dependent nature of the eccentric RR force are also missing in Eq.~\eqref{eq:fact_RRforce_QC_az}.

To motivate a RR force resummation that accounts for this information, we first write down an alternative expression for Eq.~\eqref{eq:genflux}.
Assuming that $ A _{ \ell m} =\nolinebreak A _{ \ell m}(r, p_r, p_\phi) $, we obtain
\begin{equation}
\dot A _{ \ell m} =
\frac{\partial A_{ \ell m}}{\partial r} \, \dot r 
+ \frac{\partial A_{ \ell m}}{\partial p_r} \, \dot p_r
+\frac{\partial A_{ \ell m}}{\partial p_\phi} \,\dot p_\phi.
\end{equation}
Then, substituting this relation in Eq.~\eqref{eq:genflux}, we get
\begin{align}
\label{eq:genflux2}
\Phi_J^\text{gw}
&=
\frac{1}{16 \pi}\sum^{\infty}_{\ell=2} \sum_{m=-\ell}^{\ell} 
\left\{
m^2 \Omega
- \Im \left[ \frac{\partial \ln (A _{ \ell m}^m)}{\partial r} \right] \dot r
\right. \nonumber
\\
&\quad \left.
-\, \Im \left[ \frac{\partial \ln (A _{ \ell m}^m)}{\partial p_r} \right] \dot p_{r}
- \Im \left[ \frac{\partial \ln (A _{ \ell m}^m)}{\partial p_\phi} \right] \dot p_\phi
\right\} \big |d _{ \mathcal L} \s h _{ \ell m} \big|^2 .
\end{align}
Next, we focus on the leading order piece of Eq.~\eqref{eq:genflux2}.
We take $ \ell =2 $, $ m = \pm 2 $, and neglect the dissipative part of $ \dot p_r $ and $ \dot p_\phi $,
\begin{subequations}
\begin{align}
\dot p_r
&\approx
\dot p_{r,\,\text{cons}} = - \frac{\partial H _{ \text{EOB}}}{\partial r} ,
\\
\dot p_\phi
&\approx
0 .
\end{align}
\end{subequations}
After some rearrangements, this leads to
\begin{align}
\label{eq:genflux3}
\Phi_J^\text{gw}
& \approx
\frac{M}{8 \pi \s (M\Omega)^{1/3}}
\left\{
(M\Omega)^{4/3}
- \frac{M^{4/3}\Omega^{1/3}}{4} \Im \left[ \frac{\partial \ln (A _{ 22}^2)}{\partial r} \right] \dot r
\right. \nonumber
\\
&\quad \left.
- \, \frac{M^{4/3}\Omega^{1/3}}{4} \Im \left[ \frac{\partial \ln (A _{ 22}^2)}{\partial p_r} \right] \dot p_{r,\,\text{cons}}
\right\} \, 2^2 \,  \big | \hat d _{ \mathcal L} \s h _{ 22} \big|^2 ,
\end{align}
where $ \hat d _{ \mathcal L} = d _{ \mathcal L} / M $ is the dimensionless luminosity distance.

Now, we make a comparison with Eq.~\eqref{eq:Fphi_LO}, which also contains terms proportional to $ \dot p_r $ and $ \dot r $.\footnote{
Note that we do not aim to establish a relation between the $ A _{22} $ terms in Eq.~\eqref{eq:genflux3} and the terms in Eq.~\eqref{eq:Fphi_LO}.
Rather, we are motivating the structure of a factorization that encodes the leading order RR force and has a similar form as the standard QC factorization shown in Eq.~\eqref{eq:fact_RRforce_QC_az}.
}
The main differences \emph{in the structure} of these equations are:
\begin{itemize}
\item[1)]
the presence of the $ \big |\hat d _{ \mathcal L} \, h_{22} \big|^2 $ factor in Eq.~\eqref{eq:genflux3}, and
\item[2)]
the absence of a $ M^2/r^2 $ term in Eq.~\eqref{eq:genflux3}.
\end{itemize}

The first difference can be addressed by noting that, from the leading order contribution to $ h_{22} $ for QC binaries,
\begin{equation}
h _{ 22} ^{ \text{LO, qc}} = \frac{8 \nu}{\hat d_\mathcal{L}} \sqrt{ \frac{\pi}{5} } \, (M\Omega)^{2/3} \, \e^{- i 2 \phi }  , 
\end{equation}
we get
\begin{equation}
\label{eq:prefactorQC}
\frac{M \,2^2 \, \big|\hat d _{ \mathcal L} \s h _{ 22} \big|^2}{8 \pi \s (M\Omega)^{1/3}}
\to
\frac{32 \nu^2 M^2 \Omega}{5},
\end{equation}
which is the prefactor of $ - \mathcal F_\phi ^{ \text{LO}} $ from Eq.~\eqref{eq:Fphi_LO}.
This suggests using $ \big | \hat d_\mathcal L \s h _{ \ell m} ^{ \text{qc}} \big| $ in the factorization of the eccentric RR force.

As for the absence of a $ M^2/r^2 $ term in Eq.~\eqref{eq:genflux3}, we need to introduce a special treatment.
For circular orbits, we would have $ (M\Omega)^{4/3} = M^2/r^2 $ at leading order, and the connection with Eq.~\eqref{eq:Fphi_LO} would be complete.
However, for generic planar orbits, $ \Omega $ and $ r $ are connected through the leading order relation $ \Omega = p_\phi / (\s\mu r^2) $, which complicates the association with Eq.~\eqref{eq:Fphi_LO}.
This motivates the introduction of a function which depends solely on $ r $ and whose circular-orbit limit coincides with $ \Omega $.
A natural candidate is the PN-expanded expression of $ \Omega $ for circular orbits written as a function of $ r $.
We calculate such expression up to 5PN order to include in the result the calibration parameters $ a_6 $ and $ \hat d _{ \text{SO}} $, which appear at 5PN and 4.5PN orders, respectively.
This results in the expression, $ \Omega _{ \text{c}} = \Omega _{ \text{c}}(r) $, provided in Eq.~\eqref{eq:Omega_c} of Appendix~\ref{sec:PN_expressions}.

In this way, $ \Omega _{ \text{c}} $ can be employed, together with the relation~\eqref{eq:prefactorQC}, to relate Eqs.~\eqref{eq:Fphi_LO} and \eqref{eq:genflux3} at leading order via
\begin{equation}
\frac{M^2}{r^2} \to [M \Omega _{ \text{c}} (r)]^{4/3}.
\end{equation}

We emphasize that $ \Omega _{ \text{c}} \neq \Omega = \partial H _{ \text{EOB}} / \partial p_\phi $ for non circular orbits.
The only purpose of $ \Omega _{ \text{c}} $ is to make a connection with the standard EOB factorizations of the RR force, which make use of the orbital frequency $ \Omega $.
In practice, we employ a $ (0,5) $ Pad\'e resummation, $ P^0_5[\Omega _{ \text{c}}] $, since it improves the agreement with the true orbital frequency $ \Omega $ for QC binaries and makes the model more robust.
When computing the Pad\'e approximant, we treat $ \ln r $ as a constant, as done with other elements entering into the \texttt{SEOBNRv5} Hamiltonian~\cite{Khalilv5}.


\subsubsection{Resummation proposal}
\label{sec:proposal_RR}

With these results, we now write down our proposal:
\begin{widetext}
\begin{subequations}
\label{eq:RR_force_proposal}
\begin{align}
\mF_\phi &= \mF_\phi^\text{modes}\, \mF_\phi^\text{ecc},
\label{eq:RR_force_fact_deriv_a}
\\
\mF_r &= \frac{p_{r}}{p_\phi} \ \mF_r^\text{modes} \,\mF_r^\text{ecc},
\label{eq:RR_force_fact_deriv_b}
\\
\mF_\phi^\text{modes} &= -\frac{M}{8 \pi}
\left( 
P^0_5[M \Omega _{ \text{c}}]^{4/3}
- \frac{2 \alpha + 1}{4} \,  \frac{M \dot r^2}{r}
+ \frac{ \alpha + 2}{4 } \, \frac{ M }{\mu} \dot p_{r_*, \s\text{cons}} 
\right) 
\sum_{\ell=2}^8 \sum_{m=1}^{\ell} m^2 \frac{\big| \hat d_\mathcal L \s h_{\ell m}^\text{F,\,qc}(\Omega) \big|^2}{(M\Omega)^{1/3}}\!,
\\
\mF_r^\text{modes} &= -\frac{M}{8 \pi}
\left( 
\frac{ - 3 \alpha -10}{6} \, P^0_5[M \Omega _{ \text{c}}]^{4/3}
- \frac{\alpha - \beta + 2}{2 } \, \frac{M \dot r^2}{r}
+ \frac{  \alpha - 3 \beta -1}{4 } \, \frac{ M }{\mu} \dot p_{r_*, \s\text{cons}}
\right) 
\sum_{\ell=2}^8 \sum_{m=1}^{\ell} m^2 \frac{\big| \hat d_\mathcal L \s h_{\ell m}^\text{F,\,qc}(\Omega) \big|^2}{(M\Omega)^{1/3}} \!,
\end{align}
\end{subequations}
\end{widetext}
where, as before,
\begin{equation}
\dot p_{r_*, \s\text{cons}} = - \xi(r) \frac{\partial H_\text{EOB}}{\partial r}(r,p_{r_*},p_\phi),
\end{equation}
and we add the following terms to $ f_{31} $ \big(appearing inside $ h_{31}^\text{F,\,qc} $\big)
\begin{equation}
\label{eq:f_31_addition}
(M\Omega)^{4/3} \left[
\frac{-12 \nu - 5}{2} \, \chi _{ \text{A}}^2
+(4 \nu -5) \, \frac{\chi _{ \text{A}} \chi _{ \text{S}}}{\delta }
- \frac{5}{2} \, \chi _{ \text{S}}^2
\right],
\end{equation}
to have the complete QC flux for aligned-spin BBHs at 3PN order, compared to that used in \texttt{SEOBNRv5} models \cite{Pompiliv5} [see Eq.~(B3b) of Ref.~\cite{Henry:2022dzx}].

The eccentricity corrections $ \mF_\phi^\text{ecc} $ and $ \mF_r^\text{ecc} $ are obtained by requiring that the PN expansion of Eqs.~\eqref{eq:RR_force_fact_deriv_a} and \eqref{eq:RR_force_fact_deriv_b} matches the 3PN RR force derived in Sec.~\ref{sec:PN_RR}.
The underlying QC factorization in Eqs.~\eqref{eq:RR_force_proposal} ensures that these corrections satisfy $ \mF_\phi^\text{ecc} \to 1 $ and $ \mF_r^\text{ecc} \to 1 $ in the QC-orbit limit.
For later convenience, we denote by $ \mF_\phi^\text{ecc, PN} $ and $ \mF_r^\text{ecc, PN} $ the explicit PN-expanded expressions, to distinguish them from their resummed counterparts [see Eqs.~\eqref{eq:F_ecc_resum}].

By construction, these expressions are consistent with the leading order RR force for generic planar orbits \eqref{eq:RR_LO}, since the corrections $ \mF_\phi^\text{ecc} $ and $ \mF_r^\text{ecc} $ start at relative 1PN order.
In the QC orbit limit $ [\Omega _{ \text{c}} \to \nolinebreak \Omega + \nolinebreak \mathcal O(\epsilon^{11}), \, \dot r^2 \to \nolinebreak \mathcal O(\epsilon^{10}), \, \dot p_r \to \nolinebreak \mathcal O(\epsilon^{10})] $, this resummation satisfies the gauge in Eqs.~\eqref{eq:RRforce_QCgauge_gen}, closely following the standard QC expressions~\eqref{eq:fact_RRforce_QC}.
Furthermore, using the QC modes $ h_{\ell m}^\text{F,\,qc} $ without eccentricity corrections and as functions of the instantaneous frequency $ \Omega $, makes the evaluation of the RR force highly efficient.

To improve the behavior of the eccentricity corrections $ \mF_\phi^\text{ecc} $ and $ \mF_r^\text{ecc} $, we find it important to parametrize them in terms of $ (r, \dot r, \dot p_{r_*, \s\text{cons}}) $ and to perform an additional resummation.
Without such control, previous work found significant numerical difficulties with the eccentric RR force, which led to the total (in \texttt{SEOBNRv4EHM} \cite{Ramos-Buades:2021adz}) or partial (in \texttt{TEOBResumS-Dal\'i}~\cite{Nagar:2021gss,Nagar:2023zxh}, where the radial RR force component contains a QC reduction of the 2PN results from Ref.~\cite{Bini:2012ji}) removal of eccentricity corrections.
\texttt{SEOBNRv5EHM} \cite{Gamboa:2024hli} keeps complete eccentricity corrections, but at the expense of introducing extra Keplerian parameters, which limit the model's applicability to bound orbits and introduce potential issues in the equations of motion.

The parametrization $ (r, \dot r, \dot p_{r_*, \s\text{cons}}) $ makes the RR force \eqref{eq:RR_force_proposal} fully applicable to generic planar orbits and it enables an accurate recovery of the underlying QC factorization, since $ \dot r $ and $ \dot p_{r_*, \s\text{cons}} $ are very close to zero for QC inspirals.
We find that using $ \Omega $ or $ p_\phi $ leads to worse performance, as these variables have inherent noncircular contributions, and thus rely on PN corrections to achieve a good QC-orbit limit.
In contrast, $ \dot r $ and $ \dot p_{r_*, \s\text{cons}} $ are independent variables that naturally separate the eccentricity contributions from the PN orders.
We employ $ \dot r $ instead of $ p_{r_*} $ since the former remains smaller in magnitude throughout the entire evolution.
The use of $ \dot p_{r_*, \s\text{cons}} $ is convenient, as it is directly obtained during the integration of the equations of motion \eqref{eq:EOM} and does not depend explicitly on the dissipative dynamics, thereby avoiding a circular (implicit) dependence on the RR force.
The 3PN-accurate transformation rules required to write all the relevant expressions in terms of $ (r, \dot r, \dot p_{r_*, \s\text{cons}}) $ are given in Appendix~\ref{sec:transformation_to_rrdprdst}.

\begin{figure*}
\hspace*{-15pt}
\includegraphics[width=0.9\linewidth]{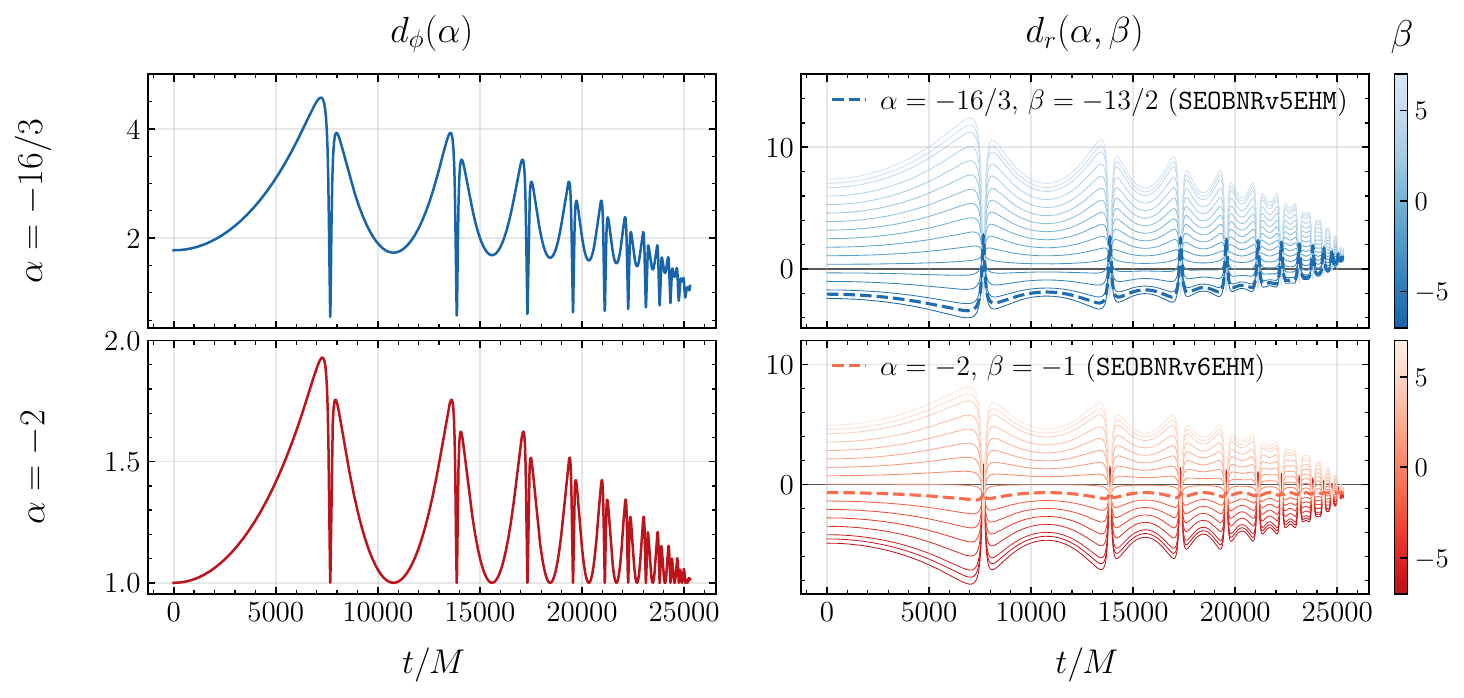}
\vspace{-8pt}
\caption{
Evolution of the functions $ d_\phi = d_\phi(r, \dot r, \dot p_{r_*, \s\text{cons}}; \alpha)$ (left panels) and $ d_r = d_r(r, \dot r, \dot p_{r_*, \s\text{cons}}; \alpha, \beta) $ (right panels), defined in Eqs.~\eqref{eq:denominators}, for different values of the leading-order RR gauge constants $ \alpha $ and $ \beta $.
Each row corresponds to a fixed value of $ \alpha $, while $ \beta $ is varied over the interval $ [-7, 7] $, as indicated by the color bars.
Dashed curves indicate specific gauge choices, including the one used in \texttt{SEOBNRv6EHM}.
The functions are evaluated on the dynamics of an equal-mass, nonspinning binary with initial orbit-averaged frequency $ \langle M \Omega \rangle = 0.0005 $, eccentricity $ e = 0.95 $, and relativistic anomaly $ \zeta  = \pi $, generated with the \texttt{SEOBNRv6EHM} model.
These functions $ d_\phi $ and $ d_r $ enter the denominators of the eccentricity corrections to the RR force [see Eqs.~\eqref{eq:F_argexp}], and are directly related to its leading-order (Newtonian) contributions [see Eqs.~\eqref{eq:RR_LO}].
}
\label{fig:ab_gauges}
\end{figure*}

The additional resummation applied to $ \mF_\phi^\text{ecc} $ and $ \mF_r^\text{ecc} $ is a novel proposal that guarantees robust numerical behavior during the plunge and highly relativistic periastron passages.
As the PN order increases, the eccentric RR force becomes more intricate, involving multiple terms with different powers of three dynamical variables [e.g., see Eq.~\eqref{eq:RRforces_EOB_1PN}].
This becomes problematic in high-velocity configurations, often leading to issues in the integration of the equations of motion.

Specifically, the resummation takes the form:
\begin{subequations}
\label{eq:F_ecc_resum}
\begin{align}
\mF_\phi^\text{ecc} &= \frac{2}{1 + \exp\big(\mF_{\phi}^{\text{arg exp} }\big)},
\\
\mF_r^\text{ecc} &= \frac{2}{1 + \exp\big(\mF_{r}^{\text{arg exp} }\big)},
\end{align}
\end{subequations}
where $ \mF_{\phi}^\text{arg exp} $ and $ \mF_{r}^\text{arg exp} $ start at 1PN order (relative to the leading 2.5PN order), and are obtained from the relations given in Appendix~\ref{sec:RR_corr_exp_resum}, which ensure that the PN expansion of Eqs.~\eqref{eq:F_ecc_resum} is consistent with the expressions $ \mF_\phi^\text{ecc, PN} $ and $ \mF_r^\text{ecc, PN} $.
The 1PN part of $ \mF_{\phi}^\text{arg exp} $ and $ \mF_{r}^\text{arg exp} $ is shown in Eqs.~\eqref{eq:F_argexp} of Appendix~\ref{sec:PN_expressions}, while the full 3PN expressions are provided in the Supplemental Material~\cite{SEOBNRv6EHM:SupplementalMaterial}.

This sigmoid resummation satisfies:
\begin{itemize}
\item
$ \mF_\phi^\text{ecc} \to 1 $ and $ \mF_r^\text{ecc} \to 1$ when $ \mF_{\phi}^{\text{arg exp} } \to 0 $ and $ \mF_{r}^{\text{arg exp} } \to 0$ in the Newtonian limit (low velocities),
\item
$ \mF_\phi^\text{ecc} \to 0 $ or $ \mF_r^\text{ecc} \to 0$ in strong periastron passages or during the plunge, if $ \mF_{\phi}^{\text{arg exp} } \to \infty $ or $ \mF_{r}^{\text{arg exp} } \to \nolinebreak \infty$,
\item
$ \mF_\phi^\text{ecc} \to 2 $ or $ \mF_r^\text{ecc} \to 2$ in strong periastron passages or during the plunge, if $ \mF_{\phi}^{\text{arg exp} } \to -\infty $ or $ \mF_{r}^{\text{arg exp} } \to \nolinebreak -\infty$,
\end{itemize}
thereby always controlling the growth of the corrections.


\subsubsection{Selecting a convenient RR gauge}
\label{sec:RR_gauge}

Choosing a gauge is important for the accuracy and numerical robustness of the RR force.
In general, the gauge constants can be mapped to specific coordinate choices, and can take arbitrary values (within a certain class of coordinate systems), as explained in Refs.~\cite{Iyer:1993xi,Iyer:1995rn,Gopakumar:1997ng}.
While all the gauges are physically equivalent in the PN regime (weak-field, slow-velocities), the situation is more complex during strong periastron passages or the plunge, where highly relativistic velocities are reached.
For example, Ref.~\cite{Fumagalli:2025rhc} showed that different RR gauges lead to significantly distinct features in the evolution, particularly when the eccentricity is high.
Indeed, the gauge constants determine the algebraic structure of the RR force at each PN order [e.g., see Eqs.~\eqref{eq:RRforces_EOB_1PN}].
Thus, the gauge has a non-negligible impact on the evolution of the binary.

Our gauge is determined by Eqs.~\eqref{eq:RRforce_QCgauge_gen} together with the ans\"atze for the Schott terms from Ref.~\cite{Gamboa:2024imd}.
This leaves two free gauge constants, $ \alpha $ and $ \beta $.
To fix them, we focus on the denominators shown in Eqs.~\eqref{eq:F_argexp}, given by
\begin{subequations}
\label{eq:denominators}
\begin{align}
d_\phi
&=
1 - \frac{2 \alpha + 1}{4} \, \frac{r \s \dot r^2}{M} + \frac{\alpha + 2}{4} \, \frac{r^2 \s \dot p_{r_*, \s\text{cons}}}{\mu M} ,
\\
d_r
&=
 \frac{ -3 \alpha -10}{6} -\frac{\alpha - \beta + 2}{2} \, \frac{r \s \dot r^2}{M} + \frac{\alpha - 3 \beta - 1}{4} \, \frac{r^2 \s \dot p_{r_*, \s\text{cons}}}{\mu M}  ,
\end{align}
\end{subequations}
which, by construction, are associated with the leading order terms of the RR force \eqref{eq:RR_LO}.

In Fig.~\ref{fig:ab_gauges}, we show the evolution of $ d_\phi $ (left panels) and $ d_r $ (right panels) across various $ \alpha $ and $  \beta $ values, for a fixed highly eccentric binary dynamics.
Particularly, we focus on the gauge choice from \texttt{SEOBNRv5EHM} \cite{Gamboa:2024imd} ($ \alpha = -16/3 $, $ \beta =\nolinebreak -13/2 $), and a gauge that removes the factor $ r^2 \s \dot p_{r_*, \s\text{cons}} $ from $ d_\phi $ and $ d_r $, such that
\begin{equation}
\label{eq:ab_gauge}
\alpha = -2, \qquad \beta = -1.
\end{equation}
We see that some gauge constants (including the ones from \texttt{SEOBNRv5EHM}) lead to vanishing denominators, making them unsuitable for robust evolutions.
In contrast, the values \eqref{eq:ab_gauge} produce a regular behavior of the denominators.
This can be inferred from Eqs.~\eqref{eq:denominators}, since for these values of $ \alpha $ and $ \beta $, $ d_\phi $ and $ d_r $ remain strictly positive and negative, respectively, for all values of $ (r, \dot r, \dot p_{r_*, \s\text{cons}}) $, thereby guaranteeing nonproblematic denominators.
Furthermore, in calculating the eccentricity corrections to the RR force [e.g., Eq.~\eqref{eq:F_argexp}], several terms acquire a dependence on the factors $ \alpha + 2 $ and $ \alpha - 3 \beta - 1$, which vanish in the gauge \eqref{eq:ab_gauge}.
This choice, therefore, leads to more compact expressions that are faster to evaluate.

For these reasons, we use the gauge \eqref{eq:ab_gauge} in \texttt{SEOBNRv6EHM}.
We emphasize that this choice is based on properties of our proposed RR force resummation \eqref{eq:RR_force_proposal}.
Other resummations may benefit from similar arguments to select an appropriate RR gauge.
As discussed later in Sec.~\ref {sec:comparison_ecc}, the gauge has important effects on the accuracy of the model, and thus is worth studying with more detail in future work.


\subsubsection{The \texttt{SEOBNRv6EHM} RR force}
\label{sec:RR_v6}

The RR force in \texttt{SEOBNRv6EHM} is obtained from Eqs.~\eqref{eq:RR_force_proposal} with the gauge condition \eqref{eq:ab_gauge}, and 1PN contributions to $ \mF_{\phi}^\text{arg exp} $ and $ \mF_{r}^\text{arg exp} \! $, as given in Eq.~\eqref{eq:F_argexp}.
Additionally, some of the factorized modes $ h_{\ell m}^\text{F,\,qc} $ are augmented with calibration parameters, as discussed in Sec.~\ref{sec:calibration_RR}.
This results in the expression \eqref{eq:overview_RR_force} presented before.

We restrict the eccentricity corrections to 1PN order, as higher-order contributions degrade agreement with NR waveforms.
A similar issue was observed during the development of \texttt{SEOBNRv4EHM}~\cite{Ramos-Buades:2021adz}.
We conjecture that this originates from the current paradigm for resumming the RR force, namely, enforcing the standard QC factorization $ \propto \Omega \sum m^2 \s \big | d_\mathcal L h_{\ell m}^\text{F,\,qc} \big |^2 $ supplemented with eccentricity corrections.
Under such construction, the corrections inherit a dependence on different powers of $ \Omega $, leading to increasingly intricate expressions at higher PN orders [e.g., the multiple terms proportional to $ p_{\phi,\,\text{N}}^{n/3} $ in Eq.~\eqref{eq:F_argexp}];
as a result, we note a reduction in waveform accuracy, particularly at low eccentricities.

To quantify this, during the development of \texttt{SEOBNRv6EHM} we performed a systematic, albeit limited, assessment of the PN order of the corrections $ \mF_\phi^\text{ecc} $ and $ \mF_r^\text{ecc} $.
For each choice of PN order, we recalibrated the model to ensure a fair comparison, and we computed waveform mismatches over a given total mass range using the metrics of Sec.~\ref{sec:metrics} (see also Sec.~\ref{sec:comparison_bound});
we restricted these tests to the nonspinning sector (see Secs.~\ref{sec:calibration_RR} and \ref{sec:calibration_nonspinning}) due to the high computational cost of the spinning calibration.
Against a representative set of 39 nonspinning QC NR simulations, the median (maximum) waveform mismatch is $ \sim 2.8 \times 10^{-5} $ ($ \sim 8 \times 10^{-5} $) with 0PN corrections, and $ \sim 1.5 \times 10^{-5} $ ($ \sim 4 \times 10^{-5} $) with 1PN corrections, a clear accuracy gain.
Beyond 1PN, however, each subsequent order reduces the accuracy;
with 3PN corrections we obtain $ \sim 2.6 \times 10^{-5} $ ($ \sim 1 \times 10^{-4} $), i.e., no better than the 0PN eccentricity corrections.
For 81 representative nonspinning eccentric NR simulations, the situation is analogous:
the median (maximum) waveform mismatch is $ \sim 1.6 \times 10^{-4} $ ($ \sim 7 \times 10^{-2} $) with 0PN corrections, and $ \sim 9.2 \times 10^{-5} $ ($ \sim 2 \times 10^{-2} $) with 1PN corrections, while for the 3PN corrections we obtain $ \sim 1 \times 10^{-4} $ ($ \sim 5 \times 10^{-2} $).
For the scattering angles (presented in Sec.~\ref{sec:comparison_gen} only for the 1PN RR force adopted in \texttt{SEOBNRv6EHM}), the dependence on PN orders is non-monotonic:
the 0PN and 3PN corrections yield the most accurate predictions, while the 1PN corrections are the least accurate.
The differences amount to a few percent in the weak-field regime, but reach tens of percent for the most relativistic configurations near the capture threshold.

Since we consider a more accurate model for bound binaries to have higher priority, we restrict the eccentricity corrections to 1PN order, and leave for future work a careful study of the effects of higher PN orders on model accuracy.
Here, we emphasize that the QC part of the RR force \eqref{eq:RR_force_proposal} includes high-PN-order information that partially captures eccentric effects via its dependence on the instantaneous frequency $ \Omega $.
Hence, it is not straightforward to quantify the impact of using 1PN corrections on the phasing of eccentric inspirals.

This reflects the complexity of modeling eccentric binaries and underscores the need for better resummations.
Relaxing the requirement to recover standard QC factorizations may prove advantageous in future work.
This would allow the RR force to naturally adapt to eccentric variables, eliminating the need for ad hoc elements, such as the $ \Omega _{ \text{c}} $ function \eqref{eq:Omega_c_overview}.
The sigmoid resummation \eqref{eq:F_ecc_resum} is promising for incorporating higher PN orders without compromising the robustness of the expressions.
Test-mass-limit studies could help assess different RR force prescriptions, as in Refs.~\cite{Chiaramello:2020ehz,Albanesi:2021rby,Albanesi:2022ywx,Albanesi:2022xge,Albanesi:2023bgi,Faggioli:2024ugn}.


\section{Initial conditions for binaries on generic planar orbits}
\label{sec:ics}

An aligned-spin binary has four degrees of freedom, as encoded in the equations of motion \eqref{eq:EOM}.
Thus, four initial values $ (r_0, \phi_0, p_{r_*0}, p_{\phi \s 0}) $ are required to uniquely determine the binary's evolution.
Here, we discuss how these initial conditions are computed from the input parameters given in Table~\ref{tab:input_values}.
Specifically, we detail the procedures for determining initial conditions 1) for \emph{bound orbits}, given values of the dimensionless orbit-averaged orbital frequency $ \langle M \Omega \rangle $, eccentricity $ e $, and relativistic anomaly $ \zeta $ (or mean anomaly~$ l \s\s$),\footnote{
Other models use a \emph{reference} frequency besides the \emph{starting} frequency (e.g., see Refs.~\cite{Planas:2025feq,Ramos-Buades:2026kbq}).
In \texttt{SEOBNRv6EHM}, the reference and starting frequencies are the same, but we provide the option to integrate backward the full equations of motion \eqref{eq:EOM}.
In the companion manuscript~\cite{Pompili:2026yxq}, we show the usefulness of this in parameter-estimation analyses.
\label{fn:bwd_int}
}
and 2) for \emph{generic planar orbits}, given values of the energy $ E/M $, total angular momentum $ J/M^2 $, and separation $ r $.


\subsection{Bound orbits}
\label{sec:ics_bound}

We determine initial conditions for bound orbits from starting values of the orbit-averaged orbital frequency $ \langle M \Omega \rangle $, eccentricity $ e $, and relativistic anomaly $ \zeta $, following a post-adiabatic approach.
Under this approximation the conservative (adiabatic) values of the dynamical variables, $ ( r_{ 0, \s \text{cons}},  \, p_{ \phi \s 0, \s \text{cons}}, \, p_{ r_* 0, \s \text{cons}} ) $,  are computed first, followed by dissipative (post-adiabatic) corrections, $ ( r_{ 0, \s \text{diss}},  \, p_{ \phi \s 0, \s \text{diss}}, \, p_{ r_* 0, \s \text{diss}} ) $, due to the effect of the RR force starting at 2.5PN order.
This results in the splitting of the initial values as
\begin{subequations}
\label{eq:PA_approx}
\begin{align}
\phi_0
&= 0 ,
\\
r_0
&= r_{ 0, \s \text{cons}} + r_{ 0, \s \text{diss}} ,
\\
p _{\phi \s 0}
&= p_{ \phi \s 0, \s \text{cons}} + p_{\phi \s 0, \s \text{diss}} ,
\\
p _{r_* 0}
&= p_{ r_* 0, \s \text{cons}} + p_{r_* 0, \s \text{diss}} ,
\end{align}
\end{subequations}
where we set $ \phi_0 $ to zero by convention.

In \texttt{SEOBNRv6EHM}, the conservative part of the initial values is determined with PN formulas and a root solving procedure.
Specifically, the radial momentum is calculated with a 3PN-accurate relation given in Eq.~(B11) of Ref.~\cite{Gamboa:2024imd}, namely
\begin{align}
\label{eq:prstar_xez}
p_{ r_* 0, \s \text{cons}} (\langle M\Omega\rangle, e, \zeta)
&= \frac{\langle M\Omega \rangle^{1/3} \s e \s \sin \zeta}{\sqrt{1-e^2}} 
+ \Order(1/c^2).
\end{align}
Next, the values of $ r_{0, \s \text{cons}} $ and $  p_{\phi \s 0, \s \text{cons}} $ are determined by numerically finding the roots of the equations
\begin{subequations}
\label{eq:root_cons}
\begin{align}
\Omega(\langle M\Omega\rangle, e, \zeta)  &= 
  \frac{\partial  H_ \text{EOB}[r, p_{ r_* 0, \s \text{cons}}, p_\phi] }{\partial p_\phi}  ,
\label{eq:root_cons_b}
\\
\dot p_{r_*} (\langle M\Omega\rangle, e, \zeta)  
&= 
- \xi (r) \, \frac{\partial  H_ \text{EOB}[r, p_{ r_* 0, \s \text{cons}} , p_\phi] }{\partial r} ,
\label{eq:root_cons_a} 
\end{align}
\end{subequations}
where $ \Omega(\langle M\Omega\rangle, e, \zeta) $ and $ \dot p_{r_*} (\langle M\Omega\rangle, e, \zeta) $ are calculated with conservative 3PN-accurate formulas, given in Appendix B of Ref.~\cite{Gamboa:2024imd}. 
The initial guesses of the root solver are calculated with Eqs.~(34b) and (34c) of Ref.~\cite{Gamboa:2024imd}.

The dissipative contributions $ r_{ 0, \s \text{diss}} $, $ p_{\phi \s 0, \s \text{diss}} $, and $ p_{r_* 0, \s \text{diss}} $, are directly calculated with the 2.5PN terms given in Eqs.~(40b-d) of Ref.~\cite{Gamboa:2024imd}.
These terms have a dependence on the leading-order RR gauge constants $ \alpha $ and $ \beta $, which we set to $ -2 $ and $ -1 $, respectively, as discussed in Sec.~\ref{sec:RR_gauge}.

This prescription for the initial conditions closely follows the one employed in the \texttt{SEOBNRv5EHM} model, but changes the way in which $ p _{r_* 0} $ is calculated.
In \texttt{SEOBNRv5EHM}, the procedure to determine $ p _{r_* 0} $ involved another root-finding procedure, and it was designed to accurately recover the method employed in the QC model \texttt{SEOBNRv5HM}.
This was important to avoid spoiling the calibration to QC NR simulations inherited from \texttt{SEOBNRv5HM}.
However, in \texttt{SEOBNRv6EHM}, we do not have such constraint, since the model has been recalibrated.
Thus, we decided to use the (simpler) prescription for the radial momentum $ p _{r_* 0} $ described above, since it is less prone to errors, compared to the method used in \texttt{SEOBNRv5EHM}, which can be problematic when applied to highly eccentric binaries.

To ensure the generation of physically meaningful waveforms, we impose constraints on the allowed initial conditions.
Specifically, we prevent waveform generation if any of the following conditions are met:
\begin{itemize}
\item[1)]
The initial angular momentum differs more than $ 1.5\% $ from its PN estimate given by Eq.~(34b) of Ref.~\cite{Gamboa:2024imd}.
\item[2)]
The initial separation differs more than $ 1\% $ from its PN estimate given by Eq.~(34c) of Ref.~\cite{Gamboa:2024imd}.
\item[3)]
The initial separation is less than $ 4M $.
\end{itemize}
These conditions were determined empirically, and they appear for very challenging configurations, e.g., when the binary is very close to merger, or when the eccentricity is very large at high frequencies (for example, see Fig.~\ref{fig:initial_separation} in Sec.~\ref{sec:robustness}).
The first two conditions are also applied in \texttt{SEOBNRv5EHM}.
Additionally, such model performs a backward integration of a set of secular evolution equations $ \{ \dot x, \s \dot e, \s \dot \zeta \}$ whenever the starting separation of the binary is less than $ 10 M $.
This allows the determination of initial conditions at earlier (less challenging) stages, and it was necessary to avoid problems with the equations of motion \cite{Gamboa:2024hli}.
However, as discussed in Sec.~\ref{sec:robustness}, \texttt{SEOBNRv6EHM} does not perform such secular backward integration, and instead it imposes a cutoff in waveform generation whenever the starting separation is less than $ 4 M $, a regime in which the approximations used to construct the initial conditions are expected to be much less accurate.

Beyond the constraints on initial conditions, \texttt{SEOBNRv6EHM} imposes additional requirements to ensure high-quality waveforms.
In particular, waveform generation is aborted if the dynamics duration is less than $ 200M $ and the final azimuthal angle is less than $ \pi $ (meaning that the binary has completed less than half an orbit), if the dynamics duration exceeds a threshold determined by a background QC evolution (see Sec.~\ref{sec:nqcs_bqc}), or if the separation during the evolution significantly exceeds (by a factor of $ 10 $) a Newtonian estimate.
The first condition prevents unphysical features arising from an inaccurate merger-ringdown attachment, while the latter two discard configurations that are inconsistent with the intended physical system represented by the input values of $ e $, $ \langle M \Omega \rangle $, and $ \zeta $.

Overall, these empirical conditions build up the robustness of \texttt{SEOBNRv6EHM}, which we discuss in Sec.~\ref{sec:robustness}.
We expect these conditions to change as better initial conditions and more accurate evolution equations are available.


\subsubsection*{Mean anomaly $ l $ as an input parameter}
\label{sec:mean_anomaly}

The default radial phase parameter in \texttt{SEOBNRv6EHM} is the relativistic anomaly $ \zeta $, but we also provide the option to use the \emph{mean anomaly} $ l $ as an input parameter.
The mean anomaly is an angle that measures (linearly in time) the fraction of the orbital period that has elapsed since the last periastron passage \cite{damour1985general,Schmidt:2017btt,Ramos-Buades:2019uvh,Islam:2021mha,Ramos-Buades:2022lgf,Shaikh:2023ypz}.
Given an input value of the mean anomaly, we employ Newtonian transformations to obtain the corresponding relativistic anomaly, which is used internally to compute EOB initial conditions as described above.
We note that this is the approach followed by the \texttt{TEOBResumS-Dal\'i} model \cite{Nagar:2024dzj}.

To determine the relativistic anomaly $ \zeta $ from the mean anomaly $ l $, first we compute the \emph{eccentric anomaly} $ u $ by numerically solving Kepler's equation,
\begin{equation}
\label{eq:kepler_equation}
l =  u - e \sin u.
\end{equation}
Then, the relativistic anomaly is obtained as \cite{Broucke:1973anomalies}
\begin{equation}
\label{eq:eccanomaly_to_relanomaly}
\zeta
=
u
+ 2 \arctan \left( \frac{\beta \sin u}{1 - \beta \cos u} \right) ,
\end{equation}
where
\begin{equation}
\beta
=
\frac{e}{1 + \sqrt{1 - e^2 }} .
\end{equation}

For completeness, we also provide the inverse transformation from relativistic anomaly to mean anomaly.
First, the eccentric anomaly is calculated through
\begin{equation}
\label{eq:relanomaly_to_eccanomaly}
u
=
\zeta
- 2 \arctan \left( \frac{\beta \sin \zeta}{1 + \beta \cos \zeta} \right).
\end{equation}
Afterwards, the mean anomaly is directly obtained from Kepler's equation \eqref{eq:kepler_equation}.

\begin{figure}
\hspace{-4pt}
\vspace{4pt}
\includegraphics[width=0.99\linewidth]{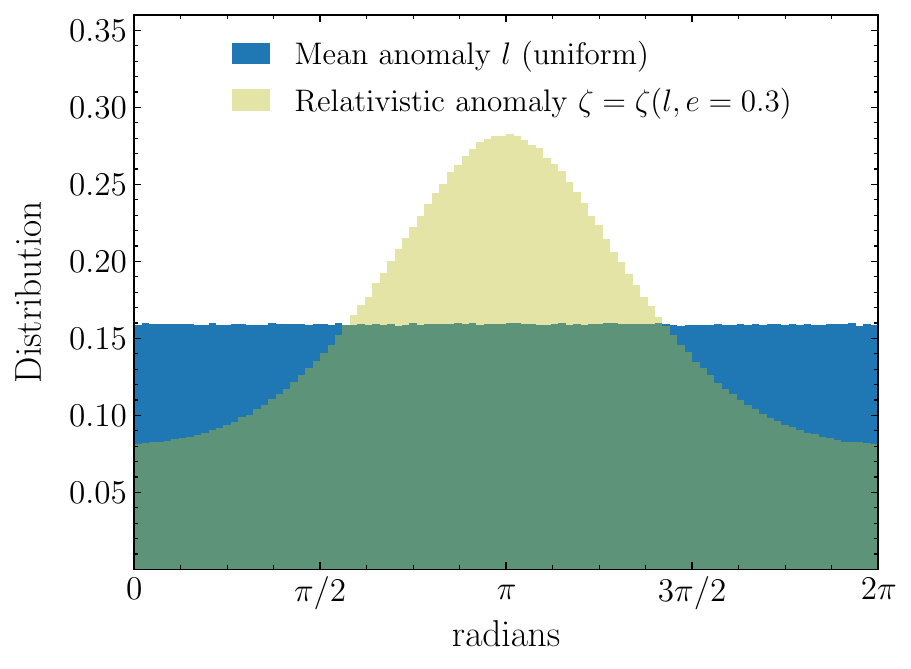}
\includegraphics[width=0.99\linewidth]{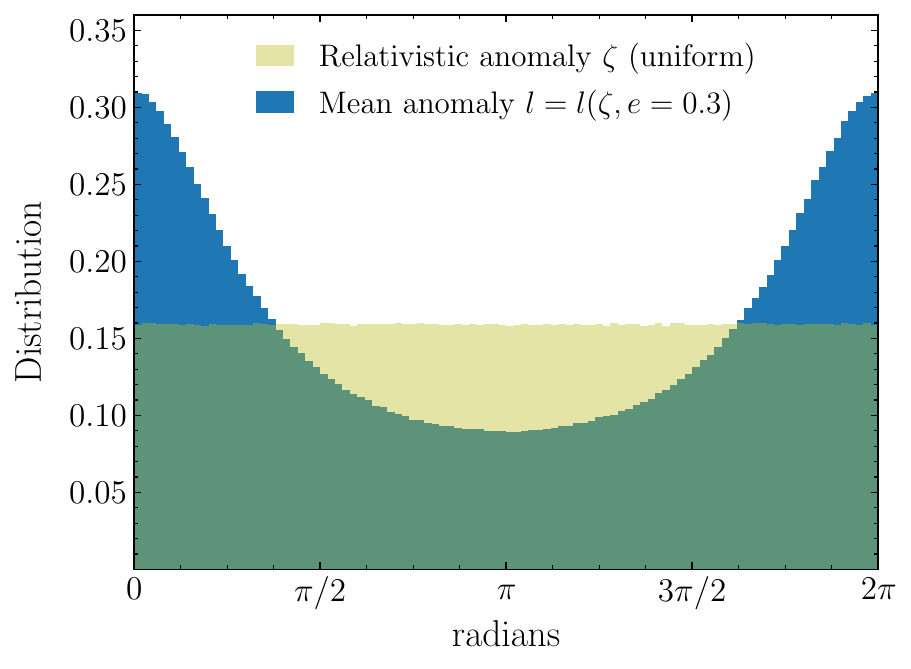}
\vspace{-15pt}
\caption{
\emph{Top panel:}
Uniform distribution in mean anomaly and its mapping to relativistic anomaly through Eqs.~\eqref{eq:kepler_equation} and \eqref{eq:eccanomaly_to_relanomaly}.
\emph{Bottom panel:}
Uniform distribution in relativistic anomaly and its mapping to mean anomaly through Eqs.~\eqref{eq:kepler_equation} and \eqref{eq:relanomaly_to_eccanomaly}.
Both uniform distributions are composed of $ 10^7 $ random samples over the interval $ [0, \s 2\pi] $.
The mappings are performed at an eccentricity of $ 0.3 $.
}
\label{fig:distribution_radial_anomalies}
\end{figure}

To illustrate the differences between uniform distributions in relativistic anomaly and mean anomaly, Fig.~\ref{fig:distribution_radial_anomalies} shows the corresponding distributions obtained by randomly drawing $ 10^7 $ samples of a given anomaly from the interval $ [0, \s 2 \pi] $ and mapping them to the other anomaly at an eccentricity of $ 0.3 $.
Specifically, the top panel shows a uniform distribution in mean anomaly and the corresponding relativistic anomaly distribution computed using Eqs.~\eqref{eq:kepler_equation} and \eqref{eq:eccanomaly_to_relanomaly}, while the bottom panel shows a uniform distribution in relativistic anomaly and the corresponding mean anomaly distribution computed using Eqs.~\eqref{eq:kepler_equation} and \eqref{eq:relanomaly_to_eccanomaly}.
We observe that a uniform distribution in mean anomaly (cf.~a uniform prior in mean anomaly, in the context of parameter estimation), leads to a concentration of the relativistic anomaly values around $ \zeta = \pi $ (apastron).
Conversely, a uniform distribution in relativistic anomaly (cf.~a uniform prior in relativistic anomaly), leads to a concentration of mean anomaly values around $ l = 0 $ (periastron).
The effects of relativistic and mean anomalies in the waveforms are illustrated in Fig.~\ref{fig:smoothness_e_z} of Sec.~\ref{sec:robustness}, where we see that the mean anomaly parametrization leads to a \emph{visually} more evenly distributed waveform amplitude (see also Ref.~\cite{Nee:2025zdy}).

%


\subsection{Generic planar orbits}
\label{sec:ics_generic}

We determine initial conditions for generic planar orbits from starting values of the energy $ E/M $, total angular momentum $ J / M^2$, and relative separation $ r_0 $, following the prescription employed in the \texttt{TEOBResumS-Dal\'i} model (e.g., see Refs.~\cite{Damour:2014afa,Hopper:2022rwo,Albanesi:2024xus}).
Specifically, we set $ \phi_0 = 0 $, we determine the initial orbital angular momentum through
\begin{equation}
\frac{p_{\phi\s 0}}{\mu M}
=
\frac{1}{\nu} \left( \frac{J}{M^2} -\chi_1 \s X_1^2 - \chi_2 \s X_2^2 \right),
\end{equation}
and we compute the initial radial momentum $ p_{r_* 0} $ by numerically finding the root of the equation
\begin{equation}
E = H _{ \text{EOB}} (r_0, p_{r_* 0}, p_{\phi\s 0}),
\end{equation}
where, as an initial guess, we use the relation
\begin{equation}
\label{eq:pr_guess}
p_{r_* 0}
\approx
\pm \sqrt{H _{ \text{eff}}^2 - A _{ \text{noS}}(r_0) \left( \mu^2 + \frac{p_{\phi\s 0}^2}{r_0^2} \right)} \,.
\end{equation}
The sign in Eq.~\eqref{eq:pr_guess} is chosen depending on the initial binary's direction, with $ + $ ($ - $) used for an outgoing (ingoing) trajectory.


\section{Multipolar waveforms for binaries on generic planar orbits}
\label{sec:waveforms}

We model the GWs emitted by binaries on generic planar orbits in terms of their \emph{waveform modes} $ h _{ \ell m} $.
These are the expansion coefficients of the $ h_+ $ and $ h_\times $ polarizations in a $-2$ spin-weighted spherical harmonic basis decomposition~\eqref{eq:gw_polarizations_decomp}.
We first numerically integrate the equations of motion~\eqref{eq:EOM}, and then compute the modes from analytical formulas evaluated on the dynamics.
For coalescing binaries, we employ a phenomenological ansatz for the merger-ringdown phase.

More specifically, we decompose the modes according to
\begin{equation}
\label{eq:h_bound_unbound}
h_{\ell m} =
\left\{
\begin{array}{ll}
h_{\ell m}^{\text {F}},
& \text{unbound orbits},
\\
h_{\ell m}^{\text {insp-plunge }} = 
N_{\ell m} \, h_{\ell m}^{\text {F}},
& \text{bound orbits, } t \leq t_{\text {match }},
\\
h_{\ell m}^{\text {merger-RD }}\hspace{-25.5pt},
& \text{bound orbits, } t \geq t_{\text {match }},
\end{array}
\right. 
\end{equation}
where the functions $  h_{\ell m}^{\text {F}} $ are the \emph{factorized} EOB modes which resum analytical PN results, $ N_{\ell m} $ are \emph{nonquasicircular} (NQC) \emph{corrections} which incorporate information from NR waveforms to improve the late-inspiral waveform of binary coalescences, and $ h_{\ell m}^{\text {merger-RD }} $ is the merger-ringdown portion of the waveform starting at the matching time $ t_{\text {match }} $ related to the peak of the $(2,2)$ mode.
We present all these elements with more detail in the following subsections.

The multipoles supported by the \texttt{SEOBNRv6EHM} model are:
\begin{equation}
\label{eq:supported_modes}
(\ell, |m|) = \{(2, 2),\, (3, 3),\, (2, 1),\, (4, 4),\, (3, 2),\, (4, 3)\},\!
\end{equation}
since for these we have information for the merger-ringdown from QC NR simulations from Ref.~\cite{Pompiliv5}.
These are the same multipoles modeled in the \texttt{SEOBNRv5EHM} model~\cite{Gamboa:2024hli}.

The waveforms are aligned ($ t = \nolinebreak 0 $) at the \emph{last} peak of the frame-invariant amplitude $ A _{ \text{inv}} $, given a threshold value of $ 10\% $ with respect to the \emph{largest} peak of $ A_{ \text{inv}} $, where 
\begin{equation}
A_{ \text{inv}} = \sqrt{\sum_{\ell, \,|m| \leq \ell} |h_{\ell m}|^2} \, .
\end{equation}
This convention allows the possibility that the last prominent peak of $ A_{ \text{inv}} $ (associated with the start of the merger-ringdown) is not the largest one, which can occur for some high-eccentricity binaries, as noted in Refs.~\cite{Gold:2012tk,Carullo:2023kvj,Albanesi:2024xus}.
The threshold of $ 10\%$ is chosen to avoid selecting amplitude peaks due to mode-mixing during the ringdown phase.

In the following, we describe the construction of multipolar waveforms for binaries on generic planar orbits.
We introduce a new mode factorization, detail improvements to the inspiral-plunge stage via NR-informed NQC corrections, and outline the procedure for attaching the merger-ringdown phase.


\subsection{Factorized EOB modes}
\label{sec:factorized_modes}

Expressions for the eccentric modes have been derived within the PN framework \cite{Mishra:2015bqa,Boetzel:2019nfw,Ebersold:2019kdc,Henry:2023tka,Paul:2024ujx}.
However, their accuracy and numerical robustness deteriorate at high velocities, e.g., during the plunge or in strong periastron passages.
To improve this, EOB models resum the modes in terms of physically and numerically motivated factors \cite{Damour:2007xr,Damour:2007yf,Damour:2008gu,Pan:2010hz}.

In \texttt{SEOBNRv6EHM}, we factorize the waveform modes as
\begin{equation}
\label{eq:model_fact}
h_{\ell m}^\mathrm{F}  = h_{\ell m}^\mathrm{F,\,hyb}(\Omega, r, \dot r, \dot p_{r_*, \s\text{cons}}) \, h_{\ell m}^\mathrm{ecc} \left(r, \dot r, \dot p_{r_*, \s\text{cons}} \right),
\end{equation}
where $ h_{\ell m}^\text{F,\,hyb} $ is a factor similar to the one employed in typical EOB QC models, and $ h_{\ell m}^\text{ecc} $ contains eccentricity contributions required for Eq.~\eqref{eq:model_fact} to have the correct PN expansion.

The term $ h_{\ell m}^\text{F,\,hyb} $ hybridizes elements employed in previous \texttt{SEOBNRv5} models \cite{Pompiliv5, Gamboa:2024hli} with a novel Newtonian prefactor.
Specifically, it takes the form
\begin{equation}
\label{eq:model_qcFactModes}
h_{\ell m}^\text{F,\,hyb} = h_{\ell m}^\text{N}  \, \hat{S}_\text{eff} \, T_{\ell m}^\text{qc} \, f_{\ell m}^\text{qc} \, \e^{i \delta_{\ell m}^\text{qc}}.
\end{equation}

Starting with the second factor in Eq.~\eqref{eq:model_qcFactModes}, $\hat{S}_\text{eff}$, this is the dimensionless effective source given by
\begin{equation}
\label{eq:Seff}
\hat{S}_{\mathrm{eff}} =
\begin{cases}
H_{\mathrm{eff}}/\mu,
& \ell + m \text{ even},
\\[0.5em]
(M\Omega)^{1/3} \, p_\phi/(M\mu),
& \ell + m \text{ odd}.
\end{cases}
\end{equation}
whose structure is motivated by solutions to the Regge-Wheeler-Zerilli equation for QC orbits.
A possible extension to generic planar orbits is discussed in Appendix~\ref{sec:ecc_corr_modes}.

The third factor, $ T_{\ell m}^\text{qc} $, resums an infinite contribution of tail terms for QC orbits~\cite{Blanchet:1997jj} and is given by~\cite{Damour:2007xr, Damour:2007yf}\footnote{
Recently, the function $ T _{ \ell m} $ was generalized in Ref.~\cite{Ivanov:2025ozg} to resum both the leading (infrared) logarithms (terms proportional to $ \Omega^n \log ^n\Omega $) and the universal ultraviolet tails (terms proportional to $ \Omega^{n+k} \log ^n\Omega $, with $ k > \nolinebreak 0 $ corresponding to dissipative tails).
}
\begin{equation}
T_{\ell m}^\text{qc} = \frac{\Gamma\left(\ell + 1 - 2 i \hat{k}\right)}{\Gamma (\ell + 1)} e^{\pi \hat{k}} e^{2i \hat{k} \ln (2m \s M\Omega \s r_0)},
\end{equation}
where $ \Gamma $ is the Euler gamma function, $\hat{k}\equiv m \s  \Omega \s H _{ \text{EOB}}$, and the value of $r_0$ is set to $ 2/\!\sqrt{\text{e}}$ to improve the agreement with waveforms in the test-body limit~\cite{Pan:2010hz}.

The amplitude $ f_{\ell m}^\text{qc} $ and phase $\delta_{\ell m}^\text{qc} $ factors in Eq.~\eqref{eq:model_qcFactModes} give consistency with the PN expanded modes for QC orbits.
We do not include calibration parameters in these functions, as done in other EOB models (particularly, in \texttt{SEOBNRv5EHM} \cite{Gamboa:2024hli});
we only add them in the RR force \eqref{eq:overview_RR_force}, as discussed in Sec.~\ref{sec:calibration}.
The amplitude factor is further resummed in terms of its $ \ell $-th root, $ f_{\ell m}^\text{qc} = (\s\s \rho_{\ell m}^\text{qc} ) ^{ \ell}$, according to \cite{Damour:2008gu,Pan:2010hz,Taracchini:2012ig,Taracchini:2013rva}
\begin{align}
\label{eq:frholm}
f_{\ell m}^\text{qc} =
\left\{
\begin{array}{ll}
(\s\s \rho_{\ell m}^\text{qc})^\ell, & \quad m \text{ even},
\\
(\s\s \rho_{\ell m}^\text{qc,\,NS})^\ell + f_{\ell m}^\text{qc,\,S}, & \quad m \text{ odd},
\end{array}
\right. 
\end{align}
where $ \rho_{\ell m}^\text{qc,\,NS} $ denotes the nonspinning part of $ \rho_{\ell m}^\text{qc} $, and $ f_{\ell m}^\text{qc,\,S} $ denotes the spinning part of $ f_{\ell m}^\text{qc} $.
The full expressions are given explicitly in Appendix B of Ref.~\cite{Pompiliv5} [for the RR force, we add more terms to $ f_{31} $ as shown in Eq.~\eqref{eq:f_31_addition}].
We remark that the $ f_{\ell m}^\text{qc} $ and $\delta_{\ell m}^\text{qc} $ factors depend on the instantaneous orbital frequency $ \Omega $ and on the Hamiltonian $ H _{ \text{EOB}} $.

The factor $h_{\ell m}^\text{N}$ in Eq.~\eqref{eq:model_qcFactModes} contains the leading-order (Newtonian) term for generic planar orbits, and is expressed as
\begin{equation}
\label{eq:model_hlm_N_generic}
h_{\ell m}^{\rm N}
=
\frac{\nu M}{d_{\mathcal L}} n_{\ell m} \, c_{\ell+\epsilon_{\ell m}}(\nu) \, \hat {h}_{\ell m} \s Y_{\ell-\epsilon_{\ell m},-m}\left(\frac{\pi}{2},\phi\right),
\end{equation}
where $\epsilon_{\ell m}$ is the parity of the mode (such that $\epsilon_{\ell m} = 0$ if $\ell+m$ is even, and $\epsilon_{\ell m} = 1$ if $\ell+m$ is odd), $Y_{\ell, m}$ is the scalar spherical harmonic, the functions $n_{\ell m}$ and $c_{k}(\nu)$ are defined by
\begin{align}
\label{eq:model_n_c}
c_{k} (\nu)
&=
\left( \frac{1 - \sqrt{1 - 4 \nu}}{2} \right)^{k - 1} \!\!+ (-1) ^{ k} \left( \frac{1 + \sqrt{1 - 4 \nu}}{2} \right) ^{k - 1}\!\!\!\!,
\\[4pt]
n_{\ell m} &=
\begin{cases}
\displaystyle
\frac{ 8\pi (i m)^\ell}{(2\ell+1)!!}
\sqrt{\tfrac{(\ell+1)(\ell+2)}{\ell(\ell-1)}}, 
& \text{$\ell+m$ even,}
\\[6pt]
\displaystyle
\frac{-16i \pi (i m)^\ell}{(2\ell+1)!!}
\sqrt{\tfrac{(2\ell+1)(\ell+2)(\ell^2-m^2)}{(2\ell-1)(\ell+1)\ell(\ell-1)}}, 
& \text{$\ell+m$ odd,}
\end{cases}
\end{align}
and the factor $ \hat {h}_{\ell m} $ is given, for each \texttt{SEOBNRv6EHM} mode, as
\begin{subequations}
\label{eq:hat_hlm_N_generic_rrdprd}
\begin{align}
\hat h_{22}
&=
P^0_5[M \Omega _{ \text{c}}]^{2/3} + \frac{1}{2} \s \frac{r \s  \dot p_{r_*, \s\text{cons}}}{\mu}  - \frac{1}{2} \s \dot r^2 + i \s \dot r \s r \s \Omega ,
\label{eq:h22_N_generic_rrdprd}
\\[2pt]
\hat h_{21}
&=
M\Omega,
\label{eq:h21_N_generic_rrdprd}
\\[2pt]
\hat h_{33}
&=
M\Omega \left( 1 + \frac{2}{9} \frac{ r^2 \dot p_{r_*, \s\text{cons}}}{\mu M} - \frac{2}{3} \frac{r \s \dot r^2}{M} \right)
\nonumber
\\
& \quad
+ i \, \dot r \left( \frac{10}{9} \frac{M}{r} + \frac{2}{3} \frac{r \dot p_{r_*, \s\text{cons}}}{\mu} - \frac{2}{9} \dot r^2 \right),
\\[2pt]
\hat h_{32}
&=
P^0_5[M\Omega _{ \text{c}}]^{4/3} + \frac{M \dot p_{r_*, \s\text{cons}}}{\mu} + \frac{i}{4} \dot r \s M\Omega,
\\[2pt]
\hat h_{44}
&=
P^0_5[M\Omega _{ \text{c}}]^{4/3}
+ \frac{63}{64} \frac{M \dot p_{r_*, \s\text{cons}}}{\mu}
+ \frac{3}{32} \frac{r^2 \dot p_{r_*, \s\text{cons}}^2}{\mu^2}
\nonumber
\\
& \quad
- \dot r^2 \left( \frac{27}{32} \frac{M}{r} + \frac{9}{16} \frac{r \dot p_{r_*, \s\text{cons}}}{\mu} \right)
+ \frac{3}{32} \dot r^4
\nonumber
\\
& \quad
+ \frac{3 i}{8} \s \dot r \s M\Omega \left( \frac{13}{4} - \frac{r \s \dot r^2}{M} + \frac{r^2 \dot p_{r_*, \s\text{cons}}}{\mu M} \right),
\\[2pt]
\hat h_{43}
&=
M\Omega \, P^0_5[M\Omega _{ \text{c}}]^{2/3}
+ \frac{1}{27} \s M\Omega \left( 23 \, \frac{r \dot p_{r_*, \s\text{cons}}}{\mu} - 2 \s \dot r^2 \right)
\nonumber
\\
& \quad
+ \frac{10 i}{27} \s \dot r \left( \frac{M \dot p_{r_*, \s\text{cons}}}{\mu}+ \frac{M^2}{r^2} \right),
\label{eq:h43_N_generic_rrdprd}
\end{align}
\end{subequations}
where $ \dot p_{r_*, \s\text{cons}} = -\xi \, \partial H _{ \text{EOB}} / \partial r $ and $ P^0_5[M\Omega _{ \text{c}}] $ is the $ (0,5) $ Pad\'e resummation of the function $ \Omega _{ \text{c}} = \Omega _{ \text{c}}(r) $ given in Eqs.~\eqref{eq:Omega_c}.

This parametrization yields a robust behavior during the plunge and a seamless transition to the QC-orbit limit.
In such limit, the Newtonian prefactor $ h_{\ell m}^\text{N,\,qc} $ is defined as in Eq.~\eqref{eq:model_hlm_N_generic}, but with $ \hat h_{\ell m}^\text{qc}  = (M \Omega)^{(\ell+\epsilon_{\ell m})/3} $ \cite{Kidder:2007rt}.
This definition is used in the RR force [see Eqs.~\eqref{eq:overview_RR_force} and \eqref{eq:overview_h_F_qc}] without any resummation to $\hat h_{\ell m}^\text{qc}$, as done, e.g., in \texttt{SEOBNRv5HM} \cite{Pompiliv5}.

The expressions \eqref{eq:hat_hlm_N_generic_rrdprd} for $ \hat h _{ \ell m} $ are motivated by our RR force prescription~\eqref{eq:RR_force_proposal}, and they are obtained from the leading-order part of the modes given in Ref.~\cite{Gamboa:2024imd}.
For example, by using the Newtonian relations, $ \Omega = p_\phi/(\s \mu r^2) $, $ \dot r = \nolinebreak p_r / \mu $, and $ p_{\phi}/(\s\mu M) = \nolinebreak \sqrt{r/M + \dot p_r \s r^3 / (\s \mu M^2)  } $, the $ (2,2) $-mode function
\begin{equation}
\label{eq:hat_h22_N_generic_rprL}
\hat h_{22}
=
\frac{M}{2 \s r}
+ \frac{p_\phi^2}{2 \s \mu^2 r^2 }
- \frac{p_r^2}{2 \mu^2}
+ i \,\frac{p_\phi p_r}{\mu^2 r} ,
\end{equation}
can be rewritten as 
\begin{equation}
\label{eq:hat_h22_N_generic_rrdOm}
\hat h_{22}
=
\frac{M}{r} + \frac{1}{2} \s \frac{r \s  \dot p_r}{\mu}  - \frac{1}{2} \s \dot r^2 + i \s \dot r \s r \s\Omega  .
\end{equation}
Then, to accurately recover the known circular-orbit limit \big[$ \hat h_{22} ^{ \text{qc}} = (M\Omega)^{2/3} $\big], we replace the $ M/r $ term in Eq.~\eqref{eq:hat_h22_N_generic_rrdOm} by its expression in terms of $ P^0_5[M\Omega _{ \text{c}}] $ and we use $ \dot p_{r_*, \s\text{cons}} $, in analogy with the procedure done in Sec.~\ref{sec:new_RR_motivation}.
This process results in Eq.~\eqref{eq:h22_N_generic_rrdprd}.
We note, however, that a degree of arbitrariness remains in this construction.
For example, the $ p_\phi $ variable in the complex term of Eq.~\eqref{eq:hat_h22_N_generic_rprL} has been replaced by $ \mu r^2 \s \Omega  $ instead of $ \mu M \sqrt{r/M + r^3 \s \dot p_{r_*, \s\text{cons}} /(\mu M^2)} $;
both options are equally valid, provided that the eccentricity corrections $ h_{\ell m}^\text{ecc} $ (discussed below) are consistent with the chosen leading-order term.
An analogous procedure (with arbitrary choices) is done for the higher-order modes \eqref{eq:h21_N_generic_rrdprd}--\eqref{eq:h43_N_generic_rrdprd}.
We defer a systematic assessment of the impact of these choices on mode accuracy to future work.

The second term in Eq.~\eqref{eq:model_fact}, $ h_{\ell m}^\text{ecc} $, represents the eccentricity corrections to the modes, provided in Eqs.~\eqref{eq:modes_ecc_corr}.
These corrections are derived by requiring a consistency between the PN expansion of Eq.~\eqref{eq:model_fact} and the EOB modes from Ref.~\cite{Gamboa:2024imd}.
The parametrization $ \left(r, \dot r, \dot p_{r_*, \s\text{cons}} \right) $ has the same advantages discussed in Sec.~\ref{sec:proposal_RR}, namely, applicability to generic planar orbits and accurate recovery of the QC-orbit limit, in which $ h_{\ell m}^\text{ecc} \to 1 $.
Furthermore, the corrections $ h_{\ell m}^\text{ecc} $ are always well behaved, eliminating the need for ad hoc treatments, such as the sigmoid functions employed in the \texttt{SEOBNRv4EHM} \cite{Ramos-Buades:2021adz} and \texttt{TEOBResumS-Dal\'i} models \cite{Nagar:2021gss}.

The factors $ h_{\ell m}^\text{ecc} $ are employed in a PN-expanded form up to 1PN order (relative to the leading-order of the $ (\ell, m) $ mode).
During the development of \texttt{SEOBNRv6EHM}, we tested higher PN corrections.
However, we found that their effect negatively overlaps with that of the NQC corrections $ N _{ \ell m} $, since the amplitudes of $ h_{\ell m}^\text{ecc} $ grow considerably during the plunge.
Such issue is not observed in \texttt{SEOBNRv5EHM}~\cite{Gamboa:2024hli}, since it employs a different prescription for $ h_{\ell m}^\text{ecc} $;
this enables the incorporation of higher PN orders, but restricts the model applicability to bound orbits and introduces nonphysical features in highly eccentric waveforms (see Fig.~\ref{fig:sxs_2527}).
This motivates adopting the new approach at 1PN order.
We emphasize, however, that the factor $ h_{\ell m}^\text{F,\,hyb}(\Omega) $ contains high-PN-order QC information, which induces additional eccentric effects through the instantaneous frequency $ \Omega $.
This may provide a sufficiently accurate approximation to the full expressions for generic planar orbits.
Future work could explore alternative resummations or parametrizations that enable the incorporation of higher PN corrections without interfering with the NQC corrections.

For scattering orbits, Eq.~\eqref{eq:model_fact} determines the full multipolar emission modeled in \texttt{SEOBNRv6EHM}.
No phenomenological corrections are included, due both to the lack of a suitable framework for their implementation and to the limited availability of NR scattering waveforms.
Future work could address this as more NR data becomes available.

For coalescing binaries, we improve the late-time waveform and attach a merger-ringdown, as described next.


\subsection{NQC corrections and background QC dynamics}
\label{sec:nqcs_bqc}

The inspiral-plunge part of the modes for coalescing binaries is computed as
\begin{equation}
h_{\ell m}^{\text {insp-plunge}} =  N_{\ell m} \, h_{\ell m}^{\text {F}},
\end{equation}
where $ N_{\ell m} $ are phenomenological NQC corrections which ensure that the amplitude and frequency of the modes agree with input values coming from fits of QC NR simulations.

\texttt{SEOBNRv6EHM} adopts the same approach as \texttt{SEOBNRv5EHM} for the computation of the NQC corrections using a \emph{background QC dynamics}.
Specifically, it employs the same functional form for the NQC corrections as in Ref.~\cite{Pompiliv5}, but evaluates them on the dynamics of a QC binary with the same values of masses, spins, and starting orbit-averaged frequency as the eccentric system.
To reduce computational cost, this QC dynamics is obtained from a post-adiabatic evolution of the equations of motion \eqref{eq:EOM} \cite{Nagar:2018gnk,Rettegno:2019tzh,Mihaylov:2021bpf,Akcay:2018yyh,Nagar:2018plt}.
For this evolution, we employ Eqs.~(19) and (20) from Ref.~\cite{Pompiliv5}, modifying only the last term in their Eq.~(20) to be used with a generic radial RR force component, $ \mathcal F_r $, to match our prescription \eqref{eq:overview_RR_force_radial}.

Thus, the \texttt{SEOBNRv6EHM} NQC corrections take the form
\begin{equation}
\label{eq:nqcs}
\begin{aligned}
N_{\ell m} &=\left[1+\frac{p_{r_*,\,\text{qc}}^2}{(\mu r_{\text{qc}}\s \Omega_{\text{qc}})^2}\left(a_1^{h_{\ell m}}
+\frac{M a_2^{h_{\ell m}}}{r_{\text{qc}}}
+\frac{M^{3/2} a_3^{h_{\ell m}}}{r^{3 / 2}_{\text{qc}}}\right)\right] \\
&\quad \times \exp \left[i\left(b_1^{h_{\ell m}} \frac{p_{r_*,\,\text{qc}}}{\mu r_{\text{qc}}\s \Omega_{\text{qc}}}
+b_2^{h_{\ell m}} \frac{p_{r_*,\,\text{qc}}^3}{\mu^3 r_{\text{qc}}\s \Omega_{\text{qc}}}\right)\right],
\end{aligned}
\end{equation}
where $ (r_{\text{qc}},  \Omega_{\text{qc}}, p_{r_*,\,\text{qc}}) $ come from the background QC evolution, and $(a_1^{h_{\ell m}}\!$, $a_2^{h_{\ell m}}\!$, $a_3^{h_{\ell m}}\!$, $b_1^{h_{\ell m}}\!$, $b_2^{h_{\ell m}})$ are coefficients determined by requiring that the amplitude, its first and second derivatives, as well as the mode frequency and its first derivative, agree for all the $ (\ell, m) $ modes~\eqref{eq:supported_modes} with fits of NR input values at the matching time $ t _{ \text{match}} $ (discussed below) \cite{Taracchini:2013rva,Bohe:2016gbl,Cotesta:2018fcv}.
The fits are given in Appendix C of Ref.~\cite{Pompiliv5}.

The NQC corrections \eqref{eq:nqcs} become relevant only during the late inspiral and plunge.
In particular, they satisfy $ N_{\ell m} \to 1$ in the early inspiral since $ |p_{r_*,\,\text{qc}}/\mu | \ll \nolinebreak  r_{\text{qc}}\s \Omega_{\text{qc}} $ during a QC evolution.
This would not be the case if Eq.~\eqref{eq:nqcs} were evaluated on an eccentric dynamics, since the quantity $ p_{r_*} / ( \mu r \s \Omega) $ oscillates throughout the inspiral.
This problem was addressed in the \texttt{SEOBNRv4EHM} model by using orbit-averaged dynamical variables $ ( \bar r, \,\bar p_{r_*}, \,\bar \Omega ) $ \cite{Ramos-Buades:2021adz}, and in the \texttt{TEOBResumS-Dal\'i} model by using a sigmoid function that tapers the effects of the NQC corrections in the early inspiral \cite{Nagar:2021gss}.

We find the use of a background QC dynamics more convenient since it is independent of the eccentric dynamics (e.g., the time to merger, or the eccentricity in the late inspiral), and it helps in the determination of the merger-ringdown matching time, as discussed in Sec.~\ref{sec:matching_time}.
Furthermore, the time to merger of the QC dynamics, $ t _{ \text{merger}} ^{ \text{qc}} $, can be used as a threshold to prevent the generation of non-physical waveforms.
At fixed initial orbit-averaged frequency, eccentric systems are expected to have a shorter time to merger, $ t _{ \text{merger}} ^{ \text{ecc}} $, due to the enhanced loss of energy and angular momentum through GW emission.
Therefore, an eccentric system with $ t _{ \text{merger}} ^{ \text{ecc}} > t _{ \text{merger}} ^{ \text{qc}} $ would indicate inaccuracies in the initial conditions of the dynamics or in its subsequent evolution.
Thus, in \texttt{SEOBNRv6EHM}, we restrict waveform generation to the systems that satisfy $ t _{ \text{merger}} ^{ \text{ecc}} \leq 1.05 \, t _{ \text{merger}} ^{ \text{qc}} $, where we added an empirically determined factor to account for different stopping conditions between the eccentric and QC dynamics.

As in \texttt{SEOBNRv5EHM}, the eccentric and QC dynamics are aligned at a common value of the binary's relative separation, $ r _{ \text{ref}} $.
The final dynamics in both systems is very similar thanks to the binary circularization \cite{Peters:1963ux,Peters:1964zz,Hinder:2007qu,Sperhake:2007gu}.
Therefore, we assume that there is a time $ t _{ *} $ such that, for all subsequent times $ t \geq t _{ *}  $, we have $ r \approx \nolinebreak r _{ \text{qc}} $, where $ r $ is the separation of the eccentric binary and $ r _{ \text{qc}} $ is the separation of the QC binary.
Thus, we align both dynamics at the times $ t _{ \text{ref,\,ecc}} $ and $ t _{ \text{ref,\,qc}} $ at which
\begin{subequations}
\label{eq:t_r_ref_background_qc}
\begin{align}
r  (t _{ \text{ref,\,ecc}}) &= r _{ \text{qc}} (t _{ \text{ref,\,qc}}) = r _{ \text{ref}},
\\
 r _{ \text{ref}} &\equiv \max \left( r_\text{final}, \, r _{ \text{final,\,qc}}\right),
 \label{eq:r_ref}
 \end{align}
\end{subequations}
where $  r _{ \text{final}} $ and $ r _{ \text{final,\,qc}} $ are the final values of the separation in the eccentric and QC dynamics, respectively, both of which are assumed to occur after $ t _{ *} $.
These approximations hold even for large eccentricities, as demonstrated by the comparisons against eccentric NR waveforms shown in Sec.~\ref{sec:comparison_bound}.
However, the assumptions may break down for extreme configurations.
Future work could focus on developing other methods to determine the NQC corrections without resorting to orbit averages, sigmoid functions, or additional binary evolutions.


\subsection{Matching time}
\label{sec:matching_time}

For coalescing binaries, the time $ t _{ \text{match}} $ at which the NQC coefficients $ \big\{ a_i^{h_{\ell m}}\!, b_j^{h_{\ell m}} \big\}$ are calculated, and at which the inspiral-plunge modes $ h_{\ell m}^{\text {insp-plunge }} $ are matched to the corresponding merger-ringdown $ h_{\ell m}^{\text {merger-RD }} \!\!$, is given by
\begin{subequations}
\label{eq:t_attach_ecc}
\begin{align}
t _{ \text{match}} &= t _{ \text{ref,\,ecc}} + \Delta t _{ \text{match,\,qc}},
\\
\Delta t _{ \text{match,\,qc}} &= t ^{22,\,\text{peak}} _{ \text{qc}} - t _{ \text{ref,\,qc}},
\end{align}
\end{subequations}
where the reference times $ t _{ \text{ref,\,qc}} $ and $ t _{ \text{ref,\,ecc}} $ are determined from Eqs.~\eqref{eq:t_r_ref_background_qc}, and $ t ^{22,\,\text{peak}} _{ \text{qc}} $ is the time at which the $ (2,2) $ mode peaks in QC binaries, and is calculated from \cite{Pompiliv5}
\begin{equation}
\label{eq:t_peak_QC}
t ^{22,\,\text{peak}} _{ \text{qc}}
= t_\text{ISCO} + \Delta t_\text{ISCO} ^{ 22},
\end{equation}
with $\Delta t_\text{ISCO}^{ 22}$ a parameter calibrated with QC NR waveforms, given in Eq.~\eqref{eq:fit_dt}, and $ t_\text{ISCO}$ being the time at which the separation of the background QC dynamics is equal to the Kerr geodesic innermost stable circular orbit (ISCO) radius \cite{Bardeen:1972fi},
\begin{equation}
\label{eq:r_rISCO}
r _{ \text{qc}}(t_\text{ISCO}) = r_\text{ISCO}.
\end{equation}
The value of $r_{\text{ISCO}}$ is calculated using the final mass and spin of the remnant BH, which are determined with fits obtained from QC NR simulations \cite{Jimenez-Forteza:2016oae, Hofmann:2016yih}.
We remark that the value of $ t_\text{ISCO} $ calculated from Eq.~\eqref{eq:r_rISCO} is independent of features in the late dynamics, and is uniquely defined since the separation decreases monotonically in a QC binary.

This method for determining the matching time $ t _{ \text{match}} $ is the same as in \texttt{SEOBNRv5EHM}.
For eccentric binaries, one cannot employ directly Eq.~\eqref{eq:t_peak_QC} to determine $ t _{ \text{match}} $ since 1) $ \Delta t_\text{ISCO} ^{ 22} $ is calibrated to QC NR waveforms, and 2) eccentricity oscillations could result in multiple roots of the equation $ r (t) = r_\text{ISCO}$, which would make the model unstable under perturbations of the input parameters.
Therefore, given the absence of a calibration to eccentric NR waveforms, the prescription in Eqs.~\eqref{eq:t_attach_ecc} for the matching time is a good strategy that leverages the background QC dynamics (used in the computation of the NQC corrections) to inherit the calibration done with QC NR waveforms.
Future investigations could focus on finding alternative methods to compute the matching time, potentially introducing a calibration to eccentric NR simulations, or by having a reference point in the late dynamics of generic coalescing systems to which one could add a calibration parameter, as in the \texttt{TEOBResumS-Dal\'i} model (e.g., see Ref.~\cite{Nagar:2024oyk}).


\subsection{Merger-ringdown waveform modes}
\label{sec:mrd}
 
We employ a QC prescription for the merger-ringdown modes of coalescing binaries.
We use the same phenomenological ansatz for the merger-ringdown as in \texttt{SEOBNRv5HM}, informed by QC NR simulations and QC test-mass limit waveforms \cite{Pompiliv5}.
The ansatz physically represents a Kerr BH that vibrates 
down to a state of equilibrium, emitting damped oscillations known as \emph{quasinormal modes} (QNMs).
Specifically, it takes the form \cite{Damour:2014yha,Bohe:2016gbl, Cotesta:2018fcv, Pompiliv5}
\begin{equation}
h_{\ell m}^{\text {merger-RD }}(\bm{\Pi}; t) =
\nu \, \tilde{A}_{\ell m}(t) \, \e^{i \tilde{\phi}_{\ell m}(t)} \, \e^{-i \sigma_{\ell m 0}\left(t-t_{\text {match}}\right)},
\label{eq:ansatz_hlm}
\end{equation}
where $ \boldsymbol{\Pi} = \nolinebreak \{q, \s \chi_1,\s \chi_2\} $ and $\sigma_{\ell m 0}$ is the complex frequency of the least-damped QNM of the remnant BH (obtained for each ($ \ell $, $ m $) mode as a function of the BH's final mass and spin, using the \texttt{qnm} Python package of Ref.~\cite{Stein:2019mop}), and where the functions $\tilde{A}_{\ell m}$ and $\tilde{\phi}_{\ell m}$ are given by \cite{Bohe:2016gbl, Cotesta:2018fcv, Pompiliv5}
\begin{subequations}
\label{eq:MRD_A_phi}
\begin{align}
\tilde{A}_{\ell m}(t) &=c_{1, c}^{\ell m} \tanh \left[c_{1, f}^{\ell m}\left(t-t_{\text {match}}\right)+c_{2, f}^{\ell m}\right]+c_{2, c}^{\ell m},
\label{eq:ansatz_a}
\\
\tilde{\phi}_{\ell m}(t) &=
\phi_{\text {match }}^{\ell m}-d_{1, c}^{\ell m} \log \left[\frac{1+d_{2, f}^{\ell m} \e^{-d_{1, f}^{\ell  m}\left(t-t_{\text {match}}\right)}}{1+d_{2, f}^{\ell m}}\right],
\label{eq:ansatz_phi}
\end{align}
\end{subequations}
with $\phi_{\text {match }}^{\ell m}$ being the phase of the inspiral-plunge $(\ell, m)$ mode at $t_{\text {match}}$, the coefficients $ c_{i, c}^{\ell m} $ ($ i \in \{ 1, 2\} $) and $ d_{1, c}^{\ell m} $ ensure that the modes are continuously differentiable at $ t _{ \text{match}} $, and the values of $ c_{i, f}^{\ell m} $ and $ d_{i, f}^{\ell m} $ ($ i \in \{ 1, 2\} $) are obtained from fits in $ \nu $ and $ \chi $ (the BH's final spin) informed by NR and test-mass limit waveforms.
The fits are given in Appendix D of Ref.~\cite{Pompiliv5}.

Following \texttt{SEOBNRv5HM}, another effect modeled in the merger-ringdown of \texttt{SEOBNRv6EHM} is the \emph{mode mixing} between the $(3, 2)$ and $(4, 3)$ modes.
Due to a mismatch between the \textit{spherical} harmonic basis and the \textit{spheroidal} harmonics basis employed in the computation of QNMs in BH perturbation theory~\cite{Buonanno:2006ui, Kelly:2012nd}, the $(3, 2)$ and $(4, 3)$ modes get mixed with other modes, and this leads to postmerger oscillations.
In contrast, the modes $ (\ell, m) = \{(2, 2), \s (3, 3), \s (2, 1), \s (4, 4) \} $ of comparable-mass binaries show a predominant monotonic amplitude and frequency evolution.

More investigation is needed to understand the biases introduced by applying a QC prescription to the merger-ringdown of eccentric binaries.
Even at relatively high eccentricities, deviations from a QC merger-ringdown remain at most at the percent level \cite{Nee:2025zdy}.
However, one naturally expects this to be a source of error for highly eccentric binaries \cite{Rao:2026lmz}, which motivates future research directions.
Additionally, it is known that QNMs can be excited by scattering encounters or highly eccentric periastron passages (e.g., see Refs.~\cite{Thornburg:2019ukt,Bae:2023sww}).
We leave the inclusion of these effects for future work.


\section{Calibration to quasi-circular waveforms}
\label{sec:calibration}

Model accuracy can be improved through calibration to waveforms generated by NR or BH perturbation theory.
This process extends the validity of models beyond the range of their underlying approximations (e.g., low-velocity or small-mass-ratio assumptions).
For EOB models, it is possible to calibrate both the binary dynamics and the corresponding waveform.
Identifying gauge-invariant quantities for the calibration process is straightforward in the QC-orbit limit, and has been done, in particular, in previous \texttt{SEOBNR} models.
In the following subsections, we describe the calibrations to QC NR waveforms performed in \texttt{SEOBNRv6EHM}.


\subsection{Calibration summary}
\label{sec:calibration_summary}

We calibrate \texttt{SEOBNRv6EHM} to 425 QC \texttt{SXS} NR waveforms following a nested-sampling approach.
Namely, we determine posterior distributions for specific calibration parameters, which are then fitted with multidimensional functions.
We employ \emph{ten} calibration parameters:\footnote{
We count the calibration parameters based on the specific place in which they enter, and not on the number of coefficients employed in their fits.
}
seven in certain $ \rho _{ \ell m} $ functions entering the RR force, two in the Hamiltonian, and one controlling the matching time of the merger-ringdown.
Specifically, these parameters are
\begin{itemize}
\item $ \Delta \rho_{\ell m}^{(1)} (\nu) $ \s, \\
representing linear-in-$ \nu $ polynomials entering the $ \rho_{\ell m} $ amplitudes in Eq.~\eqref{eq:frholm} for the seven modes $ (\ell, m) \in \nolinebreak \{ (2,2), \s (2,1), (3,3), \s(3,2), \s(4,4), \s(4,3), \s(5,5)\} $.
They enter the RR force [see Eqs.~\eqref{eq:overview_RR_force} and \eqref{eq:overview_h_F_qc}], but are not included in the output \texttt{SEOBNRv6EHM} modes.
These parameters are approximate fits of numerical second-order GSF (2GSF) flux data, except for $\Delta \rho_{22}^{(1)}$, which is obtained by calibrating to an equal-mass, nonspinning QC NR waveform.
The fits are given in Eqs.~\eqref{eq:fits_rho}.
\item $ a_6 (\nu) $\s,  \\
a 5PN parameter entering the nonspinning part of the effective Hamiltonian (it enters the $A_\text{noS}$ potential shown in Eq.~(9) of Ref.~\cite{Pompiliv5}), calibrated to QC NR waveforms.
The fit is given in Eq.~\eqref{eq:fit_a6}.
\item $ \hat d_{\text{SO}} (\nu, \s a_+, \s a _-) $\s, \\
a 4.5PN parameter entering the odd-in-spin part of the effective Hamiltonian [see Eq.~\eqref{eq:so_calib}], calibrated to QC NR waveforms.
The fit is given in Eq.~\eqref{eq:fit_hatdSO}.
\item $ \Delta t^{22}_{\text{ISCO}} = \Delta t^{22}_{\text{ISCO,\,noS}}(\nu) + \Delta t^{22}_{\text{ISCO,\,S}}(\nu, \s a_+, \s a _-) $\s, \\
a parameter that quantifies the time difference between the peak of the $(2,2)$-mode amplitude and the Kerr ISCO prediction for QC systems [see Eq.~\eqref{eq:t_peak_QC}], and that determines the merger-ringdown attachment through Eqs.~(\ref{eq:t_attach_ecc}).
The parameter $\Delta t^{22}_{\text{ISCO}}$ is split into a nonspinning part $ \Delta t^{22}_{\text{ISCO,\,noS}} $ and a spinning contribution $ \Delta t^{22}_{\text{ISCO,\,S}} $ which vanishes in the nonspinning limit.
The corresponding fits are given in Eqs.~\eqref{eq:fit_dt}.
\end{itemize}
The main differences with respect to the \texttt{SEOBNRv5} models are:
1) we employ a different calibration parameter $\big( \hat d _{ \text{SO}} \big)$ for the spinning part of the effective Hamiltonian, since it has a better behavior for \texttt{SEOBNRv6EHM} across the parameter space;
2) we use quartic polynomial fits for $ \hat d _{ \text{SO}} $ and for $ \Delta t^{22}_{\text{ISCO,\,S}} $, instead of cubic polynomial fits, since this improves the overall performance of the model;
and 3) we adopt a hierarchical fitting strategy for spinning systems, first constructing fits for equal-mass configurations and then extending them to unequal masses, resulting in improved accuracy.


\subsection{Calibration framework}
\label{sec:calibration_framework}

We employ the nested-sampling methods presented in Refs.~\cite{Pompiliv5,Bohe:2016gbl}.
Here, we specify the metrics, settings, and the data set employed for the calibration.


\subsubsection{Metrics}
\label{sec:calibration_metrics}

To quantify the similarity between two waveforms $h_1$ and $h_2$, we use the \emph{overlap}, defined as a noise-weighted inner product within the frequency band $ [f_{\text{min}}, f_{\text{max}}]$ as \cite{Sathyaprakash:1991mt,Finn:1992xs}
\begin{equation}
\label{eq:overlap}
\langle h_1 \mid h_2 \rangle \equiv 4 \,\Re \int^{f_{\text{max}}}_{f_{\text{min}}} \frac{\tilde{h}_1(f) \,\tilde{h}^*_2(f)}{S_\text{n}(f)} \, \text{d}f,
\end{equation}
where the tilde $ \, \tilde{} \, $ denotes the Fourier transform, the star $ ^{*} $ denotes the complex conjugate, and $S_\text{n}$ is the one-sided power-spectral density (PSD) of the detector's noise.
Additionally, we define the normalized waveform $\hat h$ as
\begin{equation}
\label{eq:normalization}
\hat h \equiv \frac{ h}{\sqrt{ \langle h \mid h \rangle }}.
\end{equation}

In this work, we use the \texttt{A+} PSD of LIGO detectors \cite{sensitivity_curves} (expected design sensitivity for the fifth observing run O5), we set a sampling rate such that the waveforms have a time spacing of $ 0.1 \s M$ for the lowest total mass considered, we taper the time-domain waveforms using a Planck window \cite{McKechan:2010kp} before transforming them into the frequency domain, and we set $f_{\text{min}}=10\,$Hz and $f_{\text{max}}=2048\,$Hz.
For short QC NR waveforms whose frequency content is above $ 10\, $Hz, we instead set $f_{\text{min}} = 1.35 \, f_{\text{start}}$, where $f_{\text{start}}$ is defined as the peak of the NR waveform in the frequency domain, and the factor of $1.35$ mitigates Fourier-transform artifacts that would otherwise contaminate the overlap.

Given a set of intrinsic parameters for a QC binary $ \boldsymbol{\Pi} ^{ \text{qc}}_{ 0} = \nolinebreak \{q, \s \chi_1,\s \chi_2\} $ defined at a time $ t_0 $, and a set of calibration parameters $ \boldsymbol{\theta} $, we define the \emph{calibration mismatch} of an EOB waveform $h_{\mathrm{EOB}}$ to a NR waveform $h_{\mathrm{NR}}$ as
\begin{equation}
\label{eq:metric_mm}
\mathcal{M}(M, \boldsymbol{\theta}) =
1
-  \max_{t_\text{c},\, \varphi}
\left[
\left.
\Big\langle \hat{h}_{\text{NR}} \bigm | \hat{h}_{\text{EOB}}(t_\text{c}, \varphi; \boldsymbol{\theta}) \Big\rangle \,
\right \vert_{
\substack{
\vspace{-0.3cm} \\
\boldsymbol{\Pi} _{\text{NR}, 0} ^{ \text{qc}} = \boldsymbol{\Pi} _{\text{EOB}, 0} ^{ \text{qc}}
}}
\right],
\end{equation}
where the total mass $ M $ dependence appears since it determines the physical frequency of the waveform.

Additionally, we define the \emph{difference in merger time} as
\begin{equation}
\label{eq:metric_deltat_merger}
\delta t_{\text{merger}} = \left | \, t^{22,\,\text{peak}} _{ \text{NR}} - t^{22,\,\text{peak}} _{ \text{EOB}} \, \right | ,
\end{equation}
where $ t^{22,\,\text{peak}} _{ \text{NR}} $ and $ t^{22,\,\text{peak}} _{ \text{EOB}} $ denote the times of the $(2,2)$-mode amplitude peaks of the NR and EOB waveforms, respectively, after a low-frequency phase alignment.

Our calibration procedures make use only of the $(2,2)$ mode of QC waveforms and change depending on the intrinsic parameters $ \boldsymbol{\Pi} ^{ \text{qc}}_{ 0} $ and the calibration parameters $ \boldsymbol{\theta}$.
However, all our calibration procedures use the metrics \eqref{eq:metric_mm} and \eqref{eq:metric_deltat_merger} to compare against NR waveforms.\footnote{
We note that the calibration is done using a specific PSD (in particular, we use the \texttt{A+} PSD of LIGO detectors).
Future work could benefit from using a flat PSD to remove the dependence of the calibration on specific detectors.
In this direction, one could also use \emph{time-domain} metrics to avoid the introduction of artifacts from the Fourier transforms.
This will be of particular relevance when calibrating to eccentric waveforms.
}


\subsubsection{Nested-sampling analysis}
\label{sec:calibration_nested_sampling}

Following \texttt{SEOBNRv4} \cite{Bohe:2016gbl} and \texttt{SEOBNRv5} \cite{Pompiliv5}, we adopt a calibration strategy based on stochastic sampling to obtain a posterior distribution for the calibration parameters for each NR simulation.
This approach provides an efficient and flexible calibration procedure, well suited to the dimensionality of the problem and the large number of NR simulations available.
Stochastic sampling methods allow for an exploration of high-dimensional parameter spaces, and have the advantage of providing information regarding the correlations between calibration parameters.
As in \texttt{SEOBNRv5}, we obtain the best computational performance by employing nested sampling~\cite{Skilling:2006gxv},
using the sampler \texttt{nessai}~\cite{Williams:2021qyt} through \texttt{Bilby}~\cite{Ashton:2018jfp}.

We employ a likelihood function designed to minimize the mismatch between EOB and NR waveforms while simultaneously improving the prediction of the merger time (the mismatch alone is not particularly sensitive to this quantity).
Thus, we define the likelihood function as:
\begin{equation}
\label{eq:calib_likelihood}
\mathcal L (h_\text{NR} \mid \boldsymbol{\theta}) \propto \exp \left[-\frac{1}{2}\left(\frac{{\mathcal{M}}_{\max_ M}(\boldsymbol{\theta})}{\sigma_{\mathcal{M}}}\right)^2-\frac{1}{2}\left(\frac{\delta t_{\text {merger}}(\boldsymbol{\theta})}{\sigma_t}\right)^2\right],
\end{equation}
where ${\mathcal{M}}_{\max_ M}(\boldsymbol{\theta})$ is the maximum mismatch between EOB and NR waveforms over the range $10 \, \solarmass \leq \nolinebreak M \leq \nolinebreak 300 \, \solarmass$, and $ \sigma_{\mathcal{M}} $ and $ \sigma_t $ are tolerance parameters that control the effective width of the likelihood.
For this work, we set a mismatch goal of $ \sigma_{\mathcal{M}} = 10^{-4} $ for nonspinning configurations, $ \sigma_{\mathcal{M}} = 5 \times 10^{-4} $ for spinning configurations, and $ \sigma_t = 5 M $ for all cases.
In \texttt{SEOBNRv5HM}, $ \sigma_{\mathcal{M}} $ was set to $ 10^{-3} $ for all cases \cite{Pompiliv5};
here, we used smaller values of $\sigma_{\mathcal{M}}$ to facilitate the identification of the correct modes in multimodal posteriors (see next paragraph).
Achieving a mismatch goal of $10^{-4}$ for \emph{all} binary configurations would require revisiting the calibration procedure and modifying the analytic structure of the model, since not even the maximum-likelihood points for all calibration data lie below this threshold.
Our objective here is to get a model with accuracy comparable to that of \texttt{SEOBNRv5HM}.

For each NR simulation in the calibration data set, we obtain a \emph{posterior distribution}, which is employed to construct fits for the calibration parameters as functions of the binary parameters, $\boldsymbol{\theta}(\boldsymbol{\Pi})$.
The mean and variance of these posteriors, as well as the correlations among the calibration parameters, are related to the requirements on $ \sigma_{\mathcal{M}} $ and $ \sigma_t $.
In some cases, these correlations lead to secondary modes in the posteriors (see Fig.~3 of Ref.~\cite{Pompiliv5});
for such cases, we employ continuity considerations to select only one mode to get a more regular fit across the parameter space.
For all posteriors, we discard samples that do not satisfy the calibration requirements on $ \sigma_{\mathcal{M}} $ and $ \sigma_t $.
If this would discard more than $50 \%$ of the points, we instead keep half of the original samples of the selected mode with the best likelihood values.


\subsubsection{Calibration data set}
\label{sec:calibration_data}

The calibration of \texttt{SEOBNRv6EHM} employs 425 out of the 592 QC, aligned-spin NR waveforms from the third release of the \texttt{SXS} catalog 
\cite{SXS:catalog,Boyle:2019kee,Chu:2015kft,Blackman:2017dfb,Hemberger:2013hsa,Scheel:2014ina,Lovelace:2014twa,Abbott:2016apu,Blackman:2015pia,Lovelace:2016uwp,Varma:2018mmi,Abbott:2016nmj,Varma:2019csw,Kumar:2015tha,Mroue:2013xna,Yoo:2022erv,Scheel:2025jct}, all produced with the Spectral Einstein code (\texttt{SpEC})~\cite{SpECwebsite}.
The calibration dataset comprises 21 nonspinning and 404 spinning cases, of which 78 are equal-mass spinning configurations and 326 are unequal-mass spinning configurations.
Only two out of these 425 simulations were not employed in the calibration of \texttt{SEOBNRv5HM}.
Namely, \texttt{SXS:BBH:4121} with $ q = 4$, $ \chi_1 =  0.9$, $ \chi_2 =  0.5 $, and \texttt{SXS:BBH:4123} with $ q = 4$, $ \chi_1 = 0.9 $, $ \chi_2 = -0.5 $.
These lie at parameter-space corners, enabling better extrapolation of the calibration-parameter fits.

\begin{figure}
\hspace{-5pt}
\includegraphics[width=\linewidth]{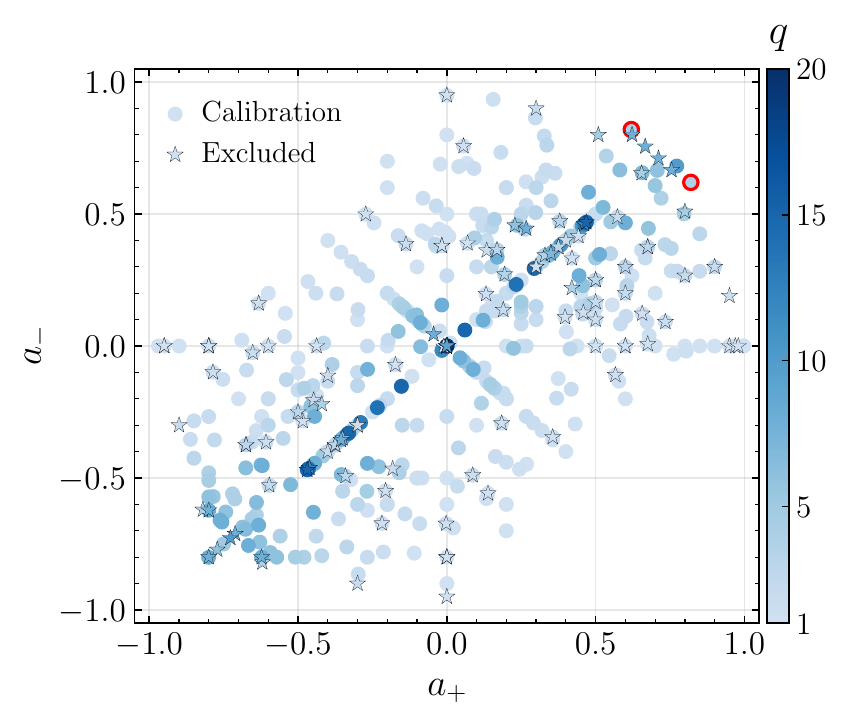}
\vspace{-5pt}
\caption{
Parameter-space distribution of the 592 QC \texttt{SXS} NR simulations employed throughout this work.
The 425 simulations used in the calibration of \texttt{SEOBNRv6EHM} are shown with round markers, while those excluded from the calibration are shown with star markers.
Two additional simulations employed in the calibration of \texttt{SEOBNRv6EHM} (not used in \texttt{SEOBNRv5HM}) are highlighted in red.
}
\label{fig:param_space_QC}
\end{figure}

The parameter-space distribution of the 592 QC \texttt{SXS} NR waveforms employed throughout this work is shown in Fig.~\ref{fig:param_space_QC}, as a function of the spin variables $ a_+ $ and $ a_- $, and the mass ratio $ q $.
This includes the 425 waveforms used for the calibration of \texttt{SEOBNRv6EHM} (round markers), and the remaining 167 waveforms (star markers).
We highlight in red the simulations \texttt{SXS:BBH:4121} and \texttt{SXS:BBH:4123} (top right corner).

\texttt{SEOBNRv6EHM} benefits from the merger-ringdown and input-value fits of \texttt{SEOBNRv5HM} \cite{Pompiliv5}.
These were constructed with additional 18 nonspinning \texttt{SXS} NR waveforms, one \texttt{Einstein} \texttt{Toolkit} waveform \cite{EinsteinToolkit:2022_11, Cotesta:2018fcv}, and 13 test-mass-limit waveforms computed using the Teukolsky equation~\cite{Barausse:2011kb,Taracchini:2014zpa}.
These add up to the 442 NR + 13 test-mass-limit waveforms quoted in Ref.~\cite{Pompiliv5}.
The number 425 for \texttt{SEOBNRv6EHM} only counts the simulations actually employed in the new calibration of the parameters $ \{a_6, \hat d _{ \text{SO}}, \Delta t _{ \text{ISCO}} ^{ 22} \} $.

Finally, we note that these 425 QC simulations form a sufficient calibration set, yielding an accuracy comparable to that of \texttt{SEOBNRv5HM} (see Sec.~\ref{sec:comparison_qc}).
We tested extending the calibration set to all 592 available QC \texttt{SXS} waveforms, but found no significant gain in accuracy (median mismatches and the number of outliers remain comparable).
This indicates that the QC parameter space is already well covered by the 425 simulations, at least within our calibration procedure.

%


\subsection{RR force calibration}
\label{sec:calibration_RR}

The first part in our calibration procedure consists of determining suitable coefficients at high PN orders for the $ \rho _{ \ell m} $ functions entering the RR force \eqref{eq:overview_RR_force} via Eqs.~\eqref{eq:overview_h_F_qc} and \eqref{eq:frholm}.

Following Ref.~\cite{VandeMeentv5}, we split the $ \rho _{ \ell m} $ functions based on their dependence on the symmetric mass ratio $ \nu $, as
\begin{align}
\rho_{\ell m}
&= \rho_{\ell m}^{(0)}+\nu \rho_{\ell m}^{(1)}+\mathcal{O}\left(\nu^2\right),
\nonumber
\\
&= \rho_{\ell m}^{(0)} + \nu \left( \rho_{\ell m}^{(1),\,\text{EOB}} + \Delta\rho_{\ell m}^{(1)} \right) +\mathcal{O}\left(\nu^2\right).
\label{eq:rholm_full}
\end{align}
The functions $ \rho_{\ell m}^{(0)} $ and $\rho_{\ell m}^{(1),\,\text{EOB}}$ only contain terms derived from PN calculations up to a certain PN order (these terms can be read from Eq.~(46) and Appendix B of Ref.~\cite{Pompiliv5}), and the functions $\Delta\rho_{\ell m}^{(1)}$ are employed as calibration parameters with a dependence on the instantaneous velocity
\begin{equation}
v_\Omega = (M \Omega)^{1/3}.
\end{equation}
In \texttt{SEOBNRv5HM} \cite{Pompiliv5}, the functions $\Delta\rho_{\ell m}^{(1)}$ were given as polynomials in $ v^2_\Omega $ {\large(}starting at the lowest order in $v^2_\Omega$ not already included in $\rho_{\ell m}^{(1),\,\text{EOB}}${\large)} whose coefficients were determined by fitting to the numerical, linear in $ \nu $, contributions to the $\rho_{\ell m}$ functions from 2GSF flux data \cite{VandeMeentv5,Warburton:2021kwk}.\footnote{
After the publication of the \texttt{SEOBNRv5} models, the authors of Ref.~\cite{Warburton:2021kwk} found an error in their numerical data, hence affecting the physical correctness of the $\Delta\rho_{\ell m}^{(1)}$ calibration done in Ref.~\cite{VandeMeentv5} for the \texttt{SEOBNRv5HM} model.
This was recently corrected in Ref.~\cite{Leather:2025nhu}, where it was pointed out that, apart from the error in Ref.~\cite{Warburton:2021kwk}, there was an inconsistency in the way the GSF calibration was propagated to the waveform modes in Ref.~\cite{VandeMeentv5}.
Namely, the correct GSF calibration should yield \emph{different} $\Delta\rho_{\ell m}^{(1)}$ contributions for the EOB RR force and for the waveform modes.
\label{fn:gsf_calibration}
}

In \texttt{SEOBNRv6EHM}, we employ the same $\Delta\rho_{\ell m}^{(1)}$ functions as derived in Ref.~\cite{Leather:2025nhu}, except for $\Delta\rho_{22}^{(1)}$, which we determine by fitting to an equal-mass, nonspinning, QC NR waveform.
Formally, the new RR force \eqref{eq:overview_RR_force} requires a different calibration to GSF data because the $ \Omega _{ \text{c}} $ function introduces a linear-in-$ \nu $ correction to the RR force [as seen in Eqs.~\eqref{eq:Omega_c}].
In particular, the functional form of Eq.~(53) for $\rho_{\ell m}^{(1),\s\text{GSF}}$ in Ref.~\cite{VandeMeentv5} would need to change.
For simplicity, we do not follow this GSF calibration route.
Instead, we calibrate the (dominant) contribution $\Delta\rho_{22}^{(1)}$ to an equal-mass, nonspinning, QC NR waveform, and we use the results for the rest of the functions $\Delta\rho_{\ell m}^{(1)}$ given in Ref.~\cite{Leather:2025nhu} as \emph{approximations} to the physically correct calibration.
Another option was to remove all the $\Delta\rho_{\ell m}^{(1)}$ functions;
however, we observed that having these functions (in particular, $\Delta\rho_{22}^{(1)}$) is highly important for the calibration to spinning, QC NR waveforms, as it gives a more regular structure to the calibration data, and improves the overall accuracy.

To determine the $\Delta\rho_{22}^{(1)}$ contribution, we decompose it as
\begin{equation}
\Delta\rho_{22}^{(1)} =
\Delta \rho_{22,\, v^8} \, v^8_\Omega
+ \Delta \rho_{22,\, v^{10}} \, v^{10}_\Omega,
\end{equation}
and perform a nested-sampling analysis over the parameters $ \boldsymbol{\theta} = \{ \Delta \rho_{22,\, v^8}, \,\Delta \rho_{22,\, v^{10}}, \, a_6, \,\Delta t^{22}_{\text{ISCO}}\} $, with uniform priors $ \Delta \rho_{22,\, v^8} \in [-20, 40] $, $ \Delta \rho_{22,\, v^{10}} \in [-200, 50] $, $a_6 \in [-100, 150]$, and $\Delta t^{22}_{\text{ISCO}} \in [-80, 20]$, and with the settings specified in Sec.~\ref{sec:calibration_framework}.
We include the parameters $ a_6 $ and $ \Delta t^{22}_{\text{ISCO}} $ since we do not know them a priori, and this gives the sampler the freedom to explore the parameter space (the resulting posteriors of $ a_6 $ and $ \Delta t^{22}_{\text{ISCO}} $ are broadly consistent with the actual values determined in Sec.~\ref{sec:calibration_nonspinning}).
We employ the equal-mass, nonspinning QC NR simulation \texttt{SXS:BBH:2325}.

Afterwards, we simply select the maximum-likelihood values, which correspond to:
\begin{subequations}
\label{eq:rho_maxL}
\begin{align}
\Delta \rho_{22,\, v^8\!, \,\text{max} \mathcal{L}}
&= 40.3,
\\
\Delta \rho_{22, \, v^{10}\!, \,\text{max} \mathcal{L}}
&= -473.
\end{align}
\end{subequations}
These values are similar (in order of magnitude and sign) to the ones employed in \texttt{SEOBNRv5HM} [namely, $ \Delta \rho_{22,\, v^8} = 21.2 $ and $ \Delta \rho_{22,\, v^{10}} = -411 $; see Eq.~(45a) of Ref.~\cite{Pompiliv5}], and to the ones determined in Ref.~\cite{Leather:2025nhu} [namely, $ \Delta \rho_{22,\, v^8} = 20.6 $ and $ \Delta \rho_{22,\, v^{10}} = -410 $; see Eq.~(56a)].

We observed that coefficients that significantly depart from those in Eqs.~\eqref{eq:rho_maxL}, lead to a fit for the nonspinning part of $ \Delta t^{22}_\text{ISCO} $ which is problematic for the spinning calibration (in particular, the medians of the calibration posteriors are not well fitted for small spins, as they lie significantly away from the predicted nonspinning value $ \Delta t^{22}_{\text{ISCO,\,noS}} $).
Thus, having coefficients similar to Eqs.~\eqref{eq:rho_maxL} leads to overall better spinning fits.
This explains why the nonspinning GSF calibration in \texttt{SEOBNRv5HM} \cite{VandeMeentv5} led to an improved accuracy for spinning systems compared to an uncalibrated model ($\Delta\rho_{\ell m}^{(1)} = 0$), despite using incorrect GSF data (see footnote \ref{fn:gsf_calibration}).

Collecting these results with the ones from Eqs.~(56b--g) of Ref.~\cite{Leather:2025nhu}, gives the values of $\Delta\rho_{\ell m}^{(1)}$, as used in \texttt{SEOBNRv6EHM}:
\begin{subequations}
\label{eq:fits_rho}
\begin{align}
\Delta\rho_{22}^{(1)} &=
\mathbf{40.3} \, v^8_\Omega
- \mathbf{473}\, v^{10}_\Omega,
\\
\Delta\rho_{21}^{(1)} &=
1.65 \,  v^6_\Omega
+\mathbf{24.3} \, v^8_\Omega
+80  \,  v^{10}_\Omega,
\\
\Delta\rho_{33}^{(1)} &=
12 		\, v^8_\Omega
-\mathbf{222} 	\, v^{10}_\Omega,
\\
\Delta\rho_{32}^{(1)} &=
\mathbf{-0.25}  \, v^6_\Omega
-6.5 \, v^8_\Omega
+98  \,  v^{10}_\Omega,
\\
\Delta\rho_{44}^{(1)} &=
-3.56 \,  v^6_\Omega
+\mathbf{15.3} \, v^8_\Omega
-216   \, v^{10}_\Omega,
\\
\Delta\rho_{43}^{(1)} &=
-0.654  \, v^4_\Omega
-\mathbf{3.73} \, v^6_\Omega
+\mathbf{18}   \, v^8_\Omega,
\\
\Delta\rho_{55}^{(1)} &=
\mathbf{-2.608}  \, v^4_\Omega
+\mathbf{1.13} \,  v^6_\Omega
-\mathbf{35.1} \,   v^8_\Omega,
\end{align}
\end{subequations}
where we highlight terms differing from \texttt{SEOBNRv5HM}.

We emphasize that these parameters enter only the RR force and not the output \texttt{SEOBNRv6EHM} modes.
We find that including them in the waveform modes leads to unphysical features in scattering waveforms, such as a small indentation near the peak of the $(2,2)$-mode amplitude.
Since a proper GSF calibration would in any case yield different coefficients (see footnote \ref{fn:gsf_calibration}), we omit them from the modes for simplicity.


\subsection{Calibration of non-spinning sector}
\label{sec:calibration_nonspinning}

With the polynomials $\Delta\rho_{\ell m}^{(1)}$ in Eqs.~\eqref{eq:fits_rho} fixed, we turn to the calibration of the $ a_6 $ parameter entering the nonspinning sector of the Hamiltonian, and the $ \Delta t^{22}_{\text{ISCO,\,noS}} $ parameter which determines the merger-ringdown attachment for nonspinning binaries.
We follow the procedure done in \texttt{SEOBNRv5HM} \cite{Pompiliv5}.

We perform a nested-sampling analysis over the parameters $ \boldsymbol{\theta} = \{ a_6, \,\Delta t^{22}_{\text{ISCO,\,noS}} \} $, with uniform priors
$a_6 \in [-500, 500]$ and $\Delta t^{22}_{\text{ISCO,\,noS}} \in [-100, 40]$.
We employ the calibration settings specified in Sec.~\ref{sec:calibration_framework}, for 21 nonspinning, QC \texttt{SXS} NR simulations \cite{Scheel:2025jct}, with mass ratios ranging from $ q=1 $ to $ q=20 $. 
This results in a set of calibration posteriors for $ a_6 $ and $ \Delta t^{22}_{\text{ISCO,\,noS}}$, which we parametrize in terms of the symmetric mass ratio $ \nu $, as in \texttt{SEOBNRv5HM}.

We complement the nonspinning NR data set with the following test-mass-limit information:
\begin{subequations}
\begin{align}
a_6 |_{\nu = 0.001}
&=
39.1161 ,
\\
\Delta t^{22}_{\text{ISCO,\,noS}} |_{\nu = 0.001}
&=
-225.4 .
\end{align}
\end{subequations}
The value of $a_6 | _{\nu = 0.001}$ is inferred by requiring that the ISCO shift predicted by the EOB Hamiltonian agrees with the 1GSF ISCO shift, and the value of $ \Delta t^{22}_{\text{ISCO,\,noS}} | _{\nu = 0.001} $ is estimated from a Teukolsky waveform \cite{Taracchini:2014zpa} (the details for these calculations can be found in Ref.~\cite{Pompiliv5}).

The simple structure of the nonspinning calibration data enables us to directly fit the maximum-likelihood values of the posteriors via least-squares regression (we do not use more information from the posteriors in constructing the nonspinning fits).
For $a_6$ we use a quartic polynomial ansatz in $\nu$, while for $ \Delta t^{22}_{\text{ISCO,\,noS}} $ we use an ansatz of the form \cite{Pompiliv5}
\begin{equation}
\Delta t^{22}_{\text{ISCO,\,noS}} = \left (\Delta_0 + \Delta_1 \s\nu + \Delta_2 \s\nu^2 + \Delta_3 \s\nu^3 \right)\,\nu^{-1/5 + \Delta_4\s \nu},
\end{equation}
where $ \Delta_i $ ($ i \in [0, 4] $) represents fitting coefficients, and the $\nu^{-1/5}$ factor ensures the expected test-mass scaling for $\big (t ^{22,\,\text{peak}} _{ \text{qc}} - t_{\text{ISCO}} \big)$~\cite{Buonanno:2000ef}, and provides a better extrapolation of the fit in the $\nu \rightarrow 0$ limit.
The maximum-likelihood values of the posteriors are fitted with the \texttt{optimize.curve\_fit} function from the \texttt{scipy} package \cite{2020SciPy-NMeth};
we assign the same uncertainty to all cases, except for the test-mass-limit data of $ a_6 $, for which we set $\texttt{sigma} = 0.3$ to give it greater weight.

The data for $ a_6 $ and $ \Delta t^{22}_{\text{ISCO,\,noS}} $, and the resulting fits are given in Eqs.~\eqref{eq:fit_a6} and \eqref{eq:fit_dt_ns}, and shown in Fig.~\ref{fig:a6_dt_v6} of Appendix~\ref{sec:fits}.
These results are used in the calibration and fitting procedure for the spinning cases discussed below.


\subsection{Calibration of spinning sector}
\label{sec:calibration_spinning}

Having determined the fits for the nonspinning parameters $\big\{ \Delta\rho_{\ell m}^{(1)}, \s a_6, \s\Delta t^{22}_{\text{ISCO,\,noS}} \big\} $, we now proceed with the calibration of $ \Delta t^{22}_{\text{ISCO,\,S}} $ which determines the merger-ringdown attachment for spinning binaries, and of $ \hat d _{ \text{SO}} $ which enters the spinning sector of the Hamiltonian, via [see Eq.~(12) of Ref.~\cite{Pompiliv5}]
\begin{equation}
\label{eq:so_calib}
H _{ \text{eff}} \,\, \propto \,\, \text{SO} _{ \text{calib}} = \nu \frac{M^5}{r^3} p_\phi \s \hat d _{ \text{SO}}.
\end{equation}
In \texttt{SEOBNRv5HM}, one has $ \hat d _{ \text{SO}} = d _{ \text{SO}} \s a_+/M  $, but here we directly calibrate $ \hat d _{ \text{SO}} $ as it yields more regular calibration posteriors.

We adopt a different procedure from that used in \texttt{SEOBNRv5HM}.
In particular, we employ a \emph{hierarchical spinning approach}, where the equal-mass cases are fitted first, followed by the unequal-mass cases (e.g., see Ref.~\cite{Nagar:2024dzj}).
This approach is particularly useful for \texttt{SEOBNRv6EHM}, as it provides greater flexibility in fitting the data observed for systems with close-to-equal masses ($ q \approx 1 $), as discussed below. 

We perform a nested sampling analysis over the parameters $ \boldsymbol{\theta}_{\text{S}} = \big \{ \hat d _{ \text{SO}}, \,\Delta t^{22}_{\text{ISCO,\,S}} \big \} $, with uniform priors
$\hat d _{ \text{SO}} \in [-500, 500]$ and $\Delta t^{22}_{\text{ISCO,\,S}} \in [-100, 100]$.
We employ the calibration settings specified in Sec.~\ref{sec:calibration_framework}, for 404 spinning, QC \texttt{SXS} NR simulations \cite{Scheel:2025jct}, with mass ratios ranging from $ q=1 $ to $ q=15 $, and dimensionless spins $ \chi_{1, \s2} \in ( -1 , 1) $.
This results in a set of calibration posteriors for $ \hat d _{ \text{SO}} $ and $ \Delta t^{22}_{\text{ISCO,\,S}}$ for different mass ratios and spin values, which we parametrize in terms of $ (\nu, \s a_+, \s a_-) $, as in \texttt{SEOBNRv5HM}.
As discussed in Sec.~\ref{sec:calibration_framework}, these posteriors are processed by removing secondary modes and excluding samples that do not satisfy the calibration requirements.
For all processed posteriors, we extract the medians $\langle\boldsymbol{\theta}_{\text{S}}\rangle_{(n)}$ and covariance matrices ${C_{\text{S}}}_{(n)}$, with $n$ labeling each of the employed NR waveforms.
We complement this data set with estimates for $ \Delta t^{22}_{\text{ISCO,\,S}}$ in the test-mass-limit from Table III of Ref.~\cite{Barausse:2011kb} (see also Sec.~IV C of Ref.~\cite{Pompiliv5}).
The estimated values, $ \Delta t^{22,\,\text{TML}}_{\text{ISCO,\,}\chi_i} $, are employed in Eq.~\eqref{eq:chi_tml} below.

Out of the 404 spinning, QC NR simulations, 78 correspond to equal-mass binaries, for which:
\begin{equation}
a_+ \, |_{\, q=1} \! = \frac{\chi_1 + \chi_2}{2}
\qquad
\text{and}
\qquad
a_- \, |_{\, q=1} \! = \frac{\chi_1 - \chi_2}{2}.
\end{equation}
This suggests a further classification of the 78 equal-mass simulations into: \emph{equal-spin} (20 cases with $ a_- = 0 $), \emph{opposite-spin} (7 cases with $ a_+ = 0 $), and \emph{different-spin} (51 cases with $ a_+ \neq 0 $ and $ a_- \neq 0 $) configurations.
The calibration data for $ \Delta t^{22}_{\text{ISCO}} = \Delta t^{22}_{\text{ISCO,\,noS}} + \Delta t^{22}_{\text{ISCO,\,S}}$ and $ \hat d _{ \text{SO}} $, corresponding to each of these spinning cases, is plotted in Fig.~\ref{fig:dtS_hatdSO} of Appendix~\ref{sec:fits}.
Notably, the data corresponding to the equal-spin and different-spin configurations display a behavior that is not possible to fit with simple, low-order polynomials, due to the structure seen for $ a_+ \in [0, 0.5] $.

This behavior motivates the use of a more flexible fitting strategy than that used in \texttt{SEOBNRv5HM}, leveraging the different spinning configurations of equal-mass binaries.
This hierarchical spinning approach proceeds as follows:
first, we fit the equal-mass, equal-spin cases as a function of $a_+$ only, and the equal-mass, opposite-spin cases as a function of $a_-$ only.
Next, we fit the equal-mass, different-spin cases using the residuals from the previous fits.
Finally, the unequal-mass cases are fitted using the results of all previous fits.

Specifically, these hierarchical spinning fits take the form:
\begin{subequations}
\begin{align}
\hat d _{ \text{SO}}
&=
\hat d _{ \text{SO}} ^{\, q=1} + \! \sqrt{1 - 4 \nu} \,\, \hat d _{ \text{SO}} ^{\, q \neq1} \!,
\\
\Delta t^{22}_{\text{ISCO,\,S}}
&=
\Delta t^{22,\, q=1}_{\text{ISCO,\,S}} + \! \sqrt{1 - 4 \nu} \,\, \Delta t^{22,\, q \neq1}_{\text{ISCO,\,S}} \s ,
\end{align}
\end{subequations}
where 
\begin{subequations}
\begin{align}
\hat d _{ \text{SO}} ^{\, q=1}
&=
\, \hat d _{ \text{SO}} ^{\, a_+} 
+ \hat d _{ \text{SO}} ^{\, a_-} 
+ \hat d _{ \text{SO}} ^{\, a_+ a_-}  ,
\\
\Delta t^{22,\, q=1}_{\text{ISCO,\,S}}
&=
\, \Delta t_{\text{ISCO,\,S}}^{22,\, a_+} 
+ \Delta t_{\text{ISCO,\,S}}^{22,\, a_-} 
+ \Delta t_{\text{ISCO,\,S}}^{22,\, a_+ a_-}  .
\end{align}
\end{subequations}
In these equations, the terms with superscript ``$ q = 1 $'' are associated with equal-mass cases, while those with superscript ``$ q \neq \nolinebreak 1 $'' are associated with unequal-mass cases (the factor $ \sqrt{1 - 4 \nu} $ allows the equal-mass cases to be fitted first).
Additionally, the terms with superscript ``$ a_+ \! $'' are fitted to equal-spin cases, those with ``$ a_-\! $'' to opposite-spin cases, and those with ``$ a_+ a_-\! $'' to different-spin cases.

An advantage of this splitting is that it allows \texttt{SEOBNRv6EHM} to satisfy the expected \emph{spin-exchange symmetry} for equal-mass binaries:
the dynamics and waveform of equal-mass systems must remain invariant under the interchange $ \chi_1 \leftrightarrow \chi_2 $.
The \texttt{SEOBNRv5} models fail to reproduce this property, because the fits for the parameters $ d _{ \text{SO}} $ and $ \Delta t^{22}_{\text{ISCO,\,S}} $ lack the required symmetry;
instead, these fits rely on the approximate behavior inherited from the calibration data.
In contrast, \texttt{SEOBNRv6EHM} respects this symmetry by enforcing it in the fitting procedure.


\subsubsection{Equal-mass spinning fits}
\label{sec:calibration_eqmass_spinning}

To fit the structures observed in the upper panels of each plot in Fig.~\ref{fig:dtS_hatdSO}, we employ the following ans\"atze:

\begin{itemize}[leftmargin=12pt]
\item
For the equal-spin contributions, $ \hat d _{ \text{SO}} ^{\, a_+}  $ and $ \Delta t_{\text{ISCO,\,S}}^{22,\, a_+}  $, we use a quartic polynomial in $a_+$ augmented with a Gaussian term,
\vspace{-10pt}
\begin{subequations}
\label{eq:equal_mass_equal_spin_ansatz}
\begin{align}
\hat d _{ \text{SO}} ^{\, a_+} 
&=
4 \s \nu \, \bigg[ d_1^{\s a_+} \, a_+ + d_2^{\s a_+} \, a_+^2 + d_3^{\s a_+} \, a_+^3 + d_4^{\s a_+} \, a_+^4
\nonumber
\\
&\qquad
+ d_5^{\s a_+} \, a_+^2 \, \mathrm{e}^{ \, d_6^{\s a_+} \, (a_+ - d_7^{\s a_+})^2 }  \bigg ] ,
\\
\Delta t_{\text{ISCO,\,S}}^{22,\, a_+} 
&=
4 \s \nu \,  \bigg[ \Delta_1^{\s a_+} \, a_+ + \Delta_2^{\s a_+} \, a_+^2 + \Delta_3^{\s a_+} \, a_+^3 + \Delta_4^{\s a_+} \, a_+^4 
\nonumber
\\
&\qquad
+ \Delta_5^{\s a_+} \, a_+^2 \, \mathrm{e}^{  \Delta_6^{\s a_+} \, (a_+ - \Delta_7^{\s a_+})^2 } \bigg ] \,  \nu^{-1/5} ,
\end{align}
\end{subequations}
where $ d_i^{\s a_+} $ and $ \Delta_j^{\s a_+} $ ($ i, j \in [1, 7] $) represent fitting coefficients, and we have introduced normalization factors of $ 4 \s \nu $ (equal to one for equal-mass cases), apart from the test-mass-limit scaling of $\nu^{-1/5}$ for $\Delta t_{\text{ISCO,\,S}}^{22}$, since this improved the fits when considering unequal-mass systems.
\item
For the opposite-spin contributions, $ \hat d _{ \text{SO}} ^{\, a_-}  $ and $ \Delta t_{\text{ISCO,\,S}}^{22,\, a_-} $, shown in the middle panels of the plots in Fig.~\ref{fig:dtS_hatdSO}, we use a quartic polynomial in $a_-$, but we keep only even powers to enforce the spin-exchange ($ a_- \to - a_- $) symmetry.
\item
For the different-spin contributions, $ \hat d _{ \text{SO}} ^{\, a_+ a_-} $ and $ \Delta t_{\text{ISCO,\,S}}^{22,\, a_+ a_-}  $, shown in the lower panels of the plots in Fig.~\ref{fig:dtS_hatdSO}, we use a quartic polynomial in both $a_+$ and $a_-$, excluding the common terms already used in the equal-spin and opposite-spin ans\"atze, and also keeping only even powers in $ a_- $.
\end{itemize}

In this way, the full expressions for $ \hat d _{ \text{SO}} ^{\, q=1} $ and $ \Delta t^{22,\, q=1}_{\text{ISCO,\,S}} $ are given by quartic polynomials in $ a_+ $ and $ a_- $, where odd powers of $ a_- $ are excluded, plus Gaussian terms.
We fit the median values of the calibration posteriors with the \texttt{optimize.curve\_fit} function from \texttt{scipy}~\cite{2020SciPy-NMeth}.
To facilitate the fitting process of $ \hat d _{ \text{SO}} ^{\, a_+} $ and $ \Delta t_{\text{ISCO,\,S}}^{22,\, a_+}  $, we first obtain the coefficients of the polynomial part of Eqs.~\eqref{eq:equal_mass_equal_spin_ansatz} by fitting to equal-spin cases with $ a_+ \leq 0 $ or $ a_+ \geq 0.55 $, and then, these results are employed to get the coefficients of the exponential part by fitting to all equal-spin cases.
In \texttt{optimize.curve\_fit}, we assign the same uncertainty to all cases, except for two equal-spin cases with $ a_+ \sim 0.25  $, for which we set $ \texttt{sigma} = 0.3 $ to give them greater weight.
The resulting fits for $ \hat d _{ \text{SO}} ^{\, q=1} $ and $ \Delta t^{22,\, q=1}_{\text{ISCO,\,S}} $ are presented in Eqs.~\eqref{eq:fit_hatdSO_equal} and \eqref{eq:fit_dt_s_equal}, and are plotted in Fig.~\ref{fig:dtS_hatdSO}.


\subsubsection{Unequal-mass spinning fits}
\label{sec:calibration_uneqmass_spinning}

The fitting process for unequal-mass cases follows the methods employed in \texttt{SEOBNRv5HM} \cite{Pompiliv5, Bohe:2016gbl}, which we summarize here.
First, we propose ans\"atze for 
$ \hat d _{ \text{SO}} ^{\, q \neq1} $ and $ \Delta t^{22,\, q \neq1}_{\text{ISCO,\,S}} $, in the form of the most generic quartic polynomials in $(\nu,\s a_{+}, \s a_{-})$, such that they vanish in the nonspinning limit, when $ a_{+} = a_{-} = 0 $.
We note that \texttt{SEOBNRv5HM} uses cubic polynomials for the spinning fits, with an extra $ a_+^4$ feature for $ \Delta t^{22}_{\text{ISCO,\,S}} $.
However, we find that quartic polynomials help reduce outlier mismatches and improve the overall accuracy.

The coefficients of the polynomial fits are determined by minimizing the function,
\begin{equation}
\label{eq:chi_s}
\chi_{\text{s}}^2 \, = \sum_{n \, \in \, \mathcal{S}_{\text{s}}} \frac{w}{2}\left(\boldsymbol{\theta}_{\text{S}}-\left\langle\boldsymbol{\theta}_{\text{S}}\right\rangle_{(n)}\right)\left(C_{\text{S}}^{-1}\right)_{(n)}\left(\boldsymbol{\theta}_{\text{S}}-\left\langle\boldsymbol{\theta}_{\text{S}}\right\rangle_{(n)}\right)^{\text{T}} +\chi_{\text{TML}}^2,
\end{equation}
where $ \boldsymbol{\theta}_{\text{S}} = \Big\{ \hat d _{ \text{SO}} ^{\, q \neq1}\!, \Delta t^{22,\, q \neq1}_{\text{ISCO,\,S}} \Big\}$, $ C_{\text{S}} $ are the covariance matrices of the calibration posteriors, and $w$ is a weighting factor given by
\begin{equation}
w = \chi_1^2+\chi_2^2+\frac{|\, \chi_S+\chi_A \delta / (1-2 \nu)|}{2 \nu}\,,
\end{equation}
which accounts for the inhomogeneous distribution of NR waveforms in the binary BH parameter space, and $\chi_{\text{TML}}^2$ is a term that penalizes deviations from the test-mass limit of $\Delta t^{22}_{\text{ISCO,\,S}}$ and takes the form
\begin{equation}
\label{eq:chi_tml}
\chi_{\text{TML}}^2 = \sum_{\chi_{i} \, \neq \, 0} \frac{\Big( \Delta t^{22}_{\text{ISCO,\,S}} - \Delta t^{22,\,\text{TML}}_{\text{ISCO,\,}\chi_i} \Big)^2}{\sigma^2_{\text{TML}}},
\end{equation}
with $\sigma_{\text{TML}} = 5 M$, as in \texttt{SEOBNRv5HM}.

The minimization of the function~\eqref{eq:chi_s} determines the fit coefficients for the unequal-mass, spinning contributions $ \hat d _{ \text{SO}} ^{\, q \neq1} $ and $ \Delta t^{22,\, q \neq1}_{\text{ISCO,\,S}} $.
This process is done with a Sequential Least Squares Programming minimization algorithm \cite{Bohe:2016gbl}.
\\

We performed extensive tests of the model and its fits to ensure physical robustness across broad regions of parameter space, including eccentricity.
These tests are discussed in Sec.~\ref{sec:robustness}.
To guard against overfitting, we also performed comparisons with additional QC and eccentric NR simulations not used in the calibration, as well as with QC NR surrogate models, as detailed in Sec.~\ref{sec:comparison_bound}.
Moreover, the reliability of the calibration pipeline for the multidimensional spinning fits was validated in Sec.~E of Ref.~\cite{Pompiliv5}.
Overall, these results indicate that the fits in \texttt{SEOBNRv6EHM} achieve the same level of robustness as the established \texttt{SEOBNRv5HM} model.

Our results highlight the complex interplay between the calibration process, fitting strategies, and model components (Hamiltonian, RR force, and waveform modes).
For example, the calibration parameters $ \Delta \rho _{ \ell m}^{(1)} $ discussed in Sec.~\ref{sec:calibration_RR} improve the model accuracy because they facilitate the overall calibration-fitting procedure, both in \texttt{SEOBNRv5HM} and \texttt{SEOBNRv6EHM}.
Future work may explore alternative calibration and fitting strategies, assess the role of individual model components in the calibration process, and extend the calibration to NR or small-mass-ratio eccentric waveforms.


\section{Model validation}
\label{sec:validation}

In this section, we assess the accuracy, speed, and robustness of \texttt{SEOBNRv6EHM} across different parameter space regions.
Accuracy is quantified through waveform and scattering-angle comparisons with NR results, efficiency through waveform-generation benchmarks, and robustness through an analysis of the regions where \texttt{SEOBNRv6EHM} yields physically meaningful and accurate waveforms.
In the companion manuscript \cite{Pompili:2026yxq}, we perform further tests within the context of Bayesian parameter estimation, confirming the high accuracy and computational efficiency of \texttt{SEOBNRv6EHM}.


\subsection{Waveform metrics}
\label{sec:metrics}

The \emph{strain} measured in a GW detector can be expressed as
\begin{align}
h(t) \, =&  \; \, F_+(\Theta,\Phi,\Psi) \, h_+(\iota, \varphi,d_\mathcal{L},\bm{\Pi},t_{ \text{c}};t)
\nonumber
\\
&+ F_\times(\Theta,\Phi,\Psi) \, h_\times(\iota, \varphi,d_\mathcal{L},\bm{\Pi},t_{ \text{c}};t),
\nonumber
\\[2pt]
=& \;\,
\mathcal{A}(\Theta,\Phi)\left ( h_+ \cos \kappa + h_+ \sin \kappa\right ) ,
\label{eq:h_detector}
\end{align}
where $F_+$ and $F_\times$ are the detector's antenna-patterns \cite{Sathyaprakash:1991mt,Finn:1992xs}, $ \,h_+ $ and $ h_\times $ are the GW polarizations defined in Eq.~\eqref{eq:gw_polarizations_decomp}, $\mathcal{A}(\Theta,\Phi)$ is given in Refs.~\cite{Cotesta:2018fcv,Ossokine:2020kjp}, and $\kappa = \kappa (\Theta,\Phi,\Psi)$ is an effective polarization angle.
The strain depends both on \emph{extrinsic} and \emph{intrinsic} parameters of the source.
The extrinsic parameters determine the location of the source with respect to a GW detector:
the angular position of the line of sight measured in the source frame ($\iota$, $\varphi$), the sky location of the source in the detector frame $(\Theta,\Phi)$, the polarization angle $\Psi$, the luminosity distance of the source $d_{\mathcal L}$, and the coalescence time $t _{ \text{c}}$.
The intrinsic parameters $ \bm{\Pi} $ determine a unique physical configuration:
the binary masses $m_1$ and $m_2$ (or, equivalently, mass ratio $ q $ and total mass $ M $), the dimensionless spin components along the $ z $ direction $\chi_1$ and $\chi_2$, as well as an orbital eccentricity $e$ and a radial phase $\zeta$ specified at a certain stage of the binary's evolution (e.g., at an orbit-averaged frequency $ \langle M \Omega \rangle $), or a set of reference values for the energy $ E/M $, angular momentum $ J/M^2 $, and relative separation $ r $ of the binary.

To quantify the agreement between a template $ h _{ \text{t}} $ (e.g., an EOB waveform) and a signal $ h _{ \text{s}} $ (e.g., a NR waveform), we employ the \emph{faithfulness} functions from Sec.~IV A of Ref.~\cite{Gamboa:2024hli}.
In general, these functions maximize the overlap~\eqref{eq:overlap} between the normalized template and signal,
\begin{equation}
\label{eq:faithfulness_gen}
\mF (M, \Sigma _{ \text{s}}) = \max_{\Sigma _{ \text{t}}} \, \big \langle \hat h _{ \text{s}} | \s \hat h _{ \text{t}} \big \rangle ,
\end{equation}
by optimizing the template parameters $ \Sigma _{ \text{t}} $ at a certain total mass $ M $ and for specific source parameters $ \Sigma _{ \text{s}} $.
For inner-product calculations, we use the settings discussed below Eq.~\eqref{eq:normalization}.
For comparisons against eccentric NR waveforms, however, we interpret $ f _{ \text{start}} $ as the initial orbit-averaged frequency of the NR $(2,2)$ mode and we trim the longest waveform to ensure that both waveforms $ h_{ \text{NR}} $ and $ h_{ \text{t}} $ have the same time to merger.
This approach ensures that only physically meaningful frequencies are considered, thereby avoiding contamination from Fourier transform artifacts \cite{Pompiliv5,RamosBuadesv5,Gamboa:2024hli}.

The template parameters optimized in Eq.~\eqref{eq:faithfulness_gen} depend on the type of binary (QC or eccentric) and on the mode content [e.g., $ (2,2) $-mode only or including higher-order modes].
For QC waveforms, the coalescence time $ t _{ \text{c}} $, azimuthal angle $ \varphi $, and effective polarization angle $ \kappa $ of the template are optimized \cite{Capano:2013raa,Harry:2017weg,Cotesta:2018fcv,Ossokine:2020kjp,Garcia-Quiros:2020qpx}.
For eccentric waveforms, additional optimizations are required due to the gauge dependence of orbital eccentricity in general relativity.

The standard approach for comparing eccentric waveforms is to optimize over the template's eccentric parameters \cite{Ramos-Buades:2021adz,Knee:2022hth,Gamboa:2024hli,Planas:2025feq,Ramos-Buades:2026kbq,Thomas:2026qrg,Morras:2026fho} (another approach, based on time shifts, is presented in Ref.~\cite{Bonino:2024xrv}).
Here, we follow the procedure described in Sec.~IV A 3 of Ref.~\cite{Gamboa:2024hli}, in which the faithfulness is numerically optimized over a grid of eccentricities and initial orbit-averaged frequencies with a fixed radial phase parameter.
Particularly, the radial phase is set to apastron, $ \zeta = \pi $, as it avoids waveform-generation issues at high eccentricities across all the considered models.\footnote{
For a highly eccentric binary, initializing in the vicinity of periastron ($ \zeta \approx 0 $) at a large starting frequency would place the BHs only a few units of $ M $ apart, close to their horizons.
In this strong-field regime, the prescriptions for the EOB initial conditions become unreliable (due to the underlying post-adiabatic and PN approximations), and the models considered here handle such configurations differently.
Initializing at apastron avoids these issues and provides a more uniform treatment across models.
}
This brute-force optimization is performed only for the $ (2,2) $ mode and for the lowest total mass under consideration.

When including higher-order modes, we use the \emph{sky-and-polarization-averaged, signal-to-noise-ratio (SNR)-weighted faithfulness} \cite{Ossokine:2020kjp,Cotesta:2018fcv},
\begin{equation}
\label{eq:faithfulness_avg_snr}
\overline{\mathcal{F}}_{\text{SNR}}(M, \iota _{ \text{s}}) = \sqrt[3]{\frac{\int_{0}^{2\pi} d\varphi_{\text{s}} \int_{0}^{2\pi} d\kappa_ {\text{s}} \ \mathcal{F}^{3} \ \text{SNR}^3}{\int_{0}^{2\pi} d\varphi_{\text{s}} \int_{0}^{2\pi} d\kappa_{\text{s}} \ \text{SNR}^3}} \, ,
\end{equation}
where $ \text{SNR} = \sqrt{\langle h_{\text{s}} | h_{\text{s}} \rangle} $.
These integrals are computed as a discrete sum over a uniform grid of $ 8 \kern-0.08em \times \kern-0.08em 8 $ values for $ \varphi_{\textrm{s}}, \kappa_{\text{s}} \in \nolinebreak{ [0, \s 2 \pi]} $ with a fixed value of $ \iota_{\text{s}} $.

In the following subsections, we quote all our results in terms of the \emph{unfaithfulness} (or \emph{mismatch}) metric
\begin{equation}
\mathcal{M} \equiv 1-\mathcal{F} .
\label{eq:mismatch}
\end{equation}
Accordingly, we denote $ \mathcal M_{\ell m} $ for individual $ (\ell, m) $-mode mismatches, and $ \overline{\mathcal{M}}_{\text{SNR}} $ for the SNR-weighted mismatches.
In this way, lower mismatch values imply better agreement between two waveforms for a given GW detector.


\subsection{Comparison against NR simulations}
\label{sec:comparison_bound}

\begin{figure}
\hspace{-5pt}
\includegraphics[width=\linewidth]{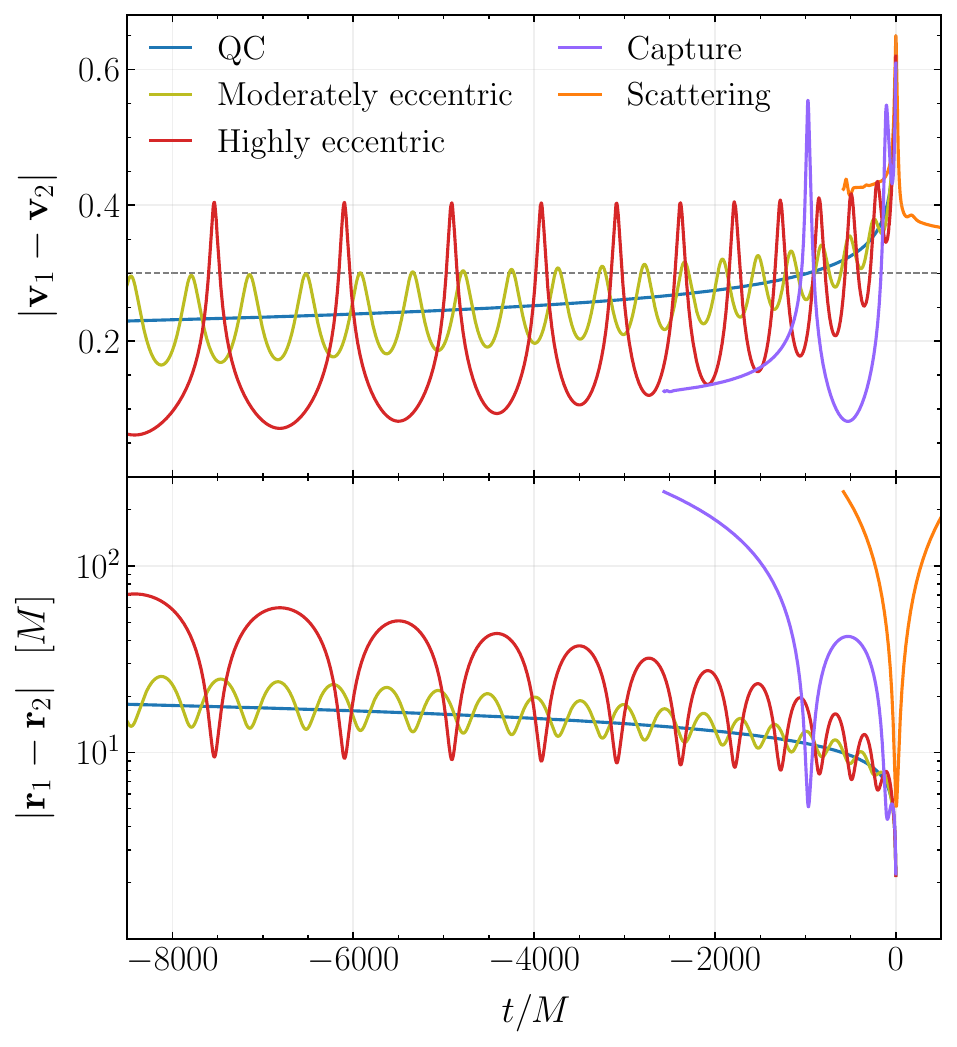}
\vspace{-10pt}
\caption{
Relative velocity (top panel) and separation (bottom panel) for a set of NR equal-mass, nonspinning BBH configurations with varying eccentricity:
QC (\texttt{SXS:BBH:2325}, blue), moderately eccentric (\texttt{SXS:BBH:2521}, yellow), highly eccentric (\texttt{SXS:BBH:2527}, red), dynamical capture (\texttt{SXS:BBH:4000}, violet), and scattering encounter (\texttt{SXS:BBH:4292}, orange).
The dashed line in the top panel marks $ 30\% $ of the speed of light, underscoring the highly relativistic nature of eccentric systems.
The systems are aligned such that $ t =0 $ corresponds to the peak of GW emission.
Here, $\mathbf{r}_i$ and $\mathbf{v}_i$ ($i\in \{1,2 \}$) denote the position and velocity vectors of the two BHs.
}
\label{fig:NR_dynamics}
\end{figure}

In general, eccentric systems present a significant modeling challenge, as their periastron passages involve small separations and high relativistic velocities.
This is illustrated in Fig.~\ref{fig:NR_dynamics}, which shows the relative velocity (top panel) and separation (bottom panel) for a subset of equal-mass, nonspinning NR simulations with varying eccentricities.
State-of-the-art approximants accurately model the GW emission from QC binaries (blue curves) up to velocities of order a few tenths of the speed of light (for reference, we indicate the value of $ 30\%$ by a dashed line), corresponding to a certain number of orbits before merger.
As the eccentricity increases, however, the binary becomes progressively more relativistic at periastron. This is evident in moderately (yellow curves) and highly eccentric (red curves) simulations, whose periastron passages reach velocities comparable to those reached by QC binaries at the very end of their inspiral. 
Importantly, these highly relativistic configurations arise even during the early inspiral of long-lived systems with moderate-to-high eccentricities, since the major change in the dynamics of these binaries occurs at apastron, while the velocity and separation at periastron remain relatively constant throughout the binaries' evolution.
Therefore, accurate modeling of eccentric waveforms requires capturing the strong-field, high-velocity binary dynamics to prevent error accumulation at periastron passages.

Here, we quantify the accuracy of the \texttt{SEOBNRv6EHM} dynamics and waveforms for different BBH systems.
Specifically, we compute waveform mismatches against sets of QC and eccentric NR simulations, we compare against NR waveforms from a dynamical capture and scattering encounters, and we compute scattering angles for different configurations for which NR data is available.
All the NR waveforms and scattering angles were computed with the \texttt{SpEC} code \cite{SpECwebsite} from the \texttt{SXS} collaboration \cite{SXS:catalog,Boyle:2019kee,Chu:2015kft,Blackman:2017dfb,Hemberger:2013hsa,Scheel:2014ina,Lovelace:2014twa,Abbott:2016apu,Blackman:2015pia,Lovelace:2016uwp,Varma:2018mmi,Abbott:2016nmj,Varma:2019csw,Kumar:2015tha,Mroue:2013xna,Yoo:2022erv,Scheel:2025jct}.
We also include predictions from the state-of-the-art eccentric, aligned-spin models \texttt{SEOBNRv5EHM} \cite{Gamboa:2024hli} and \texttt{TEOBResumS-Dal\'i}~\cite{Nagar:2024dzj}.
For waveform comparisons, we include all modes supported by each model.
Namely, for \texttt{SEOBNRv5EHM} and \texttt{SEOBNRv6EHM} we use $( \ell, |m| ) \in \nolinebreak \{(2, 2),\, (3, 3),\, (2, 1),\, (4, 4),\, (3, 2),\, (4, 3)\}$, and for \texttt{TEOBResumS-Dal\'i} we employ the modes $( \ell, |m| ) \in \nolinebreak \{(2, 2),\, (3, 3),\, (2, 1),\, (4, 4)\}$.\footnote{
In this work, we employ the \texttt{TEOBResumS-Dal\'i} model presented in Ref.~\cite{Nagar:2024dzj}, whose waveform modes $ (\ell, m) \in \nolinebreak \{ (2,2), (2,1), (3,3), (4,4)\} $ for aligned-spin BBHs have been reviewed by the LVK collaboration.
We have used the code on the \texttt{LVK-Dali} branch from the repository
\url{https://git.ligo.org/waveforms/software/teobresums},
with git hash \texttt{5868b3a965ad4227ac99a3aaf2b348206ea94f48}.
\label{fn:dali}
}
For NR waveforms, we include all modes up to $\ell = 4$.
Results for \texttt{SEOBNRv5HM} \cite{Pompiliv5} in QC waveform comparisons are identical to those of \texttt{SEOBNRv5EHM}, and thus are not included.


\subsubsection{QC binaries}
\label{sec:comparison_qc}

\begin{figure*}
\hspace{-5pt}
\includegraphics[width=0.93\linewidth]{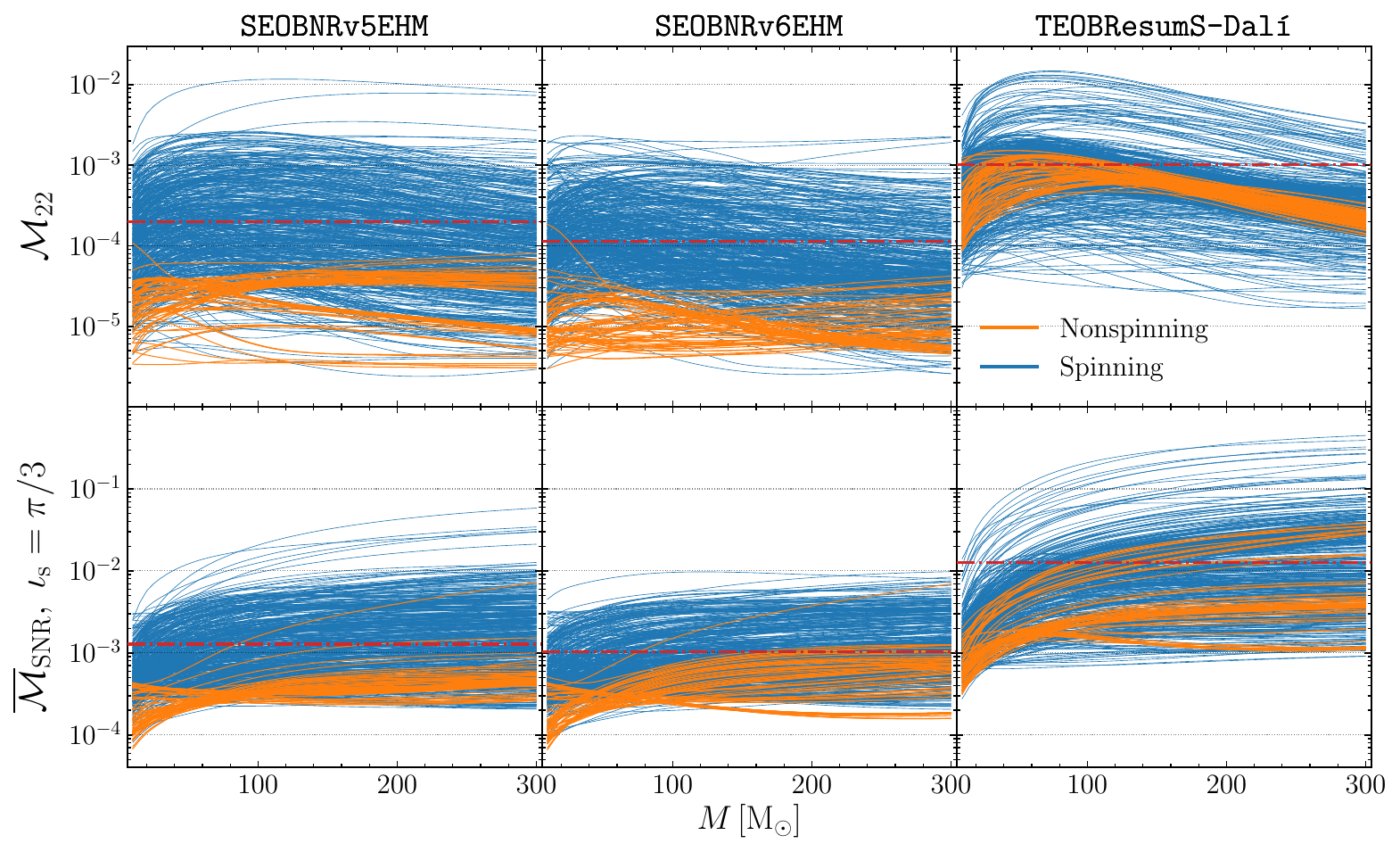}
\vspace{-5pt}
\caption{
Waveform mismatches between different approximants and the 592 QC \texttt{SXS} NR simulations used in this work, computed over the total mass range $ M \in [10, 300]\, \solarmass$.
Nonspinning cases are shown in orange and spinning cases in blue.
Columns 1--3 correspond to \texttt{SEOBNRv5EHM}, \texttt{SEOBNRv6EHM}, and \texttt{TEOBResumS-Dal\'i}, respectively.
The top panels show the $(2,2)$-mode mismatch $ \mathcal M_{22} $, while the bottom panels display the sky-and-polarization averaged, SNR-weighted mismatch $\overline{\mathcal{M}}_\mr{SNR}$ (including higher-order modes) for an inclination $\iota _{ \text{s}} = \pi/3$.
We compute the maximum mismatch $ \max_M \mathcal{M} $ across the total mass range for each NR waveform, and the median of these values, $ \text{med}[\max_M \mathcal{M}] $, is shown in each panel as a red dot-dashed line.
}
\label{fig:mismatch_spaghetti_qc}
\end{figure*}
\begin{figure}
\hspace{-7pt}
\vspace{5pt}
\includegraphics[width=\linewidth]{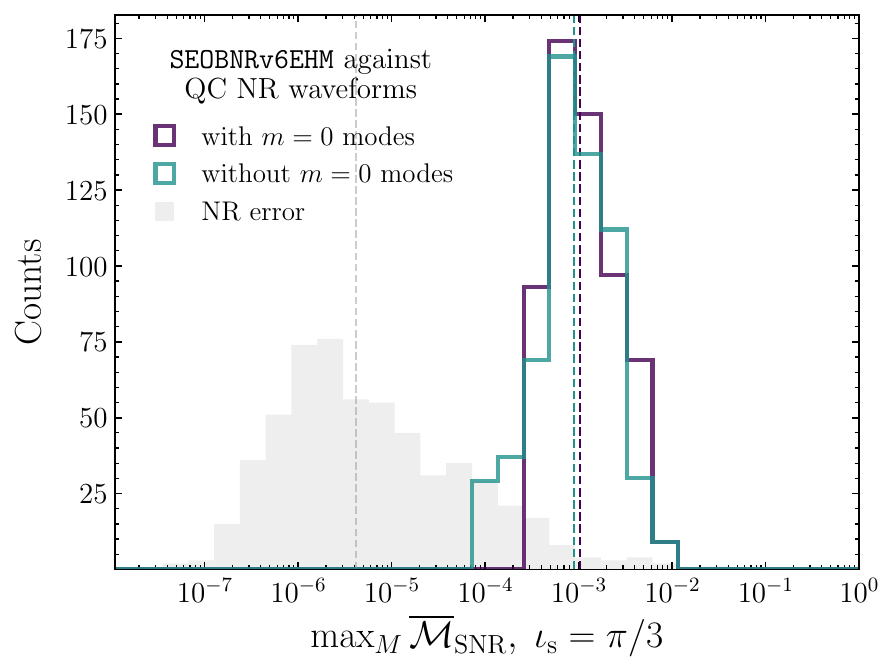}
\includegraphics[width=0.99\linewidth]{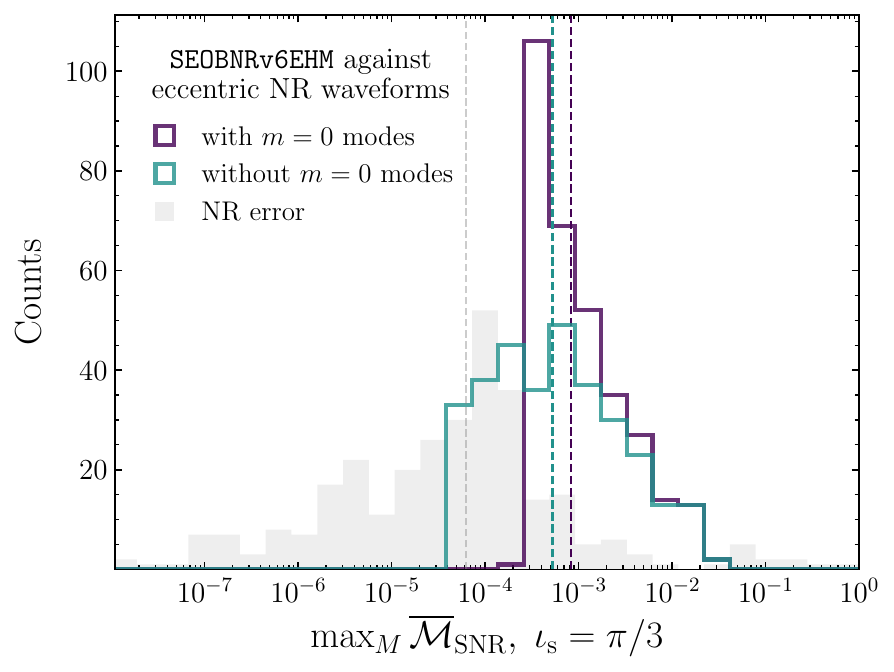}
\vspace{-5pt}
\caption{
Histograms of SNR-weighted mismatches, maximized over a specific total-mass range, for the \texttt{SEOBNRv6EHM} model against 592 QC (top panel) and 319 eccentric (bottom panel) NR \texttt{SXS} waveforms with inclination $\iota_\mathrm{s} = \pi/3$ and $ \ell \leq 4 $, including $m=0$ modes (dark purple) and excluding them (green).
The total-mass range spans $ [10, 300]\, \solarmass$ for QC waveforms, and $ [20, 200]\, \solarmass$ for eccentric waveforms.
An estimate of the NR error (gray) is also shown.
}
\label{fig:mismatch_distributions_m0}
\end{figure}
\begin{figure}
\vspace{5pt}
\includegraphics[width=\linewidth]{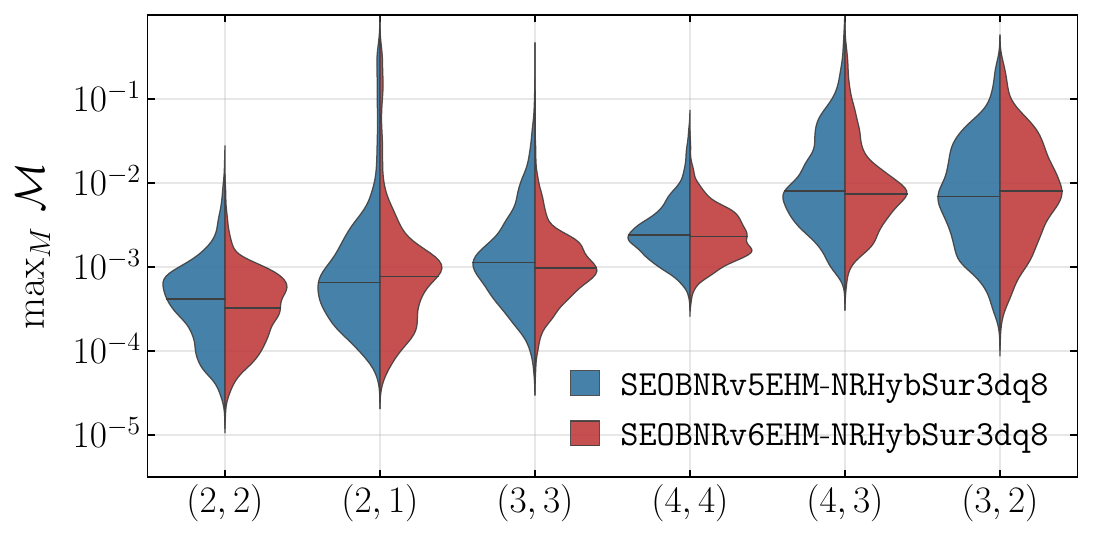}
\includegraphics[width=\linewidth]{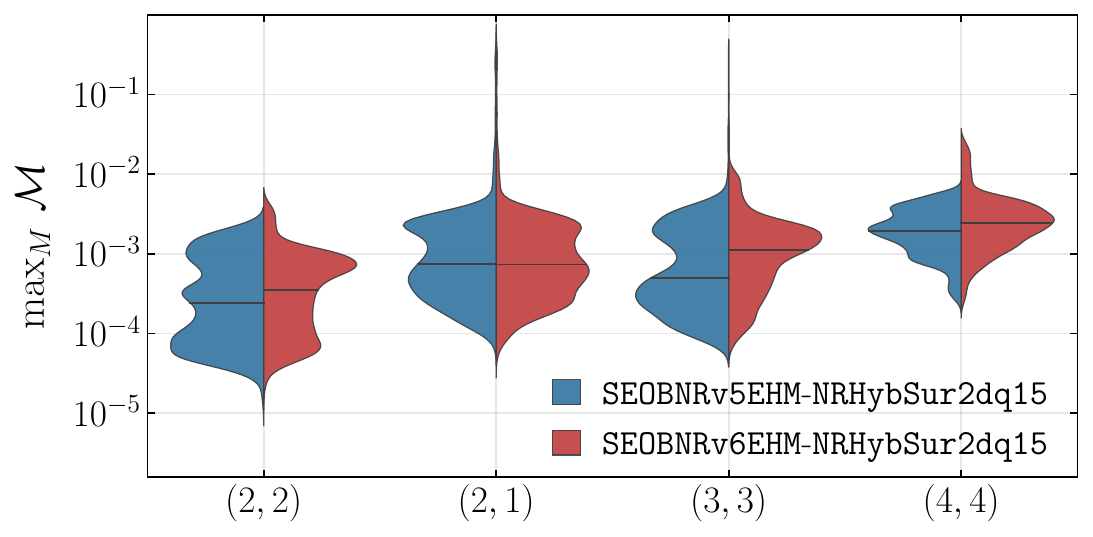}
\vspace{-15pt}
\caption{
Distribution of maximum mode-by-mode mismatches between \texttt{SEOBNRv5EHM} (blue) and \texttt{SEOBNRv6EHM} (red) against the NR surrogates \texttt{NRHybSur3dq8} (top panel) and \texttt{NRHybSur2dq15} (bottom panel), computed over the total mass range $ M \in [10, 300] \, \solarmass$ across different, randomly-distributed binary QC configurations.
For \texttt{NRHybSur3dq8}, we take $5000$ points with $ q \leq 8 $, $ |\chi_i|\leq 0.9 $ ($ i \in \nolinebreak \{1, 2\} $), and $ M \Omega = 0.015$, while for \texttt{NRHybSur2dq15}, we take $5000$ points with $ q \leq 15 $, $ |\chi_1|\leq 0.6 $, $ \chi_2 = 0 $, and $ M \Omega = 0.015$.
}
\label{fig:violins_NRsur}
\end{figure}

We compute waveform mismatches against a set of 592 BBH QC \texttt{SXS} NR waveforms \cite{Scheel:2025jct}.
The parameter space distribution of these waveforms is shown in Fig.~\ref{fig:param_space_QC}.
A summary of each NR waveform is provided in an ancillary file~\cite{SEOBNRv6EHM:SupplementalMaterial}.\footnote{
In the ancillary file, we list the mass ratio $ q $, dimensionless spin components $ \chi _1 $ and $ \chi _2 $, and the initial dimensionless orbital frequency $\langle M \Omega \rangle$.
Additionally, we provide the maximum $ (2,2) $-mode mismatch $ \max_M \mathcal M_{22} $ (calculated over the range $ M \in [10, 300] \, \solarmass $) for each waveform model used in this work.
}

The $ (2,2) $-mode mismatches are computed with Eq.~(33) of Ref.~\cite{Gamboa:2024hli} over a range of total masses $ M \in\nolinebreak {[10, 300]}\, \solarmass$, and the results are shown in the top panels of Fig.~\ref{fig:mismatch_spaghetti_qc}.
We observe an overall consistency between the accuracies of both \texttt{SEOBNR} models, with a slightly better median mismatch, a better control of mismatch outliers, and a slightly improved accuracy for nonspinning waveforms with \texttt{SEOBNRv6EHM}.
Additionally, we observe an overall accuracy improvement of one order of magnitude for both \texttt{SEOBNR} models with respect to \texttt{TEOBResumS-Dal\'i}.
The most challenging cases correspond to systems with high mass ratios and/or high spins.
However, the mismatches of \texttt{SEOBNRv6EHM} for all these systems remain below $ 3 \times 10^{-3}\! $, thanks to its analytical content, the higher-order fits of the calibration parameters in \texttt{SEOBNRv6EHM}, and the augmented calibration set with respect to \texttt{SEOBNRv5EHM}.\footnote{
The accuracy of \texttt{SEOBNRv6EHM} degrades only slightly when using analogous fits and the same calibration set as in \texttt{SEOBNRv5EHM}.
The final fits and the selected NR simulations for the calibration of \texttt{SEOBNRv6EHM} were chosen to get the best performance across different binary configurations, as discussed in Sec.~\ref{sec:calibration}.
}

The accuracy of GW polarizations (with higher-order modes) is quantified with the sky-and-polarization-averaged, SNR-weighted faithfulness \eqref{eq:faithfulness_avg_snr} [see also Eq.~(32) of Ref.~\cite{Gamboa:2024hli}], calculated for an inclination $ \iota _{ \text{s}} = \pi/3 $, over the same range of $ M \in [10, 300]\, \solarmass$.
The corresponding mismatches are shown in the bottom panels of Fig.~\ref{fig:mismatch_spaghetti_qc}.
For multipolar waveforms, \texttt{SEOBNRv6EHM} also has a better overall accuracy compared to \texttt{SEOBNRv5EHM}, and shows a significant improvement with respect to \texttt{TEOBResumS-Dal\'i}.
In particular, the latter model underperforms in many cases, even resulting in mismatches of $ \sim 40\% $.
By contrast, the mismatches of \texttt{SEOBNRv6EHM} remain always below $ 1\% $.
We note that all models exhibit reduced accuracy for systems with high mass ratios and/or high spins, as already pointed out, e.g., in Ref.~\cite{Pompiliv5}).
Furthermore, the models also show an increase in mismatch with respect to the total mass, indicating that the number of modeled higher-order modes and their accuracy (particularly in the late inspiral) need to be improved \cite{FooInPrep}.

We also observe a spread in the SNR-weighted mismatches of nonspinning waveforms (orange curves) for \texttt{SEOBNRv6EHM} relative to \texttt{SEOBNRv5EHM} (compare bottom-left and bottom-middle panels of Fig.~\ref{fig:mismatch_spaghetti_qc}).
Thus, some multipolar nonspinning cases in \texttt{SEOBNRv6EHM} are underperforming.
In contrast, for the $ (2,2) $ mode, there is an overall improvement for nonspinning cases in \texttt{SEOBNRv6EHM}.
This behavior is likely related to the arbitrary choices made when defining the Newtonian prefactors in Eqs.~\eqref{eq:hat_hlm_N_generic_rrdprd}.
Since these mismatches stay below $ \sim 10^{-3} $ and are still comparable to those of \texttt{SEOBNRv5EHM}, we defer the investigation of improved Newtonian prefactors for higher-order modes to future work.

As a complementary study, we provide in the top panel of Fig.~\ref{fig:mismatch_distributions_m0} histograms of maximum \texttt{SEOBNRv6EHM} SNR-weighted mismatches for all the QC NR waveforms with (dark purple bins) and without (green bins) the $ m = 0 $ multipoles, for $ \ell \leq 4 $.
We also include an estimate of the NR errors computed as the SNR-weighted mismatches between the highest and second-highest resolutions (always including the $ m = 0 $ modes), when available.
For the highest-mismatch cases, other sources of error (e.g., large spins) dominate over the absence of $ m=0 $ modes in \texttt{SEOBNRv6EHM} (the same is true for \texttt{SEOBNRv5EHM} and \texttt{TEOBResumS-Dal\'i}).
However, we observe that the lowest maximum mismatches are pushed toward values above $ \sim 2 \times 10^{-4} $, and we also see an increase in the number of cases with higher mismatches.
This explains why the mismatches shown in the bottom panels of Fig.~\ref{fig:mismatch_spaghetti_qc} are higher than those presented in Refs.~\cite{Pompiliv5,Gamboa:2024hli}, which did not include $ m=0 $ modes in their comparison.
Overall, these results point to the importance of including $ m=0 $ modes to achieve mismatches below $ \sim 10^{-4} $.
This direction has only recently started to be explored within EOB modeling in Refs.~\cite{Albanesi:2024fts,EstellesInPrep}.

We provide a complementary accuracy study by comparing against QC NR surrogate models.
Both \texttt{SEOBNRv5EHM} and \texttt{SEOBNRv6EHM} have been calibrated against a subset of the NR waveforms used in Fig.~\ref{fig:mismatch_spaghetti_qc}.
Thus, computing mismatches against waveforms outside the calibration set is an important step to consolidate the models' accuracy and to discard overfitting.
For this purpose, we compare \texttt{SEOBNRv6EHM} against the \texttt{NRHybSur3dq8} \cite{Varma:2018mmi} and \texttt{NRHybSur2dq15} \cite{Yoo:2022erv} QC NR surrogate models, which are hybridized using the \texttt{SEOBNRv4} \cite{Bohe:2016gbl} and \texttt{SEOBNRv4HM} \cite{Cotesta:2018fcv} models, respectively.
\texttt{NRHybSur3dq8} is a model for binaries with mass-ratios $ q \in \nolinebreak {[1, 8]} $ and spin magnitudes $ |\chi_i|\leq 0.8 $ ($ i \in \{1, 2\} $), while \texttt{NRHybSur2dq15} is a model for binaries with mass-ratios $ q \in \nolinebreak {[1, 15]} $, primary spin $ |\chi_1|\leq 0.5 $, and secondary spin $ \chi_2 = 0 $.
These surrogate models provide QC waveforms with errors comparable to the NR accuracy in the region of parameter space where the models were trained.

In Fig.~\ref{fig:violins_NRsur}, we show the mode-by-mode comparisons of \texttt{SEOBNRv5EHM} and \texttt{SEOBNRv6EHM} against the NR surrogates \texttt{NRHybSur3dq8} (top panel) and \texttt{NRHybSur2dq15} (bottom panel), for randomly distributed points in a region of parameter space slightly larger than the training region of each surrogate model, to allow for some extrapolation.
Specifically, for the \texttt{NRHybSur3dq8} comparisons, we take $5000$ points in the region determined by $ q \leq 8 $, $ |\chi_i|\leq 0.9 $ ($ i \in \{1, 2\} $), and starting dimensionless orbital frequency of $ M \Omega = 0.015$ (we choose this value to reduce the contributions from the hybridization of the NR surrogate in the early inspiral), while for the \texttt{NRHybSur2dq15} comparisons we take $5000$ points with $ q \leq 15 $, $ |\chi_1|\leq 0.6 $, $ \chi_2 = 0 $, and $ M \Omega = 0.015$.
For each of these configurations, we compute the maximum mismatch of each $ (\ell, m) $ mode across an interval of total masses $ M \in [10, 300] \, \solarmass$, denoted as $ \max_M \mathcal M $.

Overall, Fig.~\ref{fig:violins_NRsur} shows that the accuracies of \texttt{SEOBNRv5EHM} and \texttt{SEOBNRv6EHM} are very similar across the considered regions of parameter space, with the most challenging configurations being binaries with high $ q $ and large spins, as before.
For certain cases, \texttt{SEOBNRv5EHM} performs better than \texttt{SEOBNRv6EHM}, but the opposite is also true.
Notably, both models have the same tails of large mismatches, particularly for the $ (2,1) $, $ (3,3) $, and $ (4,3) $ modes, for systems with large spins.
These high mismatches originate from the small amplitudes these modes can reach near merger, which conflicts with the methods for constructing and attaching the merger-ringdown part of the waveform, as already identified in Refs.~\cite{Cotesta:2018fcv,Pompiliv5}.
Ongoing work is addressing these problems, in preparation for the accuracy requirements of next-generation GW detectors \cite{FooInPrep}.

These results show that the accuracy of \texttt{SEOBNRv6EHM} in the zero-eccentricity limit is at the same order as other state-of-the-art \emph{dedicated} QC models.\footnote{
The NR mismatches of other QC waveform models have been shown, e.g., in Fig. 7 of Ref.~\cite{Pompiliv5}.
Such results were produced for a representative subset of the NR waveforms employed in this work, using Advanced LIGO's PSD \cite{Barsotti:2018} with all NR modes up to $ \ell = 5 $.
}
We remark that developing an eccentric model with an accurate QC limit is highly important to avoid biases in parameter estimation \cite{Bonino:2022hkj, Ramos-Buades:2023yhy}.


\subsubsection{Eccentric (bound) binaries}
\label{sec:comparison_ecc}

\begin{figure}
\hspace*{-10pt}
\includegraphics[width=\linewidth]{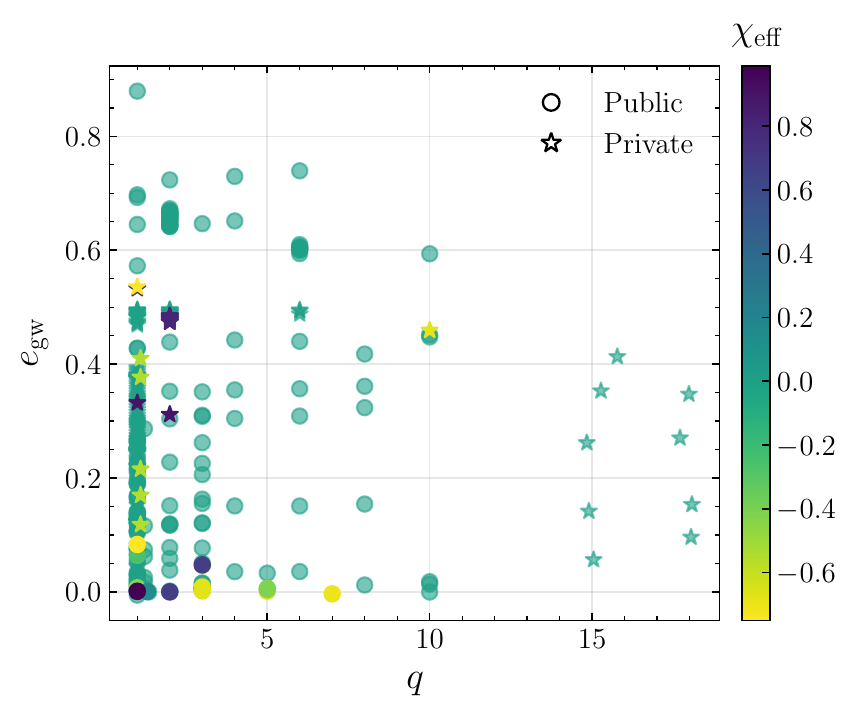}
\vspace{-8pt}
\caption{
Parameter-space distribution of the 319 eccentric \texttt{SXS} NR waveforms employed in this work, in terms of the mass ratio $ q $, dimensionless effective spin $ \chi_{\text{eff}} $, and earliest GW eccentricity $ e _{ \text{gw}} $ measured by \texttt{gw\_eccentricity}.
We represent the 184 publicly available NR waveforms with circles and the 135 private ones with stars.
}
\label{fig:param_space_NR_ecc_sims}
\end{figure}
\begin{figure}
\hspace*{-10pt}
\includegraphics[width=\linewidth]{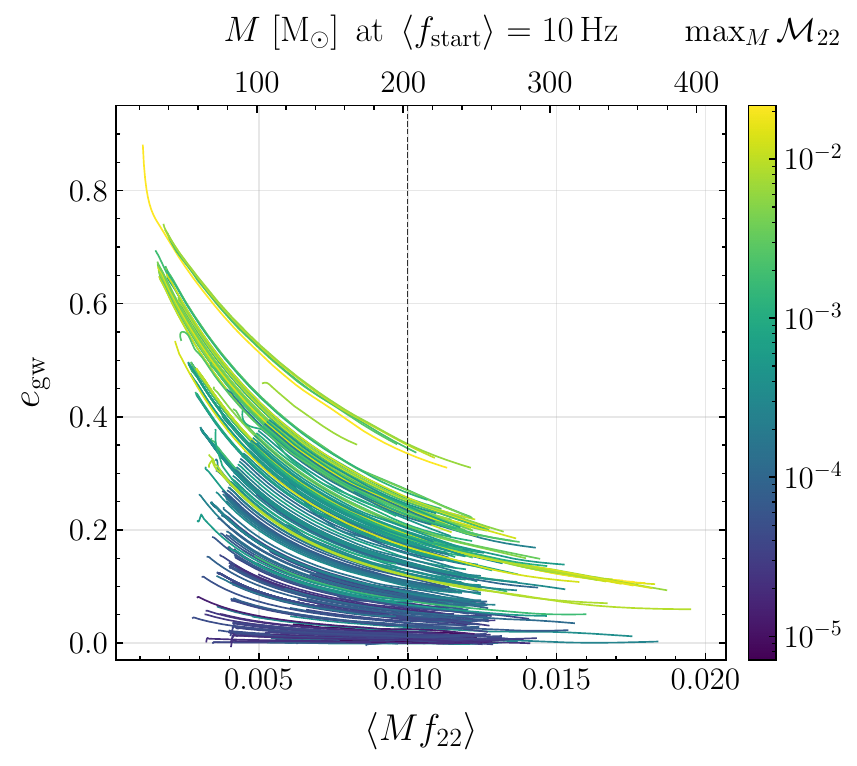}
\vspace{-8pt}
\caption{
Evolution of the GW eccentricity $ e _{ \text{gw}} $ as a function of the mass-scaled, orbit-averaged $(2,2)$-mode frequency $ \langle M f_{22} \rangle $ (in geometric units) for the 319 eccentric NR waveforms employed in this work.
The color of each curve encodes the maximum $ (2,2) $-mode mismatch $ \mathcal M_{22} $ over the total mass range $ M \in [20, 200]\, \solarmass $.
The dashed vertical line indicates a reference value $ \langle M f_{22} \rangle = 0.01 $.
}
\label{fig:mismatch_egw_frequency}
\end{figure}
\begin{figure*}
\hspace{-5pt}
\includegraphics[width=0.93\linewidth]{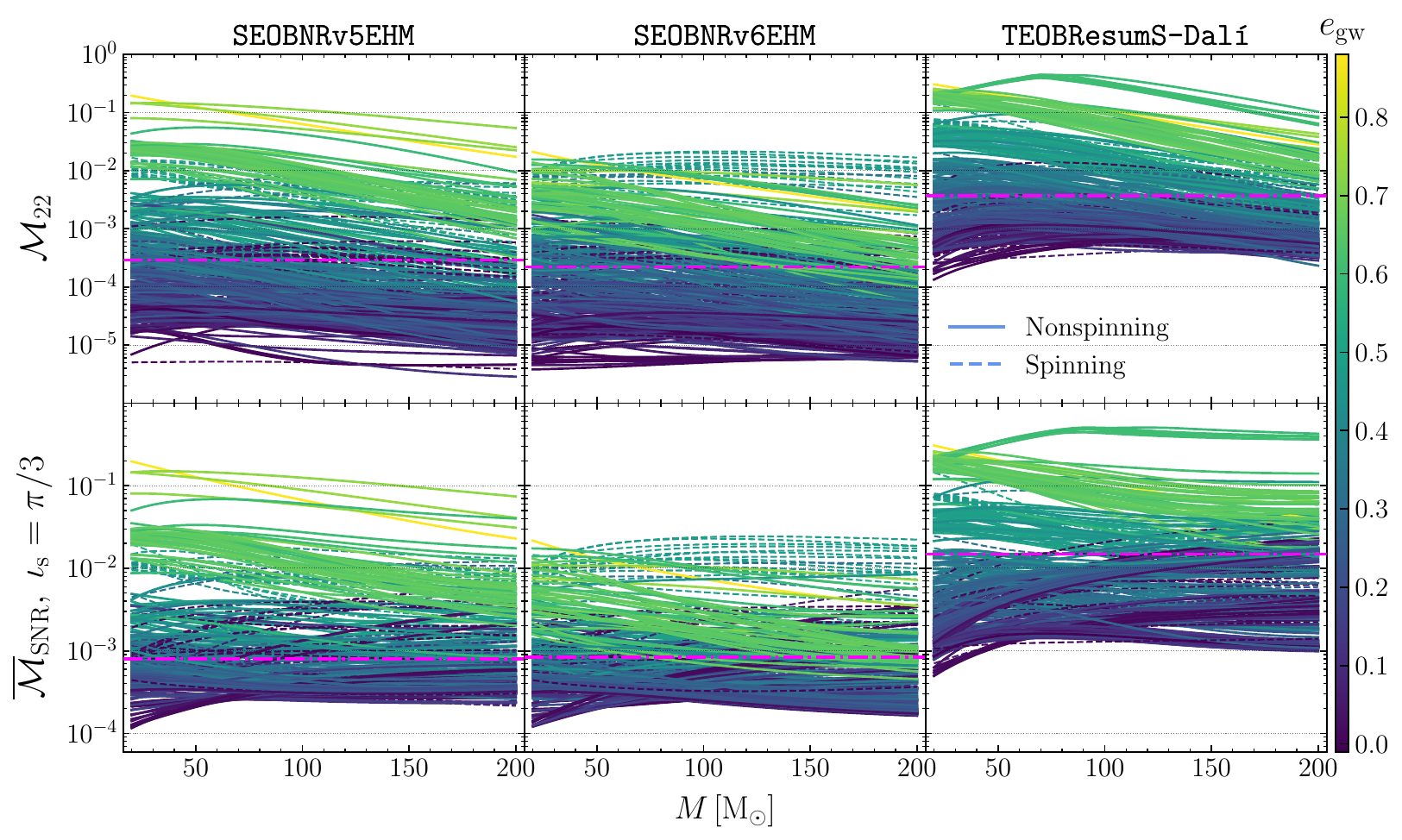}
\vspace{-5pt}
\caption{
Waveform mismatches between different approximants and the 319 eccentric \texttt{SXS} NR simulations used in this work, computed over the total mass range $ M \in [20, 200]\, \solarmass$.
Curve color represents the earliest GW eccentricity $ e_{\text{gw}} $ of each NR waveform measured by \texttt{gw\_eccentricity}.
Nonspinning cases are shown with solid curves, and spinning cases with dashed curves.
Columns 1--3 correspond to \texttt{SEOBNRv5EHM}, \texttt{SEOBNRv6EHM}, and \texttt{TEOBResumS-Dal\'i}, respectively.
The top panels display the $(2,2)$-mode mismatch $ \mathcal M_{22} $, while the bottom panels show the sky-and-polarization averaged, SNR-weighted mismatch $\overline{\mathcal{M}}_\mr{SNR}$ (including higher-order modes) for an inclination $\iota _{ \text{s}} = \pi/3$.
We compute the maximum mismatch $ \max_M \mathcal{M} $ across the total mass range for each NR waveform, and the median of these values, $ \text{med}[\max_M \mathcal{M}] $, is shown in each panel as a magenta dot-dashed line.
}
\label{fig:mismatch_spaghetti_ecc}
\end{figure*}

\begin{figure*}
\centering
\hspace{-7pt}
\subfloat{
\includegraphics[width=0.47\linewidth]{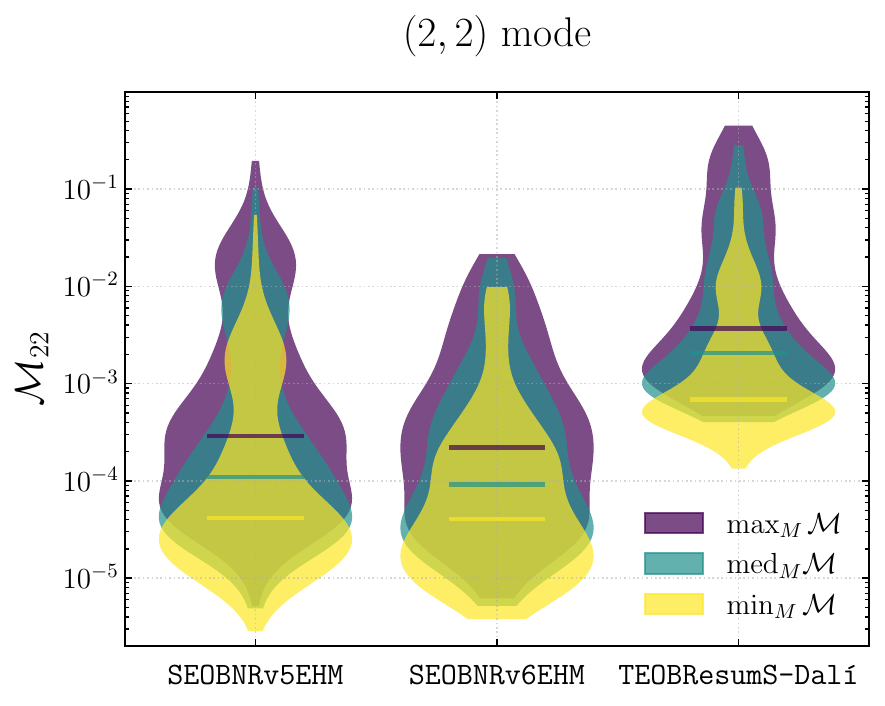}
}
\hspace{7pt}
\subfloat{
\includegraphics[width=0.47\linewidth]{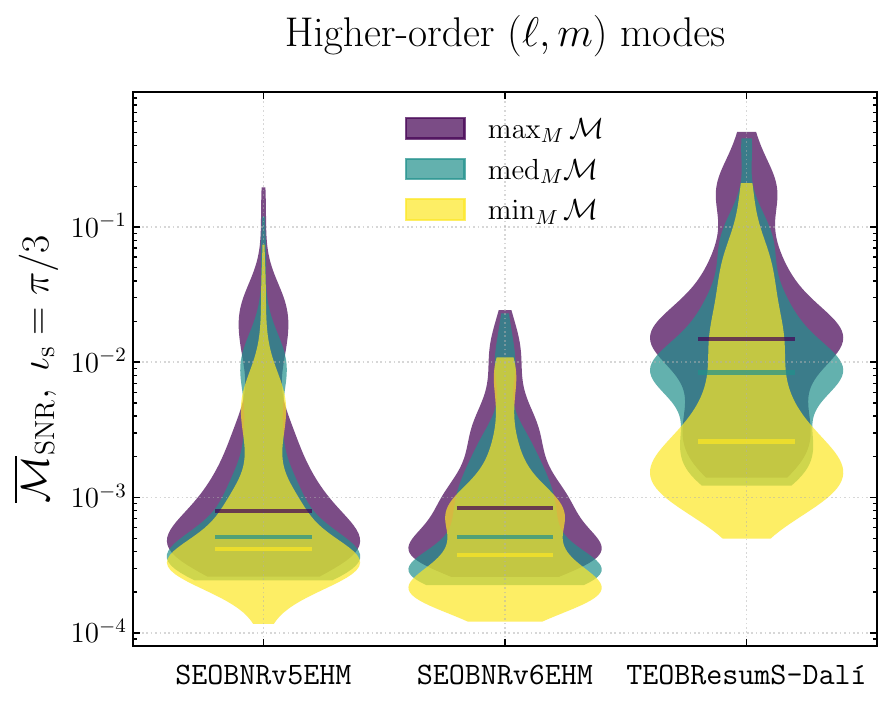}
}

\hspace{1pt}
\subfloat{
\includegraphics[width=0.46\linewidth]{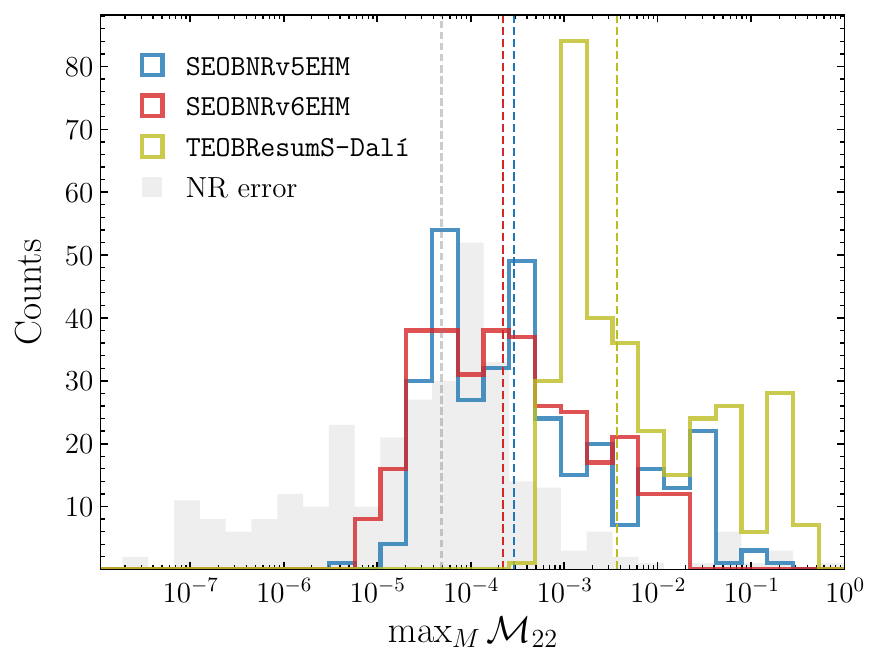}  
}
\hspace{12pt}
\subfloat{
\includegraphics[width=0.46\linewidth]{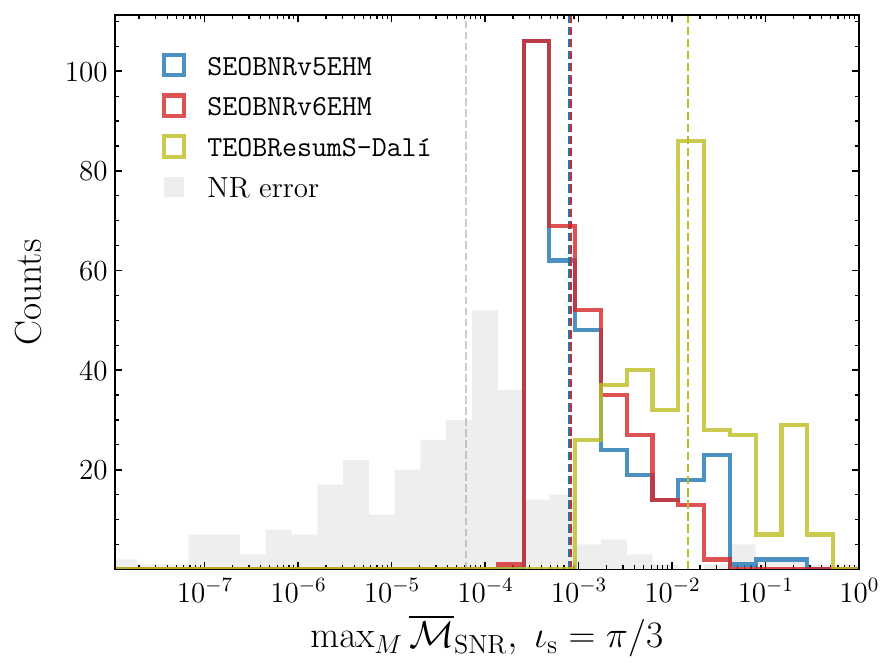}  
}
\vspace{-4pt}
\caption{
\emph{Top panels:}
Distributions of the maximum (purple), median (green), and minimum (yellow) $(2,2)$-mode mismatches $ \mathcal M_{22} $ (left) and sky-and-polarization-averaged, SNR-weighted mismatches $ \overline{\mathcal M} _{ \text{SNR}} $ (right) over the total mass range $ M \in [20, 200]\, \solarmass $, for different waveform models against the 319 eccentric \texttt{SXS} NR waveforms employed in this work.
The horizontal lines represent the corresponding medians of the maximum, median, and minimum distributions.
\emph{Bottom panels:}
Histograms of the maximum $(2,2)$-mode mismatches (left) and maximum sky-and-polarization-averaged, SNR-weighted mismatches (right) for the 319 eccentric NR waveforms against the considered models.
The NR error histograms (gray) are obtained from the waveform mismatches between the highest and second-highest NR resolutions, when available.
The vertical dashed lines display the median of the corresponding distributions.
}
\label{fig:mismatch_distributions_ecc}
\end{figure*}

We validate the accuracy of \texttt{SEOBNRv6EHM} by computing waveform mismatches against a catalog of 319 BBH eccentric \texttt{SXS} NR waveforms.
Of these, 184 (composed of 159 nonspinning and 25 aligned-spin configurations) are publicly available in the latest \texttt{SXS} catalog \cite{Scheel:2025jct}, and 135 (composed of 107 nonspinning and 28 aligned-spin configurations) are private waveforms from Refs.~\cite{Ramos-Buades:2022lgf,Nee:2025zdy}.\footnote{
Following the \texttt{SXS} catalog, we classify a NR simulation as \emph{eccentric} if its reference eccentricity exceeds $0.001$, and as \emph{nonspinning} if the dimensionless $z$-components of the spins satisfy $|\chi_i | < \nolinebreak 0.001$, with $i \in \{1,2\}$.
}
Thus, in total, we compare against 266 nonspinning and 53 aligned-spin eccentric NR waveforms, all of which are represented in Fig.~\ref{fig:param_space_NR_ecc_sims}, characterized by their mass ratio $ q $, dimensionless effective spin parameter $ \chi _{ \text{eff}} $, and their earliest GW eccentricity $ e _{ \text{gw}} $ as defined in Refs.~\cite{Ramos-Buades:2022lgf,Shaikh:2023ypz}.\footnote{
We use version \texttt{1.0.2} of the \texttt{gw\_eccentricity} package from the repository \url{https://github.com/vijayvarma392/gw_eccentricity}.
We note that the \emph{earliest} $ e _{ \text{gw}} $ value is not the \emph{initial} GW eccentricity of the waveform.
This is because $ e _{ \text{gw}} $ is calculated only at the times where periastron and apoastron frequency interpolants are defined, thus avoiding extrapolation.
Throughout this work, we always report the earliest $ e _{ \text{gw}} $ value as returned by \texttt{gw\_eccentricity}.
\label{fn:e_gw}
}
We provide a summary of each of these NR waveforms in an ancillary file~\cite{SEOBNRv6EHM:SupplementalMaterial}.\footnote{
In the ancillary file, we list the mass ratio $ q $, dimensionless spin components $ \chi _1 $ and $ \chi _2 $, the GW eccentricity $ e _{ \text{gw}} $ at the earliest $ (2,2) $-mode orbit-averaged dimensionless frequency $\langle M \omega_{22} \rangle$ calculated with \texttt{gw\_eccentricity}, number of periastron passages $ N_p $, and time to merger $ t _{ \text{merger}} / M$.
Additionally, we provide the optimum values of initial eccentricity $ e_0 $, initial dimensionless orbit-averaged orbital frequency $ \langle M \Omega \rangle_0 $, and maximum $ (2,2) $-mode mismatch $ \max_M \mathcal M_{22} $ (calculated over the range $ M \in [ 20, 200] \, \solarmass $), corresponding to the best-fitting template for each eccentric waveform model employed in this work.
\label{fn:ancillary}
}

An additional characterization of the eccentric NR waveforms employed in this work is given in Fig.~\ref{fig:mismatch_egw_frequency}.
For each NR waveform, we show the evolution of its mass-scaled (geometric) orbit-averaged $ (2,2) $-mode frequency $ \langle M f _{ 22} \rangle $ as a function of its GW eccentricity $ e _{ \text{gw}} $, and colored according to the value of the maximum $ (2,2) $-mode mismatch corresponding to the best-fitting \texttt{SEOBNRv6EHM} waveform (to be discussed below).
This figure allows one to estimate the eccentricities of the NR waveforms at a given physical frequency for a specific total mass.
For example, the dashed horizontal line indicates the value $ \langle M f_{22} \rangle  = 0.01 $ which is the one employed in the companion manuscript \cite{Pompili:2026yxq} for quoting the eccentricities in an injection-recovery study (see Fig.~3 therein).

We calculate mismatches against NR waveforms following the methods summarized in Sec.~\ref{sec:metrics}.
The three models, \texttt{SEOBNRv5EHM}, \texttt{SEOBNRv6EHM}, and \texttt{TEOBResumS-Dal\'i}, are treated equally to ensure a fair comparison.
The best-fitting parameters for each model and each eccentric NR simulation are provided in the ancillary file to this work (see footnote \ref{fn:ancillary}).

The results are presented in Fig.~\ref{fig:mismatch_spaghetti_ecc} for the three models in separate columns.
Each panel shows the mismatch as a function of the total mass ($ M \in  [20, \s 200] \, \solarmass $) corresponding to the best-fitting waveforms for each NR simulation.
The color of the curves corresponds to the initial GW eccentricity of the NR waveform, and we use dashed curves to indicate spinning configurations.
The top row shows the $ (2,2) $-mode mismatch [calculated with Eq.~(34) of Ref.~\cite{Gamboa:2024hli}], while the bottom row displays the sky-and-polarization-averaged, SNR-weighted mismatch (which includes higher-order modes) at a source inclination of $ \iota _{ \text{s}} = \pi/3 $ [calculated with Eqs.~(36) and (37) of Ref.~\cite{Gamboa:2024hli}].
In each panel, we overlay a horizontal, dot-dashed, magenta line indicating the median of the maximum mismatches, $ \text{med}[\max_M \mathcal{M}] $, obtained by first taking, for each NR waveform, the maximum mismatch over the given total mass interval, and then computing the median of these values over all the NR waveforms.

As a complement, we provide in Fig.~\ref{fig:mismatch_distributions_ecc} plots that summarize the statistical information about the mismatches for each model.
The top panels show the maximum, median, and minimum mismatches calculated for each NR waveform over the employed range of total masses, with the left panel showing the $ (2,2) $-mode mismatches, and the right panel displaying the SNR-weighted mismatches.
The bottom panels provide histograms for the maximum $ (2,2)$-mode (left) and maximum SNR-weighted (right) mismatches over the same range of total masses for all the NR waveforms, including an estimate of their numerical error computed as the mismatch between the maximum and second-maximum resolutions, when available.

The results shown in Figs.~\ref{fig:mismatch_spaghetti_ecc} and \ref{fig:mismatch_distributions_ecc} indicate that \texttt{SEOBNRv6EHM} is, overall, the most accurate model across the considered configurations.
For the $ (2,2) $-mode mismatches, the median is $ 0.029\% $ for \texttt{SEOBNRv5EHM}, $ 0.023\% $ for \texttt{SEOBNRv6EHM}, and $ 0.37\% $ for \texttt{TEOBResumS-Dal\'i}.
For the SNR-weighted mismatches, the median is $ 0.081\% $ for both \texttt{SEOBNRv5EHM} and  \texttt{SEOBNRv6EHM}, and $ 1.48\% $ for \texttt{TEOBResumS-Dal\'i}.
In the low eccentricity limit, the mismatches of \texttt{SEOBNRv6EHM} are comparable to those of \texttt{SEOBNRv5EHM}.
Since many of the employed NR waveforms have low eccentricities, the minimum, medium, and maximum mismatches across the considered range of total masses remain very similar between these models, for both the $ (2,2) $-mode and SNR-weighted mismatches (see Fig.~\ref{fig:mismatch_distributions_ecc}).
However, as eccentricity increases, \texttt{SEOBNRv5EHM} progressively loses accuracy, with mismatches rising to $\sim \nolinebreak 20\%$ at the highest eccentricities.
In contrast, \texttt{SEOBNRv6EHM} keeps the maximum mismatches below $ \sim 2\% $ for \emph{all} eccentric NR waveforms in the \texttt{SXS} catalog.
The difference in accuracy becomes even more pronounced when compared to \texttt{TEOBResumS-Dal\'i}, for which \texttt{SEOBNRv6EHM} achieves mismatches that are an order of magnitude smaller across all configurations.

Another aspect to note in Fig.~\ref{fig:mismatch_spaghetti_ecc} is that some spinning configurations are better modeled in \texttt{SEOBNRv5EHM}.
This behavior originates from the choice of the RR gauge (discussed in Sec.~\ref{sec:RR_gauge}).
During the development of \texttt{SEOBNRv6EHM}, we observed that using the \texttt{SEOBNRv5EHM} RR gauge (i.e., $ \alpha = -16/3 $ and $ \beta = -13/2 $) leads to the same improved mismatch behavior for these spinning configurations.
However, the mismatch for the highly eccentric configurations increases significantly (for example, the mismatch for the highly eccentric simulation \texttt{SXS:BBH:2527} goes from $ \sim \nolinebreak 2\% $ to $ \sim \nolinebreak 15\% $, compared to the $ \sim \nolinebreak 20\% $ of \texttt{SEOBNRv5EHM}).
Furthermore, as commented in Sec.~\ref{sec:RR_gauge}, the \texttt{SEOBNRv5EHM} gauge could lead to numerical problems with our resummation of the RR force~\eqref{eq:RR_force_proposal} (see Fig.~\ref{fig:ab_gauges}).
For these reasons, and considering that the mismatches of \texttt{SEOBNRv6EHM} for the most demanding spinning configurations are still within the $ \sim \nolinebreak 2\% $ level, we decided to keep a different RR gauge (namely, $ \alpha = -2 $ and $ \beta = -1 $).
We additionally note that this behavior is not related to the QC spinning calibration.
For instance, one of the spinning simulations with high mismatch in \texttt{SEOBNRv6EHM} 
has a mass ratio $ q = 2 $, dimensionless spins $ \chi_1 = 0.8 $ and $ \chi_2 = 0.8 $, and initial GW eccentricity $ e _{ \text{GW}} =  0.48 $.
The QC counterpart \texttt{SXS:BBH:0618} has the same intrinsic parameters, but the maximum mismatch for \texttt{SEOBNRv6EHM} is $\sim 0.073 \%$, while \texttt{SEOBNRv5EHM} has $ \sim 0.084 \%$ (i.e., \texttt{SEOBNRv6EHM} models slightly better this QC configuration).
Overall, these results point to the need for further investigation into the impact of the RR gauge on the accuracy of EOB models.

\begin{figure*}
\hspace{-5pt}
\includegraphics[width=0.93\linewidth]{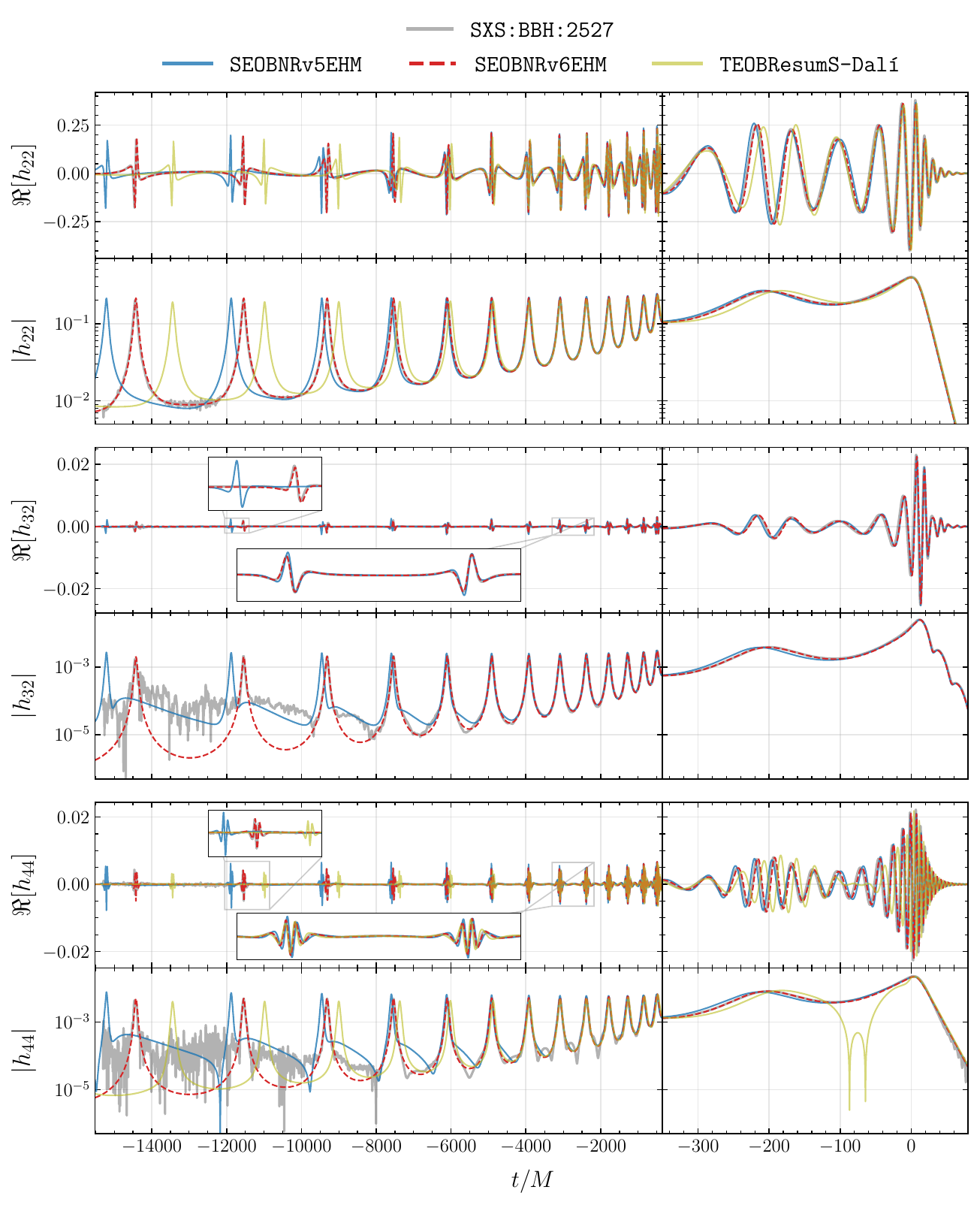}
\vspace{-10pt}
\caption{
Real parts and amplitudes of the waveform modes for a highly eccentric, equal-mass, nonspinning BBH, shown in geometric units.
We plot the waveform modes of the NR simulation \texttt{SXS:BBH:2527} (gray) together with the best-fitting \texttt{SEOBNRv5EHM} (blue), \texttt{SEOBNRv6EHM}~(red, dashed), and \texttt{TEOBResumS-Dal\'i} (yellow) waveforms.
The NR waveform has a GW eccentricity of $ e _{ \text{gw}} \approx 0.88 $ at a dimensionless orbit-averaged $ (2,2) $-mode frequency of $ \langle M \omega_{22} \rangle = 0.0069 $ (corresponding to the second apoastron passage at $ t/M \approx -10500 $), and the best-fit parameters for each model are provided in the ancillary file.
Only the nonvanishing $(\ell, m)$ modes supported by the models are included (the $(3,2)$ mode of \texttt{TEOBResumS-Dal\'i} is not reviewed, see footnote \ref{fn:dali}).
Zoomed insets of the $(3,2)$ and $(4,4)$ modes are shown for clarity.
NR modes are plotted from the relaxation time onward, yet significant noise is present in the higher-order modes.
}
\label{fig:sxs_2527}
\end{figure*}

We exemplify in Fig.~\ref{fig:sxs_2527} the accuracy of \texttt{SEOBNRv6EHM} when tested against the most eccentric simulation in the catalog, \texttt{SXS:BBH:2527}.
This corresponds to an equal-mass, nonspinning binary with a GW eccentricity of $ e _{ \text{gw}} \approx 0.88 $ at a dimensionless orbit-averaged $ (2,2) $-mode frequency of $ \langle M \omega_{22} \rangle = 0.0069 $, corresponding to a time near the second apoastron passage, approximately $t/M \sim -10500$ before merger.
We display the real part and amplitude of the nonvanishing multipoles $\{ (2,2), (3,2), (4,4) \} $ for the best-fitting waveforms of the three models under consideration (the corresponding input parameters are given in the ancillary file; see footnote \ref{fn:ancillary}).
For a total mass of $20 \,\solarmass$, \texttt{SEOBNRv6EHM} yields a mismatch of $\mathcal{M}_{22} = 2.1\%$, while \texttt{SEOBNRv5EHM} and \texttt{TEOBResumS-Dal\'i} give significantly larger values of $\mathcal{M}_{22} = 19.5\%$ and $\mathcal{M}_{22} = 30.6\%$, respectively.
\texttt{SEOBNRv6EHM} is therefore able to accurately capture the bursts of radiation at each periastron passage, where the relative velocity reaches up to $ 40\% $ of the speed of light (see Fig.~\ref{fig:NR_dynamics}).
In contrast, both \texttt{SEOBNRv5EHM} and \texttt{TEOBResumS-Dal\'i} predict a less faithful evolution, characterized by wrong intervals of time between each periastron passage.
Moreover, these two models exhibit unphysical features in the higher-order modes:
\texttt{SEOBNRv5EHM} shows an incorrect amplitude behavior during the early inspiral, while \texttt{TEOBResumS-Dal\'i} develops zero-crossings in the $ (4,4) $ mode near merger.
In comparison, the higher-order modes of \texttt{SEOBNRv6EHM} display a well-behaved evolution, even qualitatively outperforming the NR modes, which are affected by numerical noise due to their small amplitudes.

The accurate modeling of highly eccentric binaries is enabled thanks to the new analytical ingredients employed in \texttt{SEOBNRv6EHM}.
Particularly, the novel resummation of the RR force, the choice of the RR gauge, and the use of a suitable parametrization, played a key role in achieving small mismatches.
\texttt{SEOBNRv6EHM} employs 1PN eccentricity corrections to the RR force and waveform modes, compared to the 3PN expressions used in \texttt{SEOBNRv5EHM}.
Despite this, \texttt{SEOBNRv6EHM} yields a clear improvement for high eccentricities.
As discussed in Sec.~\ref{sec:RR_v6}, better resummations and parametrizations are needed to include higher PN orders in a way which benefits the model accuracy.

Apart from the RR force, the modeling of higher-order modes also requires improvement in future developments.
When transitioning from the $ (2,2) $-mode metric to the SNR-weighted metric, the median of maximum mismatches increases from $ 0.030\% $ to $ 0.081\% $ (a factor of $ \sim 2.7 $) for \texttt{SEOBNRv5EHM}, from $ 0.023\% $ to $ 0.084\% $ (a factor of $ \sim 3.6 $) for \texttt{SEOBNRv6EHM}, and from $ 0.37\% $ to $ 1.48\% $ (a factor of $ \sim 4 $) for \texttt{TEOBResumS-Dal\'i}.
Since the degradation of mismatches when including higher-order modes is larger in \texttt{SEOBNRv6EHM} than in \texttt{SEOBNRv5EHM}, this suggests that the prescription for higher-order modes in \texttt{SEOBNRv6EHM} may not be the most adequate.
As discussed in Secs.~\ref{sec:factorized_modes} and \ref{sec:comparison_qc}, there is a significant freedom in how the higher-order modes are modeled;
future work could focus on studying more accurate prescriptions.
However, we emphasize that the improvement for large eccentricities is also observed when higher-order modes are included.
Thus, \texttt{SEOBNRv6EHM} provides an overall better approximation for the full GW signal of eccentric binaries.

\begin{figure}
\hspace*{-5pt}
\includegraphics[width=\linewidth]{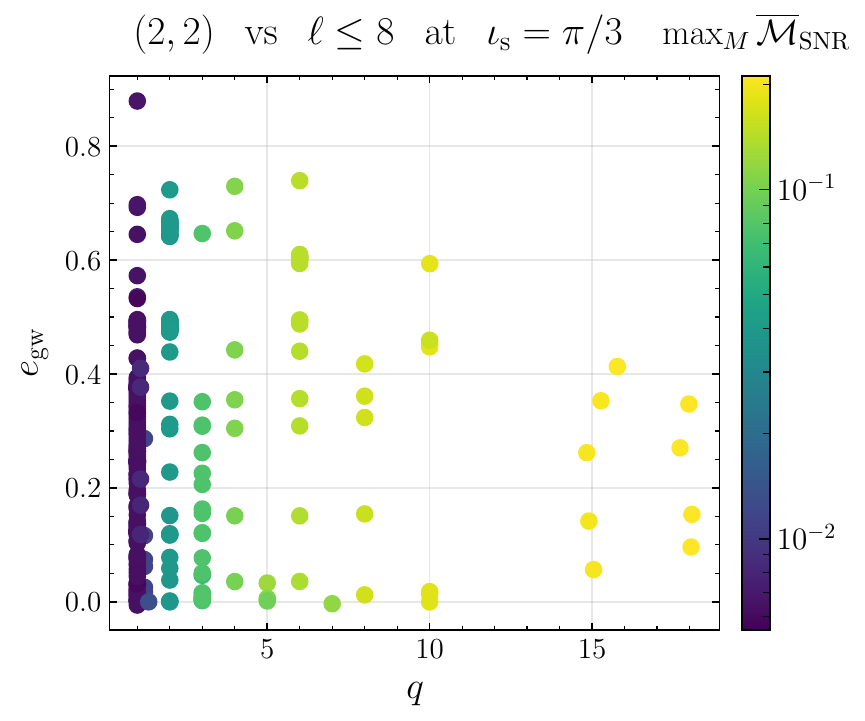}
\vspace{-8pt}
\caption{
Mismatch between NR waveforms with only the $(2,2)$ mode and those including all modes up to $\ell = 8$, for the 319 eccentric \texttt{SXS} waveforms used in this work, shown as a function of the mass ratio $ q $ and earliest GW eccentricity $ e _{ \text{gw}} $ measured by \texttt{gw\_eccentricity}.
The color indicates the sky-and-polarization-averaged SNR-weighted mismatch maximized over the total mass range $ M \in [20, 200] \s \solarmass $ for an inclination $\iota _{ \text{s}} = \pi/3$.
}
\label{fig:param_space_NR_ecc_sims_22_HM}
\end{figure}

The contribution of higher-order modes to the full waveform is, however, independent of eccentricity.
Figure~\ref{fig:param_space_NR_ecc_sims_22_HM}, shows the sky-and-polarization-averaged SNR-weighted mismatch between NR waveforms including only the $(2,2)$ mode and those including all modes up to $\ell = 8$, for an inclination $\iota _{ \text{s}} = \pi/3$, as a function of the mass ratio and initial GW eccentricity (only NR waveforms are used to disentangle the results from inaccuracies in waveform models).
It is well known that large mass ratios and non-negligible inclinations enhance the contributions of higher-order modes in QC binaries (e.g.,~\cite{Cotesta:2018fcv}).
This is also the case for eccentric binaries, as shown in Fig.~\ref{fig:param_space_NR_ecc_sims_22_HM}, where we observe a baseline mismatch of $ \sim 1\% $, increasing with mass ratio up to $ \sim 20\% $.
However, the mismatch shows no dependence on eccentricity.\footnote{
See also Fig.~11 of Ref.~\cite{Islam:2026blk} for analogous results restricted to certain modes up to $ \ell = 5 $ and mass ratios up to $ q = 4 $, obtained with a waveform approximant constructed from a NR surrogate model.
Additionally, the impact of neglecting higher-order modes in eccentric parameter-estimation analyses has recently been explored in Ref.~\cite{Tang:2026jvl}, where high mass ratios and inclinations lead to larger biases.
The role of eccentricity, however, is less clear and requires further investigation.
} 
Indeed, orbital eccentricity does not introduce asymmetries on the sphere $ (\iota, \varphi) $ in which $ h_+ $ and $ h_\times $ are expanded in spherical-harmonic modes $ h_{\ell m} $.
Rather, eccentric binaries introduce an additional timescale characterized by the radial oscillations.\footnote{
The higher-order modes of waveform decompositions based on \emph{radial frequencies} do become more important as eccentricity increases.
This is the case for the typical decompositions employed in extreme-mass-ratio eccentric waveform models (e.g., see Fig.~4 in Ref.~\cite{Chapman-Bird:2025xtd}).
}
Therefore, the increase in mismatch with eccentricity observed in the bottom panels of Fig.~\ref{fig:mismatch_spaghetti_ecc} is entirely driven by inaccuracies in the binary phasing, rather than by the impact of higher-order modes or the absence of additional multipoles.
These results further suggest that improvements in the modeling of higher-order modes for QC binaries will be beneficial regardless of the eccentricity.

To complement this discussion about higher-order modes, we provide in the bottom panel of Fig.~\ref{fig:mismatch_distributions_m0} histograms of maximum \texttt{SEOBNRv6EHM} SNR-weighted mismatches for all the eccentric NR waveforms with (dark purple bins) and without (green bins) the $ m = 0 $ multipoles, for $ \ell \leq 4 $.
We also display the estimated NR errors computed as the SNR-weighted mismatches between the highest and second-highest resolutions (always including the $ m =0 $ modes), when available.
Similar to the case of QC binaries discussed in Sec.~\ref{sec:comparison_qc}, the inclusion of NR $ m = 0 $ modes degrades the comparisons against NR, but the highest-mismatch cases are dominated by other sources of error (e.g., large spins, large eccentricities), rather than the absence of $ m=0 $ modes in \texttt{SEOBNRv6EHM}.
Although other sources of error (e.g., inaccuracies in the RR force) currently play a more significant role in limiting the accuracy for moderate-to-high eccentricities, low-eccentricity systems would immediately benefit from the modeling of $ m = 0 $ modes.
We leave this for future work.

We conclude by summarizing our results as follows:
\begin{itemize}
\item
\texttt{SEOBNRv6EHM} matches the accuracy of \texttt{SEOBNRv5EHM} at low eccentricities, while at high eccentricities it can be up to an order of magnitude more accurate.
\item
\texttt{SEOBNRv6EHM} is overall an order of magnitude more accurate than \texttt{TEOBResumS-Dal\'i} for all eccentricities.
\item
The improvements of \texttt{SEOBNRv6EHM} at high eccentricities come solely from new analytical results, and not from calibrations to eccentric NR simulations.\footnote{
While no explicit numerical calibration is performed, it is important to note that different analytical choices were made based on comparisons against eccentric NR waveforms.
}
\item
The RR gauge adopted in \texttt{SEOBNRv6EHM} is key to its accuracy at high eccentricities, but, compared to \texttt{SEOBNRv5EHM}, degrades the accuracy for some spinning configurations.
However, even in the most demanding spinning cases, the mismatches of \texttt{SEOBNRv6EHM} remain within $ \sim 2\% $.
\item
When including higher-order modes, the median of the maximum mismatches of \texttt{SEOBNRv6EHM} increases by a factor of $ \sim 3.6 $ with respect to the $ (2,2) $ mode, a larger degradation than the factor of $ \sim 2.7 $ observed for \texttt{SEOBNRv5EHM} (and smaller than the factor of $ \sim 4 $ for \texttt{TEOBResumS-Dal\'i}), suggesting that alternative prescriptions for higher-order modes should be explored.
\end{itemize}
%


\subsubsection{Generic planar binaries}
\label{sec:comparison_gen}

Modeling the GW emission from dynamically captured binaries or from scattering encounters poses a great challenge, since they probe higher velocities than QC or moderately-eccentric systems, as observed in the dynamics from NR simulations shown Fig.~\ref{fig:NR_dynamics}.
The relative velocity of the considered dynamically captured binary (violet) reaches $ \sim 55\% $ the speed of light during the first and second periastron passages, at which the binary's separation is $ \sim 5 M $.
Similarly, the relative velocity of the shown scattered binary (orange) reaches $ \sim 65\% $ the speed of light during the encounter at a separation of $ \sim 5 M $.
By contrast, the QC and eccentric (bound) systems reach $ \sim 60\% $ the speed of light \emph{only} at the end of the plunge.

To quantify the accuracy of \texttt{SEOBNRv6EHM} for generic planar-orbit configurations, we compare with NR simulations.
Specifically, we compare with the NR waveforms from a dynamical capture (\texttt{SXS:BBH:4000}), a $ 90^\circ $ scattering encounter (\texttt{SXS:BBH:3999}), and a $ 180^\circ $ scattering encounter (\texttt{SXS:BBH:4292}), all publicly-available, equal-mass, nonspinning simulations from the \texttt{SXS} catalog \cite{Scheel:2025jct}.
Additionally, we employ data (predicted scattering angles) from Ref.~\cite{Long:2025nmj}, comprising 61 \texttt{SXS} NR scattering encounters spanning a wide range of binary configurations, with varying energy, angular momentum, spins, and mass ratios.
In all the comparisons, we also include the predictions from the \texttt{TEOBResumS-Dal\'i} model \cite{Nagar:2024dzj}.

The waveforms comparisons are provided in Fig.~\ref{fig:sxs_scatters}.
The first set of panels (left column of the page) is dedicated to the dynamical capture, while the second and third sets of panels (right column of the page) focus on the $ \sim 90^\circ $ (top) and $ \sim 180^\circ $ (bottom) scattering encounters.
In each set of panels, we display the real part and amplitude of the non-vanishing modes, for the corresponding NR simulation (gray), along with the \texttt{SEOBNRv6EHM} (red) and \texttt{TEOBResumS-Dal\'i} (yellow) predictions.
To generate the waveforms, both models are initialized with the ADM energy and total angular momentum from the NR metadata, at an initial separation of $ 10^3 M $.

\begin{figure}
\hspace{-6pt}
\includegraphics[width=0.942\linewidth]{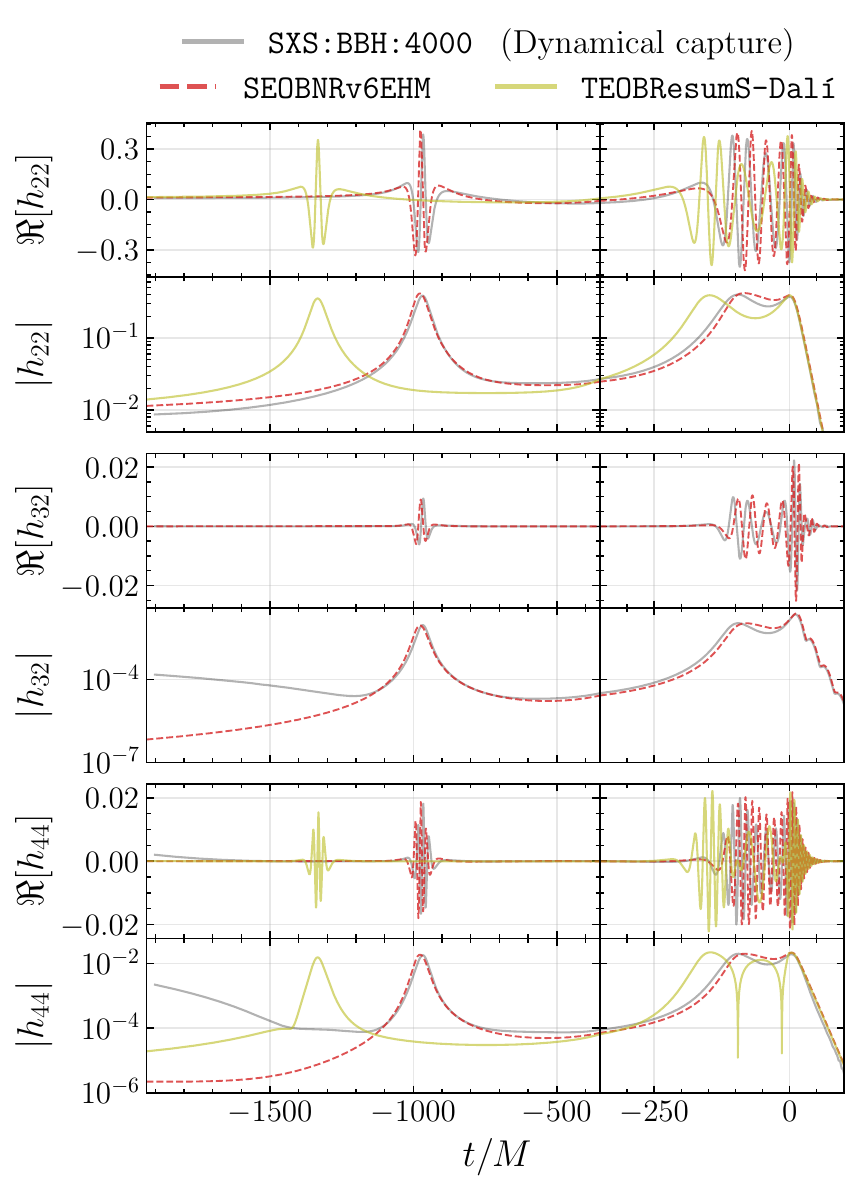}
\vspace{-5pt}
\caption{
Real part and amplitude of the waveform modes for an equal-mass, nonspinning BBH in different configurations:
a dynamical capture with $ E/M = 1.001 $ and $ L/(\mu M) = 4.02 $ (left panels), a scattering encounter of $ \sim 90^\circ $ with $ E/M = 1.02264 $ and $ L/(\mu M) = 6.4 $ (top-right panels), and a scattering encounter of $ \sim \! 180^\circ $ with $ E/M = 1.02264 $ and $ L/(\mu M) = 4.8 $ (bottom-right panels), all shown in geometric units.
For each case, we show a \texttt{SXS} NR waveform (gray) alongside the corresponding \texttt{SEOBNRv6EHM} (red) and \texttt{TEOBResumS-Dal\'i} (yellow) predictions.
Only the nonvanishing $(\ell, m)$ modes supported by the models are shown (the $(3,2)$ mode of \texttt{TEOBResumS-Dal\'i} is not reviewed, see footnote \ref{fn:dali}).
Both models are initialized with the same energy and angular momentum as the corresponding NR simulation, at an initial separation of $r = 10^3 M$.
}
\label{fig:sxs_4000}
\end{figure}
\begin{figure}\ContinuedFloat
\hspace{-11pt}
\vspace{9pt}
\includegraphics[width=0.96\linewidth]{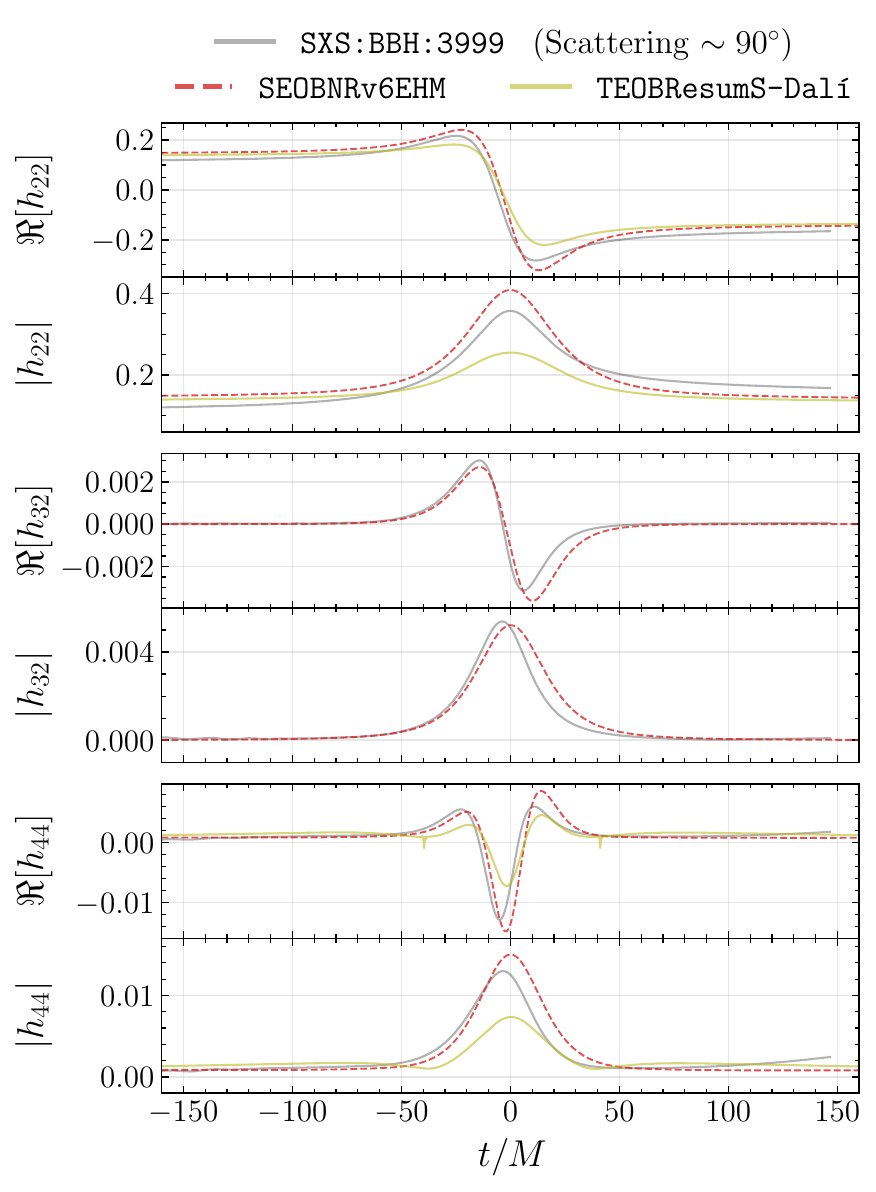}
\includegraphics[width=0.953\linewidth]{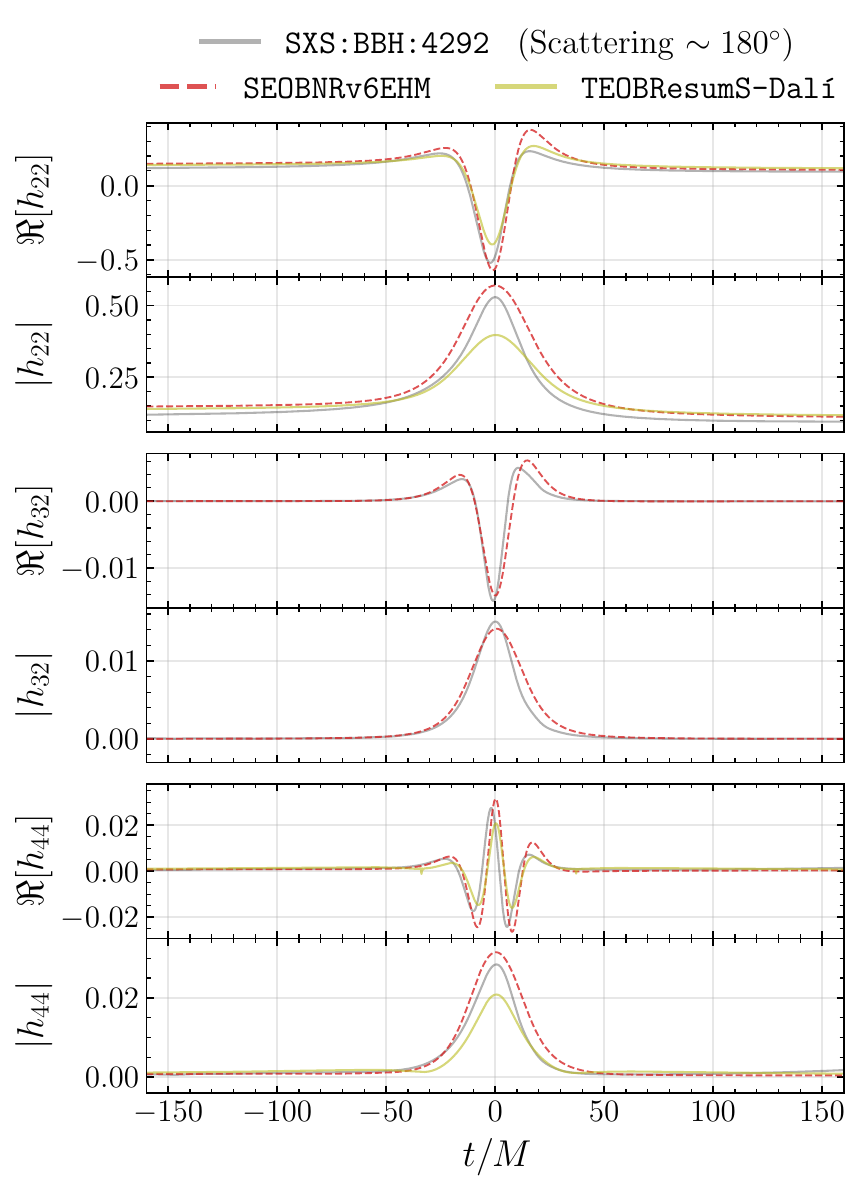}
\vspace{-5pt}
\caption{
\emph{(Continued)}
}
\label{fig:sxs_scatters}
\end{figure}
\begin{figure*}
\centering
\hspace{-8pt}
\subfloat{
\includegraphics[width=0.48\linewidth]{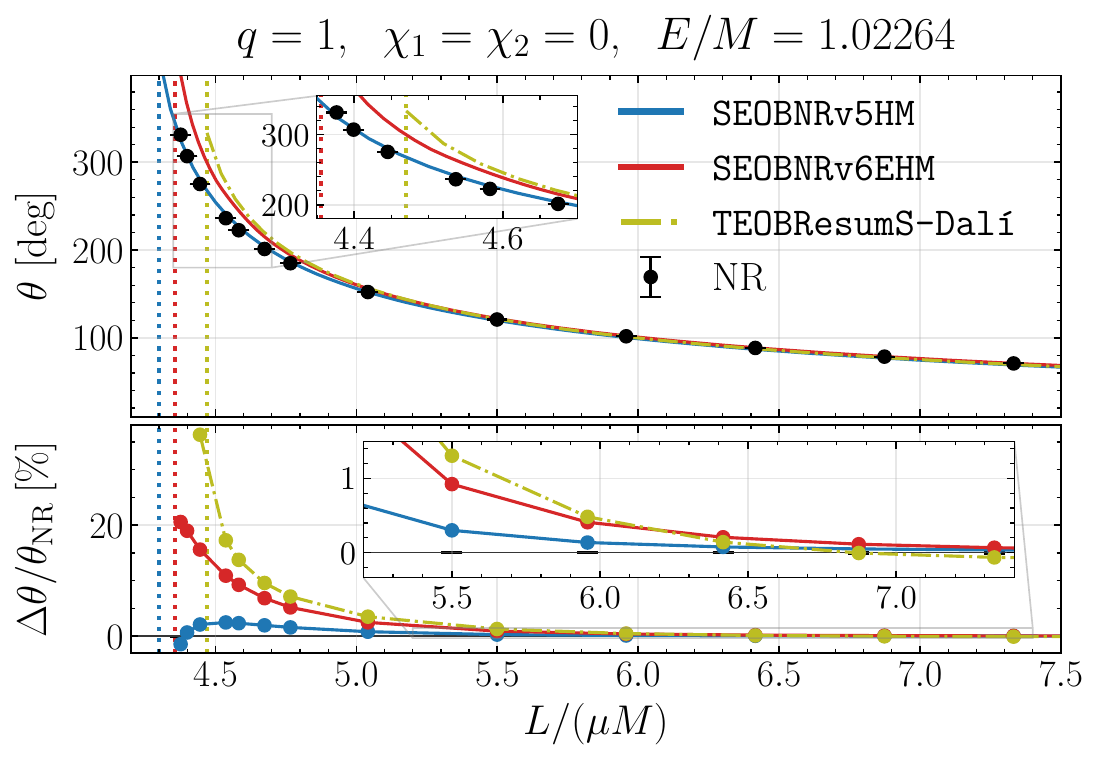}
}
\hspace{12pt}
\subfloat{
\includegraphics[width=0.48\linewidth]{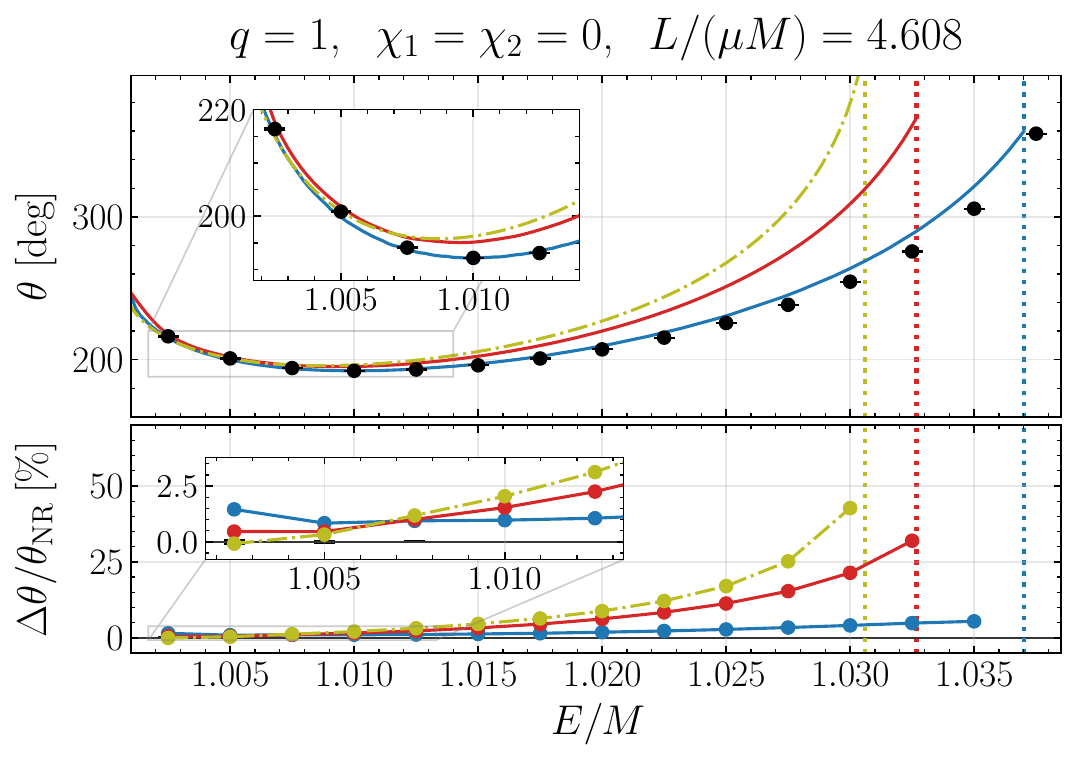}
}

\hspace{-8pt}
\subfloat{
\includegraphics[width=0.48\linewidth]{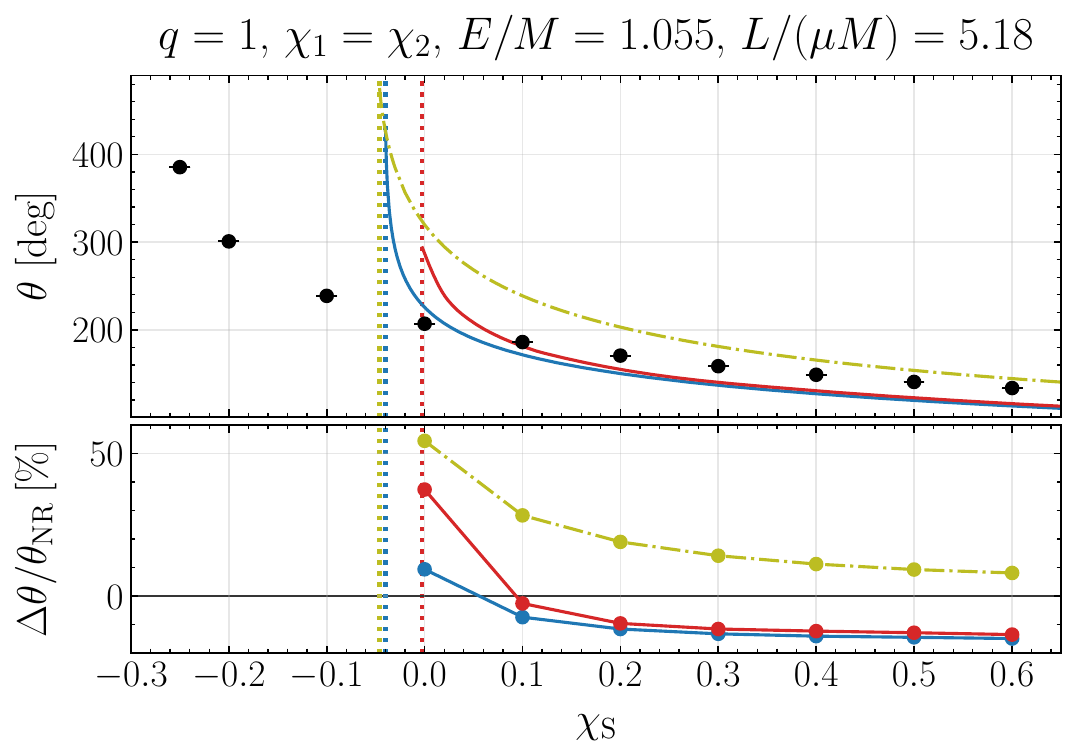}  
}
\hspace{12pt}
\subfloat{
\includegraphics[width=0.48\linewidth]{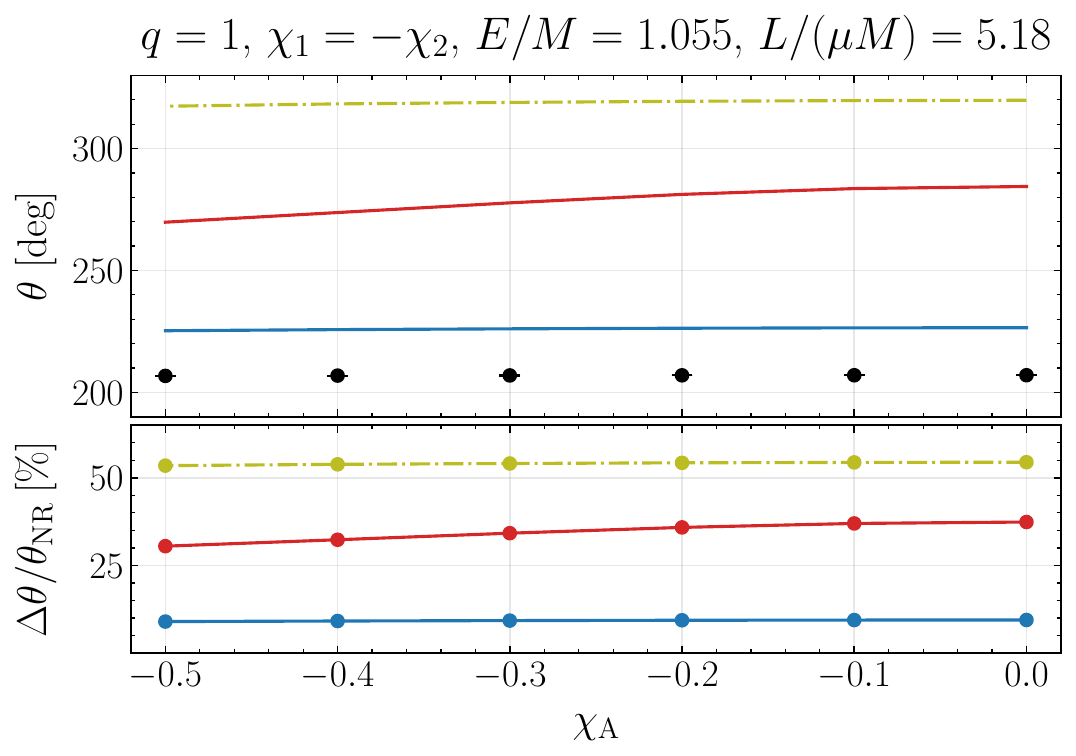}  
}

\hspace{-8pt}
\subfloat{
\includegraphics[width=0.48\linewidth]{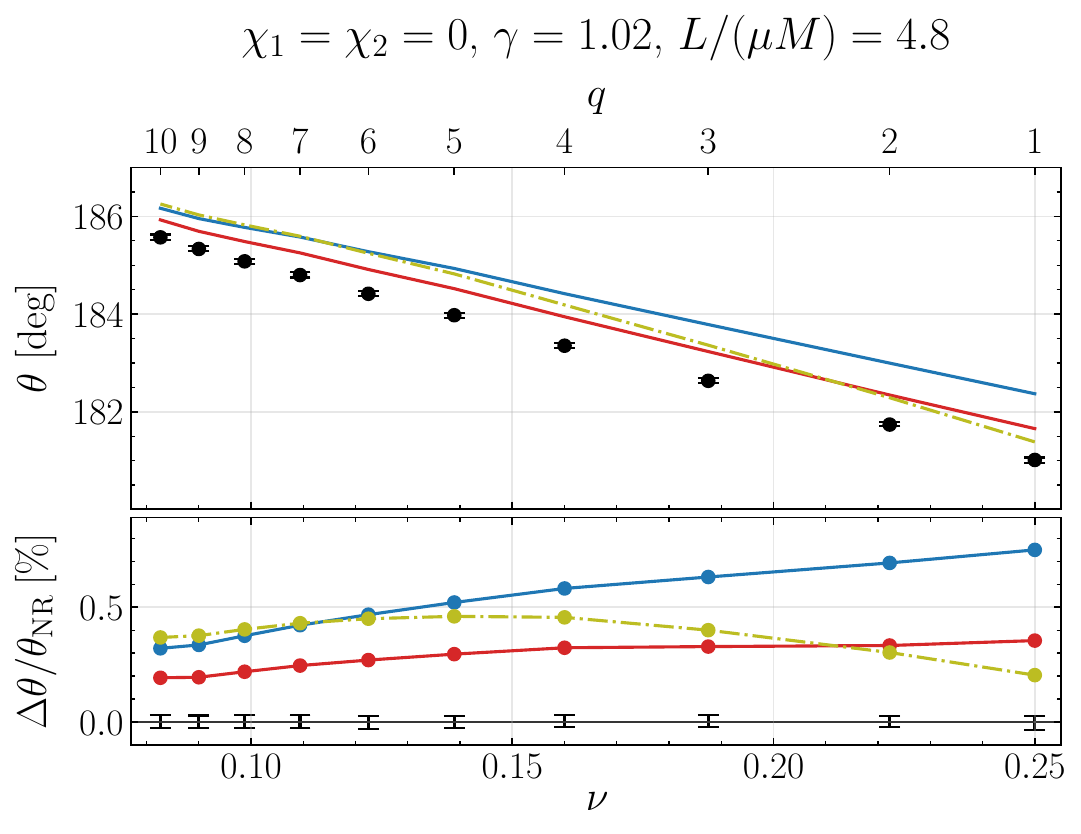}  
}
\hspace{12pt}
\subfloat{
\includegraphics[width=0.48\linewidth]{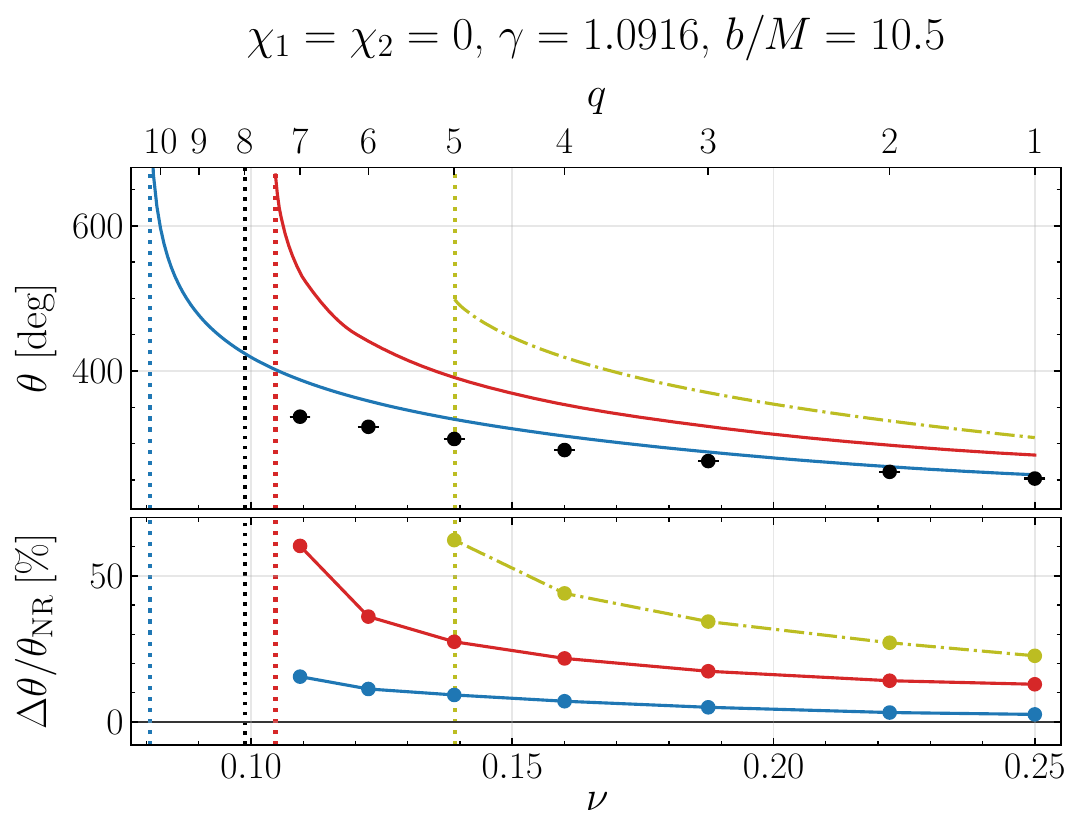}  
}
\caption{
Scattering angles $ \theta $ for six sets of binary configurations as functions of mass ratio, spins, energy $ E/M $, reduced orbital angular momentum $ L/(\mu M) $, effective energy $ \gamma $, and impact parameter $ b/M $.
Data are extracted from NR simulations \cite{Long:2025nmj} (black dots with error bars) and compared with the predictions of \texttt{SEOBNRv5HM} (blue), \texttt{SEOBNRv6EHM} (red), and \texttt{TEOBResumS-Dal\'i} (yellow);
vertical dotted lines mark the first known dynamical capture for each model.
In each panel, the upper plot shows $\theta$, while the lower plot displays the relative difference with respect to NR;
zoomed insets are included where useful for clarity.
From top left to bottom right:
equal-mass, nonspinning binaries varying $ L/(\mu M) $ at $ E/M = 1.02264 $;
equal-mass, nonspinning binaries varying $ E/M $ at $ L/(\mu M) = 4.608 $;
equal-mass, equal-spin binaries varying $\chi_\text{S}$ at $ E/M = 1.055 $ and $ L/(\mu M) = 5.18 $;
equal-mass, opposite-spin binaries varying $ \chi_\text{A} $ at $ E/M = 1.055 $ and $ L/(\mu M) = 5.18 $;
nonspinning binaries varying mass ratio at $ \gamma =1.02 $ and $ L/(\mu M) = 4.8 $;
and nonspinning binaries varying mass ratio at $\gamma = 1.0916$ and $b/M = 10.5$.
}
\label{fig:scattering_angles}
\end{figure*}

The results in Fig.~\ref{fig:sxs_scatters} suggest a better qualitative performance of \texttt{SEOBNRv6EHM} with respect to \texttt{TEOBResumS-Dal\'i}.
For the dynamical capture case, we find that \texttt{SEOBNRv6EHM} yields better predictions for the time between periastron passages and the merger.
This results in an improved overall agreement with the NR waveforms, including the higher-order modes.
We note, however, that the NR higher-order modes exhibit unphysical features in the amplitude at early times, associated with the waveform extrapolation procedure used in the simulation.
In contrast, the higher-order modes of \texttt{SEOBNRv6EHM} show a smooth decay at early times and remain qualitatively consistent with the NR results close to merger, even reproducing the postmerger oscillations of the $(3,2)$ mode, which arise from the mode mixing between spherical and spheroidal harmonics~\cite{Buonanno:2006ui, Kelly:2012nd} and are modeled following the \texttt{SEOBNRv5HM} prescription (see Sec.~\ref{sec:mrd}).
This is not the case for \texttt{TEOBResumS-Dal\'i}, where we see that the $ (4,4) $ mode displays nonsmooth behavior around $ t = -1500 M$, as well as amplitude zero crossings near merger.
For scattering waveforms, \texttt{TEOBResumS-Dal\'i} also develops unphysical features in the higher-order modes around $ t \approx \pm 40 M $, whereas \texttt{SEOBNRv6EHM} reproduces more closely the NR waveforms.

The predicted scattering angles for a wide range of binary configurations are shown in Fig.~\ref{fig:scattering_angles}.
Each panel shows the measured scattering angles from NR simulations (black dots with error bars), along with the predictions of \texttt{SEOBNRv6EHM} and \texttt{TEOBResumS-Dal\'i}.
Additionally, we include the predictions by the \texttt{SEOBNRv5HM} model \cite{Pompiliv5} when given generic planar-orbit initial conditions \cite{Long:2025nmj}.
The models are initialized with the ADM energy and total angular momentum from the NR metadata at an initial separation of $10^6 M$, to ensure smaller errors in the EOB angles than in the NR data.
The scattering angle is computed as $ \theta = \phi _{ \text{f}} - \phi _{ 0} - \pi $, where $ \phi _{ \text{f}} $ and $ \phi _{ 0} $ are the final and initial azimuthal orbital angles, respectively, returned by the models.

The results shown in Fig.~\ref{fig:scattering_angles} indicate that \texttt{SEOBNRv6EHM} provides slightly more accurate predictions of the scattering angle than \texttt{TEOBResumS-Dal\'i}.
In most cases, the scattering angles and predicted captures (dotted vertical lines) from \texttt{SEOBNRv6EHM} are closer to the NR data.
However, this improvement is less pronounced than in QC and eccentric (bound) configurations.
Remarkably, the best performance is achieved by \texttt{SEOBNRv5HM}, as already reported in Ref.~\cite{Long:2025nmj}.
\texttt{SEOBNRv5HM} shares the same Hamiltonian structure as \texttt{SEOBNRv6EHM}, but has different calibration parameters and employs a QC RR force (which is a function of the instantaneous $ \Omega $ and $ H _{ \text{EOB}} $ and thus can be numerically evaluated on a generic-orbit dynamics).
One might naively expect that employing an eccentric RR force would improve the modeling of scattering encounters, but this is not the case (at least not with our chosen resummation).
During the development of \texttt{SEOBNRv6EHM}, we observed that the 0PN and 3PN eccentricity corrections to the RR force yield more accurate scattering angles than the 1PN corrections (see Sec.~\ref{sec:RR_v6}).
However, both choices degrade the waveform accuracy for bound systems compared to the 1PN corrections adopted in the model.

Taken together, our findings for the generic planar-orbit waveforms and scattering angles underscore the need for further investigation into alternative resummations of the waveform modes and the RR force that accurately apply to both bound and unbound binaries.
This, together with a more systematic investigation of model accuracy for unbound orbits, is left for future work.
For now, conclusions drawn from generic planar-orbit results obtained with \texttt{SEOBNRv6EHM} and \texttt{TEOBResumS-Dal\'i} should be interpreted with caution.

We conclude that \texttt{SEOBNRv6EHM} demonstrates a strong performance across all configurations, marking an important step toward accurate models for generic-orbit binaries.


\subsection{Computational performance}
\label{sec:comp_perf}

Waveform generation in eccentric models is computationally more expensive than their QC counterparts due to the extra equations and functions required to model eccentricity.
Furthermore, eccentric binaries depend on two extra parameters with respect to QC binaries (e.g., eccentricity and a radial phase).
Thus, strategies and codes that reduce the overall computational cost of eccentric models need to be developed.

\texttt{SEOBNRv6EHM} is implemented within the high-performance and flexible \texttt{pySEOBNR} Python package \cite{Mihaylovv5}, which has been optimized to reduce waveform generation time.
The optimizations consisted mainly of reducing redundancy in the original code (as used in \texttt{SEOBNRv5} models) and using \texttt{Cython} \cite{behnel2011cython} to speed up the evaluation of cost-intensive expressions.
Furthermore, \texttt{SEOBNRv6EHM} evolves only four equations of motion \eqref{eq:EOM} and employs compact expressions for the RR force and modes, compared to \texttt{SEOBNRv5EHM}, which evolves six equations of motion and uses longer expressions.

\begin{figure}
\hspace{-5pt}
\includegraphics[width=\linewidth]{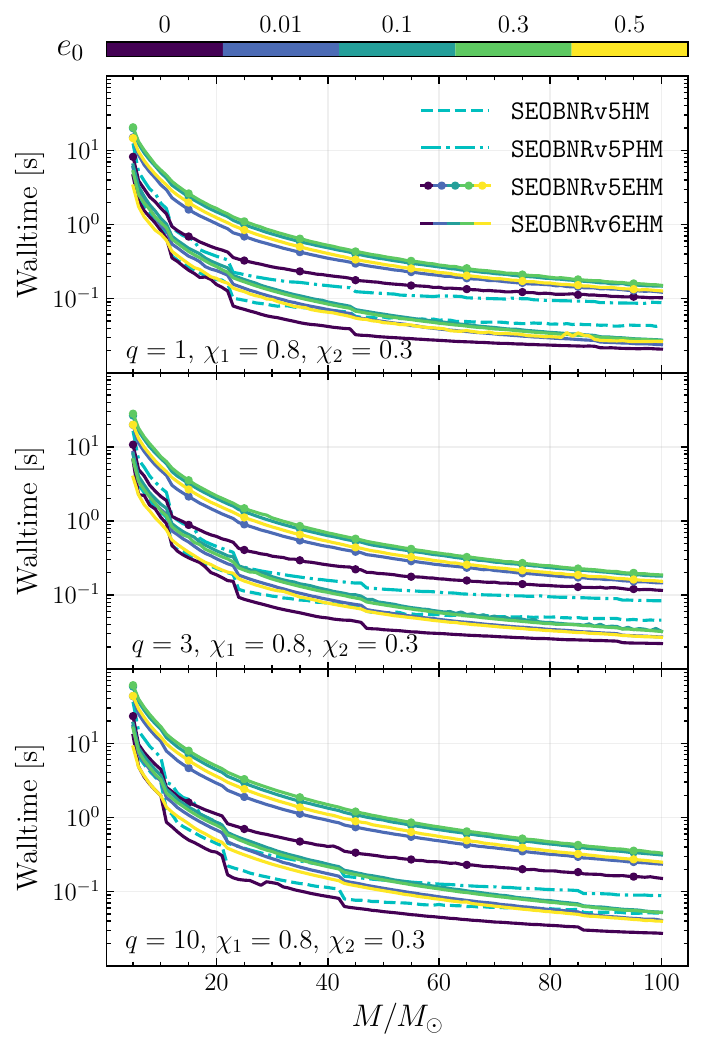}
\vspace{-7pt}
\caption{
Walltime for waveform generation of \texttt{SEOBNRv5HM} (dashed), \texttt{SEOBNRv5PHM} (dash-dotted), \texttt{SEOBNRv5EHM} (solid-dotted), and \texttt{SEOBNRv6EHM} (solid), as function of the total mass $ M \in \nolinebreak {[5, \, 100]\, \solarmass} $.
The systems are characterized by a starting orbit-averaged frequency $ \langle f _{ \text{start}} \rangle = 10\, $Hz, dimensionless spin components $ \chi_{1} = 0.8 $ and $ \chi_{2} = 0.3 $, and three different mass ratios $ q \in \{1, 3, 10 \} $.
For \texttt{SEOBNRv5EHM} and \texttt{SEOBNRv6EHM}, the systems are initialized at apastron ($ \zeta_0 = \pi $) with different starting eccentricities $ e_0 \in \{0, 0.01, 0.1, 0.3, 0.5\} $, each represented by a different color.
}
\label{fig:benchmarks_seobnr}
\end{figure}
\begin{figure}
\hspace{-5pt}
\vspace{5pt}
\includegraphics[width=\linewidth]{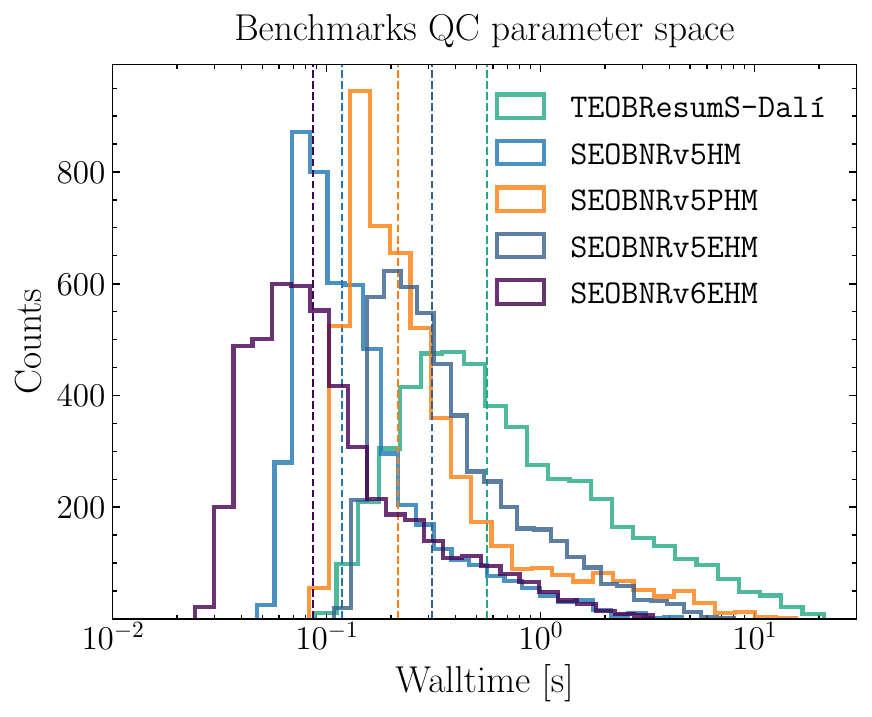}
\includegraphics[width=\linewidth]{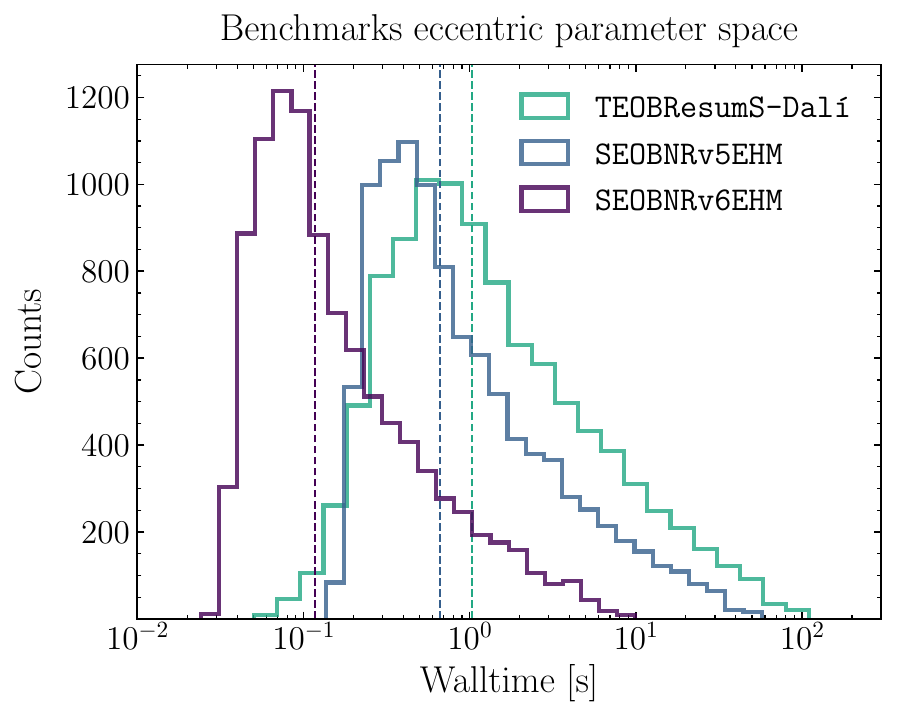}
\vspace{-15pt}
\caption{
Histograms of waveform-evaluation walltimes for different models and for randomly distributed QC (top panel) and eccentric (bottom panel) binary configurations, initialized at an orbit-averaged frequency of $ \langle f _{ \text{start}} \rangle = 10\, $Hz with a fixed sampling rate of $ 2048 \,$Hz.
For the QC parameter space, we sample $5000$ points with $ q \in [1,10] $, $ M \in [5,100]\,\solarmass $, and $ |\chi_i| \leq 0.9 $ ($ i \in \{1,2\} $, aligned-spins). For the eccentric parameter space, we sample $10000$ points with the same mass and spin ranges, and with $ e \in [0,0.5] $ and $ \zeta \in [0,2\pi] $.
The vertical dashed lines indicate the median of the corresponding histogram.
}
\label{fig:benchmarks_param_space}
\end{figure}

Here, we quantify the speed of \texttt{SEOBNRv6EHM} and compare it with other state-of-the-art models across various binary configurations typically expected for current LVK detectors.
For all the models below, we include all their supported modes up to $ \ell = 4 $, with the remaining settings kept as default.
The benchmarks were performed with the \texttt{Hypatia} computer cluster at the Max Planck Institute for Gravitational Physics in Potsdam, on a compute node equipped with a dual-socket 64-core AMD EPYC (Rome) 7742 CPU.

In Fig.~\ref{fig:benchmarks_seobnr}, we show the walltimes for waveform generation (time-domain GW polarizations) as a function of the binary's total mass $ M \in [5, \, 100]\, \solarmass$ for three mass ratios $ q \in \{ 1, 3, 10\} $, and for four waveform approximants:
the QC aligned-spin \texttt{SEOBNRv5HM} model \cite{Pompiliv5}, the QC precessing-spin \texttt{SEOBNRv5PHM} model \cite{RamosBuadesv5}, and the eccentric aligned-spin models \texttt{SEOBNRv5EHM} \cite{Gamboa:2024hli} and \texttt{SEOBNRv6EHM}.
The considered configurations correspond to aligned-spin binaries with dimensionless spin components $ \chi_{1} = 0.8 $ and $ \chi_{2} = 0.3 $, starting at an orbit-averaged frequency $ \langle f _{ \text{start}} \rangle = 10\, $Hz.
For the eccentric models, the binaries are initialized with different eccentricities $ e_0 \in\nolinebreak \{0, 0.01, 0.1, 0.3, 0.5 \} $ at apastron $ \zeta_0 = \pi $.
For each total mass, we choose a sampling rate that satisfies the Nyquist criterion for the $ \ell = 4 $ multipoles.
This is reflected as discontinuities in the curves shown in Fig.~\ref{fig:benchmarks_seobnr}, since smaller sampling rates lead to faster waveform generations.
For systems with $ e_0 > 0 $, the discontinuities are less pronounced, as the computational cost of modeling eccentricity dominates over the differences caused by the change in sampling rate.
Related to this, for both \texttt{SEOBNRv5EHM} and \texttt{SEOBNRv6EHM}, we observe in Fig.~\ref{fig:benchmarks_seobnr} an increase of walltimes as the eccentricity is augmented, until a certain point at which a further increase of eccentricity decreases the generation time.
This is counterintuitive to the fact that, at fixed orbit-averaged frequency, an increase in eccentricity leads to a decrease in waveform duration due to larger losses of energy and angular momentum at periastron passages.
This effect was documented in Ref.~\cite{Gamboa:2024hli} and, in summary, can be explained by the balance between the numerical integrator resolving the eccentricity timescales and the waveforms becoming shorter as the eccentricity increases.

Additionally, we show in Fig.~\ref{fig:benchmarks_param_space} histograms of the walltimes for waveform generation (time-domain GW polarizations) for randomly distributed configurations in the QC (top panel) and eccentric (bottom panel) parameter spaces, for different waveform approximants.
For the QC configurations, we take $ 5000 $ random points with $ q \in [1, 10] $, $ M \in \nolinebreak {[5, \, 100]\, \solarmass} $, $ |\s\s \chi_i| \leq 0.9 $ ($ i \in \{1, 2\} $, aligned-spins), while for the eccentric configurations, we take $ 10000 $ random points with the same bounds for masses and spins but with $ e \in [0, 0.5] $ and $ \zeta \in [0, 2 \pi] $.
For both cases, we start the waveforms at $ \langle f _{ \text{start}} \rangle = 10\, $Hz with a fixed sampling rate of $ 2048 \,$Hz.

In the zero-eccentricity limit, both figures show that
\texttt{SEOBNRv6EHM}
can be slightly faster than
\texttt{SEOBNRv5HM}, and 
outperforms \texttt{SEOBNRv5PHM} and \texttt{SEOBNRv5EHM} by a wide margin.
These are remarkable properties, as the \texttt{SEOBNRv6EHM} code includes several additional functions for modeling generic planar-orbit waveforms.
Particularly, \texttt{SEOBNRv6EHM} evolves \emph{two} binary dynamics (a main dynamics and a background QC evolution that helps in the construction of inspiral-merger-ringdown waveforms; see Sec.~\ref{sec:nqcs_bqc}),
and does not use any post-adiabatic acceleration for the main
evolution (it only uses a post-adiabatic prescription for the background QC evolution).
This encourages the development of analytical methods to accelerate the EOB dynamics of eccentric binaries \cite{Lynch:2026ibo}, since they will bring QC and eccentric models to a common computational performance.
We also observe in the top panel of Fig.~\ref{fig:benchmarks_param_space} a considerable improvement with respect to \texttt{TEOBResumS-Dal\'i}:
the median walltime of \texttt{SEOBNRv6EHM} is $ \sim 6 $ times smaller, and the tail of large evaluation times (low total mass and high mass ratios) is markedly reduced.\footnote{
We also benchmarked the QC model \texttt{TEOBResumS-GIOTTO}~\cite{Nagar:2023zxh} under identical settings, finding that its median walltime is $ \sim 2 $ times smaller than that of \texttt{SEOBNRv5HM} and \texttt{SEOBNRv6EHM}, in consistency with Ref.~\cite{Pompiliv5}.
}

For eccentric configurations, the computational efficiency of \texttt{SEOBNRv6EHM} outperforms that of \texttt{SEOBNRv5EHM} and \texttt{TEOBResumS-Dal\'i}.
Figure \ref{fig:benchmarks_seobnr} shows explicitly that, for \emph{all} cases, \texttt{SEOBNRv6EHM} is significantly faster than \texttt{SEOBNRv5EHM}.
Both models have a similar definition of eccentricity (differing only in the RR gauge), yielding nearly equal-duration waveforms for the same input parameters (with the exception of high-eccentricity systems, not included in these tests).
This justifies a fair timing comparison between the two models when evaluated at the same point in parameter space.
In contrast, due to a different eccentricity treatment, waveforms generated with \texttt{TEOBResumS-Dal\'i} are typically shorter than those of \texttt{SEOBNRv6EHM} (and \texttt{SEOBNRv5EHM}) when given the same input parameters.
This precludes a comparison similar to that in Fig.~\ref{fig:benchmarks_seobnr}, unless the eccentric parameters of one of the models are optimized to find the most adequate waveform for each considered case.
For this reason, we find it more convenient to compare models by evaluating them at random points across a given region of parameter space, as shown in Fig.~\ref{fig:benchmarks_param_space}.
There, it is straightforward to see the waveform-evaluation improvement of \texttt{SEOBNRv6EHM} over \texttt{SEOBNRv5EHM} (by factors of $ \sim 2 - 5 $) and \texttt{TEOBResumS-Dal\'i} (by factors of $ \sim 2 - 8 $) across all configurations, including binaries with high mass ratios and small total masses, which compose the tail of high walltimes in the histograms of Fig.~\ref{fig:benchmarks_param_space}.\footnote{
We note that \texttt{TEOBResumS-Dal\'i} supports only four multipoles, $ \{(2,2),\, (2,1), \,(3,3), \,(4,4)\} $, whereas both \texttt{SEOBNR} models additionally include the $ (3,2) $ and $ (4,3) $ modes.
Thus, Fig.~\ref{fig:benchmarks_param_space} shows that \texttt{SEOBNRv6EHM} is faster, even though it returns additional multipoles.
}

These results indicate that \texttt{SEOBNRv6EHM} is faster than all EOB eccentric models.
Our benchmarks are complemented by results from the companion manuscript \cite{Pompili:2026yxq}.
There, it is shown that \texttt{SEOBNRv6EHM} can be three to five times faster than  \texttt{SEOBNRv5EHM} in GW parameter estimation.
Hence, we expect \texttt{SEOBNRv6EHM} to open the possibility of cost-intensive investigations that require accurate eccentric waveforms.


\subsection{Robustness across parameter space}
\label{sec:robustness}

\begin{figure}
\hspace{-5pt}
\vspace{1pt}
\includegraphics[width=\linewidth]{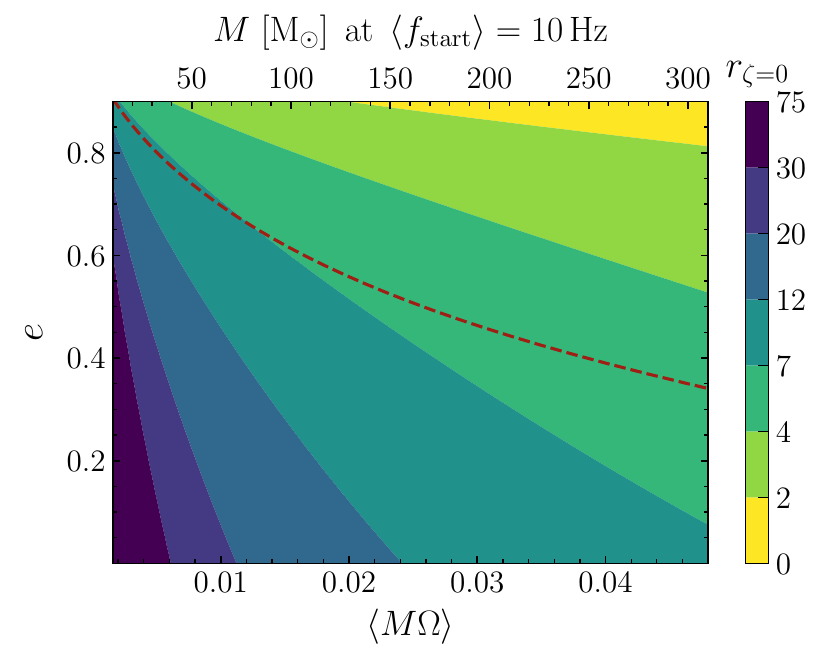}
\hspace*{-3pt}
\vspace{2pt}
\includegraphics[width=1.02\linewidth]{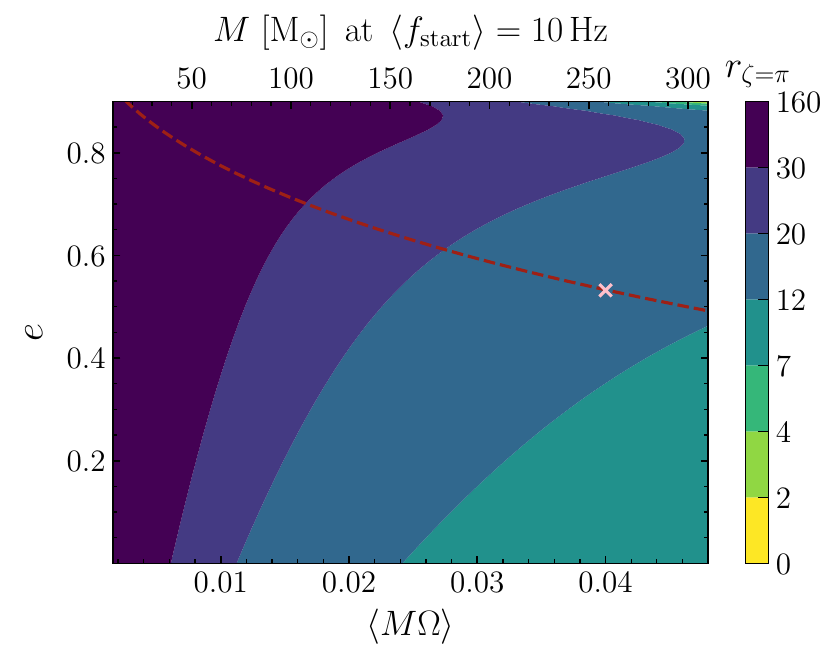}
\hspace*{-8pt}
\includegraphics[width=0.98\linewidth]{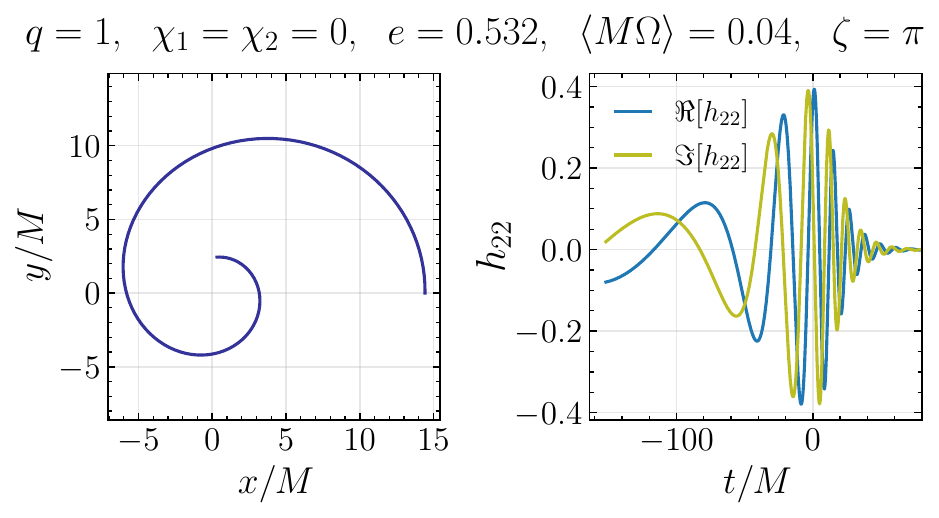}
\vspace{-10.5pt}
\caption{
Contour plots of the initial separation as a function of the eccentricity and frequency for an equal-mass, nonspinning BBH, as predicted by \texttt{SEOBNRv6EHM} and Eq.~(34c) of Ref.~\cite{Gamboa:2024imd}, along with a particular EOB dynamics and waveform.
The values in the top panel are computed for binaries initialized at periastron ($ \zeta = 0 $), while those in the middle panel correspond to initialization at apastron ($\zeta = \pi$).
The red dashed curves separate regions where waveform evaluation is successful (below) from those where it fails (above), due to constraints imposed to ensure physically sensible waveforms.
For reference, we also show the corresponding total mass $M$ at a GW orbit-averaged frequency of $ \langle f _{ \text{start}} \rangle = 10\,$Hz.
The bottom panels display the trajectory (left) and $(2,2)$ mode (right) for a representative boundary configuration (pink cross in the middle panel).
}
\label{fig:initial_separation}
\end{figure}

Studying the behavior of a waveform model across the binary parameter space is crucial for establishing its reliability.
Here, we analyze the robustness of \texttt{SEOBNRv6EHM} by identifying the parameter space regions where we expect the model to generate physically meaningful waveforms, and by studying waveform changes under variations of the input parameters.

Determining initial conditions for generic planar orbits is challenging, as one must know precise instantaneous values of the relative separation and momenta of the binary.
To ensure physically accurate initial conditions, \texttt{SEOBNRv6EHM} imposes a range of constraints, as discussed in Sec.~\ref{sec:ics}.
The excluded initial conditions are always associated with challenging configurations, for example, very small separations or states very close to merger.
As an example, we show in Fig.~\ref{fig:initial_separation} the predicted initial separation for an equal-mass, nonspinning binary as a function of the starting eccentricity and frequency, as predicted by \texttt{SEOBNRv6EHM}, complemented with a PN formula (see below).
The top (middle) panel shows the predicted separation when the binary is initialized at periastron ($\zeta = 0$) [apastron ($\zeta = \pi$)].
For reference, besides the dimensionless orbit-averaged orbital frequency $ \langle M \Omega \rangle $, we also indicate the total mass $ M $ of the binary when the physical orbit-averaged GW frequency is equal to $ 10\, $Hz.
The red dashed curves delineate the regions where \texttt{SEOBNRv6EHM} waveform generation is successful (below the curves) from those where it fails (above the curves); these curves were calculated numerically, by iterating model evaluations over the eccentricity.
The separation is calculated using Eq.~(34c) of Ref.~\cite{Gamboa:2024imd},\footnote{
Intuitively, one would expect the apastron distance to increase with eccentricity at fixed frequency.
However, this is not observed in the contours of the middle panel of Fig.~\ref{fig:initial_separation} for, e.g., $ \langle M \Omega \rangle \approx 0.04 $ and $ e \approx 0.8 $.
This unintuitive behavior arises from the 3PN terms of Eq.~(34c) of Ref.~\cite{Gamboa:2024imd};
removing such terms gives the contours the expected behavior.
We interpret this as a manifestation of the asymptotic nature of the PN series, since at such high eccentricities and frequencies the 3PN contributions become comparable to the lower-order terms.
Consequently, we cannot assess whether this is a true feature of the apastron distance or an artifact of the PN truncation at high eccentricities.
}
but we verified that plotting the true \texttt{SEOBNRv6EHM} predicted separation results in the same plots (for the regions below the red dashed curves).
Additionally, in the bottom panels, we include the EOB trajectory and the $ (2,2) $ mode for one representative boundary case (indicated by a pink cross in the middle panel).
For other configurations (unequal mass ratios, nonzero spins, or different frequencies), the qualitative behavior of the initial separation is analogous to the one observed in Fig.~\ref{fig:initial_separation}.

It is crucial to note that not all input values $ (\langle M \Omega \rangle, \, e,\, \zeta) $ will produce a successful waveform generation.
Some combinations are not physically meaningful (e.g., initial separations below $ 2M $), or correspond to configurations resembling head-on collisions.
This is illustrated in the bottom panels of Fig.~\ref{fig:initial_separation}, which show that higher eccentricities result in shorter-duration systems that fall outside the model's assumptions.
QC models overcome this issue by exploiting the system's circular symmetry, thereby determining initial conditions at lower frequencies (which still correspond to the \emph{same} physical system).
An eccentric binary, however, cannot be trivially evolved backward as its dynamical state depends not only on the frequency, but on the eccentricity and radial anomaly.

This is a problem common to all eccentric models, and it is important to take it into account when drawing conclusions from their results (e.g., when observing posterior railing in parameter estimation \cite{Gupte:2026whi}).
The \texttt{SEOBNRv5EHM} model addresses this problem by integrating backward a set of PN secular evolution equations whenever the starting separation is smaller than $ 10M $.
In this way, EOB initial conditions can be determined at an earlier (less challenging) stage of the evolution.
This approach was needed in \texttt{SEOBNRv5EHM} to avoid issues with its equations of motion.
Nonetheless, a study assessing its impact (e.g., in parameter estimation) has not yet been performed.
Such method, however, turned out to be highly non-user-friendly, as it gives the model an unexpected behavior from the user perspective, and its purpose is not obvious to someone outside the eccentric waveform modeling field.
For these reasons, and thanks to the simpler analytical structure of the \texttt{SEOBNRv6EHM} equations of motion, we do not adopt the secular backward integration approach (not to be confused with the backward evolution of the \emph{full} equations of motion, as explained in footnote \ref{fn:bwd_int}).
This results in a smoother model behavior across the parameter space.
As demonstrated in the companion manuscript \cite{Pompili:2026yxq}, this choice does not negatively affect parameter estimation, as all the runs there were executed seamlessly (some even with fewer waveform generation failures than \texttt{SEOBNRv5EHM}), including high-eccentricity injection-recovery analyses with upper priors on eccentricity of $ 0.9 $ (we also used an upper prior on eccentricity of $ 0.99 $ for some additional analyses, finding no issues with the runs or their convergence).
Thus, \texttt{SEOBNRv6EHM} enables accurate high-eccentricity parameter estimation studies.

Apart from the excluded challenging initial configurations, the parameter-space coverage of \texttt{SEOBNRv6EHM} has \emph{no} inherent restrictions.
The \texttt{SEOBNRv5EHM} model had to be constrained to a conservative region of parameter space to avoid a ``desynchronization'' of its equations of motion that can potentially appear for highly-eccentric, long-lived systems (see Appendix E of Ref.~\cite{Gamboa:2024hli}).
In \texttt{SEOBNRv6EHM}, no such constriction is needed, as the equations of motion \eqref{eq:EOM} are well-defined and self-consistent (this is one of the major improvements with respect to \texttt{SEOBNRv5EHM}).
Thus, \texttt{SEOBNRv6EHM} can be used anywhere in parameter space, with the usual caveats of models.
For example, we expect reduced accuracy for configurations that differ significantly from those used to calibrate the model (e.g., high mass ratios or large spins; see Sec.~\ref{sec:calibration}),
and reduced accuracy as eccentricity increases.

\begin{figure*}
\hspace{-9pt}
\includegraphics[width=\textwidth]
{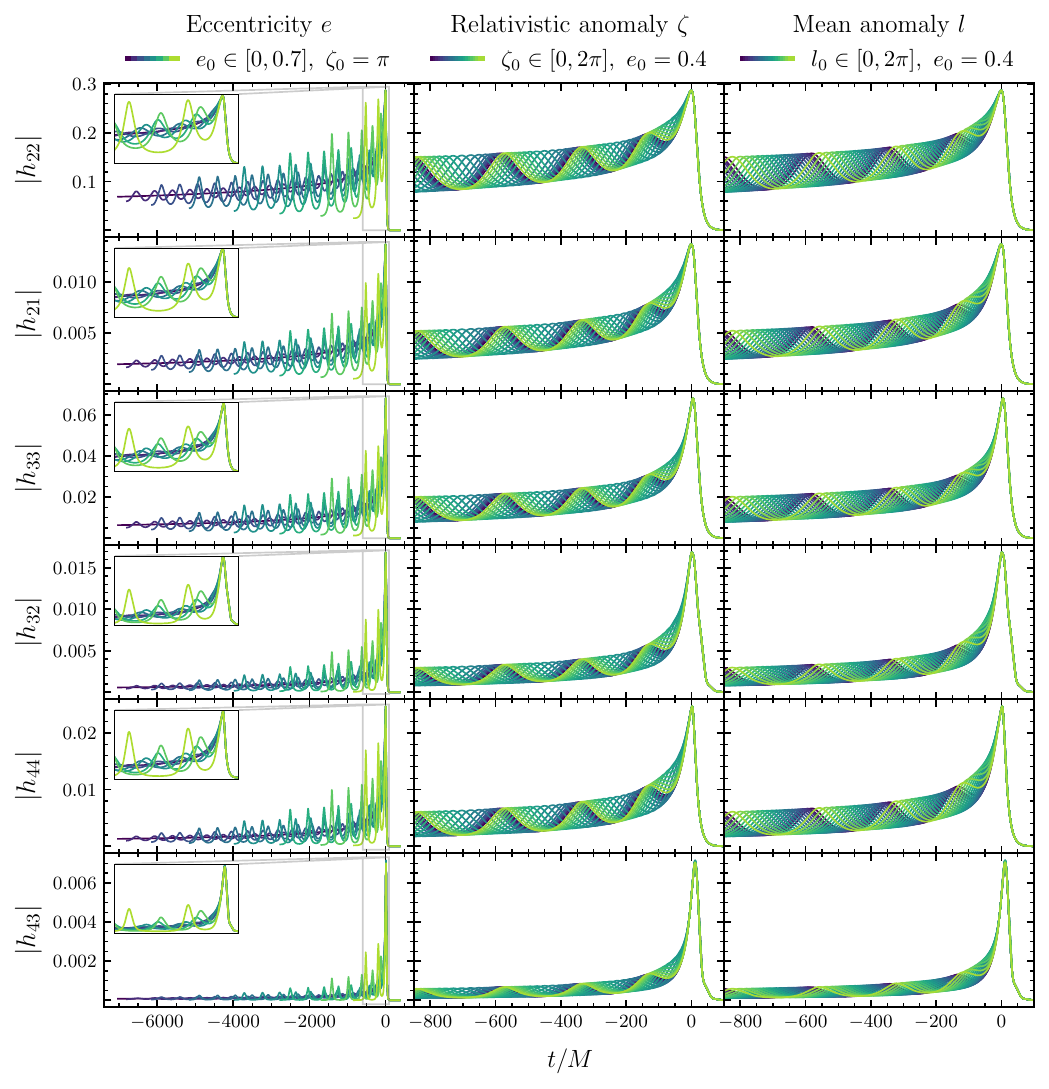}
\vspace{-15pt}
\caption{
Uniform variation of eccentricity (left panels), relativistic anomaly (middle panels), and mean anomaly (right panels) for the different \texttt{SEOBNRv6EHM} waveform mode amplitudes (in geometric units).
Each of these waveforms is characterized by a mass ratio $ q = 3 $, dimensionless spin components $ \chi_{1} = 0.6 $ and $ \chi_{2} = -0.3 $, and a dimensionless orbit-averaged orbital frequency of $ \langle M \omega \rangle \approx 0.01547 $ (equivalent, e.g., to a starting GW frequency $ \langle f _{ \text{start}} \rangle = 20\, $Hz for a total mass $ M = 50 \, \solarmass $).
The systems in the left panels are initialized at apastron ($ \zeta_0 = \pi $), and the eccentricity is varied uniformly from $ e_0 = 0 $ (dark blue) to $ e_0 = 0.7 $ (yellow).
The systems in the middle and right panels are initialized with eccentricity $ e_0 = 0.4 $, and the anomalies are varied uniformly from $ 0 $ (dark blue) to $ 2 \pi $ (yellow).
In the left panels, we include insets of the zoomed-in merger-ringdown portions of the waveform modes to ease visualization, while in the middle and right panels, we directly show the ending portions to better illustrate the dependence on the radial anomaly.
The transformation from mean anomaly to relativistic anomaly for the waveforms shown in the right panels is done under a Newtonian approximation using Eqs.~\eqref{eq:kepler_equation} and \eqref{eq:eccanomaly_to_relanomaly}, as the intrinsic (and default) radial phase parameter in \texttt{SEOBNRv6EHM} is the relativistic anomaly.
}
\label{fig:smoothness_e_z}
\end{figure*}
\begin{figure*}
\hspace{-5pt}
\vspace{2pt}
\includegraphics[width=0.925\linewidth]{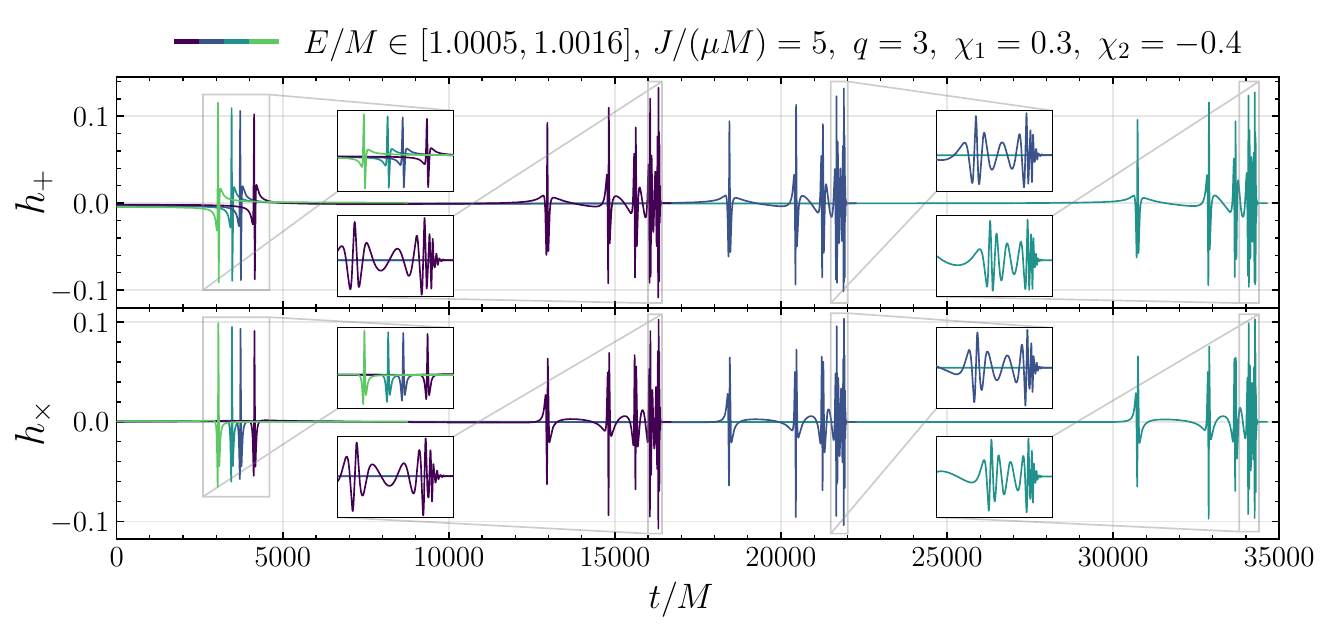}
\hspace*{-15pt}
\includegraphics[width=0.9\linewidth]{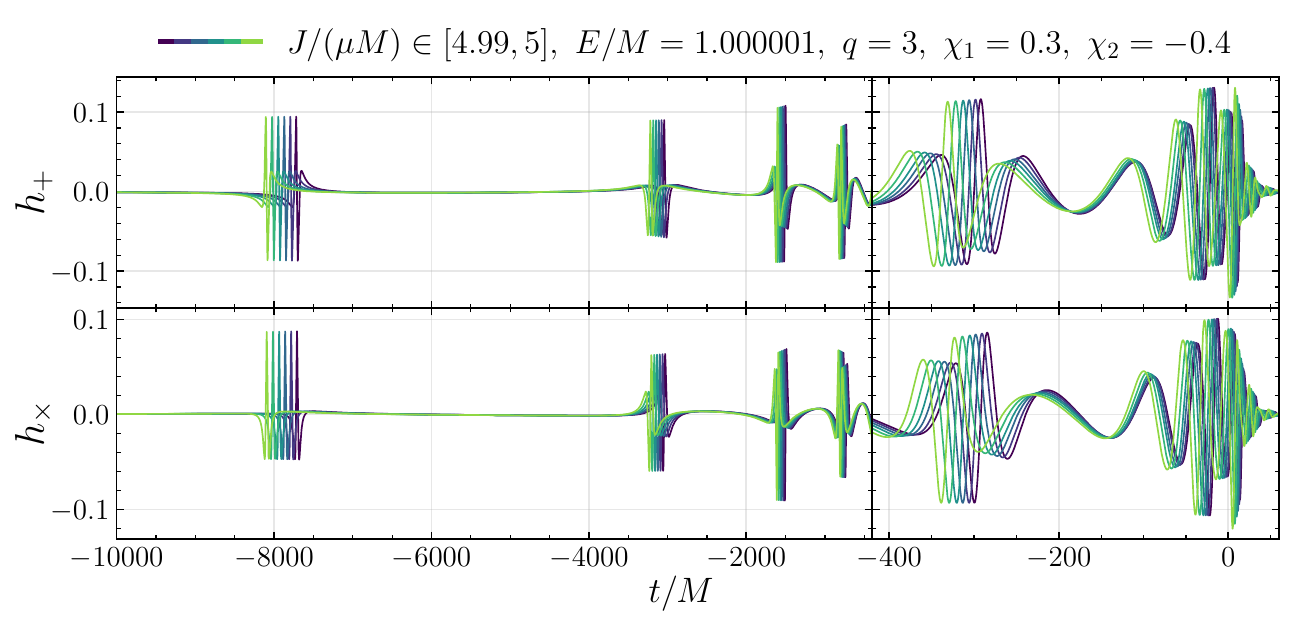}
\vspace{-10pt}
\caption{
Variation of the GW polarizations, $ h_+ $ and $ h_\times $ (in geometric units), with respect to the energy $ E/M $ (top panel) and total angular momentum $ J/M^2 $ (bottom panel), as predicted by \texttt{SEOBNRv6EHM}.
Each of these waveforms belongs to a BBH system with mass ratio $ q = 3 $, dimensionless spin components $ \chi_{1} = 0.3 $ and $ \chi_{2} = -0.4 $, initial separation $ r = 500 M $, and a line-of-sight inclination of $ \iota = \pi/3 $.
The specific energy values employed in the top panel are $E/M \in \{1.0005, 1.0008, 1.00105, 1.0016\}$ at fixed total angular momentum $ J/(\mu M) = 5 $, while the specific total angular momentum values used in the bottom panel are $J/(\mu M) \in \{4.99, 4.992, 4.994, 4.996, 4.998, 5\}$ at fixed energy $ E/M = 1.000001 $.
We also include insets of zoomed-in portions of the waveforms in the top panel to ease visualization.
}
\label{fig:smoothness_generic}
\end{figure*}

Building on the LVK review tests carried out for \texttt{SEOBNRv5EHM}, we successfully assessed the robustness of \texttt{SEOBNRv6EHM} across broad parameter space regions covering mass ratios $ q \in [1,100] $, the full spin range, and eccentric parameters $ (e, \s \zeta)$ for bound orbits, as well as $ (r, \s E/M, \s J/M^2) $ for generic planar orbits.
Namely, we performed visual inspection of the waveforms, stability under perturbations of the input parameters, checks of the mode hierarchy (ensuring that the $ (2, 2) $ mode has the largest amplitude), and verification of the monotonicity of the inspiral amplitude and frequency evolution for the QC $ (2, 2) $ mode, as well as of the envelope defined by the maxima and minima of the eccentric $ (2, 2) $ mode.

For example, in Fig.~\ref{fig:smoothness_e_z}, we show the smooth changes of the amplitude of all the \texttt{SEOBNRv6EHM} waveform modes, $( \ell, |m| ) = \{(2, 2),\, (3, 3),\, (2, 1),\, (4, 4),\, (3, 2),\, (4, 3)\}$, under uniform variations of the eccentric parameters $ e \in [0, \s 0.7] $ (left panels), $ \zeta \in [0, \s 2\pi] $ (relativistic anomaly, middle panels), and $ l \in [0, \s 2\pi] $ (mean anomaly, right panels), at a dimensionless orbit-averaged orbital frequency of $ \langle M \omega \rangle \approx 0.01547 $ (equivalent, e.g., to an initial GW frequency $ \langle f _{ \text{start}} \rangle = 20\, $Hz for a total mass $ M = 50 \, \solarmass $) for a system with mass ratio $ q = 3 $, and dimensionless spin components $ \chi_{1} = 0.6 $ and $ \chi_{2} = -0.3 $.

In the left panels of Fig.~\ref{fig:smoothness_e_z}, we observe the expected decrease in the time to merger with increasing eccentricity at fixed starting frequency, while in the middle and right panels, we observe the amplitude oscillations induced by the radial anomaly.
The middle panels show that a uniform distribution on relativistic anomaly leads to a clustering of the amplitude around periastron at $ \zeta \in \{ 0, 2\pi \} $.
In contrast, the right panels show that a uniform distribution on mean anomaly leads to evenly distributed waveform amplitudes.
These observations are complemented by Fig.~\ref{fig:distribution_radial_anomalies} in Sec.~\ref{sec:ics_bound}, where one can see the effects of uniformly sampling in either of the two anomalies (see also Fig.~5 of Ref.~\cite{Nee:2025zdy} for the effect of mean anomaly in NR waveforms).
In the context of surrogate modeling of eccentric waveforms, the mean anomaly approach has been quite fruitful \cite{Nee:2025nmh,Maurya:2025shc}.
However, the impact of using either anomaly on parameter estimation still needs to be addressed.
For the present, we chose the relativistic anomaly $ \zeta $ as the default radial phase parameter in \texttt{SEOBNRv6EHM}.

In Fig.~\ref{fig:smoothness_e_z}, all cases show analogous behavior during merger-ringdown due to the employed QC prescription.
The true behavior at merger has a dependence on eccentricity and radial anomaly, as illustrated in Fig.~5 of Ref.~\cite{Nee:2025zdy} (see also Refs.~\cite{Radia:2021hjs,Wang:2023vka,Wang:2023wol,Wang:2024jro,Carullo:2023kvj,Carullo:2024smg,Rao:2026lmz}).
These effects are relatively small, so we neglect them and leave their inclusion for future work.
Further investigation is required to characterize the impact of eccentric merger-ringdown effects on waveform accuracy and parameter estimation.

Going beyond bound orbits, we show in Fig.~\ref{fig:smoothness_generic} the changes of the waveform polarizations, $ h_+ $ and $ h_\times $, under variations of the input parameters $ E/M $ and $ J/M^2 $ for generic planar orbits, for systems with mass ratio $ q = 3 $, dimensionless spins $ \chi_{1} = 0.3 $ and $ \chi_{2} = -0.4 $, initial separation $ r = 500 M $, and a line-of-sight inclination of $ \iota = \pi/3 $.
Specifically, the top panel shows the changes with respect to a set of energy values $E/M \in \{1.0005, 1.0008, 1.00105, 1.0016\}$ at fixed total angular momentum $ J/(\mu M) = 5 $, while the bottom panel shows the changes with respect to different total angular momentum values $J/(\mu M) \in \{4.99, 4.992, 4.994, 4.996, 4.998, 5\}$ at fixed energy $ E/M = 1.000001 $.
To ease visualization, we show insets of zoomed-in portions of the waveforms in the top panel.

A particularly interesting behavior observed in the top panel of Fig.~\ref{fig:smoothness_generic} is the transition from dynamical capture to a scattering encounter.
Starting from the dynamically captured system with $E/M = 1.0005 $ (dark blue curves), we observe that increasing the energy delays the subsequent periastron passages to later times, until the system becomes unbound (light green curves, with $ E/M = 1.0016 $).
The precise threshold value between scattering and dynamical capture for this system lies between $ E/M = 1.00105 $ and $ E/M = 1.0016 $.
In contrast, the bottom panel of Fig.~\ref{fig:smoothness_generic} focuses on the smooth changes of waveform morphology under small variations of $ J $, for a system just above the parabolic limit ($ E/M = 1 $) that is captured after its first periastron passage.
We remark that these smoothness properties exemplified in Fig.~\ref{fig:smoothness_generic} are important for studies analyzing dynamical-capture and scattering waveforms \cite{East:2012xq,Mukherjee:2020hnm,Gamba:2021gap,Morras:2021atg,Bini:2023gaj,Henshaw:2025arb,Smith:2025rnp,Lange:2026eqx}.
As a proof of principle, in the companion manuscript \cite{Pompili:2026yxq}, we have successfully performed parameter estimation over different high-mass events using generic planar-orbit initial conditions.

Overall, the behavior of \texttt{SEOBNRv6EHM} remains robust across different parameter space regions.
This robustness ensures the model can be used straightforwardly in different GW studies.
Better prescriptions for initial conditions are needed to expand the parameter space of eccentric waveform models.
However, new data analysis techniques are also needed to enable robust eccentric GW inference.
We expect \texttt{SEOBNRv6EHM} to play a role in supporting these future investigations.


\section{Conclusions}
\label{sec:conclusions}

Eccentric GW models will play a key role in discriminating compact binary formation channels and in achieving unbiased parameter estimation.
Here, we presented \texttt{SEOBNRv6EHM}: a new multipolar, time-domain, EOB model for the complete GW radiation of comparable-mass BBHs orbiting in generic planar orbits.
The model leverages new resummations of the RR force and waveform modes, and it provides the multipoles $( \ell, |m| ) \in \{(2, 2),\, (3, 3),\, (2, 1),\, (4, 4),\, (3, 2),\, (4, 3)\}$ for the full inspiral-merger-ringdown evolution of coalescing binaries, dynamical captures, and scattering encounters.
The model has been calibrated to QC \texttt{SXS} NR simulations, and it constructs the merger-ringdown under a QC approximation.

\texttt{SEOBNRv6EHM} is \emph{accurate}.
In the low-eccentricity limit, its performance is comparable to that of \texttt{SEOBNRv5EHM} \cite{Gamboa:2024hli}, while for the largest eccentricities it improves the accuracy by up to an order of magnitude.
\texttt{SEOBNRv6EHM} achieves a median of $ \sim 0.02 \%$ for the maximum mismatches across the total mass interval $ [20, 200] \, \solarmass$ when comparing against 319 eccentric \texttt{SXS} NR waveforms \cite{Scheel:2025jct,Ramos-Buades:2022lgf,Nee:2025zdy}.
Importantly, mismatches remain below $\sim 2\%$ even for the most eccentric NR configurations.
Relative to \texttt{TEOBResumS-Dal\'i} \cite{Nagar:2024dzj}, we observe an overall improvement of about one order of magnitude across all eccentricities.
In the generic planar-orbit regime, we compare against NR waveforms from one dynamical capture and two scattering encounters, as well as against 61 scattering-angle results from Ref.~\cite{Long:2025nmj}, finding improved agreement relative to \texttt{TEOBResumS-Dal\'i}.
The high accuracy of \texttt{SEOBNRv6EHM} is further demonstrated in the companion manuscript \cite{Pompili:2026yxq}, where its capability to accurately infer the parameters of highly eccentric binaries is established through injection-recovery studies with synthetic NR signals.

\texttt{SEOBNRv6EHM} is \emph{computationally efficient}.
We performed benchmarks to assess the computational speed of \texttt{SEOBNRv6EHM} across a broad range of binary configurations.
Thanks to code optimizations and to a simpler analytical structure, \texttt{SEOBNRv6EHM} can be $ 2 - 8 $ times faster than \texttt{SEOBNRv5EHM} and \texttt{TEOBResumS-Dal\'i} in single waveform evaluations, depending on the specific binary parameters.
The improved efficiency of \texttt{SEOBNRv6EHM} in Bayesian parameter estimation is demonstrated in the companion manuscript~\cite{Pompili:2026yxq}, where the model is shown to be approximately three to five times faster than \texttt{SEOBNRv5EHM}.

\texttt{SEOBNRv6EHM} is \emph{robust}.
The parameter space covered by \texttt{SEOBNRv6EHM} has no inherent restrictions, apart from systems with very high eccentricity near merger.
This is a large improvement over \texttt{SEOBNRv5EHM}, whose parameter space coverage had to be restricted to avoid unphysical features in the waveforms, mainly occurring for long-lived, highly eccentric systems.
Leveraging this robustness and the improved computational efficiency, the companion manuscript \cite{Pompili:2026yxq} presents analyses of \NEVENTS GW events spanning a wide range of configurations, including high mass ratios, low total masses (including NSs), and large eccentricities. These results show that \texttt{SEOBNRv6EHM} is readily applicable to standard GW analyses.

Several subsequent studies can be made to improve eccentric EOB modeling.
Further research on the RR force will be essential, as it is one of the main accuracy bottlenecks.
The current paradigm of enforcing a QC factorization in resummations of the RR force and waveform modes may, in fact, be introducing limitations in the modeling of eccentric systems.
The sigmoid resummation introduced in this work may prove useful for developing new approaches.
Studying different RR gauges could be important for achieving better accuracy across different binary configurations.
The final portion of the waveform could be improved by incorporating recent developments in the modeling of the eccentric merger-ringdown phase (e.g., \cite{Carullo:2023kvj,Carullo:2024smg,Nee:2025zdy,Faggioli:2026alx,Albanesi:2026qtx,Rao:2026lmz}).
Developing a framework for calibrating EOB models to eccentric NR simulations is a challenging but promising avenue that will require careful control over the eccentric parameters.
Finally, testing the impact of post-Minkowskian results on the accuracy of inspiral-merger-ringdown waveforms remains an open problem that has been addressed only in the QC-orbit limit~\cite{Buonanno:2024byg,Damour:2025uka}.

Our next goal is to develop an accurate and efficient waveform model for generic-orbit, \emph{precessing-spin} BBHs, using \texttt{SEOBNRv6EHM} as baseline model \cite{Gamboa:2026ybr}.
In this context, there has been recent work in developing inspiral-only PN waveform models \cite{Klein:2018ybm, Klein:2021jtd, Arredondo:2024nsl,Morras:2025nlp,Morras:2026fho}, as well as inspiral-merger-ringdown EOB waveform models \cite{Liu:2023ldr, Gamba:2024cvy}.
Such generic models will be essential to allow for \emph{unbiased} BBH parameter estimation.
\\\\


\section*{Acknowledgments}

This work benefited from many interesting discussions with
Aurora Abbondanza,
H\'ector Estell\'es,
Guglielmo Faggioli,
Cheng Foo,
Nihar Gupte,
Marcus Haberland,
Mohammed Khalil,
Benjamin Leather,
Philip Lynch,
Maarten van de Meent,
Gonzalo Morras,
Nami Nishimura,
Lluc Planas,
and Luca Sebastiani.
We also thank Steffen Grunewald for his support with scientific computing.
A.B. would like to thank the Institute for Advanced Study in Princeton for its hospitality during the final stages of this work.

The computational work for this manuscript was carried out on the \texttt{Hypatia} computer cluster at the Max Planck Institute for Gravitational Physics in Potsdam.
Numerical-relativity simulations were performed at the Max Planck Computing and Data Facility on the Minerva HPC system of the ``Astrophysical and Cosmological Relativity'' division, and at the Dutch national e-infrastructure with the support of the SURF Cooperative (NWO-2024.002), and in Picasso Supercomputer from the Red Espa\~nola de Supercomputaci\'on (RES) (AECT-2025-1-0017, AECT-2025-2-0004, AECT-2025-3-0015, AECT-2026-1-0004).

This work is funded by the European Union (ERC grant GWSky/101167314).
Views and opinions expressed are however those of the authors only and do not necessarily reflect those of the European Union or the European Research Council Executive Agency. Neither the European Union nor the granting authority can be held responsible for them.
L.~P. is supported by a UKRI Future Leaders Fellowship (grant number MR/Y018060/1).
A.~R.-B. is supported by the Veni research programme which is (partly) financed by the Dutch Research Council (NWO) under the grant VI.Veni.222.396; acknowledges support from the Spanish Agencia Estatal de Investigaci\'on grant PID2024-157460NA-I00; and the Spanish Ministerio de Ciencia, Innovaci\'on y Universidades (Beatriz Galindo, BG23/00056), co-financed by UIB.


\section*{Data availability}

\texttt{SEOBNRv6EHM} is implemented in the \texttt{pyseobnr} package~\cite{Mihaylovv5} and will be made publicly available after its upcoming review within the LVK collaboration.
We provide ancillary files~\cite{SEOBNRv6EHM:SupplementalMaterial} that list the NR simulations used here, the binary parameters, and the mismatch information for each model.


\appendix

\onecolumngrid

\section{Expressions at high PN orders}
\label{sec:PN_expressions}

In this appendix, we provide expressions for various quantities derived throughout this work.
Section~\ref{sec:transformation_to_rrdprdst} derives rules to transform EOB variables to the parametrization employed in the eccentricity corrections to the RR force and waveform modes.
Section~\ref{sec:ap_PN_RR} presents the PN-expanded RR force components derived from flux-balance laws.
Section~\ref{sec:RR_corr_exp_resum} discusses the sigmoid resummation and its use in the eccentricity corrections of the RR force, $ \mF_\phi^\text{ecc} $ and $ \mF_r^\text{ecc} $.
Section~\ref{sec:ecc_corr_modes} presents the eccentricity corrections $ h_{\ell m}^\mathrm{ecc} $ to the factorized waveform modes, and discusses a possible generalization of the source term $ \hat S _{ \text{eff}} $.
Finally, Section~\ref{sec:orbital_frequency} presents a PN-expanded expression of the orbital frequency for circular orbits $ \Omega _{ \text{c}}(r) $.
Throughout this section, we employ dimensionless variables to simplify notation.
To restore units, one can employ the relations in Eqs.~\eqref{eq:dimlessVars}.

We define the conservative part of $ \dot p _{r_*} $ as
\begin{equation}
\dot p_{r_*,\s\text{cons}}
\equiv
- \xi(r) \frac{\partial H_\text{EOB}}{\partial r}(r,p_{r_*},p_\phi).
\end{equation}

We also define
\begin{equation}
\label{eq:p_phiN}
p_{\phi,\s\text{N}} \equiv \sqrt{r + r^3 \s \dot p_{r_*, \s\text{cons}}  } \, ,
\end{equation}
to simplify some expressions below.
Note that $ p_{\phi,\s\text{N}} $ reduces to the orbital angular momentum in the Newtonian limit.


\subsection{Transformation to $ (r, \dot r, \dot p_{r_*,\s \mathrm{cons}} ) $}
\label{sec:transformation_to_rrdprdst}

Here, we write the 3PN relations employed to transform the EOB variables $ (p_r, p_\phi, \Omega) $ that appear in the RR force and waveform modes, into  $ (r, \dot r, \dot p_{r_*, \s\text{cons}} = -\xi \, \partial H _{ \text{EOB}} / \partial r ) $.
The latter is a convenient parametrization of the eccentricity corrections to the RR force and modes, since it favors the recovery of an underlying QC structure of the model in the zero-eccentricity limit.

First, we consider the PN expanded equations of motion in terms of $ (r, p_r, p_\phi) $:
\begin{subequations}
\begin{align}
\dot r
&=
\frac{\partial H _{ \text{EOB}}}{\partial p_r} (r, p_r, p_\phi)
= p_r + 
\epsilon^2 \bigg[
 \frac{p_r }{r}(\nu -3)
- \frac{p_\phi^2 \s p_r  }{ r^2}\frac{\nu +1}{2}
-  p_r^3 \frac{\nu + 1}{2}
\bigg]
+ \text{3PN expansion},
\\
\Omega
&=
\frac{\partial H _{ \text{EOB}}}{\partial p_\phi} (r, p_r, p_\phi)
= \frac{p_\phi}{r^2} + 
\epsilon^2 \bigg[
\frac{p_\phi }{r^3} (\nu -1) 
- \frac{p_\phi^3 }{r^4} \frac{\nu + 1}{2}  
- \frac{p_\phi \s p_r^2}{r^2} \frac{\nu + 1}{2}  
\bigg]
+ \text{3PN expansion},
\label{eq:omega_pnexp}
\\
\dot p_r
&=
- \frac{\partial H _{ \text{EOB}}}{\partial r} (r, p_r, p_\phi) + \mathcal F_r (r, p_r, p_\phi)
\nonumber \\
&= \frac{p_\phi^2}{r^3} -\frac{1}{r^2}
-\epsilon ^2 \bigg[
\frac{\nu+1}{r^3}
+ \frac{p_\phi^4}{r^5}\frac{ \nu + 1}{2 }
+ \frac{p_\phi^2 \s p_r^2}{r^3}  \frac{\nu +1}{2 }
- \frac{p_r^2}{r^2} \frac{\nu -3}{2 }
- \frac{p_\phi^2}{r^4} \frac{3  (\nu -1)}{2 }\bigg]
+ \text{3PN expansion},
\end{align}
\end{subequations}
where $ \mathcal F_r $ is given by Eq.~\eqref{eq:Fr_LO_rprL}.
These expressions can be inverted, respectively, to get:
\begin{subequations}
\begin{align}
p_r (r, \dot r, p_\phi)
&=
\dot r
+\frac{\dot{r}\epsilon^2}{2r^2}
\Bigg[
p_\phi^2(1+\nu)
+r(6-2\nu)
+r^2\dot{r}^2(1+\nu)
\Bigg]
+ \text{3PN expansion},
\label{eq:pr_rrdL}
\\
p_\phi (r, p_r, \Omega)
&=
r^2 \s \Omega
+\frac{r\epsilon^2\Omega}{2}
\Bigg[
2
+r p_r^2 (1+\nu)
+r^3\Omega^2
+\nu(-2+r^3\Omega^2)
\Bigg]
+ \text{3PN expansion},
\label{eq:L_rprom}
\\
p_\phi (r, p_r, \dot p_r)
&=
\sqrt{r + r^3 \s \dot p_r } 
+\frac{\epsilon^2}{4\sqrt{r + r^3 \s \dot p_r } }
\Bigg[
6
+\dot{p}_r r^2(5-\nu)
+\dot{p}_r^2 r^4(1+\nu)
+4 r p_r^2  +\dot{p}_r r^3 p_r^2(1+\nu)
\Bigg]
+ \text{3PN expansion}.
\label{eq:L_rprprd}
\end{align}
\end{subequations}
Next, we substitute Eq.~\eqref{eq:L_rprom} into Eq.~\eqref{eq:pr_rrdL} and solve iteratively for $ p_r $ to get
\begin{align}
p_r (r, \dot r, \Omega)
&= 
\dot r
+\frac{\dot{r}\epsilon^2}{2r}
\Bigg[
6-2\nu
+r\dot{r}^2(1+\nu)
+r^3(1+\nu)\Omega^2
\Bigg]
+ \text{3PN expansion}.
\label{eq:pr_rrdom}
\end{align}
Afterwards, we substitute Eq.~\eqref{eq:L_rprprd} into Eq.~\eqref{eq:omega_pnexp} to get:
\begin{equation}
\label{eq:omega_rprprd}
\Omega(r, p_r, \dot p_r)
=
\frac{\sqrt{r + r^3 \s \dot p_r } }{r^2}
+\frac{\epsilon^2}{4r^2\sqrt{r+\dot{p}_r r^3}}
\Bigg[
2\nu
-\dot{p}_r r^2(3+\nu)
-\dot{p}_r^2 r^4(1+\nu)
+2r p_r^2(1-\nu)
-\dot{p}_r p_r^2r^3(1+\nu)
\Bigg]
+ \text{3PN expansion}.
\end{equation}
Then we substitute Eq.~\eqref{eq:pr_rrdom} into Eq.~\eqref{eq:omega_rprprd} and solve iteratively for $ \Omega $, leading to
\begin{equation}
\Omega(r, \dot r, \dot p_r)
=
\frac{\sqrt{r + r^3 \s \dot p_r } }{r^2}
+\frac{\epsilon^2}{4r^2\sqrt{r+\dot{p}_r r^3}}
\Bigg[
2\nu
+2r\dot{r}^2(1-\nu)
-\dot{p}_r r^2(3+\nu) -\dot{p}_r \dot{r}^2 r^3(1+\nu)
-\dot{p}_r^2 r^4(1+\nu)
\Bigg]
+ \text{3PN expansion}.
\label{eq:om_rrdprd}
\end{equation}
The next step is to substitute Eq.~\eqref{eq:om_rrdprd} into Eq.~\eqref{eq:pr_rrdom} to get
\begin{equation}
p_r(r, \dot r, \dot p_r)
=
\dot r
+\frac{\dot{r}\epsilon^2}{2r}
\Bigg[
7-\nu
+\dot{p}_r r^2(1+\nu)
+r\dot{r}^2(1+\nu)
\Bigg]
+ \text{3PN expansion}.
\label{eq:pr_rrdprd}
\end{equation}
Finally, substituting Eq.~\eqref{eq:pr_rrdprd} into Eq.~\eqref{eq:L_rprprd} gives us
\begin{align}
p_\phi(r, \dot r, \dot p_r)
&=
\sqrt{r + r^3 \s \dot p_r }
+\frac{\epsilon^2}{4\sqrt{r+\dot{p}_r r^3}}
\Bigg\{
6
+\dot{p}_r r^2[5-\nu+2r\dot{r}^2(1+\nu)]
+\dot{p}_r^2 r^4(1+\nu)
+2r\dot{r}^2(1+\nu)
\nonumber \\
&\qquad
-p_r^2\big[2r(-1+\nu)+\dot{p}_r r^3(1+\nu)\big]
\Bigg\}
+ \text{3PN expansion}.
\label{eq:L_rrdprd}
\end{align}
Thus, Eqs.~\eqref{eq:om_rrdprd}--\eqref{eq:L_rrdprd} transform $ (p_r, p_\phi, \Omega) $ into $ (r, \dot r, \dot p_r) $.

The final process is to find a transformation from $ \dot p_r $ to $ \dot p_{r_*} $.
This is convenient because $ \dot p_{r_*} $ is directly associated with the equations of motion~\eqref{eq:EOM}.
First, we consider the relation
\begin{equation}
\label{eq:def_prst}
p_{r_*} = p_r \, \xi(r),
\end{equation}
where the tortoise function $ \xi(r) $ is defined in Eq.~(44) of Ref.~\cite{Khalilv5} and whose 3PN expansion is given by
\begin{equation}
\xi(r) =
1
-\frac{2\epsilon^2}{r}
+\frac{3\epsilon^4\nu}{r^2}
+\frac{\epsilon^6}{r^3}
\Bigg[
-3\nu^2
+\nu(22-8\chi_{\text{A}}^2)
+2(\chi_{\text{A}}^2+2\delta\chi_{\text{A}}\chi_{\text{S}}+\chi_{\text{S}}^2)
\Bigg].
\end{equation}
Then, taking a time derivative of 
$p_{r_*}$ and substituting Eq.~\eqref{eq:pr_rrdprd} yields
\begin{equation}
\dot p_{r_*}
=
\dot p_r \s \xi + p_r \frac{\di \xi }{\di r} \s \dot r
=
\dot p_r  + \epsilon ^2 \left(\frac{2\s \dot r^2}{r^2}-\frac{2 \s \dot p_r}{r}\right)
+ \text{3PN expansion}.
\end{equation}
A PN inversion of this relation gives the desired formula
\begin{align}
\dot p_r
&=
\dot p_{r_*}
+\frac{2 \epsilon^2}{r^2}
\Big(
\dot{p}_{r_*}r-\dot{r}^2
\Big)
- \frac{\epsilon^4}{r^3}
\Bigg\{
\dot{p}_{r_*}r(-4+3\nu)
+\dot{r}^2\Big[
11-7\nu
+\dot{p}_{r_*}r^2(1+\nu)
\Big]
+\dot{r}^4
r(1+\nu)
\Bigg\}
\nonumber \\
&\quad
- \frac{\epsilon^6}{4 r^4}
\Bigg\{
-4\dot{p}_{r_*}r(8-34\nu+3\nu^2)
+\dot{r}^2\Big[
171-487\nu+47\nu^2
+6\dot{p}_{r_*}r^2(6+3\nu-2\nu^2)
+\dot{p}_{r_*}^2 r^4(1+5\nu+\nu^2)
\Big]
\nonumber \\
&\qquad
+\dot{r}^4\Big[
2r(19+9\nu-7\nu^2)
+2\dot{p}_{r_*}r^3(2+7\nu+2\nu^2)
\Big]
+3 \dot{r}^6
r^2(1+3\nu+\nu^2)
\Bigg\}
\nonumber \\
&\quad
+\frac{2\epsilon^6}{r^4}
\Big(
\dot{p}_{r_*}r-4\dot{r}^2
\Big)
\Big[
(-1+4\nu)\chi_{\text{A}}^2
-2\delta\chi_{\text{A}}\chi_{\text{S}}
-\chi_{\text{S}}^2
\Big],
\end{align}
which we substitute in Eqs.~\eqref{eq:om_rrdprd}--\eqref{eq:L_rrdprd} to finally obtain:
%
\begin{subequations}
\label{eq:omprL_to_rrdprdst}
\begin{align}
\Omega (r, \dot r, \dot p_{r_*})
&=
\frac{p_{\phi,\s\mathrm{N}}}{r^2}
-\frac{\epsilon^2}{4 p_{\phi,\mathrm{N}} r^4}
\Bigg[
p_{\phi,\s\mathrm{N}}^4(1+\nu)
+2r^2(1-\nu)
-p_{\phi,\s\mathrm{N}}^2 r(3+\nu)
+\dot{r}^2(1+\nu)r^2(p_{\phi,\mathrm{N}}^2+r)
\Bigg]
+\frac{\epsilon^3}{2r^3}
\Bigg[(\nu - 2)\chi_{\text{S}} -2\delta\chi_{\text{A}}\Bigg]
\nonumber \\
&\quad
-\frac{\epsilon ^4}{32 p_{\phi,\s\mathrm{N}}^3 r^6}
\Bigg\{
4r^4(-1+\nu)^2
+4p_{\phi,\s\mathrm{N}}^2 r^3(5-7\nu^2)
+p_{\phi,\s\mathrm{N}}^8(-1+4\nu-\nu^2)
+2p_{\phi,\s\mathrm{N}}^6 r(16+9\nu+2\nu^2)
\nonumber \\
&\qquad \quad
+p_{\phi,\s\mathrm{N}}^4 r^2(-55+46\nu+9\nu^2)
\nonumber \\
&\qquad
+\dot{r}^2\Big[
-4r^5(-1+\nu^2)
-4p_{\phi,\s\mathrm{N}}^4 r^3(-8-4\nu+\nu^2)
+2p_{\phi,\s\mathrm{N}}^6 r^2(1+8\nu+\nu^2)
+2p_{\phi,\s\mathrm{N}}^2 r^4(13+8\nu+7\nu^2)
\Big]
\nonumber \\
&\qquad
+\dot{r}^4\Big[
6p_{\phi,\s\mathrm{N}}^2 r^5\nu
+r^6(1+\nu)^2
+3p_{\phi,\s\mathrm{N}}^4 r^4(1+4\nu+\nu^2)
\Big]
\Bigg\}
\nonumber \\
&\quad
+\frac{\dot{r}^2\epsilon^4}{2 p_{\phi,\s\mathrm{N}} r^2}
\Bigg[
(-1+4\nu)\chi_{\text{A}}^2
-2\delta\chi_{\text{A}}\chi_{\text{S}}
-\chi_{\text{S}}^2
\Bigg]
\nonumber \\
&\quad
-\frac{4\epsilon_\text{RR}^5\nu}{15 p_{\phi,\s\mathrm{N}} r^4}
\dot{r}
\Bigg\{
r(17+9\alpha-9\beta)
+p_{\phi,\s\mathrm{N}}^2(3-3\alpha+9\beta)
+\dot{r}^2\Big[
6r^2(2+\alpha-\beta)
\Big]
\Bigg\}
\nonumber \\
&\quad
+\frac{\epsilon^5}{32 r^5}
\Bigg\{
p_{\phi,\s\mathrm{N}}^2\Big[
\delta(48+5\nu)\chi_{\text{A}}
+(48-49\nu+2\nu^2)\chi_{\text{S}}
\Big]
-2r\Big[
\delta(24+25\nu)\chi_{\text{A}}
+(24+7\nu-26\nu^2)\chi_{\text{S}}
\Big]
\nonumber \\
&\qquad
+8r^2\dot{r}^2\nu\Big[
2\delta\chi_{\text{A}}
-(-2+\nu)\chi_{\text{S}}
\Big]
\Bigg\}
\nonumber \\
&\quad
+\frac{\epsilon^6}{384 p_{\phi,\s\mathrm{N}}^5 r^8}
\Bigg\{
3p_{\phi,\s\mathrm{N}}^{12}(1+\nu-3\nu^2+\nu^3)
+3p_{\phi,\s\mathrm{N}}^{10}r(25-115\nu+35\nu^2-\nu^3)
-3p_{\phi,\s\mathrm{N}}^8 r^2(411-45\nu-97\nu^2-13\nu^3)
\nonumber \\
&\qquad \quad
+3p_{\phi,\s\mathrm{N}}^6 r^3(575-2045\nu+9\nu^2-107\nu^3)
+p_{\phi,\s\mathrm{N}}^4 r^4(-462+(-6470+492\pi^2)\nu-834\nu^2+534\nu^3)
\nonumber \\
&\qquad \quad
-12p_{\phi,\s\mathrm{N}}^2 r^5(7-5\nu-15\nu^2+13\nu^3)
+24r^6(-1+\nu)^3
\nonumber \\
&\qquad
+\dot{r}^2\Big[
9p_{\phi,\s\mathrm{N}}^{10}r^2(1-11\nu-15\nu^2+\nu^3)
-9p_{\phi,\s\mathrm{N}}^8 r^3(23+133\nu-19\nu^2+3\nu^3)
\nonumber \\
&\qquad \quad
+3p_{\phi,\s\mathrm{N}}^6 r^4(-343+217\nu+269\nu^2+21\nu^3)
-3p_{\phi,\s\mathrm{N}}^4 r^5(283-115\nu-83\nu^2+75\nu^3)
\nonumber \\
&\qquad \quad
+24p_{\phi,\s\mathrm{N}}^2 r^6(-8+\nu+3\nu^2+6\nu^3)
-36r^7(-1+\nu)^2(1+\nu)
\Big]
\nonumber \\
&\qquad
+\dot{r}^4\Big[
-3p_{\phi,\s\mathrm{N}}^8 r^4(5+93\nu+105\nu^2+5\nu^3)
+9p_{\phi,\s\mathrm{N}}^6 r^5(-31-95\nu+11\nu^2+3\nu^3)
\nonumber \\
&\qquad \quad
-3p_{\phi,\s\mathrm{N}}^4 r^6(-3+653\nu-387\nu^2+9\nu^3)
-9p_{\phi,\s\mathrm{N}}^2 r^7(9+17\nu+3\nu^2+3\nu^3)
+18r^8(-1+\nu)(1+\nu)^2
\Big]
\nonumber \\
&\qquad
+\dot{r}^6\Big[
-3p_{\phi,\s\mathrm{N}}^6 r^6(7+59\nu+63\nu^2+7\nu^3)
+3p_{\phi,\s\mathrm{N}}^4 r^7(1-17\nu-21\nu^2+\nu^3)
-3p_{\phi,\s\mathrm{N}}^2 r^8(1+9\nu+9\nu^2+\nu^3)
\nonumber \\
&\qquad \quad
-3r^9(1+\nu)^3
\Big]
\Bigg\}
\nonumber \\
&\quad
+\frac{\epsilon^6}{16 p_{\phi,\s\mathrm{N}}^3 r^6}
\Bigg\{
4p_{\phi,\s\mathrm{N}}^2\Big[
p_{\phi,\s\mathrm{N}}^4(1+\nu)(-1+4\nu)
+p_{\phi,\s\mathrm{N}}^2 r(1+\nu)(-1+4\nu)
+6r^2-28r^2\nu
\Big]\chi_{\text{A}}^2
\nonumber \\
&\qquad \quad
+8p_{\phi,\s\mathrm{N}}^2\delta\Big[
-p_{\phi,\s\mathrm{N}}^4(1+\nu)
-p_{\phi,\s\mathrm{N}}^2 r(1+\nu)
+3r^2(2-7\nu)
\Big]\chi_{\text{A}}\chi_{\text{S}}
\nonumber \\
&\qquad \quad
+\Big[
-4p_{\phi,\s\mathrm{N}}^6(1+\nu)
-4p_{\phi,\s\mathrm{N}}^4 r(1+\nu)
+2p_{\phi,\s\mathrm{N}}^2 r^2(12-76\nu+25\nu^2)
\Big]\chi_{\text{S}}^2
\nonumber \\
&\qquad
+2 \dot{r}^2 r^2\Big[
(1-4\nu)\Big(
3p_{\phi,\s\mathrm{N}}^4(1+\nu)
+p_{\phi,\s\mathrm{N}}^2 r(7-13\nu)
+2r^2(-1+\nu)
\Big)\chi_{\text{A}}^2
\nonumber \\
&\qquad \quad
+2\Big(
3p_{\phi,\s\mathrm{N}}^4(1+\nu)
+7p_{\phi,\s\mathrm{N}}^2 r(1-7\nu)
+2r^2(-1+\nu)
\Big)\delta \chi_{\text{A}}\chi_{\text{S}}
\nonumber \\
&\qquad \quad
+\Big(
3p_{\phi,\s\mathrm{N}}^4(1+\nu)
+p_{\phi,\s\mathrm{N}}^2 r(7-85\nu+18\nu^2)
+2r^2(-1+\nu)
\Big)\chi_{\text{S}}^2
\Big]
\nonumber \\
&\qquad
+2 \dot{r}^4 r^4 (p_{\phi,\s\mathrm{N}}^2+r) \Big[
 (-1+3\nu+4\nu^2)\chi_{\text{A}}^2
-2 \delta(1+\nu)\chi_{\text{A}}\chi_{\text{S}}
- (1+\nu)\chi_{\text{S}}^2
\Big]
\Bigg\}
,
\label{eq:omega_rrdprdst}
\\[8pt]
p_r (r, \dot r, \dot p_{r_*})
&=
\dot r
+\frac{\dot{r}\epsilon^2}{2r}
\Big[
7-\nu
+\dot{p}_{r_*} r^2(1+\nu)
+r\dot{r}^2(1+\nu)
\Big]
+\frac{\dot{r}\epsilon^4}{8r^2}
\Big\{
83-59\nu-\nu^2
+2\dot{p}_{r_*} r^2(14+11\nu)
+\dot{p}_{r_*}^2 r^4(1+5\nu+\nu^2)
\nonumber \\
&\qquad
+\dot{r}^2\Big[
-2r(-15-11\nu+\nu^2)
+2\dot{p}_{r_*} r^3(2+7\nu+2\nu^2)
\Big]
+3 \dot{r}^4
r^2(1+3\nu+\nu^2)
\Big\}
\nonumber \\
&\quad
+\frac{\dot{r}\epsilon^4}{r^2}
\Big[
(-1+4\nu)\chi_{\text{A}}^2
-2\delta\chi_{\text{A}}\chi_{\text{S}}
-\chi_{\text{S}}^2
\Big]
+\frac{p_{\phi,\s\mathrm{N}}\dot{r}\epsilon^5(3+\nu)}{2r^3}
\Big(
-2\delta\chi_{\text{A}}
+(-2+\nu)\chi_{\text{S}}
\Big)
\nonumber \\
&\quad
+\frac{\dot{r}\epsilon^6}{16r^3}
\Bigg\{
467-1275\nu+132\nu^2-\nu^3
+3\dot{p}_{r_*} r^2(88+39\nu-24\nu^2)
+\dot{p}_{r_*}^2 r^4(21+97\nu+14\nu^2+\nu^3)
+\dot{p}_{r_*}^3 r^6\nu(5+6\nu)
\nonumber \\
&\qquad
+\dot{r}^2\Big[
-r(-291+169\nu-74\nu^2+\nu^3)
+2\dot{p}_{r_*} r^3(44+139\nu+26\nu^2)
+3\dot{p}_{r_*}^2 r^5(1+11\nu+12\nu^2+\nu^3)
\Big]
\nonumber \\
&\qquad
+\dot{r}^4\Big[
3r^2(23+61\nu+12\nu^2-\nu^3)
+\dot{p}_{r_*} r^4(8+57\nu+60\nu^2+8\nu^3)
\Big]
+\dot{r}^6\Big[
r^3(5+29\nu+30\nu^2+5\nu^3)
\Big]
\Bigg\}
\nonumber \\
&\quad
+\frac{\dot{r}\epsilon^6}{2 r^3}
\Bigg\{
2(-5+18\nu+8\nu^2)\chi_{\text{A}}^2
-4\delta(5+8\nu)\chi_{\text{A}}\chi_{\text{S}}
+(-10-28\nu+6\nu^2)\chi_{\text{S}}^2
\nonumber \\
&\qquad
+\dot{r}^2\Big[
3r(1+\nu)(-1+4\nu)\chi_{\text{A}}^2
-6r\delta(1+\nu)\chi_{\text{A}}\chi_{\text{S}}
-3r(1+\nu)\chi_{\text{S}}^2
\Big]
\Bigg\}
,
\label{eq:pr_rrdprdst}
\\[8pt]
p_\phi (r, \dot r, \dot p_{r_*})
&=
p_{\phi,\s\mathrm{N}}
+
\frac{\epsilon^2}{4 p_{\phi,\s\mathrm{N}} r^2}
\Bigg\{
p_{\phi,\s\mathrm{N}}^4(1+\nu)
+p_{\phi,\s\mathrm{N}}^2 r(7-3\nu)
+2r^2(-1+\nu)
+\dot{r}^2 r^2 (1+\nu)(
p_{\phi,\s\mathrm{N}}^2
-r
)
\Bigg\}
\nonumber \\[3pt]
&\quad
+\frac{\epsilon^3}{2r}
\Big[
-6\delta\chi_{\text{A}}
+3(-2+\nu)\chi_{\text{S}}
\Big]
+
\frac{\epsilon^4}{32 p_{\phi,\s\mathrm{N}}^3 r^4}
\Bigg\{
p_{\phi,\s\mathrm{N}}^8(1+8\nu+\nu^2)
+2p_{\phi,\s\mathrm{N}}^6 r(22+7\nu)
+p_{\phi,\s\mathrm{N}}^4 r^2(103-110\nu-9\nu^2)
\nonumber \\[3pt]
&\qquad \quad
+4p_{\phi,\s\mathrm{N}}^2 r^3(-9+8\nu+3\nu^2)
-4r^4(-1+\nu)^2
\nonumber \\[3pt]
&\qquad
+\dot{r}^2\Big[
6p_{\phi,\s\mathrm{N}}^6 r^2(1+4\nu+\nu^2)
-4p_{\phi,\s\mathrm{N}}^4 r^3(-10+2\nu+3\nu^2)
+2p_{\phi,\s\mathrm{N}}^2 r^4(-21-8\nu+\nu^2)
+4r^5(-1+\nu^2)
\Big]
\nonumber \\[3pt]
&\qquad
+\dot{r}^4\Big[
p_{\phi,\s\mathrm{N}}^4 r^4(5+16\nu+5\nu^2)
-2p_{\phi,\s\mathrm{N}}^2 r^5(2+7\nu+2\nu^2)
-r^6(1+\nu)^2
\Big]
\Bigg\}
\nonumber \\[3pt]
&\quad
-\frac{\epsilon^4}{2p_{\phi,\s\mathrm{N}} r^2}
\Big(
2p_{\phi,\s\mathrm{N}}^2-r^2\dot{r}^2
\Big)
\Big[
(-1+4\nu)\chi_{\text{A}}^2
-2\delta\chi_{\text{A}}\chi_{\text{S}}
-\chi_{\text{S}}^2
\Big]
\nonumber \\[3pt]
&\quad
-\frac{4\dot{r}\epsilon_\text{RR}^5\nu}{15p_{\phi,\s\mathrm{N}} r^2}
\Big\{
r(17+9\alpha-9\beta)
+p_{\phi,\s\mathrm{N}}^2(3-3\alpha+9\beta)
+6 \dot{r}^2
r^2(2+\alpha-\beta)
\Big\}
\nonumber \\[3pt]
&\quad
+
\frac{\epsilon^5}{32 r^3}
\Bigg\{
\Big[
p_{\phi,\s\mathrm{N}}^2\delta(-96+119\nu)
-4r\delta(36+13\nu)
\Big]\chi_{\text{A}}
+\Big[
p_{\phi,\s\mathrm{N}}^2(-96+317\nu-82\nu^2)
+4r(-36+11\nu+20\nu^2)
\Big]\chi_{\text{S}}
\nonumber \\[3pt]
&\qquad
+\dot{r}^2\Big[
-48r^2\delta\chi_{\text{A}}
+24r^2(-2+\nu)\chi_{\text{S}}
\Big]
\Bigg\}
\nonumber \\[3pt]
&\quad
-
\frac{\epsilon^6}{384 p_{\phi,\s\mathrm{N}}^5 r^6}
\Bigg\{
3p_{\phi,\s\mathrm{N}}^{12}(1-11\nu-15\nu^2+\nu^3)
-3p_{\phi,\s\mathrm{N}}^{10}r(41+237\nu-29\nu^2+7\nu^3)
\nonumber \\[3pt]
&\qquad \quad
+3p_{\phi,\s\mathrm{N}}^8 r^2(-691+273\nu+93\nu^2+5\nu^3)
+3p_{\phi,\s\mathrm{N}}^6 r^3(-707+2777\nu-373\nu^2+31\nu^3)
\nonumber \\[3pt]
&\qquad \quad
+p_{\phi,\s\mathrm{N}}^4 r^4(918+(5726-492\pi^2)\nu+762\nu^2-174\nu^3)
+12p_{\phi,\s\mathrm{N}}^2 r^5(11-17\nu-3\nu^2+9\nu^3)
-24r^6(-1+\nu)^3
\nonumber \\[3pt]
&\qquad
+\dot{r}^2\Big[
-3p_{\phi,\s\mathrm{N}}^{10}r^2(5+93\nu+105\nu^2+5\nu^3)
+3p_{\phi,\s\mathrm{N}}^8 r^3(-223-597\nu+147\nu^2+5\nu^3)
\nonumber \\[3pt]
&\qquad \quad
+3p_{\phi,\s\mathrm{N}}^6 r^4(-483+1001\nu+269\nu^2+9\nu^3)
+3p_{\phi,\s\mathrm{N}}^4 r^5(611-267\nu-251\nu^2+3\nu^3)
\nonumber \\[3pt]
&\qquad \quad
-24p_{\phi,\s\mathrm{N}}^2 r^6(-11+4\nu+6\nu^2+3\nu^3)
+36r^7(-1+\nu)^2(1+\nu)
\Big]
\nonumber \\[3pt]
&\qquad
+\dot{r}^4\Big[
-9p_{\phi,\s\mathrm{N}}^8 r^4(7+59\nu+63\nu^2+7\nu^3)
+9p_{\phi,\s\mathrm{N}}^6 r^5(-57-89\nu+77\nu^2+13\nu^3)
\nonumber \\[3pt]
&\qquad \quad
-3p_{\phi,\s\mathrm{N}}^4 r^6(-147-887\nu+417\nu^2+9\nu^3)
-9p_{\phi,\s\mathrm{N}}^2 r^7(-13-21\nu+\nu^2+\nu^3)
-18r^8(-1+\nu)(1+\nu)^2
\Big]
\nonumber \\[3pt]
&\qquad
+\dot{r}^6\Big[
-3p_{\phi,\s\mathrm{N}}^6 r^6(15+95\nu+99\nu^2+15\nu^3)
+3p_{\phi,\s\mathrm{N}}^4 r^7(11+77\nu+81\nu^2+11\nu^3)
\nonumber \\[3pt]
&\qquad \quad
+9p_{\phi,\s\mathrm{N}}^2 r^8(1+5\nu+5\nu^2+\nu^3)
+3r^9(1+\nu)^3
\Big]
\Bigg\}
\nonumber \\[3pt]
&\quad
+
\frac{\epsilon^6}{16 p_{\phi,\s\mathrm{N}}^3 r^4}
\Bigg\{
\Big[
-8p_{\phi,\s\mathrm{N}}^6(-1+3\nu+4\nu^2)
+8p_{\phi,\s\mathrm{N}}^4 r(7-30\nu+8\nu^2)
-8p_{\phi,\s\mathrm{N}}^2 r^2(-2+9\nu+4\nu^2)
\Big]\chi_{\text{A}}^2
\nonumber \\[3pt]
&\qquad \quad
+8p_{\phi,\s\mathrm{N}}^2\Big[
2p_{\phi,\s\mathrm{N}}^4(1+\nu)
+2p_{\phi,\s\mathrm{N}}^2 r(7-2\nu)
+r^2(4-19\nu)
\Big]\delta\chi_{\text{A}}\chi_{\text{S}}
\nonumber \\[3pt]
&\qquad \quad
+\Big[
8p_{\phi,\s\mathrm{N}}^6(1+\nu)
+8p_{\phi,\s\mathrm{N}}^4 r(7-2\nu)
+2p_{\phi,\s\mathrm{N}}^2 r^2(8-72\nu+25\nu^2)
\Big]\chi_{\text{S}}^2
\nonumber \\[3pt]
&\qquad
+\dot{r}^2\Big[
2r^2(-1+4\nu)\Big(
5p_{\phi,\s\mathrm{N}}^4(1+\nu)
+p_{\phi,\s\mathrm{N}}^2 r(-1+11\nu)
-2r^2(-1+\nu)
\Big)\chi_{\text{A}}^2
\nonumber \\[3pt]
&\qquad \quad
+4r^2\Big(
-5p_{\phi,\s\mathrm{N}}^4(1+\nu)
+p_{\phi,\s\mathrm{N}}^2 r(1-47\nu)
+2r^2(-1+\nu)
\Big)\delta \chi_{\text{A}}\chi_{\text{S}}
\nonumber \\[3pt]
&\qquad \quad
+2r^2\Big(
-5p_{\phi,\s\mathrm{N}}^4(1+\nu)
+p_{\phi,\s\mathrm{N}}^2 r(1-83\nu+18\nu^2)
+2r^2(-1+\nu)
\Big)\chi_{\text{S}}^2
\Big]
\nonumber \\
&\qquad
+\dot{r}^4\Big[
2r^4(3p_{\phi,\s\mathrm{N}}^2+r)(-1+3\nu+4\nu^2)\chi_{\text{A}}^2
-4r^4(3p_{\phi,\s\mathrm{N}}^2+r)(1+\nu) \delta\chi_{\text{A}}\chi_{\text{S}}
-2r^4(3p_{\phi,\s\mathrm{N}}^2+r)(1+\nu)\chi_{\text{S}}^2
\Big]
\Bigg\}
,
\label{eq:L_rrdprdst}
\end{align}
\end{subequations}
%
These expressions are also provided in the Supplemental Material~\cite{SEOBNRv6EHM:SupplementalMaterial}.

To obtain a dependence on $ \dot p_{r_*, \s\text{cons}}   = - \xi(r) \, \partial H_\text{EOB}/\partial r$, one just needs to remove the dissipative terms appearing at 2.5PN order in the nonspinning part of Eqs.~\eqref{eq:omega_rrdprdst} and \eqref{eq:L_rrdprdst}, flagged with the parameter $ \epsilon_\text{RR} $.
Since $ \dot p_{r_*} $ appears at 1PN order in Eq.~\eqref{eq:pr_rrdprdst}, then one can just make the replacement $ \dot p_{r_*}   \to \dot p_{r_*, \s\text{cons}} $ at 3PN accuracy.


\subsection{PN-expanded RR force}
\label{sec:ap_PN_RR}

The 1PN expressions of the PN-expanded RR force components derived in Sec.~\ref{sec:PN_RR}, in terms of the leading-order gauge constants $ \alpha $ and $ \beta $, are given by:
%
\begin{subequations}
\label{eq:RRforces_EOB_1PN}
\begin{align}
\mF_\phi ^{ 1\text{PN}}( r, p_r, p_\phi )  &=
\frac{8\nu p_\phi  }{5 r^3} \left[ 
(2 \alpha +1 ) \, p_r^2  
- \frac{ \alpha + 2}{r^2} \s p_\phi^2
+ \frac{\alpha - 2}{r}
\right]
+ \frac{\epsilon ^2 \nu p_\phi   }{r^3}  \bigg\{
p_r^4 \left[ \left(-\frac{8 \alpha }{5}-\frac{236}{105}\right) \nu -\frac{8 \alpha }{5}-\frac{548}{105} \right]
\\
&\qquad
+\frac{p_{\phi }^2}{r^3}\left[\left(\frac{6 \alpha }{5}+\frac{188}{105}\right) \nu +\frac{2423 \alpha }{210}+\frac{1577}{63}\right] 
+\frac{p_r^2 p_{\phi }^2}{r^2}\left[\left(\frac{548}{105}-\frac{4 \alpha }{5}\right) \nu -\frac{4 \alpha }{5}-\frac{88}{105}\right] \nonumber 
\\
&\qquad
+ \frac{p_{\phi }^4}{r^4}\left[\left(\frac{4 \alpha }{5}+\frac{424}{105}\right) \nu +\frac{4 \alpha }{5}-\frac{22}{21}\right]
+\frac{p_r^2}{r}\left[\left(\frac{724}{21}-\frac{42 \alpha }{5}\right) \nu -\frac{2423 \alpha }{70}-\frac{3821}{105}\right]
\\
&\qquad
+\frac{1}{r^2} \left[\left(\frac{188}{35}-2 \alpha \right) \nu -\frac{1583 \alpha }{210}+\frac{1195}{63}\right]
\bigg\} , 
\\[8pt]
\mF_r ^{1\text{PN}} ( r, p_r, p_\phi ) &=
\frac{8 \nu p_r }{15 r^3}\left[6 \s (\alpha -\beta + 2)\, p_r^2 
-  \frac{3 \s (\alpha - 3 \beta-1 ) }{r^2} \s p_\phi^2
+ \frac{9 \alpha -9 \beta + 17 }{r}\right]
\nonumber\\
&\quad
+ \frac{\epsilon^2 \nu p_r}{r^3} \Bigg\{
p_r^4\left[\left(-\frac{8 \alpha }{5}-\frac{848}{105}\right) \nu -\frac{8 \alpha }{5}+\frac{44}{21}\right]
+\frac{p_{\phi }^2}{r^3}\left[\left(-\frac{2 \alpha }{5}-\frac{3256}{105}\right) \nu +\frac{71 \alpha }{210}+\frac{24 \beta }{5}+\frac{89}{63}\right] 
\nonumber\\
&\qquad
+\frac{p_r^2 p_{\phi }^2}{r^2}\left[\left(-\frac{4 \alpha }{5}-\frac{148}{15}\right) \nu -\frac{4 \alpha }{5}+\frac{878}{105}\right]
+\frac{p_{\phi }^4}{r^4}\left[\left(\frac{4 \alpha }{5}-\frac{548}{105}\right) \nu +\frac{4 \alpha }{5}+\frac{88}{105}\right]
\nonumber\\
&\qquad
+\frac{p_r^2}{r}\left[\left(\frac{40}{21}-10 \alpha \right) \nu -\frac{1863 \alpha }{70}-\frac{48 \beta }{5}-\frac{1057}{15}\right]
+\frac{1}{r^2} \left[\left(\frac{1844}{105}-6 \alpha \right) \nu -\frac{911 \alpha }{70}-\frac{96 \beta }{5}-\frac{6577}{105}\right]
\Bigg\} . 
\end{align}
\end{subequations}
%
The complete 3PN expressions and the corresponding Schott terms are provided in the Supplemental Material~\cite{SEOBNRv6EHM:SupplementalMaterial}.


\subsection{Eccentricity corrections to the RR force and the sigmoid resummation}
\label{sec:RR_corr_exp_resum}

PN expressions for the RR force tend to grow significantly during plunges or strong periastron passages, spoiling the numerical integration of the equations of motion.
Here, we introduce an \emph{sigmoid resummation} which limits their growth.

We consider a PN expanded function $ f ^{ \text{PN}}$ which has been factorized as
\begin{subequations}
\label{eq:F_fact_original}
\begin{align}
f ^{ \text{F}}
&\equiv
 f ^{ \text{fact}} \, f ^{ \text{corr}}  ,
\qquad \quad
f ^{ \text{corr}}
\equiv
1 + \sum_{i=1}^{n} \epsilon^i  \Delta f_i ,
\end{align}
\end{subequations}
where $ f ^{ \text{fact}} $ is another factorization containing the leading-order (Newtonian) term of $ f ^{ \text{PN}} $ (and, potentially, additional higher PN terms), and the functions $ \Delta f_i $ are defined in such a way that the PN expansion of $ f ^{ \text{F}} $ coincides with $ f ^{ \text{PN}}$ at a given $ n $-th PN order.

Then, the sigmoid resummation of $ f ^{ \text{corr}} $ is defined as
\begin{subequations}
\label{eq:ecc_resummation}
\begin{align}
f ^{ \text{exp}}
&\equiv
\frac{2}{1 + \exp(f ^{ \text{arg exp}})} ,
\qquad \quad 
f ^{ \text{arg exp}}
\equiv
\sum_{i=1}^{n} \epsilon^i  \Delta f_i ^{ \text{arg exp}} ,
\end{align}
\end{subequations}
where the corrections $  \Delta f_i ^{ \text{arg exp}} $ make the PN expansion of $ f ^{ \text{exp}} $ to coincide with $ f ^{ \text{corr}} $.
At 3PN order, a straightforward calculation (a Taylor expansion in the PN parameter $ \epsilon $) yields
\begin{subequations}
\label{eq:ecc_resummation_PNorders}
\begin{align}
\Delta f ^{ \text{arg exp}}_1
&=
- 2 \Delta f_1 ,
\label{eq:ecc_resummation_PNorders_1}
\\
\Delta f ^{ \text{arg exp}}_2
&=
- 2 \Delta f_2 ,
\label{eq:ecc_resummation_PNorders_2}
\\
\Delta f ^{ \text{arg exp}}_3
&=
- 2 \Delta f_3 - \frac{2}{3} \Delta f_1^3 ,
\\
\Delta f ^{ \text{arg exp}}_4
&=
-2 \Delta f_4 - \frac{6}{3} \Delta f_1^2 \Delta f_2 ,
\\
\Delta f ^{ \text{arg exp}}_5
&=
-2 \Delta f_5 - \frac{2}{5} \Delta f_1^5 - 2 \Delta f_1 \Delta f_2^2 - 2 \Delta f_1^2 \Delta f_3,
\\
\Delta f ^{ \text{arg exp}}_6
&=
-2 \Delta f_6 - \frac{2}{3} \Delta f_2^3  - 2 \Delta f_1^4 \Delta f_2 - 2 \Delta f_1^2 \Delta f_4 - 4 \Delta f_1 \Delta f_2 \Delta f_3 .
\end{align}
\end{subequations}
Thus, given the PN coefficients $ \Delta f_i $, these formulas give the elements entering into the argument of the exponential.
Importantly, if corrections to $ f ^{ \text{PN}} $ start at 1PN order ($ n=2 $), then $ \Delta f_1 =\nolinebreak 0 $, and Eqs.~\eqref{eq:ecc_resummation_PNorders} simplify significantly.

In this way, we can resum the PN-expanded function $ f ^{ \text{PN}} $ as
\begin{equation}
f ^{ \text{F, exp}} \equiv f ^{ \text{fact}} \; \frac{2}{1 + \exp(f ^{ \text{arg exp}})} ,
\end{equation}
from which we can observe the following properties:
\begin{itemize}
\item
$ f ^{ \text{F, exp}} \to f ^{ \text{fact}} \,$ when $ f^{\text{arg exp} } \to 0 $, e.g., in the Newtonian, low-velocity limit.
\item
$  f ^{ \text{F, exp}} \to 0 \,$ or $  f ^{ \text{F, exp}} \to 2  f ^{ \text{fact}} \,$  when $ f^{\text{arg exp} }  \to \pm \infty $, respectively, e.g., during strong periastron passages or the plunge.
\end{itemize}

The connection with the RR force~\eqref{eq:RR_force_proposal} is immediate:
after extracting a factor which contains the leading order RR force \big($\s f ^{ \text{fact}} \in \nolinebreak {\big\{ \mF_\phi^\text{modes}, \,  p_r/p_\phi \ \mF_r^\text{modes} \big\} } $\big), the corresponding corrections \big($\s f ^{ \text{corr}} \in \big\{ \mF_\phi^\text{ecc}, \, \mF_r^\text{ecc} \big\}$\big) can be resummed with the method discussed above.
This will result in corrections to the RR force which have a controlled behavior during high-velocity configurations.
Particularly, the fact that $  f ^{ \text{F, exp}} \to \{ 0, \, 2  f ^{ \text{fact}} \}$ is the key property that enables the RR force to stay constrained during the plunge, thus avoiding a premature end of the dynamics.

As discussed in Sec.~\ref{sec:proposal_RR}, we first calculate the PN-expanded eccentricity corrections, $ \mF_\phi^\text{ecc, PN} $ and $ \mF_r^\text{ecc, PN} $, from the PN expression of the RR force [e.g., see \eqref{eq:RRforces_EOB_1PN}] and the proposed resummation \eqref{eq:RR_force_proposal}.
Then, these are transformed to the parametrization $ (r, \dot r, \dot p_{r*,\s \mathrm{cons}} ) $ with the rules derived in Appendix~\ref{sec:transformation_to_rrdprdst}.
The resulting PN-expanded expressions are then resummed with the rules \eqref{eq:ecc_resummation} and Eq.~\eqref{eq:ecc_resummation_PNorders}.
This process leads to
\begin{subequations}
\label{eq:F_ecc_resum_appendix}
\begin{align}
\mF_\phi^\text{ecc} &= \frac{2}{1 + \exp\big(\mF_{\phi}^{\text{arg exp} }\big)},
\qquad \quad
\mF_r^\text{ecc} = \frac{2}{1 + \exp\big(\mF_{r}^{\text{arg exp} }\big)},
\end{align}
\end{subequations}
where
%
\begin{subequations}
\label{eq:F_argexp}
\begin{align}
\mF_{\phi}^{\text{arg exp} }
&=
- 2 \epsilon^2 \Bigg\{
\frac{1}{r} \left( \frac{ (273 \alpha -506) \nu}{336} +\frac{717 \alpha -7918}{4032} \right)
- \frac{p_{\phi,\s\text{N}}^2}{r^2} \left(\frac{5 (63 \alpha +86) \nu}{336} +  \frac{213 \alpha +3674}{4032}  \right)
\nonumber \\
& \qquad
+ \frac{p_{\phi,\s\text{N}}^4 }{ r^3} \left( \frac{(21 \alpha -22) \nu}{168}  +\frac{55-42 \alpha }{336}  \right)
- \frac{p_{\phi,\s\text{N}}^{2/3}}{ r^{4/3}} \frac{(\alpha -2)  (2828 \nu +2761)}{4032}
+ \frac{p_{\phi,\s\text{N}}^{16/3}}{r^{11/3}} \frac{(\alpha +2)  (4 \nu -1)}{144 }
\nonumber \\
& \qquad
+ \frac{p_{\phi,\s\text{N}}^{8/3} }{r^{7/3}} \frac{(\alpha +2) (2828 \nu +2761)}{4032 }
- \frac{p_{\phi,\s\text{N}}^{10/3}}{r^{8/3}} \frac{ (\alpha -2) (4 \nu -1)}{144 }
\nonumber \\
& \qquad
+\dot r^2 \Bigg[
\frac{41(21 \alpha -38) \nu}{336}  +\frac{1221 \alpha +3946}{1344}
+ \frac{p_{\phi,\s\text{N}}^2}{r} \left(  -\frac{(63 \alpha +116) \nu}{168}  +\frac{1-21 \alpha}{168} \right)
- \frac{p_{\phi,\s\text{N}}^{10/3} }{r^{5/3}} \frac{(2 \alpha +1) (4 \nu -1)}{144 }
\nonumber \\
& \qquad \quad
- \frac{p_{\phi,\s\text{N}}^{2/3} }{r^{1/3}} \frac{(2 \alpha +1) (2828 \nu +2761)}{4032 }
\Bigg]
- r \s \dot r^4 \Bigg[ \frac{(21 \alpha +1) \nu}{42} - \frac{29}{42}   \Bigg]
\Bigg\}
\;
\Bigg( 1 - \frac{2 \alpha +1}{4} \, r \s \dot r^2 + \frac{\alpha +2}{4} \, r^2 \s \dot p_{r_*, \s\text{cons}}   \Bigg)^{-1},
\label{eq:F_argexp_phi}
\\[8pt]
\mF_{r}^{\text{arg exp} } 
&=
- 2 \epsilon^2 \Bigg\{
\frac{1}{r} \left( \frac{  (2289 \alpha -1512 \beta -470)\nu}{1008}+\frac{-627 \alpha +8064 \beta +7778}{1344} \right)
+ \frac{p_{\phi,\s\text{N}}^4}{r^3} \left( \frac{  (21 \alpha -126 \beta +95)\nu}{168} -\frac{21 \alpha +22}{168}   \right)
\nonumber \\
& \qquad
+ \frac{p_{\phi,\s\text{N}}^2 }{r^2} \left( \frac{   (-315 \alpha +756 \beta +1600)\nu}{336}+\frac{7851 \alpha -24192 \beta -4586}{4032} \right)
- \frac{p_{\phi,\s\text{N}}^{2/3} }{r^{4/3}}\frac{(2828 \nu +2761) (9 \alpha -9 \beta +17)}{12096 }
\nonumber \\
& \qquad
+ \frac{p_{\phi,\s\text{N}}^{8/3}}{r^{7/3}}\frac{ (2828 \nu +2761) (\alpha -3 \beta -1)}{4032 }
- \frac{p_{\phi,\s\text{N}}^{10/3} }{r^{8/3}}\frac{(4 \nu -1) (9 \alpha -9 \beta +17)}{432 }
+ \frac{p_{\phi,\s\text{N}}^{16/3}}{r^{11/3}}\frac{ (4 \nu -1) (\alpha -3 \beta -1)}{144 }
\nonumber \\
& \qquad
+\dot r^2 \Bigg[
\frac{  (861 \alpha -252 \beta +712)\nu}{336} +\frac{-41 \alpha +2688 \beta +1274}{448} 
- \frac{p_{\phi,\s\text{N}}^{10/3}}{r^{5/3}}\frac{ (4 \nu -1) (\alpha -\beta +2)}{72 }
\nonumber \\
& \qquad \quad
- \frac{p_{\phi,\s\text{N}}^{2/3}}{ r^{1/3}}\frac{ (2828 \nu +2761) (\alpha -\beta +2)}{2016}
- \frac{p_{\phi,\s\text{N}}^2 }{r} \left(  \frac{(9 \alpha +5) \nu}{24}  - \frac{-42 \alpha +84 \beta -607}{336} \right)\Bigg]
\nonumber \\
& \qquad 
+r \s \dot r^4 \Bigg[ \frac{  (-42 \alpha +63 \beta -20) \nu}{84} +\frac{42 \beta -139}{168} \Bigg]
\Bigg\}
\;
\Bigg( \frac{ - 3 \alpha -10}{6} -\frac{\alpha -\beta +2}{2} \, r \s \dot r^2 + \frac{\alpha -3 \beta -1}{4} \, r^2 \s \dot p_{r_*, \s\text{cons}}   \Bigg)^{-1}.
\label{eq:F_argexp_r}
\end{align}
\end{subequations}
%
These expressions are employed in \texttt{SEOBNRv6EHM}.
The 3PN generalizations are provided in the Supplemental Material~\cite{SEOBNRv6EHM:SupplementalMaterial}.


\subsection{Eccentricity corrections to the waveform modes}
\label{sec:ecc_corr_modes}

As discussed in Sec.~\ref{sec:factorized_modes}, we calculate the eccentricity corrections, $ h_{\ell m}^\mathrm{ecc} $, to the factorized waveform modes \eqref{eq:model_fact} using the Newtonian prefactor given in Eq.~\eqref{eq:model_hlm_N_generic} and the 1PN (relative to the leading-order of the $ (\ell, m) $ mode) expressions for the modes given in Ref.~\cite{Gamboa:2024imd}.
Then, we transform the resulting expressions to the parametrization $ (r, \dot r, \dot p_{r*,\s \mathrm{cons}} ) $ with the rules derived in Appendix~\ref{sec:transformation_to_rrdprdst}.
This process yields the following corrections, as used in \texttt{SEOBNRv6EHM}:
%
\begin{subequations}
\label{eq:modes_ecc_corr}
\begin{align}
h_{22}^\mathrm{ecc} &= 
1
+ \epsilon^2 \Bigg\{
56 (\nu -3) \, r^2
+ (86-55 \nu ) \, r^{5/3} p_{\phi,\s\mathrm{N}}^{2/3}
+ (86-55 \nu ) \, r^{2/3} p_{\phi,\s\mathrm{N}}^{8/3}
+ 3 (8 \nu -5) \, p_{\phi,\s\mathrm{N}}^4
+ (30 \nu +11) \, r p_{\phi,\s\mathrm{N}}^2 
\nonumber
\\
& \qquad
+ \dot r \,
\bigg[
-2 i (55 \nu -86) \, r^{5/3} p_{\phi,\s\mathrm{N}}^{5/3}
+ 6 i (15 \nu +2) \, r p_{\phi,\s\mathrm{N}}^3
-2 i (96 \nu -25) \, r^2 p_{\phi,\s\mathrm{N}}
\bigg] 
\nonumber
\\
& \qquad
+\dot r^2
\bigg[
(55 \nu -86) \, r^{8/3} p_{\phi,\s\mathrm{N}}^{2/3}
- 21 (\nu +1) \, r^2 p_{\phi,\s\mathrm{N}}^2
+ (72 \nu -129) \, r^3
\bigg]
+ \dot r^3 \, 6 i (15 \nu +2) \, r^3 p_{\phi,\s\mathrm{N}}
\nonumber
\\
& \qquad
+ \dot r^4 \, (-45 \nu -6) \, r^4
\Bigg\}
\;
\Bigg[
42 \, r^2 \left(p_{\phi,\s\mathrm{N}}^2+2 i  r p_{\phi,\s\mathrm{N}} \s \dot r-r^2 \dot r^2+r\right)
\Bigg]^{-1} \!,
\label{eq:h22_ecc_1PN}
\\[6pt]
h_{21}^\mathrm{ecc} &=
\frac{r^{2/3}}{p_{\phi,\s\mathrm{N}} ^{ 4/3}}
+ \epsilon^2 \frac{1}{84  r^{4/3} p_{\phi,\s\mathrm{N}}^{10/3}} \Bigg\{
4 (26 \nu -69) \, r p_{\phi,\s\mathrm{N}}^2 -56 (\nu -1) \, r^2 + (177-46 \nu ) \, r^{2/3} p_{\phi,\s\mathrm{N}}^{8/3} + (43-2 \nu ) p_{\phi,\s\mathrm{N}}^4
\nonumber
\\
& \qquad
- \dot r \, 6 i (12 \nu -83) \, r  p_{\phi,\s\mathrm{N}}^3 
+ \dot r^2 \bigg[ (-38 \nu -212) \, r^2 p_{\phi,\s\mathrm{N}}^2 +28 (\nu +1) \, r^3\bigg]
\Bigg\},
\\[6pt]
h_{33}^\mathrm{ecc} &=
1
+ \epsilon^2 \Bigg\{
18 i (7 - 4 \nu) \, r^{2/3} p_{\phi,\s\mathrm{N}}^{11/3} - 63 i (4 \nu -7) \, r^{5/3} p_{\phi,\s\mathrm{N}}^{5/3} + 6 i (5 \nu -1) \, p_{\phi,\s\mathrm{N}}^5 +9 i (14 \nu -5) \, r p_{\phi,\s\mathrm{N}}^3 +12 i  (14 \nu -43) \, r^2 p_{\phi,\s\mathrm{N}}
\nonumber
\\[4pt]
& \qquad
+\dot r \, \bigg[36 (4 \nu -7) \, r^{8/3} p_{\phi,\s\mathrm{N}}^{2/3} + 54 (4 \nu -7) \, r^{5/3} p_{\phi,\s\mathrm{N}}^{8/3} + (18-90 \nu ) \, r p_{\phi,\s\mathrm{N}}^4 + 270 (\nu -1) \, r^2 p_{\phi,\s\mathrm{N}}^2 + (182-100 \nu ) \, r^3\bigg]
\nonumber
\\[4pt]
& \qquad
+\dot r^2 \bigg[54 i (4 \nu -7) \, r^{8/3} p_{\phi,\s\mathrm{N}}^{5/3} - 6 i (19 \nu +7) \, r^2 p_{\phi,\s\mathrm{N}}^3 + 54 i (6 \nu -7) \, r^3 p_{\phi,\s\mathrm{N}} \bigg]
\nonumber
\\[4pt]
& \qquad
+\dot r^3 \bigg[(126-72 \nu ) \, r^{11/3} p_{\phi,\s\mathrm{N}}^{2/3} + (30-42 \nu ) \, r^3 p_{\phi,\s\mathrm{N}}^2 +(222-66 \nu ) \, r^4\bigg]
-\dot r^4 \, 36 i (4 \nu +1) \, r^4  p_{\phi,\s\mathrm{N}}
\nonumber
\\[4pt]
& \qquad
+ \dot r^5 \, (48 \nu +12) \, r^5 \Bigg\}
\;
\Bigg[
36 i r^2 p_{\phi,\s\mathrm{N}}^3 +126 i  r^3 p_{\phi,\s\mathrm{N}} +\dot r \left(-108  r^3 p_{\phi,\s\mathrm{N}}^2 -72 r^4\right)-\dot r^2 \, 108 i  r^4 p_{\phi,\s\mathrm{N}} + \dot r^3 \, 36 r^5 
\Bigg]^{-1} \!,
\\[6pt]
h_{32}^\mathrm{ecc} &=
\frac{r^{2/3}}{p_{\phi,\s\mathrm{N}} ^{ 4/3}}
+ \epsilon^2 \Bigg\{
-4 \left(320 \nu ^2-1115 \nu +328\right) r^{2/3} p_{\phi,\s\mathrm{N}}^{11/3}-6 \left(25 \nu ^2+205 \nu -53\right) p_{\phi,\s\mathrm{N}}^5 -120 \left(3 \nu ^2+2 \nu -1\right) r^2 p_{\phi,\s\mathrm{N}}
\nonumber
\\
& \qquad \quad
+ \left(1790 \nu ^2-2990 \nu +874\right) r p_{\phi,\s\mathrm{N}}^3
\nonumber
\\
& \qquad
+\dot r \; \bigg[ -i \left(320 \nu ^2-1115 \nu +328\right) r^{5/3} p_{\phi,\s\mathrm{N}}^{8/3}-15 i  \left(72 \nu ^2-325 \nu +95\right) r p_{\phi,\s\mathrm{N}}^4 -10 i \left(35 \nu ^2-215 \nu +64\right) r^2 p_{\phi,\s\mathrm{N}}^2 
\nonumber
\\
& \qquad \quad
- 60 i \left(3 \nu ^2-4 \nu +1\right) r^3 \bigg]
+\dot r^2 \bigg[-12 \left(90 \nu ^2+250 \nu -89\right) r^2 p_{\phi,\s\mathrm{N}}^3 -60 \left(3 \nu ^2+2 \nu -1\right) r^3 p_{\phi,\s\mathrm{N}} \bigg]
\nonumber
\\
& \qquad
+\dot r^3 \bigg[30 i \left(3 \nu ^2+2 \nu -1\right) r^4-15 i \left(4 \nu ^2+43 \nu -14\right) r^3 p_{\phi,\s\mathrm{N}}^2\bigg]
\Bigg\}
\;
\Bigg[
90 (3 \nu -1) (4 p_{\phi,\s\mathrm{N}}+i \, r \s \dot r) \,  r^{4/3}  p_{\phi,\s\mathrm{N}}^{10/3} 
\Bigg]^{-1}\!,
\\[6pt]
h_{44}^\mathrm{ecc} &=
1
+ \epsilon^2 \Bigg\{
-14 \left(2625 \nu ^2-5870 \nu +1614\right) r^{8/3} p_{\phi,\s\mathrm{N}}^{2/3} -102 \left(2625 \nu ^2-5870 \nu +1614\right) r^{5/3} p_{\phi,\s\mathrm{N}}^{8/3}
\nonumber
\\
& \qquad \quad
+40 \left(1344 \nu ^2-4699 \nu +1437\right) r^3
-12 \left(2625 \nu ^2-5870 \nu +1614\right) r^{2/3} p_{\phi,\s\mathrm{N}}^{14/3} 
\nonumber
\\[4pt]
& \qquad \quad
+180 \left(39 \nu ^2-50 \nu +10\right) p_{\phi,\s\mathrm{N}}^6+9  \left(9420 \nu ^2-15580 \nu +3127\right) r p_{\phi,\s\mathrm{N}}^4 +9  \left(21160 \nu ^2-46020 \nu +13241\right) r^2 p_{\phi,\s\mathrm{N}}^2
\nonumber
\\[4pt]
& \qquad
+\dot r \; \bigg[ 720 i  \left(72 \nu ^2-28 \nu -1\right) r p_{\phi,\s\mathrm{N}}^5 -108 i  \left(2625 \nu ^2-5870 \nu +1614\right) r^{8/3} p_{\phi,\s\mathrm{N}}^{5/3} -24 i  \left(345 \nu ^2+5320 \nu -2076\right) r^3 p_{\phi,\s\mathrm{N}}
\nonumber
\\[4pt]
& \qquad \quad
-36 i \left(3765 \nu ^2-4700 \nu +1237\right) r^2 p_{\phi,\s\mathrm{N}}^3
-48 i  \left(2625 \nu ^2-5870 \nu +1614\right) r^{5/3} p_{\phi,\s\mathrm{N}}^{11/3}
\bigg]
\nonumber
\\[4pt]
& \qquad
+\dot r^2 \bigg[36 \left(2625 \nu ^2-5870 \nu +1614\right) r^{11/3} p_{\phi,\s\mathrm{N}}^{2/3}+72\left(2625 \nu ^2-5870 \nu +1614\right) r^{8/3} p_{\phi,\s\mathrm{N}}^{8/3}
\nonumber
\\[4pt]
& \qquad \quad
+180 \left(-393 \nu ^2+118 \nu +16\right) r^2 p_{\phi,\s\mathrm{N}}^4 +18  \left(10740 \nu ^2-30320 \nu +8561\right) r^3 p_{\phi,\s\mathrm{N}}^2-120 \left(819 \nu ^2+61 \nu -87\right) r^4 \bigg]
\nonumber
\\[4pt]
& \qquad
+\dot r^3 \bigg[48 i  \left(2625 \nu ^2-5870 \nu +1614\right) r^{11/3} p_{\phi,\s\mathrm{N}}^{5/3} -7920 i \left(3 \nu ^2+2 \nu -1\right) r^3 p_{\phi,\s\mathrm{N}}^3 +36 i  \left(5415 \nu ^2-10090 \nu +2568\right) r^4 p_{\phi,\s\mathrm{N}} \bigg]
\nonumber
\\[4pt]
& \qquad
+\dot r^4 \bigg[-12  \left(2625 \nu ^2-5870 \nu +1614\right) r^{14/3} p_{\phi,\s\mathrm{N}}^{2/3} -540 \left(109 \nu ^2-54 \nu +2\right) r^4 p_{\phi,\s\mathrm{N}}^2 -540 \left(43 \nu ^2-230 \nu +68\right) r^5 \bigg]
\nonumber
\\[4pt]
& \qquad
-\dot r^5 \, 2160 i  \left(35 \nu ^2-2 \nu -4\right) r^5 p_{\phi,\s\mathrm{N}}
+ \dot r^6 \, 540 \left(35 \nu ^2-2 \nu -4\right) r^6
\nonumber
\\[4pt]
& \qquad
\Bigg\}
\;
\Bigg\{
660 (3 \nu -1) \, r^2 \bigg[ 6 p_{\phi,\s\mathrm{N}}^4 +51  r p_{\phi,\s\mathrm{N}}^2 +7 r^2
+\dot r \, i  r p_{\phi,\s\mathrm{N}} \left(24 p_{\phi,\s\mathrm{N}}^2 +54  r \right)
-\dot r^2 18 r^2 \left(2 p_{\phi,\s\mathrm{N}}^2+ r\right) - \dot r^3 \, 24 i r^3 p_{\phi,\s\mathrm{N}} +\dot r^4 \, 6 r^4 \bigg]
\Bigg\}^{-1}\!,
\\[2pt]
h_{43}^\mathrm{ecc} &=
\frac{r^{2/3}}{p_{\phi,\s\mathrm{N}} ^{ 4/3}}
+ \epsilon^2 \Bigg\{
-352 \left(2 \nu ^2-3 \nu +1\right) r^3 -12 \left(160 \nu ^2-547 \nu +222\right) r^{5/3} p_{\phi,\s\mathrm{N}}^{8/3}  + \left(1064 \nu ^2-12701 \nu +3596\right) p_{\phi,\s\mathrm{N}}^6
\nonumber
\\
& \qquad \quad
+\left(9980 \nu ^2-25084 \nu +11254\right) r p_{\phi,\s\mathrm{N}}^4  +\left(2620 \nu ^2-7578 \nu +3484\right) r^2 p_{\phi,\s\mathrm{N}}^2 -69 \left(160 \nu ^2-547 \nu +222\right) r^{2/3} p_{\phi,\s\mathrm{N}}^{14/3}
\nonumber
\\
& \qquad
+\dot r \, \bigg[
440 i  \left(2 \nu ^2-3 \nu +1\right) r^3 p_{\phi,\s\mathrm{N}} -4 i \left(2093 \nu ^2-5999 \nu +2168\right) r  p_{\phi,\s\mathrm{N}}^5 -2 i \left(2438 \nu ^2-17569 \nu +7398\right) r^2 p_{\phi,\s\mathrm{N}}^3
\nonumber
\\
& \qquad \quad
-30 i \left(160 \nu ^2-547 \nu +222\right) r^{5/3} p_{\phi,\s\mathrm{N}}^{11/3} \bigg]
\nonumber
\\
& \qquad
+\dot r^2 \bigg[6  \left(160 \nu ^2-547 \nu +222\right) r^{8/3} p_{\phi,\s\mathrm{N}}^{8/3} -18  \left(348 \nu ^2+900 \nu -467\right) r^2 p_{\phi,\s\mathrm{N}}^4 -2  \left(566 \nu ^2+5035 \nu -2433\right) r^3 p_{\phi,\s\mathrm{N}}^2
\nonumber
\\
& \qquad \quad
+352 \nu  (2 \nu -1) r^4 \bigg]
\nonumber
\\
& \qquad
+\dot r^3 \bigg[-4 i  \left(458 \nu ^2+1846 \nu -907\right) r^3 p_{\phi,\s\mathrm{N}}^3 -220 i  \left(2 \nu ^2+\nu -1\right) r^4 p_{\phi,\s\mathrm{N}} \bigg]
\nonumber
\\
& \qquad
- \dot r^4 r^4 \bigg[ \left(8 \nu ^2-1316 \nu +560\right)  p_{\phi,\s\mathrm{N}}^2 + 88 \left(2 \nu ^2+\nu -1\right) r \bigg]
\nonumber
\nonumber
\\
& \qquad
\Bigg\}
\Bigg[
132 (2 \nu -1) \, r^{4/3} p_{\phi,\s\mathrm{N}}^{10/3}  \left(23 p_{\phi,\s\mathrm{N}}^2+10 i p_{\phi,\s\mathrm{N}} r \s \dot r- 2 r^2 \dot r^2+4r \right)
\Bigg]^{-1}\!.
\end{align}
\end{subequations}
%
These expressions are also provided in the Supplemental Material~\cite{SEOBNRv6EHM:SupplementalMaterial}.
\\

We note that the eccentricity corrections with $ \ell + m $ odd have the leading-order contribution $ h _{ \ell m} ^{ \text{ecc},\,\text{LO}} = r^{2/3} / p_{\phi,\s\text{N}}^{4/3} $, instead of $ h _{ \ell m} ^{ \text{ecc},\,\text{LO}} = 1 $ as the $ \ell + m $ even modes.
This discrepancy comes from the definition of $ \hat{S}_\text{eff} $ in Eq.~\eqref{eq:Seff}.
With such definition, one has $ \hat{S}_\text{eff} \sim 1 + \mathcal O (1/c^2) $ in the $ \ell + m $ odd case \emph{only} for QC orbits, for which $ p_\phi \to 1/v + \mathcal O (1/c^2) $.
This is not surprising considering that the factor $\hat{S}_{\text{eff}}$ was introduced for QC binaries in Refs.~\cite{Damour:2007xr,Damour:2007yf,Damour:2008gu}.
If one keeps the idea of having corrections in the mode factorization [Eqs.~\eqref{eq:model_fact} and \eqref{eq:model_qcFactModes}] that become unity at Newtonian order, then one could consider the definition
\begin{equation}
\label{eq:new_Seff}
\hat{S}_\text{eff}= \left\{
\begin{array}{ll}
H_\text{eff} , & \quad \ell + m \text{ even},
\\
p_\phi/(r^2  \s \Omega), & \quad \ell + m \text{ odd},
\end{array}
\right.
\end{equation}
which satisfies $ \hat{S}_\text{eff}  \sim  1 + \mathcal O(1/c^2) $ for generic planar orbits.
We do not implement this new definition in \texttt{SEOBNRv6EHM} since a change in $ \hat{S}_\text{eff} $ would propagate to the $ f _{ \ell m} ^{ \text{qc}} $ factors.
We leave for future work the derivation of expressions consistent with the definition in Eq.~\eqref{eq:new_Seff} and the study of their impact on the accuracy of the modes.


\subsection{Orbital frequency for circular orbits}
\label{sec:orbital_frequency}

The 5PN expression for the orbital frequency of circular orbits $ \Omega _{\text{c}} $ expressed as function of the separation $ r $, discussed in Sec.~\ref{sec:new_RR_motivation} and used in the RR force \eqref{eq:RR_force_proposal} and waveform modes \eqref{eq:model_fact}, is calculated by perturbatively solving the equation $ \dot p_r = 0 = \partial H _{ \text{EOB}} / \partial p_\phi \, | _{ p_r = 0}$ for $ p_\phi = p_\phi(r) $, and then replacing the solution into $ \Omega _{ \text{c}} \equiv \partial H _{ \text{EOB}} / \partial p_\phi \, | _{ p_r = 0} $, followed by a further expansion.
This results in the following PN expression:
%
\begin{subequations}
\label{eq:Omega_c}
\begin{align}
\Omega _{\text{c}}
& \equiv
\Omega _{\text{c},\s S^0} + \Omega _{\text{c},\s S^1} + \Omega _{\text{c},\s S^2} + \Omega _{\text{c},\s S^3} + \Omega _{\text{c},\s S^4} \,,
 \\[10pt]
\Omega _{\text{c},\s S^0}
&=
\frac{1}{r^{3/2}}
+\frac{\epsilon ^2 }{r^{5/2}} \frac{\nu}{2}
+ \frac{ \epsilon ^4}{r^{7/2}}
\left(\frac{3 \nu ^2}{8}-\frac{15 \nu }{8}\right)
+ \frac{\epsilon ^6}{r^{9/2}}
\bigg[\frac{5 \nu ^3}{16}-\frac{13 \nu ^2}{16}+\left(\frac{41 \pi ^2}{32}-\frac{1585}{48}\right) \nu \bigg]
\nonumber \\[4pt]
& \quad 
+ \frac{\epsilon ^8 }{r^{11/2}}
\bigg[\frac{35 \nu ^4}{128} -\frac{33 \nu ^3}{64}
 +\left(\frac{17821}{384}-\frac{205 \pi ^2}{128}\right) \nu ^2+  \left(\frac{153211}{1920}-\frac{11375 \pi ^2}{2048}-32 \gamma_\text E -64 \ln 2 \right) \nu +16 \nu  \ln r \bigg]
 \nonumber \\[4pt]
& \quad 
+\frac{\epsilon ^{10}}{r^{13/2}}
\bigg[
\frac{63 \nu ^5}{256}
-\frac{45 \nu ^4}{128} +\left(\frac{6415}{768}-\frac{41 \pi ^2}{128}\right) \nu ^3
+ \left(\frac{171571}{3840}-\frac{3629 \pi ^2}{4096}+\frac{16 \gamma_\text E }{5}+\frac{32 \ln 2}{5}\right) \nu ^2
\nonumber \\[4pt]
& \qquad 
+ \left(\frac{31511}{26880}-\frac{3 \, a_6}{2}\right) \nu
- \left(\frac{224 \nu ^2}{5}+\frac{3502 \nu }{35}\right) \ln r
\bigg],
\\[10pt]
\Omega _{\text{c},\s S^1}
&=
\frac{\epsilon ^3 }{r^3}
\Bigg[-\delta  \chi_A+ \bigg( \frac{\nu }{2}-1 \bigg)\chi_S \Bigg]
+\frac{\epsilon ^5 }{r^4}
\left[\frac{27 \nu ^2 \chi_S}{16} -  \left(\frac{45 \delta  \chi_A}{32} + \frac{63 \chi_S}{32}\right)\nu\right]
\nonumber \\[4pt]
& \quad 
+\frac{\epsilon ^7 }{r^5}
\bigg[\frac{279 \nu ^3 \chi_S}{128} - \left(\frac{63 \delta  \chi_A}{128} +\frac{357 \chi_S}{128}\right) \nu ^2
+  \left(\frac{1413 \delta  \chi_A}{128}+\frac{357 \chi_S}{32}\right)\nu \bigg]
\nonumber \\[4pt]
& \quad 
+\frac{\epsilon ^9}{r^6}
\bigg[\frac{1061 \nu ^4 \chi_S}{768} -\left(\frac{359 \delta  \chi_A}{384} +\frac{1333 \chi_S}{192}\right)\nu ^3 -  \left(\frac{75 \delta  \chi_A }{128} + \frac{575 \chi_S}{256}\right) \nu ^2
+  \left(\frac{837 \delta  \chi_A}{64}+\frac{621 \chi_S}{32} -2 \,\hat d_\text{SO} \right)\nu
\bigg],
\\[10pt]
\Omega _{\text{c},\s S^2}
&=
\frac{\epsilon ^6 }{r^{9/2}}
\Bigg[ \left(2 \chi_A^2+\frac{25 \chi_S^2}{8}\right) \nu ^2 + \left(-\frac{23 \delta  \chi_A \chi_S}{2}-\frac{11 \chi_A^2}{2}-10 \chi_S^2\right) \nu  +2 \delta  \chi_A \chi_S +\chi_A^2+\chi_S^2\Bigg]
\nonumber \\[4pt]
& \quad 
+\frac{\epsilon ^8 }{r^{11/2}}
\Bigg[ \left(3 \chi_A^2+\frac{247 \chi_S^2}{32}\right) \nu ^3 - \left(\frac{1139 \delta  \chi_A \chi_S}{64}+\frac{131 \chi_A^2}{8}+\frac{1049 \chi_S^2}{64}\right)\nu ^2 +  \Bigg(\frac{71 \delta  \chi_A \chi_S}{4}+\frac{185 \chi_A^2}{32}
+\frac{383 \chi_S^2}{32}\bigg)\nu\Bigg]
\nonumber \\[4pt]
& \quad 
+ \frac{\epsilon ^{10} }{r^{13/2}}
\Bigg\{\left(\frac{15 \chi_A^2}{4}-\frac{3761 \chi_S^2}{512}\right) \nu ^4 
+ \Bigg[\frac{79843 \delta  \chi_A \chi_S}{1024} +\frac{222349 \chi_S^2}{2048}
+\left(\frac{415701}{2048}+\frac{123 \pi ^2}{16}\right) \chi_A^2\Bigg] \nu ^2 
\nonumber \\[4pt]
& \qquad 
+  \left(\frac{11267 \delta  \chi_A \chi_S}{512}+\frac{28235 \chi_A^2}{512}+\frac{497 \chi_S^2}{512}\right) \nu ^3 
- \Bigg[\left(\frac{17155}{128}+\frac{123 \pi ^2}{32}\right) \delta  \chi_A \chi_S +\left(\frac{6911}{128}+\frac{123 \pi ^2}{64}\right) \chi_A^2
\nonumber \\[4pt]
& \qquad \quad
+\left(\frac{2561}{32}+\frac{123 \pi ^2}{64}\right) \chi_S^2\Bigg] \nu \Bigg\},
\\[10pt]
\Omega _{\text{c},\s S^3}
&=
\frac{\epsilon ^7 }{r^5}
\left[  \left(3 \delta  \chi_A \chi_S^2+\frac{3 \chi_A^2 \chi_S}{2}+\frac{3 \chi_S^3}{2}\right) \nu -6 \nu ^2 \chi_A^2 \chi_S\right]
- \frac{\epsilon ^9 }{r^6}
\Bigg[ \left(\frac{51 \delta  \chi_A^3}{4} +\frac{21 \delta  \chi_A \chi_S^2 }{4} +\frac{403 \chi_A^2 \chi_S}{8}+\frac{21 \chi_S^3}{8}\right) \nu ^2
\nonumber \\[4pt]
& \qquad 
-  \left(\frac{115 \delta  \chi_A^3}{16}+\frac{229 \delta  \chi_A \chi_S^2}{16} +\frac{383 \chi_A^2 \chi_S}{16}+\frac{89 \chi_S^3}{16}\right) \nu 
 - \frac{21}{2} \nu ^3 \chi_A^2 \chi_S +\delta  \chi_A^3+3 \delta  \chi_A \chi_S^2+3 \chi_A^2 \chi_S+\chi_S^3\Bigg],
\\[10pt]
\Omega _{\text{c},\s S^4}
&=
\frac{ \epsilon ^{10} }{r^{13/2}}
\Bigg[ 3 \nu ^3 \chi_A^2 \chi_S^2 - \left(30 \delta  \chi_A^3 \chi_S+\frac{3 \delta  \chi_A \chi_S^3}{2} +6 \chi_A^4+\frac{339 \chi_A^2 \chi_S^2}{4}+\frac{3 \chi_S^4}{4}\right) \nu ^2
\nonumber \\[4pt]
& \qquad
+ \Bigg(\frac{21  \delta  \chi_A^3 \chi_S}{2}+\frac{39 \delta  \chi_A \chi_S^3}{2} 
 +\frac{3 \chi_A^4}{2} +\frac{45 \chi_A^2 \chi_S^2}{2}+6 \chi_S^4\Bigg)\nu  \Bigg],
\end{align}
\end{subequations}
%
where $\gamma_\text E \approx 0.577$ is Euler's constant.
In the Supplemental Material~\cite{SEOBNRv6EHM:SupplementalMaterial}, we provide this expression along with its $ (0,5) $ Pad\'e resummation, $ P^0_5[\Omega _{ \text{c}}] $, as used in \texttt{SEOBNRv6EHM}.
Since the \texttt{SEOBNRv5} Hamiltonian includes complete spin effects only up to 4PN order, along with partial 5PN nonspinning contributions~\cite{Khalilv5}, the 5PN expression for $ \Omega_{\text{c}} $ \eqref{eq:Omega_c} is only partially determined.


\section{Fits for the calibration parameters $ \big (a_6 , \hat d _{ \mathrm{SO}}, \Delta t _{ \mathrm{ISCO}}^{22} \big )$}
\label{sec:fits}

Here, we present the fits for the calibration parameters $ \big (a_6 , \hat d _{ \mathrm{SO}}, \Delta t _{ \mathrm{ISCO}}^{22} \big )$ employed in \texttt{SEOBNRv6EHM}, as described in Sec.~\ref{sec:calibration}.

The fit for the $ a_6 $ parameter entering the nonspinning sector of the effective Hamiltonian is given by (see Sec.~\ref{sec:calibration_nonspinning}):
%
%
%
%
%
\begin{equation}
\label{eq:fit_a6}
a_6 (\nu) =
41.618273
-2651.44213 \, \nu
+ 36594.739 \, \nu^{2}
- 213765.281 \, \nu^{3}
+ 466493.941 \, \nu^{4}.
\end{equation}

\begin{figure}[h]
\hspace{-5pt}
\includegraphics[width=1\linewidth]{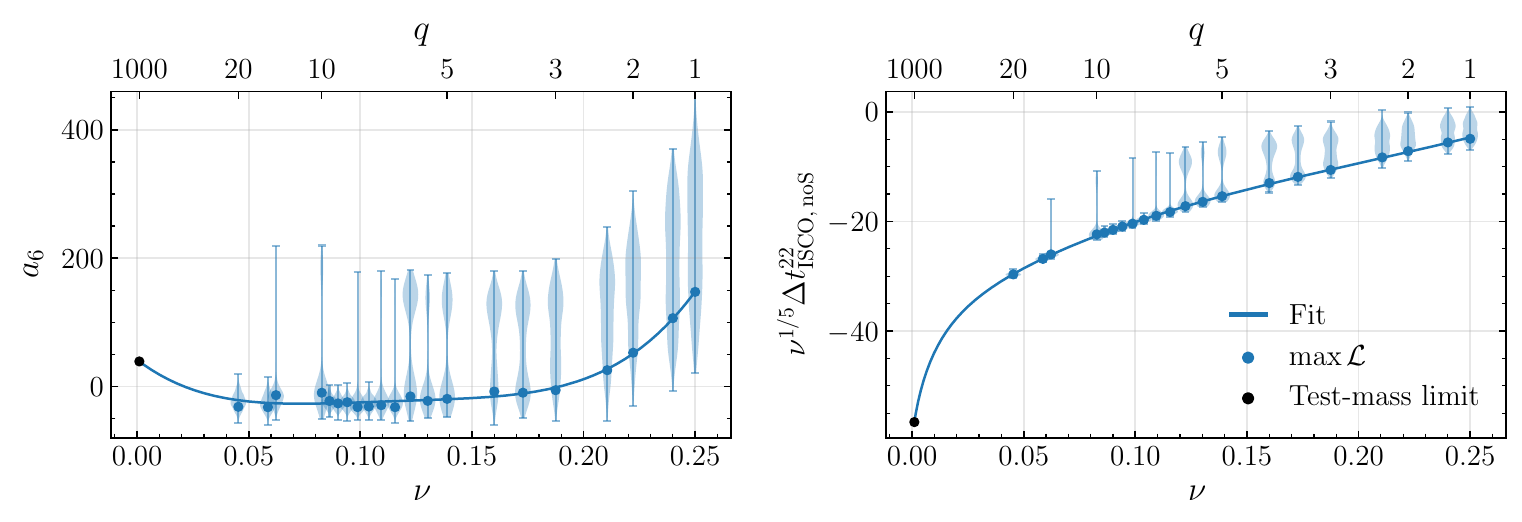}
\vspace{-5pt}
\caption{
Fits for the $ a_6 $ (left panel) and $ \Delta t^{22} _{ \text{ISCO,\,noS}} $ (right panel) calibration parameters, as functions of the symmetric mass ratio $ \nu $.
The fits are given by Eqs.~\eqref{eq:fit_a6} and \eqref{eq:fit_dt_ns}, respectively, and are obtained by least-squares regression of the maximum-likelihood values ($ \max \mathcal L $, blue dots) of the calibration posteriors (shaded violins) for a set of 21 QC \texttt{SXS} NR simulations, including test-mass limit estimates (black dots).
}
\label{fig:a6_dt_v6}
\end{figure}

\begin{figure}[H]
\centering
\hspace{-5pt}
\includegraphics[width=0.49\linewidth]{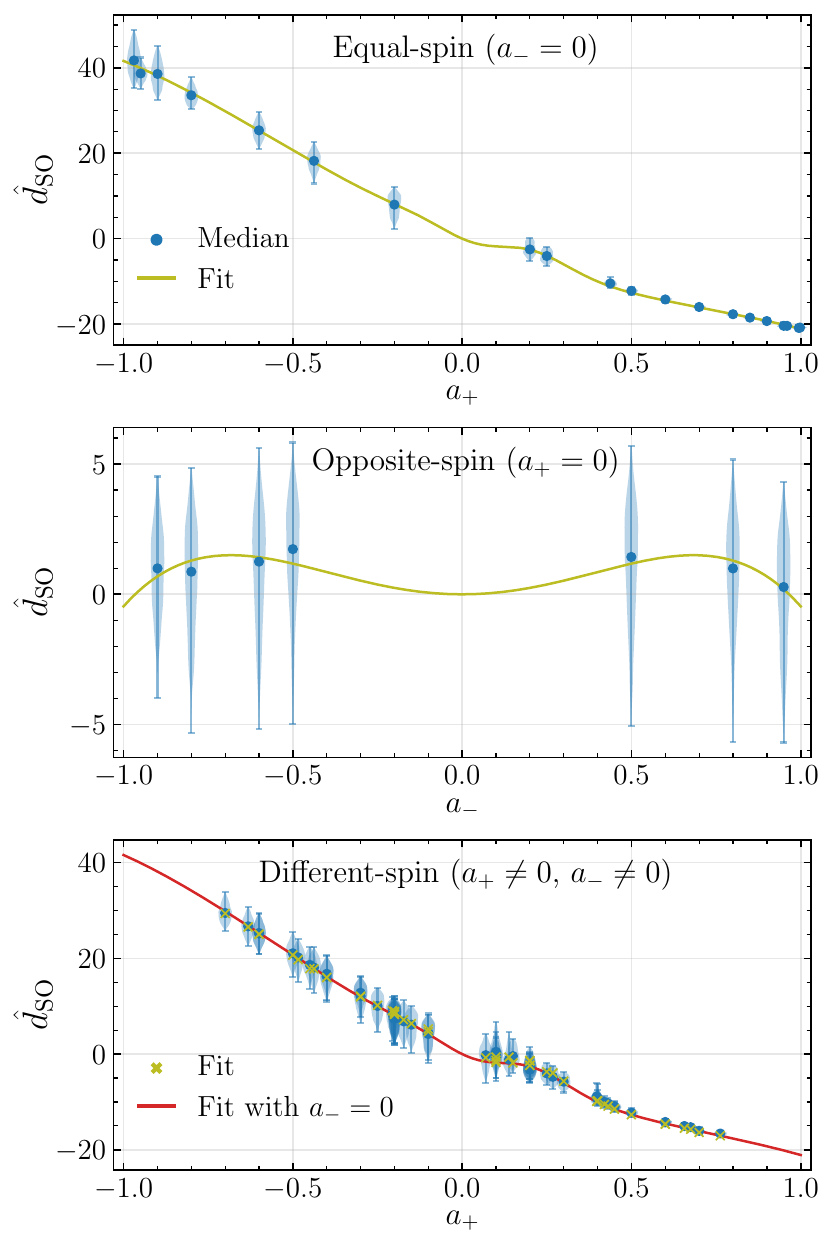}
\hspace{6pt}
\includegraphics[width=0.49\linewidth]{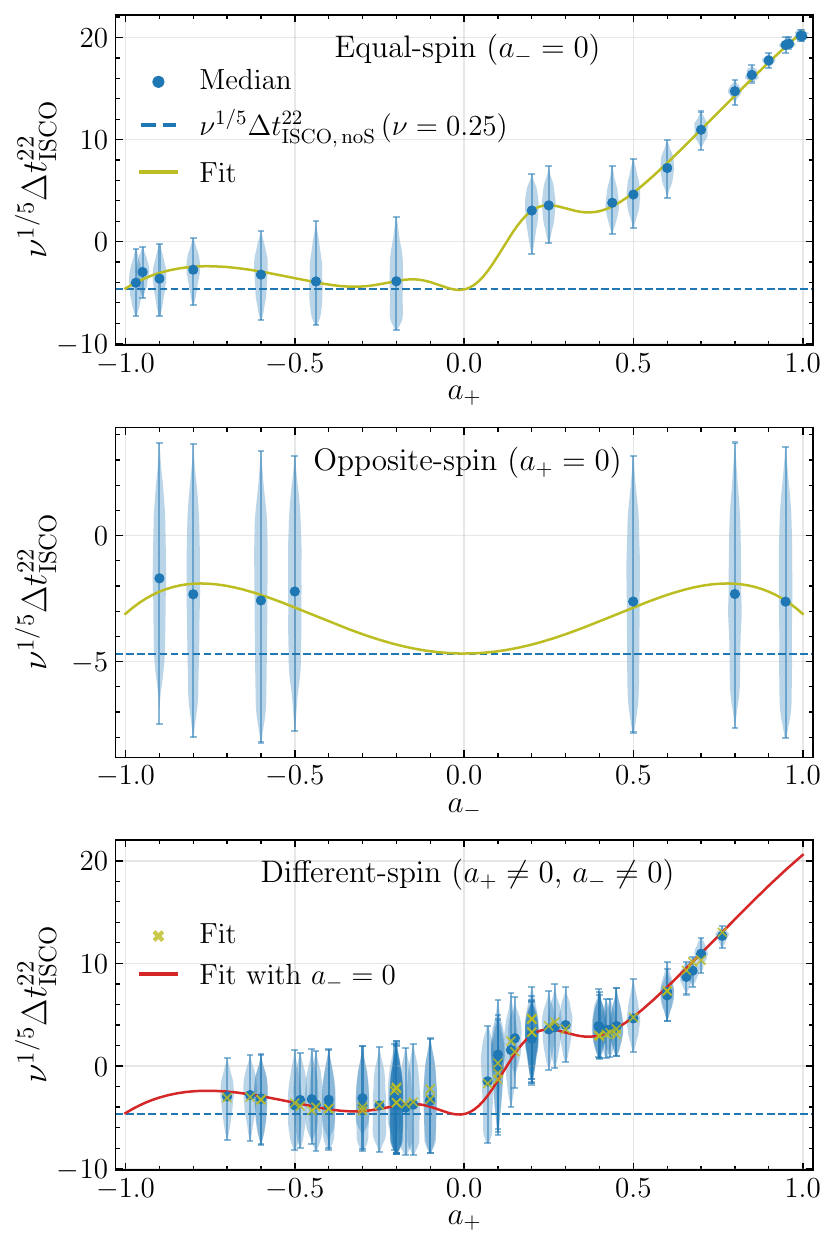}
\vspace{-5pt}
\caption{
Fits for the $ \hat d _{ \text{SO}}$ (left plot) and $ \Delta t^{22} _{ \text{ISCO}} $ (right plot) calibration parameters, as functions of the spin variables $ a_+ $ and $ a_- $.
The fits are given by Eqs.~\eqref{eq:fit_hatdSO_equal} and \eqref{eq:fit_dt_addition}--\eqref{eq:fit_dt_s_equal}, respectively, and are obtained by least-squares regression of the median values (blue dots) of the calibration posteriors (shaded violins) for a set of 78 QC, equal-mass \texttt{SXS} NR simulations.
In the right panels, a dashed horizontal line indicates the nonspinning value $ \Delta t^{22} _{ \text{ISCO,\,noS}} (\nu = 0.25)$.
In the bottom panels, yellow crosses show the fit predictions for cases with $ a_+ \neq 0 $ and $ a_- \neq 0 $, while the red curves show the corresponding fits with $ a_- = 0 $, for reference.
}
\label{fig:dtS_hatdSO}
\end{figure}

The fit for the $ \hat d _{ \text{SO}} $ parameter entering the spinning sector of the effective Hamiltonian, and the fit for the parameter $ \Delta t _{ \text{ISCO}} ^{ 22} $ determining the matching time for the merger-ringdown phase, are given by (see Sec.~\ref{sec:calibration_spinning}):
%
%
\begin{subequations}
\label{eq:fit_hatdSO}
\begin{align}
\hat d _{ \text{SO}}
&=
\hat d _{ \text{SO}} ^{\, q=1}
+ \! \sqrt{1 - 4 \nu} \,\, \hat d _{ \text{SO}} ^{\, q \neq1} \!,
\\[4pt]
%
\hat d _{ \text{SO}} ^{\, q=1} &=
4 \s \nu \, \Big(
-34.2344583 \, a_+
+ 17.6449319 \, a_+^2
+ 2.89013625 \, a_+^3
- 7.38751602 \, a_+^4 
+ 148.342235 \, a_+^2 \,\s \mathrm{e}^{-30.6665632 \, (a_+ - 0.070349849)^2}
\nonumber
\\
&\qquad \quad
+ 6.44220076 \, a_-^2
- 6.90516533 \, a_-^4
+ 3.16175018 \, a_+ a_-^2
- 31.4547453 \, a_+^2 a_-^2
\Big)
\,,
\label{eq:fit_hatdSO_equal}
\\[4pt]
%
\hat d _{ \text{SO}} ^{\, q \neq1} &=
32.8293601\,a_-^{4} -189.507823\,a_-^{3} a_+ +353.171620\,a_-^{3} \nu -78.1113444\,a_-^{3} +295.713452\,a_-^{2} a_+^{2} 
\nonumber
\\
& \quad
-33.8165178\,a_-^{2} a_+ \nu +6.76432040\,a_-^{2} a_+ -704.757568\,a_-^{2} \nu^{2} -797.461100\,a_-^{2} \nu +220.520869\,a_-^{2}   
\nonumber
\\
& \quad
-183.251252\,a_- a_+^{3} -315.081442\,a_- a_+^{2} \nu +145.700522\,a_- a_+^{2} +148.246293\,a_- a_+ \nu^{2} +1929.01011\,a_- a_+ \nu 
\nonumber
\\
& \quad
-444.742645\,a_- a_+ +87.4640195\,a_- \nu^{3} -17.1272746\,a_- \nu^{2} -236.178099\,a_- \nu +37.6064650\,a_- +30.8600291\,a_+^{4} 
\nonumber
\\
& \quad
+423.190055\,a_+^{3} \nu -117.259103\,a_+^{3} -132.828461\,a_+^{2} \nu^{2} -1058.53790\,a_+^{2} \nu +257.598409\,a_+^{2} +524.463266\,a_+ \nu^{3} 
\nonumber
\\
& \quad
+984.738126\,a_+ \nu^{2} +70.8014313\,a_+ \nu -89.7683214\,a_+
,
\end{align}
\end{subequations}
\\[-20pt]
%
%
\begin{subequations}
\label{eq:fit_dt}
\begin{align}
\Delta t^{22}_{\text{ISCO}}
&=
\Delta t^{22}_{\text{ISCO, noS}}
+ \Delta t^{22,\, q=1}_{\text{ISCO,\,S}}
+ \! \sqrt{1 - 4 \nu} \,\, \Delta t^{22,\, q \neq1}_{\text{ISCO,\,S}} \s ,
\label{eq:fit_dt_addition}
\\[4pt]
%
\Delta t^{22}_{\text{ISCO, noS}} &=
\left(
- 60.106301
- 1491.53128 \, \nu
- 22623.9596 \, \nu^{2}
+ 97254.969 \, \nu^{3}
\right) \,\s
\nu^{- 1/5 + 12.2560628 \, \nu },
\label{eq:fit_dt_ns}
\\[4pt]
%
\Delta t^{22,\, q=1}_{\text{ISCO,\,S}} &=
4 \s \nu \, \Big(
6.76727194 \, a_+
+ 23.5274911 \, a_+^2
+ 5.80829523 \, a_+^3
-10.8600381 \, a_+^4 
+ 250.474064 \, a_+^2  \,\s \mathrm{e}^{-30.680603 \, (a_+ - 0.05778969)^2}
\nonumber
\\
&\qquad \quad
+ 9.22003581 \, a_-^2
- 7.64273673 \, a_-^4
- 0.960589896 a_+ a_-^2
- 50.0695267 a_+^2 a_-^2
\Big)
\, \nu^{-1/5},
\label{eq:fit_dt_s_equal}
\\[4pt]
%
\Delta t^{22,\, q \neq1}_{\text{ISCO,\,S}} &=
\Big(
9.60527471\,a_-^{4} +64.8349363\,a_-^{3} a_+ -172.824977\,a_-^{3} \nu +23.7096189\,a_-^{3} -102.206641\,a_-^{2} a_+^{2}
\nonumber
\\
& \quad
+453.708285\,a_-^{2} a_+ \nu -34.0991808\,a_-^{2} a_+ -710.833783\,a_-^{2} \nu^{2} +311.250601\,a_-^{2} \nu -29.8797295\,a_-^{2}
\nonumber
\\
& \quad
+89.4999966\,a_- a_+^{3} +459.145193\,a_- a_+^{2} \nu -68.1529604\,a_- a_+^{2} -696.807052\,a_- a_+ \nu^{2} +47.3287461\,a_- a_+ \nu
\nonumber
\\
& \quad
+8.31004725\,a_- a_+ -24.6145296\,a_- \nu^{3} +883.819784\,a_- \nu^{2} -458.548126\,a_- \nu +56.7387561\,a_-
\nonumber
\\
& \quad
+1.65844573\,a_+^{4} -327.093040\,a_+^{3} \nu +62.0376531\,a_+^{3} -654.429154\,a_+^{2} \nu^{2} +335.530606\,a_+^{2} \nu
\nonumber
\\
& \quad
-32.4621685\,a_+^{2} +59.1310283\,a_+ \nu^{3} +952.142993\,a_+ \nu^{2} -212.059066\,a_+ \nu +2.65377616\,a_+
\Big) \, \nu^{-1/5},
\label{eq:fit_dt_s_unequal}
\end{align}
\end{subequations}
%
The nonspinning data and the corresponding fits of $ a_6 $ and $ \Delta t^{22} _{ \text{ISCO,\,noS}} $ are displayed in Fig.~\ref{fig:a6_dt_v6}.
The spinning data for equal-mass cases and the corresponding fits of $ \hat d _{ \text{SO}} $ and $ \Delta t^{22} _{ \text{ISCO,\,noS}} $ are shown in Fig.~\ref{fig:dtS_hatdSO}.

%



\twocolumngrid

\bibliography{../references/references}





\end{document}